%% file: Formulation.tex
\documentclass[DIV14,BCOR5mm]{scrreprt} 
\usepackage[english]{babel}
\usepackage{amsmath,amssymb}
\usepackage{comment}
\usepackage{listings}
\usepackage{graphicx}
\usepackage[
	colorlinks=true, 
	linkcolor=blue, 
	citecolor=cyan, 
	urlcolor=red, 
	filecolor=red, 
	setpagesize=false, 
	pdftitle={SMatrAn formulation}, 
	pdfauthor={Tatsuya Usuki}, 
	pdfsubject={formulation of simulation}, 
	pdfkeywords={S-matrix, electromagnetic field, optics}, 
]{hyperref}
\newcommand{\usukihome}{http://www.smatran.org/}
\newcommand{\usukimail}{usuki-tty@smatran.org}
\newcommand{\gototable}{[\hyperlink{contents}{Go to table of contents.}]\,
				[\href{\usukihome}{Go to home.}]}
\title{Wave scattering in frequency domain
	\begin{figure}[b!]
		\centering
	 	\includegraphics[height=15mm]{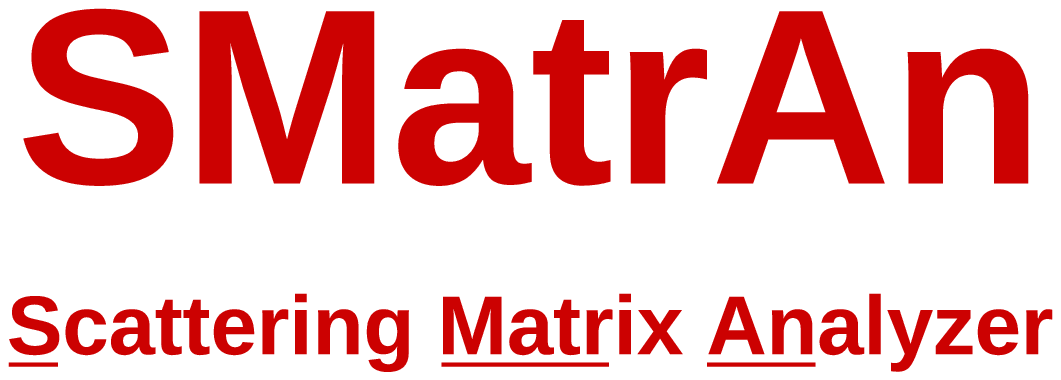}
	\end{figure}
}
\author{Tatsuya Usuki}
\date{\today}
\begin{document}
	\maketitle
	\input{./abst.tex}
	\hypertarget{contents}{\tableofcontents	}
	\input{./Chap/1_Introduction_v6.tex}~\\~\\ \gototable
	\input{./Chap/2_CoordinateTrans_v2}~\\~\\ \gototable
	\input{./Chap/3_Mode_v9.tex}~\\~\\ \gototable
	\input{./Chap/4_NonAdiabaticTransition_v2}~\\~\\ \gototable
	\input{./Chap/5_Roughness_v9}~\\~\\ \gototable
	\input{./Chap/6_Discretization_v12.tex}~\\~\\ \gototable
	\input{./Chap/7_PropagationEqYee_v2.tex}~\\~\\ \gototable
	\input{./Chap/8_FDTD_v2.tex}~\\~\\ \gototable
	
	\appendix
	
	\input{./App/A_GeneralWaveguides_v7.tex}~\\~\\ \gototable
	\input{./App/B_PeriodicWG_v2.tex}~\\~\\ \gototable
	\input{./App/C_ShroedingerEq_v1.tex}~\\~\\ \gototable
	\input{./App/D_GeneralMaxwellEq_v2.tex}\input{./App/Dsub_CheckGenMaxEq_v1.tex}~\\~\\ \gototable
	\input{./App/E_NewtonEq_v3.tex}~\\~\\ \gototable
	\input{./App/F_OpticalScattering_v1.tex}~\\~\\ \gototable
	\input{./App/G_NonUniformMesh_v1.tex}~\\~\\ \gototable
	\input{./App/H_DiscDepend_v2.tex}~\\~\\ \gototable
	\input{./App/I_DiscEquations_v2.tex}~\\~\\ \gototable
	
	\bibliographystyle{unsrt}
	\bibliography{./ref}
	\addcontentsline{toc}{chapter}{\bibname}
	~\\~\\ \gototable   
\end{document}

%% file: abst.tex
\begin{abstract}
This report shows formulation of wave scattering in frequency domain.
The formulation provides an understanding of S-matrix solver which is named as SMatrAn.
It is an abbreviation for `$\mathrm{\underline{S}cattering}$ $\mathrm{\underline{Matr}ix}$ $\mathrm{\underline{An}alyzer}$' or `$\mathrm{\underline{S}cattering}$ $\mathrm{\underline{Matr}ix}$ $\mathrm{\underline{An}alysis}$'.

The S-matrix has the whole of amplitude and phase information for reflected and transmitted waves from a complicated scatterer.
Accurate S-matrix leads to quantitative evaluation of scattering wave.
SMatrAn can provide useful information to engineers and scientists.

We can understand all formulas for the SMatrAn after reading this report.
The S-matrix provided by SMatrAn will give us detailed analysis on complicated scattering in optical structure.
I can hardly shorten its update period, but I will respond to needs from users as much as possible.
Please let \href{mailto:\usukimail}{me} know if you have any questions and comments.
\end{abstract}

%% file: Chap/1_Introduction_v6.tex
\chapter{Introduction\label{ch: intro}}
Analysis of wave scattering will start from a propagation equation \eqref{eq: propagation-equation} in frequency domain.
The equation can be derived from any fundamental equation, \textit{e.g.} Maxwell equations, Shr{\" o}dinger equation or Newton equation of motion.
Chapter \ref{ch: GeneralWaveguides}, \ref{ch: Mode} and \ref{ch: S-matrix BornApprox} define analytical  formulas for coordinates, modes and S-matrix.
Chapter \ref{ch: Roughness} focuses on roughness scattering by using the formulas.
Chapter \ref{ch: discretization}, \ref{ch: propagation Yee} and \ref{ch: FDTD} give us discrete formulation for numerical calculation.
Appendix \ref{ch: Details of two steps}, \ref{ch: Periodic WG}, \ref{ch: Shroedinger}, \ref{ch: Generalized Maxwell}, \ref{ch: Newton}, \ref{ch: optical scattering}, \ref{ch: Non-uniform mesh}, \ref{ch: RelaxDisc.}  and \ref{ch: details of disc. eq.} show detailed derivations of the formulas.
The following section \ref{sec: notation} shows notation for the SMatrAn formulas before explaining details of the formulas.

\section{Notation for the formulas\label{sec: notation}}
The report consists of many formulas.
This section will show notation for the formulas.
\begin{enumerate}
\item Three dimensional coordinates are represented as $\left(x,y,z\right)$ or $\left(x_{0},x_{1},x_{2}\right)$, 
because cyclic permutation of the latter is directly related to modulo 3 operation.
The report will also use $\left(r_{0},r_{1},r_{2}\right)$ or $\left(u_{0},u_{1},u_{2}\right)$.
The $z$, $x_{2}$, $r_{2}$ or $u_{2}$ is propagating axis.
The report chooses the above simpler notation in order to avoid mistakes in operating and programming physical models, and 
it will never use the general notation $\left(x_{1},x_{2}, x_{3}\right)$.

\item Partial derivative by $u_{\nu}$ is represented as $\partial/\partial u_{\nu}$ or $\partial_{u\nu}$.
Derivative of function is sometimes noted as ``${\,}^{'}$''.

\item The report frequently uses matrix and vector notation, and it does not distinguish between the notations.
Matrix and column vector are represented by bold symbol fonts as $\boldsymbol A$ for example.
Especially, $\boldsymbol{\Phi}_{n}$ means column vector of the $n$-th mode.
Transpose and Hermite operators are notated as ``$\,^{\mathrm T}\,$'' and ``$\,^{\dagger}\,$'', respectively.
The ``$\,^{*}\,$'' means complex conjugate, and then ${{\boldsymbol A}^{\mathrm T}}^{\dagger}={\boldsymbol A}^{*}$.
Inner product $\boldsymbol{k}{\cdot}\boldsymbol{r}$ is also represented as $\boldsymbol{k}^{\mathrm T}\boldsymbol{r}$, where $\boldsymbol{k}$ and $\boldsymbol{r}$ are wavenumber vector and position vector, respectively.

\item ``$\left(\partial/\partial x\right)^{\dagger}$'' means differential operator to the left side as $\boldsymbol{f}^{\dagger}\left(\partial/\partial x\right)^{\dagger}\boldsymbol{g} = \left(\partial \boldsymbol{f}/\partial x\right)^{\dagger}\boldsymbol{g}$. 
Integration by parts gives us a relation: $\left(\partial/\partial x\right)^{\dagger}=-\partial/\partial x$ when $\boldsymbol{f}^{\dagger}\boldsymbol{g}$ satisfies a periodic boundary condition.

\item Dual basis of $\boldsymbol{\Phi}_{n}$ is represented as $\widetilde{\boldsymbol{\Phi}}_{n}$.
The product $\boldsymbol{\Phi}_{m}^{\dagger}\widetilde{\boldsymbol{\Phi}}_{n}$ includes integral for cross section.
We will not explicitly show $\iint du_{0}du_{1}$.

\item Imaginary unit is denoted as ``$\,i\,$''. The ``$\,j\,$'' will be only used as integer parameter, and ``$\,l\,$'', ``$\,m\,$'', ``$\,n\,$'' and their capital letters are also integer.
As an exception, the ``$\,L_{\mathrm{s}}$'' has a special meaning for system length, \textit{e.g.} scattering region along the $u_{2}$-axis is defined as $-L_{\mathrm{s}}/2 < u_{2} < +L_{\mathrm{s}}/2$.
The ``$\,L_{\mathrm{c}}$'' means correlation length for other exception.
The modulo operation is used only for non-negative integer in order to avoid confusion caused by several definitions, and it is represented as ``$\,\mathrm{mod}\,$'' or ``$\,\% \,$''.
A notation ``$\,\% 3 \,$'' is frequently omitted in the modulo 3 operation.

\item Plane wave is defined as $\exp\left(i\boldsymbol{k}^{\mathrm T}\boldsymbol{r}-i\omega t\right)$,
where $\omega$ and $t$ are angular frequency and time, respectively.
The definition is well known in physics as eq. (34) in Chapter 1 of \cite{BornWolf} and in other scientific society~\cite{Treyssede}.
However it is different from the manner of engineering\,\cite{Marcuse,Auld}.
If you usually use $\exp\left(j\omega t - j\boldsymbol{k}^{\mathrm T}\boldsymbol{r}\right)$ as the plane wave, the ``$\,j\,$'' has to be replaced to ``$\,-i\,$'' in the report.
$\beta_{n}$ is also used as a propagation constant of the $n$-th mode instead of wavenumber ``$\,k\,$.'' 

\item Fourier-transform (FT) of $f\left(z\right)$ is defined as $\widehat{f}\left(k\right) = \int^{\infty}_{-\infty}f\left(z\right)e^{-ikz}dz$.
The inverse transformation is that $f\left(z\right) = \left(2\pi\right)^{-1}\int^{\infty}_{-\infty}\widehat{f}\left(k\right)e^{ikz}dk$.
Discrete Fourier transform (DFT) is also defined as $\widehat{f}_{\,\mathrm{DFT}}$ in Section \ref{sec: DFT definition}.

\item The $f\left[l\right]$ represents the discretization of a continuous function $f\left(x\right)$, which also means an array. 

\item The ``$\,\tilde{\,}\,$'' and ``$\,\hat{\,}\,$'', which are not wide as ``$\,\widetilde{\,}\,$'' nor ``$\,\widehat{\,}\,$'', mean temporary modification of parameters. 

\end{enumerate}


%% file: Chap/2_CoordinateTrans_v2.tex
\chapter{Coordinate transformation for waveguide\label{ch: GeneralWaveguides}}
\begin{figure}[h!]
\begin{centering}
\includegraphics[clip,width=0.5\columnwidth]{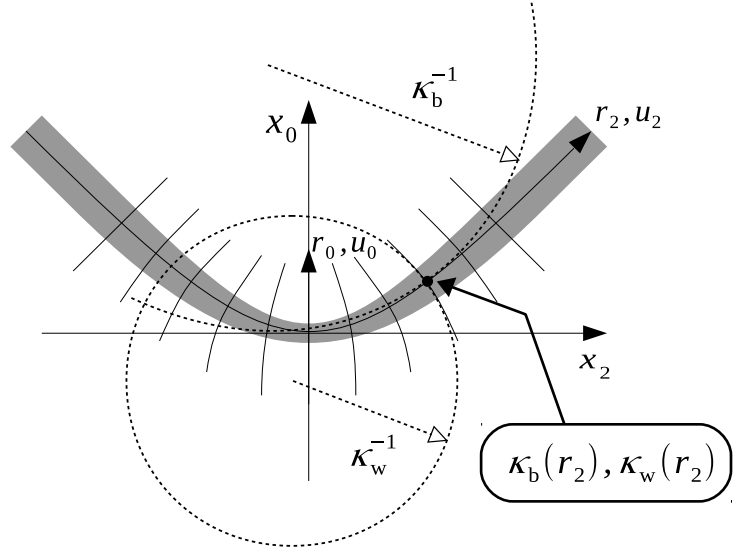}
\par\end{centering}
\caption{Coordinate transfer. 
The region of $r_{0}$ is at least limited by $r_{0}\kappa_{b} < 1$ and $\left|r_{0}\kappa_{w}\right| < 1$.
\label{fig: general waveguide}}
\end{figure}

This chapter reports coordinate transformation which can be applied
to general waveguides. 
The goal is that we obtain the scale factors after transforming coordinates as shown in Fig. \ref{fig: general waveguide}.

Let us consider two step transformation for orthogonal curvilinear coordinates such as $\boldsymbol{x}\Rightarrow\boldsymbol{r}\Rightarrow\boldsymbol{u}$.
Considered waveguide is laid on the $x_{2}x_{0}$-plane, and then
$x_{1} \Rightarrow r_{1} \Rightarrow u_{1}$ is identity transformation.
We will ignore the $x_{1}$-axis  in the following sections.
The $r_{2}$-axis and the $u_{2}$-axis are propagation axis, and $\left|r_{2}\right|, \left|u_{2}\right| < L_{\mathrm{s}}/2$ (see section \ref{sec: notation}).
Note that we will introduce two curvatures $\kappa_{b}$ and $\kappa_{w}$ to the transformation, and the region of $r_{0}$ (and $u_{0}$) is limited by $\kappa_{b}$ and $\kappa_{w}$.

\section{Two steps of the transformation: $\boldsymbol{x}\Rightarrow\boldsymbol{r}\Rightarrow\boldsymbol{u}$ \label{Coordinate trans}}

The first step $\left(x_{0},\,x_{2}\right)\Rightarrow\left(r_{0},\,r_{2}\right)$ is that 
\begin{equation}
\left\{ \begin{aligned}
x_{2}\left(r_{0},\,r_{2}\right) & = \int^{r_{2}}_{0}f\left(r\right)dr-r_{0}g\left(r_{2}\right),\\
x_{0}\left(r_{0},\,r_{2}\right) & = \int^{r_{2}}_{0}g\left(r\right)dr+r_{0}f\left(r_{2}\right),\\
f\left(r_{2}\right) & =\cos\left(\int^{r_{2}}_{0}\kappa_{b}\left(r\right)dr\right),\\
g\left(r_{2}\right) & =\sin\left(\int^{r_{2}}_{0}\kappa_{b}\left(r\right)dr\right).
\end{aligned} \right.
\label{eq: 1st-step trans}
\end{equation}
Section \ref{sec: details of x to r} in Appendix \ref{ch: Details of two steps} shows details of the above transformation.
Equation (\ref{eq: x-r ortho}) ensures that the coordinates $\left(r_{0},\,r_{2}\right)$ are orthogonal curvilinear coordinates.
Note that $\kappa_{b}\left(r_{2}\right)$ is a signed curvature for waveguide bending as shown in Fig. \ref{fig: general waveguide}, since eq. (\ref{eq: bending kappa}) gives us the definition of signed curvature:
\[
\kappa_{b} = \left.
\frac{\frac{\partial x_{2}}{\partial r_{2}}\frac{\partial^{2}x_{0}}{\partial r_{2}^{2}}-\frac{\partial^{2}x_{2}}{\partial r_{2}^{2}}\frac{\partial x_{0}}{\partial r_{2}}}{\left(\left(\frac{\partial x_{0}}{\partial r_{2}}\right)^{2}+\left(\frac{\partial x_{2}}{\partial r_{2}}\right)^{2}\right)^{3/2}}
\right|_{r_{0}=0}
\,.
\]

The second step $\left(r_{0},\,r_{2}\right)\Rightarrow\left(u_{0},\,u_{2}\right)$ is defined by 
\begin{equation}
\left\{\begin{aligned}
r_{2}\left(u_{0},\,u_{2}\right) & =
u_{2} - \int_{0}^{u_{0}} F_{\mathrm{2D}}\left(r_{0}\left(u,\,u_{2}\right),\,r_{2}\left(u,\,u_{2}\right)\right) du\,,\\
r_{0}\left(u_{0},\,u_{2}\right) & =u_{0}\zeta\left(r_{2}\left(u_{0},\,u_{2}\right)\right),\\
F_{\mathrm{2D}}\left(r_{0},\,r_{2}\right) & = 
\frac{r_{0}\kappa_{w}\left(r_{2}\right)\zeta\left(r_{2}\right)}{\left(1-r_{0}\kappa_{b}\left(r_{2}\right)\right)^{2}+\left(r_{0}\kappa_{w}\left(r_{2}\right)\right)^{2}}
\,,\\
\zeta\left(r_{2}\right) & =\exp\left(\int^{r_{2}}_{0}\kappa_{w}\left(r\right)dr\right).
\end{aligned}\right.
\label{eq: 2nd-step trans}
\end{equation}
Equations (\ref{eq: mod. ortho}), (\ref{eq: partial r2 r0 by u0 u2}) and (\ref{eq: F definition})  in Section \ref{sec: details of r to u} ensure that the coordinates $\left(u_{0},\,u_{2}\right)$ are orthogonal curvilinear coordinates.
Note that $\kappa_{w}\left(r_{2}\left(0,\,u_{2}\right)\right)$ is a signed curvature for waveguide broadening as shown in Fig. \ref{fig: general waveguide}, since eq. (\ref{eq: width kappa}) give us the definition of signed curvature:
\[
\kappa_{w} = \left.
\frac{\frac{\partial r_{2}}{\partial u_{0}}\frac{\partial^{2}r_{0}}{\partial u_{0}^{2}}-\frac{\partial^{2}r_{2}}{\partial u_{0}^{2}}\frac{\partial r_{0}}{\partial u_{0}}}{\left(\left(\frac{\partial r_{0}}{\partial u_{0}}\right)^{2}+\left(\frac{\partial r_{2}}{\partial u_{0}}\right)^{2}\right)^{3/2}}
\right|_{u_{0}=0}
\,.
\]
The region of $u_{0}$ must be numerically checked by eqs. (\ref{eq: 1st-step trans}) and (\ref{eq: 2nd-step trans}) after setting an initial condition:
\[
\left|u_{0}\right| < \min_{-L_{\mathrm{s}}/2<r<L_{\mathrm{s}}/2}\left(\frac{1}{\zeta\left(r\right)\max\left(\left|\kappa_{b}\left(r\right)\right|, \left|\kappa_{w}\left(r\right)\right|\right)}\right)
\,.
\]

\section{Scale factor}

Let us obtain scale factor $h_{0}$ and $h_{2}$ from eqs. (\ref{eq: h0 with curvatures}) and (\ref{eq: h2 with curvatures}) in Section \ref{sec: details of scale factor}. 
\begin{equation}
\left\{\begin{aligned}
h_{0} & \triangleq 
\sqrt{
	\sum_{j=0}^{2}\left(\frac{\partial x_{j}}{\partial u_{0}}\right)^{2}
}
= \sqrt{
	\left(\frac{\partial x_{0}}{\partial u_{0}}\right)^{2}
	+\left(\frac{\partial x_{2}}{\partial u_{0}}\right)^{2}
}
= \frac{\left(1-r_{0}\kappa_{b}\right)\zeta}
{\sqrt{\left(1-r_{0}\kappa_{b}\right)^{2}+\left(r_{0}\kappa_{w}\right)^{2}}}
\,,\\
h_{2}  & \triangleq 
\sqrt{
	\sum_{j=0}^{2}\left(\frac{\partial x_{j}}{\partial u_{2}}\right)^{2}
}
= \sqrt{
	\left(\frac{\partial x_{0}}{\partial u_{2}}\right)^{2}
	+\left(\frac{\partial x_{2}}{\partial u_{2}}\right)^{2}
}
= \frac{\partial r_{2}}{\partial u_{2}}
\sqrt{\left(1-r_{0}\kappa_{b}\right)^{2}+\left(r_{0}\kappa_{w}\right)^{2}}\,.
\end{aligned}\right.\label{eq: h2-h0}
\end{equation}
Note that $\kappa_{b}$, $\kappa_{w}$ and $\zeta$ are functions of $r_{2}$.
The $\partial r_{2}/\partial u_{2}$ for $h_{2}$ can be numerically solved by 
\begin{equation}\left\{
\begin{aligned}
\frac{\partial r_{2}\left(u_{0},\,u_{2}\right)}{\partial u_{2}} 
& = 1 - \int_{0}^{u_{0}}\left[
\frac{\partial}{\partial u_{2}}
F_{\mathrm{2D}}\left(r_{0}\left(u,\,u_{2}\right),\,r_{2}\left(u,\,u_{2}\right)\right)
\right] du \,,\\
\frac{\partial F_{\mathrm{2D}}\left(r_{0},\,r_{2}\right)}{\partial u_{2}}
 & = \frac{\partial r_{2}}{\partial u_{2}}
			\frac{
				r_{0}\zeta\left[
					\kappa_{w}^{'}\left(1-r_{0}\kappa_{b}\right)^{2}
					+ 2\kappa_{w}\left(1-r_{0}\kappa_{b}\right)
						\left(\kappa_{w}+r_{0}\kappa_{b}^{'}\right)
					- r_{0}^{2}\kappa_{w}^{2}\kappa_{w}^{'}
				\right]
			}{
			 \left[ 
			 \left(1-r_{0}\kappa_{b}\right)^{2}
			 +\left(r_{0}\kappa_{w}\right)^{2} 
			 \right]^{2}
			} \,,
 \end{aligned}\right.\label{eq: r2u2}
\end{equation}
where $\kappa_{b}^{'}= d\kappa_{b}/d r_{2}$ and $\kappa_{w}^{'}= d\kappa_{w}/d r_{2}$.
See eq. (\ref{eq: partial F partial u2}) for details of $\partial F_{\mathrm{2D}} / \partial u_{2}$.

When $\left|u_{0}\kappa_{b}\right|, \left|u_{0}\kappa_{w}\right| \ll 1$ and $\left|u_{0}^{2}\kappa_{b}^{'}\right|, \left|u_{0}^{2}\kappa_{w}^{'}\right| \ll 1$, eq. (\ref{eq: h2-h0}) is approximated to
\begin{equation}
\left\{\begin{aligned}
h_{0} & \simeq \zeta\left(u_{2}\right)\,,\\
\mathit{\Delta}h_{2}  & \triangleq h_{2} -1 \simeq  - u_{0}\zeta\kappa_{b} - \frac{u_{0}^{2}}{2}\zeta^{2}\kappa_{w}^{'}\,.
\end{aligned}\right.\label{eq: approx h2-h0}
\end{equation}
Note that $r_{2}$ can be replaced to $u_{2}$, since $r_{2}=u_{2}-O\left(u_{0}^{2}\kappa_{w}\right)$ from eq. (\ref{eq: 2nd-step trans}).

When a condition $\zeta \simeq 1$, \textit{i.e.} $\left|\kappa_{w}\right| \ll 2/L$ is applied to eq. (\ref{eq: approx h2-h0}), $h_{0}$ and $h_{2}$ are furthermore approximated to
\begin{equation}
\left\{\begin{aligned}
\mathit{\Delta}h_{0} & \triangleq h_{0} -1 \simeq \int^{u_{2}}_{0}\kappa_{w}\left(u\right)du
\,,\\
\mathit{\Delta}h_{2}  & = h_{2} -1  \simeq  - u_{0}\kappa_{b} - \frac{u_{0}^{2}}{2}\kappa_{w}^{'}\,.
\end{aligned}\right.\label{eq: approx2 h2-h0}
\end{equation}

%% file: Chap/3_Mode_v9.tex
\chapter{Propagation equation\label{ch: Mode}}

Fundamental equations, which are Maxwell equation, Shr{\" o}dinger equation and Newton equation of motion, can be unified to an equation focused on wave propagation in frequency domain $\omega$.
All analysis in the report derives from the propagation equation.

\section{Basic relations from propagation equation \label{sec: basic relation of prop eq}}
Wave-function $\boldsymbol{\Psi}=\left(\boldsymbol{\psi}_{a}^{\mathrm T}\,\boldsymbol{\psi}_{b}^{\mathrm T}\right)^{\mathrm T}$ as a column vector satisfies the following propagation-equation along propagation axis $u_{2}$:
\begin{equation}
\boldsymbol{M}\boldsymbol{\Psi}=-i
\frac{\partial}{\partial u_{2}}
\left(
	\begin{array}{cc}
		\boldsymbol{0} & \boldsymbol{1}\\
		\boldsymbol{1} & \boldsymbol{0}
	\end{array}
\right)
\boldsymbol{\Psi}\,.\label{eq: propagation-equation}
\end{equation}
The $2\times2$ matrix elements in the right hand side are square submatrices.
Each submatrix is the same size as continuous (or discretized) fields in 2D cross-section of the modes.
The square matrix $\boldsymbol{M}$ does not have $\partial_{2}$, that is, $\boldsymbol{M}u_{2}=u_{2}\boldsymbol{M}$.
We introduce generalized power-flow $P$ along the $u_{2}$ axis, and conservation of the power flow is derived from eq. (\ref{eq: propagation-equation}):
\begin{equation}\left\{
\begin{aligned}
P\left(u_{2}\right) &\triangleq \boldsymbol{\Psi}^{\dagger}\left(u_{2}\right)\left(\begin{array}{cc}
\boldsymbol{0} & \boldsymbol{1}\\
\boldsymbol{1} & \boldsymbol{0}
\end{array}\right)\boldsymbol{\Psi}\left(u_{2}\right)\,,
\\
\frac{\partial P}{\partial u_{2}} & =
i\boldsymbol{\Psi}^{\dagger}\left(\boldsymbol{M}-\boldsymbol{M}^{\dagger}\right)\boldsymbol{\Psi}\,.
\end{aligned}\right.
\label{eq: power flow and the conservation}
\end{equation}
The right hand side of the second in eq. \eqref{eq: power flow and the conservation} means power dissipation or power gain.
When $\boldsymbol{M}=\boldsymbol{M}^{\dagger}$, the system has no power-loss for wave propagation.

We also consider the following eigen-mode equation with replacing $-i\partial_{2}$ to propagation constant $\beta_{n}$ as the $n$-th eigenvalue for eq. (\ref{eq: propagation-equation}).
\begin{eqnarray}
\boldsymbol{M}\,\boldsymbol{\Phi}_{n} & = & \beta_{n}\left(\begin{array}{cc}
\boldsymbol{0} & \boldsymbol{1}\\
\boldsymbol{1} & \boldsymbol{0}
\end{array}\right)\boldsymbol{\Phi}_{n}\,.\label{eq: matrix equation}
\end{eqnarray}
The orthogonality of modes is derived from eq. \eqref{eq: matrix equation} with $\boldsymbol{M}=\boldsymbol{M}^{\dagger}$.
\begin{equation}
\boldsymbol{\Phi}_{m}^{\dagger}\left(\begin{array}{cc}
\boldsymbol{0} & \boldsymbol{1}\\
\boldsymbol{1} & \boldsymbol{0}
\end{array}\right)\boldsymbol{\Phi}_{n}=0\quad \mathrm{for} \quad \beta_{m}^{*}\neq\beta_{n}\,.
\label{eq: orthogonality}
\end{equation}
Previous work has already discussed orthogonality relation of propagation modes (see equations (3.2-24) in \cite{Marcuse} and (10.120) in \cite{Auld}). 
We can set power-flow normalization and mode numbering for real $\beta_{n}$ with satisfying that 
\begin{equation}
\boldsymbol{\Phi}_{n}^{\dagger}\left(\begin{array}{cc}
\boldsymbol{0} & \boldsymbol{1}\\
\boldsymbol{1} & \boldsymbol{0}
\end{array}\right)\boldsymbol{\Phi}_{n}=\frac{n}{\left|n\right|} \quad \mathrm{for} \quad n \neq 0\,.
\label{eq: propagating wave normalization}
\end{equation}
The above numbering physically means that power flow for positive (negative) $n$ is always positive (negative).
If we happen to have a mode with exactly zero power-flow and real  $\beta_{n}$, we can number the mode as $ n = 0 $ and normalize it as $ \boldsymbol {\Phi}_{0} ^ {\dagger} \boldsymbol {\Phi}_{0} = 1$.
There is a maximum number $n_{\max} \triangleq \max\left| n \right|$ for real $\beta_{n}$.
Non-real $\beta_{n}$ is numbered $n \gtrless \pm n_{\max}$ with satisfying $n\mathrm{Im}\beta_{n} > 0$, that is,
the $n > n_{\max}$ ($n < -n_{\max}$) means evanescent (divergent) wave along the $u_{2}$ axis.
A pair of complex conjugate $\beta_{n}$ is also set as $\beta_{n}=\beta_{-n}^{*}$, and the pair is normalized as 
\begin{equation}
\boldsymbol{\Phi}_{n}^{\dagger}\left(\begin{array}{cc}
\boldsymbol{0} & \boldsymbol{1}\\
\boldsymbol{1} & \boldsymbol{0}
\end{array}\right)\boldsymbol{\Phi}_{-n}=1\quad\mathrm{for}\quad n\gtrless \pm n_{\max}\,.
\label{eq: evanescent wave normalization}
\end{equation}
All of $\beta_{n}$ consists of real numbers and/or complex conjugate pairs when $\boldsymbol{M}=\boldsymbol{M}^{\dagger}$,
 because a secular equation for generalized eigenvalue problem $\boldsymbol{A}\boldsymbol{X}=\lambda\boldsymbol{B}\boldsymbol{X}$ can be deformed into
 $\left|\boldsymbol{A}-\lambda\boldsymbol{B}\right|=
 \left|\boldsymbol{A}^{\dagger}-\lambda^{*}\boldsymbol{B}^{\dagger}\right|=0$.
Note that the numbering of the complex conjugate pairs except for pure imaginary $\beta_{n}$ is different from one of Fig. 10.14 and eq. (10.125) in \cite{Auld}.
Here, we try to define dual basis $\widetilde{\boldsymbol{\Phi}}_{n}$ for the basis $\boldsymbol{\Phi}_{n}$:
\begin{eqnarray}
\widetilde{\boldsymbol{\Phi}}_{n} = \begin{cases}
\frac{n}{\left| n \right|}
\tiny 
\left(\begin{array}{cc}
\boldsymbol{0} & \boldsymbol{1}\\
\boldsymbol{1} & \boldsymbol{0}
\end{array}\right)
\normalsize
\boldsymbol{\Phi}_{n} & \mathrm{for} \quad 0 < \left| n \right| \leq n_{\max}\, ,\\
\tiny 
\left(\begin{array}{cc}
\boldsymbol{0} & \boldsymbol{1}\\
\boldsymbol{1} & \boldsymbol{0}
\end{array}\right)
\normalsize
\boldsymbol{\Phi}_{-n} & \mathrm{for} \quad \left| n \right| > n_{\max}\, ,\\
\boldsymbol{\Phi}_{0} - \sum_{m\neq 0}\left(\boldsymbol{\Phi}_{m}^{\dagger}\boldsymbol{\Phi}_{0}\right)\widetilde{\boldsymbol{\Phi}}_{m} & \mathrm{for} \quad n = 0\, .
\end{cases}
\label{eq: dual basis}
\end{eqnarray}
Equation (\ref{eq: orthogonality}) and the above normalization rules give us  bi-orthogonality~\cite{Treyssede,Spencer} and p. 285 of \cite{Shrikhande} as 
\[
\boldsymbol{\Phi}_{m}^{\dagger}\widetilde{\boldsymbol{\Phi}}_{n}=\widetilde{\boldsymbol{\Phi}}_{m}^{\dagger}\boldsymbol{\Phi}_{n}=\delta_{mn}\,,
\quad\mathrm{and}\quad
\sum_{n}\boldsymbol{\Phi}_{n}\widetilde{\boldsymbol{\Phi}}_{n}^{\dagger}=1\,.
\]
The $n = 0$ will be void below, 
because practical calculations can avoid it by slightly shifting the frequency $\omega$.

The mode equation \eqref{eq: matrix equation} can be extended to periodic system using the $\boldsymbol{M}_{p}$ of eq. (\ref{eq: periodic mode eq}) in Appendix \ref{ch: Periodic WG}.
Note that the representation basis with non-real eigenvalue of the $\boldsymbol{M}_{p}$ have to be redefined to satisfy eq. \eqref{eq: propagating wave normalization} as shown in eq. \eqref{eq: redefined evanescent mode PeriodicWG}.

\section{Special cases of mode equation\label{sec: special mode eq}}
Appendices \ref{ch: Shroedinger}, \ref{ch: Generalized Maxwell} and \ref{ch: Newton} show details of eq. \eqref{eq: propagation-equation} for the Shr{\"o}dinger equation, Maxwell's equations and Newton's equation of motion, respectively.

This section focuses on two special cases: Shr{\"o}dinger equation \eqref{eq: Shroedinger eq} when $A_{z}=0$,
 and Maxwell equation \eqref{eq: Generalized Maxwell eq} when $\boldsymbol{\varepsilon}$ and $\boldsymbol{\mu}$ are diagonal hermitian and $\boldsymbol{\alpha}=\boldsymbol{\gamma}=0$.
From eqs. \eqref{eq: Shroedinger eq. Az=0} and \eqref{eq: normal Maxwell equation}, the propagation equation \eqref{eq: propagation-equation} becomes that
\[
\boldsymbol{M}\boldsymbol{\Psi} =
\left(\begin{array}{cc}
\boldsymbol{m}_{aa} & \boldsymbol{0}\\
\boldsymbol{0} & \boldsymbol{m}_{bb}
\end{array}\right)\boldsymbol{\Psi}=-i
\frac{\partial}{\partial u_{2}}\left(\begin{array}{cc}
\boldsymbol{0} & \boldsymbol{1}\\
\boldsymbol{1} & \boldsymbol{0}
\end{array}\right)\boldsymbol{\Psi}\,.
\]
Then,
\[
\boldsymbol{m}_{aa} \boldsymbol{\psi}_{a} = -i\frac{\partial \boldsymbol{\psi}_{b}}{\partial u_{2}}\,,\quad
\boldsymbol{m}_{bb} \boldsymbol{\psi}_{b} = -i\frac{\partial \boldsymbol{\psi}_{a}}{\partial u_{2}}\,.
\]
For eq. \eqref{eq: Shroedinger eq. Az=0}, $\boldsymbol{m}_{aa}$ and $\boldsymbol{m}_{bb}$ are Hermite matrix and a real number, respectively.
For eq. \eqref{eq: normal Maxwell equation}, $\boldsymbol{m}_{aa}$ and $\boldsymbol{m}_{bb}$ are real symmetric
matrices. 
The mode equation \eqref{eq: matrix equation} is also that
$\boldsymbol{m}_{aa} \boldsymbol{\phi}_{an} = \beta_{n}\boldsymbol{\phi}_{bn}$ and 
$\boldsymbol{m}_{bb} \boldsymbol{\phi}_{bn} = \beta_{n}\boldsymbol{\phi}_{an}$
as $\boldsymbol{\Phi}_{n}=\left(\boldsymbol{\phi}_{an}^{\mathrm T}\,\boldsymbol{\phi}_{bn}^{\mathrm T}\right)^{\mathrm T}$.
Therefore, eigenvalue problem for $\beta_{n}$ is solved by 
\[
\boldsymbol{m}_{bb}\boldsymbol{m}_{aa} \boldsymbol{\phi}_{an} = \beta_{n}^{2}\boldsymbol{\phi}_{an}
\quad \mathrm{or} \quad
\boldsymbol{m}_{aa}\boldsymbol{m}_{bb} \boldsymbol{\phi}_{bn} = \beta_{n}^{2}\boldsymbol{\phi}_{bn}\,.
\]

For the special case of Shr{\"o}dinger equation \eqref{eq: Shroedinger eq. Az=0}, square of the eigenvalue $\beta_{n}^{2}$ is always real, since $\left(\boldsymbol{m}_{bb}\boldsymbol{m}_{aa}\right)^{\dagger}=\boldsymbol{m}_{aa}\boldsymbol{m}_{bb}=\boldsymbol{m}_{bb}\boldsymbol{m}_{aa}$.
Furthermore, the $\boldsymbol{\Phi}_{n}$ satisfies the simpler orthogonality:
\begin{equation}
\boldsymbol{\phi}_{am}^{\dagger}\boldsymbol{\phi}_{an} = \boldsymbol{\phi}_{bm}^{\dagger}\boldsymbol{\phi}_{bn} = 0
\quad \textrm{when} \quad
\beta_{m}^{2} \neq  \beta_{n}^{2}\,.
\label{eq: simple orthogonality}
\end{equation}
The special case \eqref{eq: case2 eigenvalue eq} for 2D tight-binding model \eqref{eq: Ando model} has the same properties as the above.

For the special case of Maxwell equation \eqref{eq: Generalized Maxwell eq}, properties of the $\boldsymbol{\Phi}_{n}$ are different from them for Shr{\"o}dinger equation.
The $\beta_{n}^{2}$ is real or complex, since $\left(\boldsymbol{m}_{bb}\boldsymbol{m}_{aa}\right)^{\dagger} = \boldsymbol{m}_{aa}\boldsymbol{m}_{bb} \neq \boldsymbol{m}_{bb}\boldsymbol{m}_{aa}$ for eq. (\ref{eq: normal Maxwell equation}).
The $\boldsymbol{\Phi}_{n}$ does not always satisfy eq. \eqref{eq: simple orthogonality}.

\section{Perfectly matched layer (PML) method\label{sec: PML}}
The dominating way for treating unbounded problems in numerical simulation is with the PML method~\cite{Berenger}. 
From eq. (1) with $\zeta=1$ of \cite{Shyroki} and eq. (2.7) of \cite{Shin}, equation \eqref{eq: matrix equation} can be modified to
\begin{equation}
\left[1+i\tilde{\sigma}_{2}\left(u_{2}\right)\right]\boldsymbol{M}\,e^{i\beta_{n}\int^{u_{2}}\left[1+i\tilde{\sigma}_{2}\left(z\right)\right]dz}\,\boldsymbol{\Phi}_{n}  =  -i
\frac{\partial}{\partial u_{2}} \left(\begin{array}{cc}
\boldsymbol{0} & \boldsymbol{1}\\
\boldsymbol{1} & \boldsymbol{0}
\end{array}\right)\,e^{i\beta_{n}\int^{u_{2}}\left[1+i\tilde{\sigma}_{2}\left(z\right)\right]dz}\,\boldsymbol{\Phi}_{n}\,.\label{eq: PML equation}
\end{equation}
Note that $n \beta_{n} \tilde{\sigma}_{2} \geq 0$ when $0 < \left| n \right| \leq n_{\max}$. 
The theoretical reflectance $\left| R_{n} \right|^{2}$ of PML with thickness $d$ and power number $M$ is defined as eq. (3) of \cite{Oskooi}: $\left| R_{n} \right|^{2} 
= \exp \left[-4\left|\beta_{n} d \right|\max\left(\tilde{\sigma}_{2}\right)/\left(M+1\right)\right]$,
where $d \gtrsim 2\lambda_{0}/3$ for vacuum wavelength $\lambda_{0}$, and $M = 2,\, 3,\,\mathrm{or}\,4$~\cite{Shin, Oskooi}.
Details of PML formulation for Maxwell's equations will be shown in eq. \eqref{eq: epsilon_l mu_l}.

%% file: Chap/4_NonAdiabaticTransition_v2.tex
\chapter{Non-adiabatic transition in frequency domain\label{ch: S-matrix BornApprox}}
This chapter discusses non-adiabatic transition by using adiabatic picture.
Equation \eqref{eq: propagation-equation} is slightly modified to
\begin{equation}
\left[\boldsymbol{M}^{\left(0\right)}\left( z \right) + \boldsymbol{M}^{\left(1\right)}\left( z \right) \right] \boldsymbol{\Psi}\left( z \right) 
= -i \frac{\partial}{\partial z}
\left(
	\begin{array}{cc}
		\boldsymbol{0} & \boldsymbol{1}\\
		\boldsymbol{1} & \boldsymbol{0}
	\end{array}
\right)
\boldsymbol{\Psi}\left( z \right)\,,\label{eq: propagation-equation M'}
\end{equation}
where the $\boldsymbol{M}^{\left(0\right)}$ always satisfies that $\boldsymbol{M}^{\left(0\right)}=\boldsymbol{M}^{\left(0\right)\dagger}$, but the additional term $\boldsymbol{M}^{\left(1\right)}$ has no limit of Hermitian property. 
Equation \eqref{eq: propagation-equation M'}  has  boundaries at $a$ and $b$ as $ -L_{\mathrm{s}}/2\leq a < b\leq L_{\mathrm{s}}/2$ (see section \ref{sec: notation}).
The $\boldsymbol{M}^{\left(0\right)}$ and $\boldsymbol{M}^{\left(1\right)}$ satisfy the following conditions as  
\begin{equation}
\frac{\partial\boldsymbol{M}^{\left(0\right)}}{\partial z}= 0\;\;\mathrm{and}\;\;\boldsymbol{M}^{\left(1\right)}=0\quad\mathrm{when}\quad z\leq a\;\;\mathrm{or}\;\; b\leq z\,.
\label{eq: propagation-eq. boundary conditions}
\end{equation}
The eigen-mode equation \eqref{eq: matrix equation} is rewritten as 
\[
\boldsymbol{M}^{\left(0\right)}\left( z \right)\boldsymbol{\Phi}_{m}\left( z \right) 
= \beta_{m}\left( z \right)
\left(
	\begin{array}{cc}
		\boldsymbol{0} & \boldsymbol{1}\\
		\boldsymbol{1} & \boldsymbol{0}
	\end{array}
\right)
\boldsymbol{\Phi}_{m}\left( z \right)\,.
\]

\section{Adiabatic picture}

To discuss the $\boldsymbol{\Psi}\left(z\right)$ in eq. \eqref{eq: propagation-equation M'},
we introduce adiabatic picture:
\begin{equation}
\boldsymbol{\Psi}_{m}^{\left(0\right)}\left(z\right) \triangleq \exp\left(i\int_{0}^{z}\beta_{m}\left(u\right)du\right)\boldsymbol{\Phi}_{m}\left(z\right)\quad\mathrm{and}\quad
\widetilde{\boldsymbol{\Psi}}_{m}^{\left(0\right)\dagger}\left(z\right) \triangleq \widetilde{\boldsymbol{\Phi}}_{m}^{\dagger}\left(z\right)\exp\left(-i\int_{0}^{z}\beta_{m}\left(u\right)du\right)\,,
\label{eq: adiabatic picture}
\end{equation}
where eqs. \eqref{eq: dual basis} and \eqref{eq: adiabatic picture}
give us 
\begin{equation}
\widetilde{\boldsymbol{\Psi}}_{m}^{\left(0\right)\dagger}\left(z\right)\boldsymbol{\Psi}_{n}^{\left(0\right)}\left(z\right)=\delta_{mn}\quad\mathrm{and}\quad
\sum_{m\neq 0}\boldsymbol{\Psi}_{m}^{\left(0\right)}\left(z\right)\widetilde{\boldsymbol{\Psi}}_{m}^{\left(0\right)\dagger}\left(z\right)=1\,.\label{eq: adiabatic dual vector}
\end{equation}
The $\widetilde{\boldsymbol{\Psi}}_{m}^{\left(0\right)\dagger}\left(z\right)$
and $\boldsymbol{\Psi}_{m}^{\left(0\right)}\left(z\right)$ can also
represent $\left\langle \widetilde{\boldsymbol{\Psi}}_{m}^{\left(0\right)}\left(z\right)\right|$
and $\left|\boldsymbol{\Psi}_{m}^{\left(0\right)}\left(z\right)\right\rangle $
using the bra-ket notation, respectively. 
Equations \eqref{eq: adiabatic picture} and \eqref{eq: adiabatic dual vector}
describe the operators $\boldsymbol{M}^{\left(0\right)}\left(z\right)$ and $\boldsymbol{M}^{\left(1\right)}\left(z\right)$
in eq. \eqref{eq: propagation-equation M'} as
\begin{equation}
\boldsymbol{M}^{\left(0\right)}\left(z\right) = \sum_{m\neq 0}\beta_{m}\left(z\right)\left(\begin{array}{cc}
\boldsymbol{0} & \boldsymbol{1}\\
\boldsymbol{1} & \boldsymbol{0}
\end{array}\right)\boldsymbol{\Psi}_{m}^{\left(0\right)}\left(z\right)\widetilde{\boldsymbol{\Psi}}_{m}^{\left(0\right)\dagger}\left(z\right)\quad\mathrm{and}\quad
\boldsymbol{M}^{\left(1\right)}\left(z\right) = \sum_{m,n\neq 0}{M}^{\left(1\right)}_{mn}\boldsymbol{\Psi}_{m}^{\left(0\right)}\left(z\right)\widetilde{\boldsymbol{\Psi}}_{n}^{\left(0\right)\dagger}\left(z\right)\,.
\label{eq: M and M'}
\end{equation}
From eqs. \eqref{eq: M and M'} and \eqref{eq: adiabatic picture},
the $\boldsymbol{\Psi}_{m}^{\left(0\right)}$ satisfies that 
\begin{equation}
\left[\frac{\partial}{\partial z}-i\left(\begin{array}{cc}
\boldsymbol{0} & \boldsymbol{1}\\
\boldsymbol{1} & \boldsymbol{0}
\end{array}\right)\left(\boldsymbol{M}^{\left(0\right)}+\boldsymbol{M}^{\left(1\right)}\right)\right]\boldsymbol{\Psi}_{m}^{\left(0\right)}\left(z\right) = \left[
	\boldsymbol{D}\left(z\right)-i\left(\begin{array}{cc}
	\boldsymbol{0} & \boldsymbol{1}\\
	\boldsymbol{1} & \boldsymbol{0}
	\end{array}\right)\boldsymbol{M}^{\left(1\right)}\left(z\right)
\right]\boldsymbol{\Psi}_{m}^{\left(0\right)}\left(z\right)
+\boldsymbol{\Psi}_{m}^{\left(0\right)}\left(z\right)\frac{\partial}{\partial z}\,,
\label{eq: adiabatic formulation}
\end{equation}
where 
\begin{equation}
\begin{split}\boldsymbol{D}\left(z\right) & \triangleq\sum_{n\neq 0}\exp\left(i\int_{0}^{z}\beta_{n}\left(u\right)du\right)\frac{\partial\boldsymbol{\Phi}_{n}}{\partial z}\widetilde{\boldsymbol{\Psi}}_{n}^{\left(0\right)\dagger}\left(z\right)=\sum_{m,n\neq 0}\boldsymbol{\Psi}_{m}^{\left(0\right)}\left(z\right)D_{mn}\left(z\right)\widetilde{\boldsymbol{\Psi}}_{n}^{\left(0\right)\dagger}\left(z\right)\,,\\
D_{mn} & =\widetilde{\boldsymbol{\Phi}}_{m}^{\dagger}\left(z\right)\exp\left(-i\int_{0}^{z}\left(\beta_{m}\left(u\right)-\beta_{n}\left(u\right)\right)du\right)\frac{\partial\boldsymbol{\Phi}_{n}}{\partial z}\,.
\end{split}
\label{eq: operator D}
\end{equation}

\section{Lippmann-Schwinger equation}

We show the following Lippmann-Schwinger equation for the $n$-th mode with outgoing ``$+$'' (incoming ``$-$'') scattered wave:
\begin{equation}
\begin{split}\boldsymbol{\Psi}_{n}^{\left(\pm\right)}\left(z\right) & =\boldsymbol{\Psi}_{n}^{\left(0\right)}\left(z\right)+\int_{a}^{b}\boldsymbol{G}_{0}^{\left(\pm\right)}\left(z,\,z'\right)\boldsymbol{\hat{M}}\left(z'\right)\boldsymbol{\Psi}_{n}^{\left(\pm\right)}\left(z'\right)dz'\,.\end{split}
\label{eq: LS equation}
\end{equation}
Here we introduce an adiabatic Green operator $\boldsymbol{G}_{0}^{\left(\pm\right)}\left(z,\,z'\right)$ to eq. \eqref{eq: LS equation}:
\begin{equation}
\boldsymbol{G}_{0}^{\left(\pm\right)}\left(z,\,z'\right)\triangleq\pm\sum_{m\neq 0}\boldsymbol{\Psi}_{m}^{\left(0\right)}\left(z\right)\frac{m}{\left|m\right|}H\left(\pm\frac{m}{\left|m\right|}\left(z-z'\right)\right)\widetilde{\boldsymbol{\Psi}}_{m}^{\left(0\right)\dagger}\left(z'\right)
\label{eq: Green operator 0}
\end{equation}
with using the Heaviside step function
\[
\begin{cases}
H\left(z\right)=1, & z>0\,,\\
H\left(z\right)=0, & z<0\,.
\end{cases}
\]
The $\boldsymbol{\hat{M}}$ in eq. \eqref{eq: LS equation} is also defined by
\begin{equation}
\begin{split}\boldsymbol{\hat{M}}\left(z\right) & \triangleq 
-\boldsymbol{D}\left(z\right)+i\left(\begin{array}{cc}
\boldsymbol{0} & \boldsymbol{1}\\
\boldsymbol{1} & \boldsymbol{0}
\end{array}\right)\boldsymbol{M}^{\left(1\right)}\left(z\right)
=\sum_{m,n\neq 0}\boldsymbol{\Psi}_{m}^{\left(0\right)}\left(z\right)\hat{M}_{mn}\left(z\right)\widetilde{\boldsymbol{\Psi}}_{n}^{\left(0\right)\dagger}\left(z\right),\\
\hat{M}_{mn}\left(z\right) & = \widetilde{\boldsymbol{\Psi}}_{m}^{\left(0\right)\dagger}\left(z\right)\left[-\boldsymbol{D}\left(z\right)+i\left(\begin{array}{cc}
\boldsymbol{0} & \boldsymbol{1}\\
\boldsymbol{1} & \boldsymbol{0}
\end{array}\right)\boldsymbol{M}^{\left(1\right)}\left(z\right)\right]\boldsymbol{\Psi}_{n}^{\left(0\right)}\left(z\right).
\end{split}
\label{eq: interaction operator}
\end{equation}
We can consider outgoing and evanescent (incoming and divergent) scattered-waves
to outside by using $\boldsymbol{G}_{0}^{\left(+\right)}$ ($\boldsymbol{G}_{0}^{\left(-\right)}$)
of eq. \eqref{eq: Green operator 0}, and the $\boldsymbol{G}_{0}^{\left(\pm\right)}$ satisfy that
\[
\left[\frac{\partial}{\partial z}-\boldsymbol{D}\left(z\right)-i\left(\begin{array}{cc}
\boldsymbol{0} & \boldsymbol{1}\\
\boldsymbol{1} & \boldsymbol{0}
\end{array}\right)\boldsymbol{M}^{\left(0\right)}\left(z\right)\right]\boldsymbol{G}_{0}^{\left(\pm\right)}\left(z,\,z'\right)=\delta\left(z-z'\right).
\]
We can check that eq. \eqref{eq: LS equation} satisfies eq. \eqref{eq: propagation-equation M'} with eqs. \eqref{eq: propagation-eq. boundary conditions} by considering eq. \eqref{eq: interaction operator} and the above equation.

When $\widetilde{\boldsymbol{\Phi}}_{n}^{\dagger}\left(z\right)\partial_{z}\boldsymbol{\Phi}_{n}\left(z\right)=0$, the $\boldsymbol{\hat{M}}$ in eq. \eqref{eq: interaction operator} can be also deformed
into
\begin{eqnarray}
\begin{split}
\hat{M}_{mn}\left(z\right) & =  \widetilde{\boldsymbol{\Psi}}_{m}^{\left(0\right)\dagger}\left(z\right)\left(\begin{array}{cc}
\boldsymbol{0} & \boldsymbol{1}\\
\boldsymbol{1} & \boldsymbol{0}
\end{array}\right)\left[\frac{\left(1-\delta_{mn}\right)\frac{\partial\boldsymbol{M}^{\left(0\right)}}{\partial z}}{\beta_{m}\left(z\right)-\beta_{n}\left(z\right)+0}+i\boldsymbol{M}^{\left(1\right)}\right]\boldsymbol{\Psi}_{n}^{\left(0\right)}\left(z\right) \\
 & =  \frac{m}{\left|m\right|}\boldsymbol{\Psi}_{m}^{\left(0\right)\dagger}\left(z\right)\left[\frac{\left(1-\delta_{mn}\right)\frac{\partial\boldsymbol{M}^{\left(0\right)}}{\partial z}}{\beta_{m}\left(z\right)-\beta_{n}\left(z\right)+0}+i\boldsymbol{M}^{\left(1\right)}\right]\boldsymbol{\Psi}_{n}^{\left(0\right)}\left(z\right)\quad\mathrm{for}\quad 0<\left|m\right|\leq n_{\max}\left(z\right).
\end{split} \label{eq: Mtild_lm}
\end{eqnarray}
If $\widetilde{\boldsymbol{\Phi}}_{n}^{\dagger}\left(z\right)\left(\partial\boldsymbol{\Phi}_{n}/\partial z\right)=\phi_{n}\left(z\right)\neq0$, a renormalization of $\boldsymbol{\Phi}_{n}$, which $\exp\left(-\int^{z}\phi_{n}du\right)\boldsymbol{\Phi}_{n}\left(z\right)\Rightarrow\boldsymbol{\Phi}_{n}\left(z\right)$,
can always remove the non-zero $\phi_{n}$ from the case with maintaining $\widetilde{\boldsymbol{\Phi}}_{n}^{\dagger}\boldsymbol{\Phi}_{n}=1$.
Here note that the $\phi_{n}$ always satisfies $\phi_{n}+\phi_{n}^{*}=0$ ( $\phi_{n}+\phi_{-n}^{*}=0$ ) when $\left|n\right|\leq n_{\max}$ ( $\left|n\right|>n_{\max}$ ).

\section{Dyson equation and T-matrix equation}

We can show the Dyson equation for the non-adiabatic transition.
\[\begin{split}
\boldsymbol{G}^{\left(\pm\right)}\left(z_{0},\,z'\right) & \triangleq  \boldsymbol{G}_{0}^{\left(\pm\right)}\left(z_{0},\,z'\right)+\sum_{N=2}^{\infty}\left[\prod_{j=1}^{N-1}\int_{a}^{b}dz_{j}\boldsymbol{G}_{0}^{\left(\pm\right)}\left(z_{j-1},\,z_{j}\right)\boldsymbol{\hat{M}}\left(z_{j}\right)\right]\boldsymbol{G}_{0}^{\left(\pm\right)}\left(z_{N-1},\,z'\right)\\
&=\boldsymbol{G}_{0}^{\left(\pm\right)}\left(z_{0},\,z'\right) + \int_{a}^{b}dz_{1}\boldsymbol{G}_{0}^{\left(\pm\right)}\left(z_{0},\,z_{1}\right)\boldsymbol{\hat{M}}\left(z_{1}\right)\boldsymbol{G}^{\left(\pm\right)}\left(z_{1},\,z'\right)\,.
\end{split}\]
The  T-matrix called as the transition matrix can be also introduced into the non-adiabatic transition as
\begin{equation}
\begin{split}\boldsymbol{T}^{\left(\pm\right)}\left(z_{0}\right) 
 & \triangleq\boldsymbol{\hat{M}}\left(z_{0}\right)\left[\boldsymbol{1} + \sum_{N=1}^{\infty}\prod_{j=1}^{N}\int_{a}^{b}dz_{j}\boldsymbol{G}_{0}^{\left(\pm\right)}\left(z_{j-1},\,z_{j}\right)\boldsymbol{\hat{M}}\left(z_{j}\right)\right]
=\boldsymbol{\hat{M}}\left(z_{0}\right)\left[\boldsymbol{1} + \int_{a}^{b}dz_{1}\boldsymbol{G}_{0}^{\left(\pm\right)}\left(z_{0},\,z_{1}\right)\boldsymbol{T}^{\left(\pm\right)}\left(z_{1}\right)\right]\,.
\end{split}
\label{eq: T-matrix equation}
\end{equation}
The perturbed Green operator $\boldsymbol{G}^{\left(\pm\right)}$ or the T-matrix $\boldsymbol{T}^{\left(\pm\right)}$ represents eq.\eqref{eq: LS equation} as
\begin{equation}
\begin{split}
\boldsymbol{\Psi}_{n}^{\left(\pm\right)}\left(z_{0}\right) - \boldsymbol{\Psi}_{n}^{\left(0\right)}\left(z_{0}\right)
& = \sum_{N=1}^{\infty}\left[\prod_{j=1}^{N}\int_{a}^{b}dz_{j}\boldsymbol{G}_{0}^{\left(\pm\right)}\left(z_{j-1},\,z_{j}\right)\boldsymbol{\hat{M}}\left(z_{j}\right)\right]\boldsymbol{\Psi}_{n}^{\left(0\right)}\left(z_{N}\right)\\
& = \int_{a}^{b}dz'\boldsymbol{G}^{\left(\pm\right)}\left(z_{0},\,z'\right)\boldsymbol{\hat{M}}\left(z'\right)\boldsymbol{\Psi}_{n}^{\left(0\right)}\left(z'\right)
 = \int_{a}^{b}dz'\boldsymbol{G}_{0}^{\left(\pm\right)}\left(z_{0},\,z'\right)\boldsymbol{T}^{\left(\pm\right)}\left(z'\right)\boldsymbol{\Psi}_{n}^{\left(0\right)}\left(z'\right)\,.
\end{split}
\label{eq: LS expansion}
\end{equation}
Note that the T-matrix of eq. \eqref{eq: T-matrix equation} is different from transfer matrix discussed in Section \ref{sec: transfer matrix}.

\section{S-matrix and its Born approximation\label{sec: BornApprox}}
We define elements of S-matrix which is discussed in Section \ref{sec: scattering matrix}:
\begin{equation}
S_{mn} \triangleq 
\begin{cases}
\widetilde{\boldsymbol{\Psi}}_{m}^{\left(0\right)\dagger}\left(b\right)
\boldsymbol{\Psi}_{n}^{\left(+\right)}\left(b\right) = \delta_{mn} + \int^{b}_{a}dz\,T^{\left(+\right)}_{mn}\left(z\right) & \mathrm{for}\;\; m>0\,,\\
\widetilde{\boldsymbol{\Psi}}_{m}^{\left(0\right)\dagger}\left(a\right)\boldsymbol{\Psi}_{n}^{\left(+\right)}\left(a\right)  = \delta_{mn} - \int^{b}_{a}dz\,T^{\left(+\right)}_{mn}\left(z\right)
& \mathrm{for}\;\; m<0\,.
\end{cases}
\label{eq: S-matrix}\end{equation}
The S-matrix shows the outgoing waves scattered in the region: $-L_{\mathrm{s}}/2 \leq	 a < z < b \leq +L_{\mathrm{s}}/2$.
The SMatrAn can numerically create the S-matrix of eq. \eqref{eq: S-matrix} by directly solving the propagation equation \eqref{eq: propagation-equation}.

If we consider the $\boldsymbol{\hat{M}}$ as a small perturbation term, we can apply the Born approximation to eq. \eqref{eq: T-matrix equation}:
\[
\boldsymbol{T}\left(z\right)=\boldsymbol{\hat{M}}\left(z\right)+O\left(\left\Vert \boldsymbol{\hat{M}}\right\Vert ^{2}\right).{}
\]
The $\boldsymbol{S}$ of eq. \eqref{eq: S-matrix} is approximated
to
\begin{equation}
S_{mn}\simeq\begin{cases}
\delta_{mn} + \int_{a}^{b}\hat{M}_{mn}\left(z\right)dz & \mathrm{for}\;\;m>0\,,\\
\delta_{mn} - \int_{a}^{b}\hat{M}_{mn}\left(z\right)dz & \mathrm{for}\;\;m<0\,.
\end{cases}\label{eq: S-matrix app}
\end{equation}
Equations \eqref{eq: Mtild_lm} and \eqref{eq: S-matrix app} give us
two pictures for wave-scattering. When $\boldsymbol{M}^{\left(1\right)}=0$, 
\[
\left\Vert \boldsymbol{S}-\boldsymbol{1}\right\Vert \propto\left\Vert \frac{\partial\boldsymbol{M}^{\left(0\right)}}{\partial z}\right\Vert
\]
which shows non-adiabatic transition in adiabatic structure.
When $\partial\boldsymbol{M}^{\left(0\right)}/\partial z=0$ and $\Im\boldsymbol{M}^{\left(1\right)}\neq0$,
\[
\left|S_{nn}\right|^{2}=1-O\left(\left\Vert i\boldsymbol{M}^{\left(1\right)}\right\Vert \right)\quad\mathrm{and}\quad\left|S_{mn}\right|^{2}=O\left(\left\Vert i\boldsymbol{M}^{\left(1\right)}\right\Vert ^{2}\right)\quad\mathrm{for}\quad m\neq n\,.
\]
Then, we can ignore scattered power from weak absorber as compared to absorbed power in it.

The Born approximation of eq. \eqref{eq: S-matrix app} is suit for not only understanding scattering process but also evaluating roughness scattering.
The following section shows the roughness scattering in the framework of the Born approximation.

%% file: Chap/5_Roughness_v9.tex
\chapter{Roughness scattering\label{ch: Roughness}}
We introduce parameter diagonal matrix $\boldsymbol{V}\left(u_{0},u_{1},u_{2}\right)$ into $\boldsymbol{M}$ in eq. (\ref{eq: propagation-equation}),
 and the $\boldsymbol{V}$ consists of several scalar functions of $u_{0}$, $u_{1}$ and $u_{2}$ for the media.
The $\boldsymbol{M}$ consists of $\partial_{u0}$, $\partial_{u1}$ and $\boldsymbol{V}$,
and it is linear with $\boldsymbol{V}$, \textit{i.e.} $ \boldsymbol{M}\left(\eta\boldsymbol{V}\right)=\eta\boldsymbol{M}\left(\boldsymbol{V}\right) $. 
We separate the $\boldsymbol{V}$ into adiabatic part $\boldsymbol{V}^{\left(0\right)}$ and non-adiabatic part $\boldsymbol{V}^{\left(1\right)}$ to use eq. \eqref{eq: propagation-equation M'} as
\begin{equation}
\boldsymbol{M}^{\left(0\right)}=\boldsymbol{M}\left(\boldsymbol{V}^{\left(0\right)}\right) \quad \mathrm{and} \quad \boldsymbol{M}^{\left(1\right)}=\boldsymbol{M}\left(\boldsymbol{V}^{\left(1\right)}\right).
\label{eq: V0V1}\end{equation}

\section{Edge roughness of straight waveguide \label{sec: roughness}}
When the $\boldsymbol{M}^{\left(0\right)}\left(u_{2}\right)$ does not depend on $u_{2}$, $\beta_{n}$ and $\boldsymbol{\Phi}_{n}$ are constant for $u_{2}$ (or $z$), and then this case simplifies eq. \eqref{eq: Mtild_lm} for $0<\left|m\right|\leq n_{\max}$ into 
\[
\hat{M}_{mn}\left(u_{2}\right)  = i
\frac{m}{\left|m\right|}\boldsymbol{\Psi}_{m}^{\left(0\right)\dagger}\left(u_{2}\right)\boldsymbol{M}^{\left(1\right)}\left(u_{2}\right)\boldsymbol{\Psi}_{n}^{\left(0\right)}\left(u_{2}\right) =
 i\frac{m}{\left|m\right|}e^{-i\left(\beta_{m}-\beta_{n}\right)u_{2}\,}\boldsymbol{\Phi}_{m}^{\dagger}\boldsymbol{M}^{\left(1\right)}\left(u_{2}\right)\boldsymbol{\Phi}_{n}\,.
\]
Fourier transform (FT) of the above $\boldsymbol{M}^{\left(1\right)}$ can be used as
\begin{equation}
\int_{-L_{\mathrm{s}}/2}^{L_{\mathrm{s}}/2}\hat{M}_{mn}\left(u_{2}\right) d u_{2}
	= i\frac{m}{\left|m\right|}\boldsymbol{\Phi}_{m}^{\dagger}\left[
	\int_{-\infty}^{\infty}\boldsymbol{M}^{\left(1\right)}\left(u_{2}\right)e^{-i\left(\beta_{m}-\beta_{n}\right)u_{2}\,} d u_{2}
	\right]\boldsymbol{\Phi}_{n}
	= i\frac{m}{\left|m\right|}\boldsymbol{\Phi}_{m}^{\dagger}\widehat{\boldsymbol{M}^{\left(1\right)}}\left(\beta_{m}-\beta_{n}\right)\boldsymbol{\Phi}_{n}
	\,,
\label{eq: Fgmn}\end{equation}
since $\boldsymbol{M}^{\left(1\right)}=0$ for $\left|u_{2}\right|>L_{\mathrm{s}}/2$.
Then equation \eqref{eq: S-matrix app} can be deformed by eqs. \eqref{eq: V0V1} and \eqref{eq: Fgmn}:
\begin{equation}
S_{mn} - \delta_{mn} \simeq  
i\boldsymbol{\Phi}_{m}^{\dagger}\widehat{\boldsymbol{M}^{\left(1\right)}}\left(\beta_{m}-\beta_{n}\right)\boldsymbol{\Phi}_{n}
 = i\boldsymbol{\Phi}_{m}^{\dagger}
	\boldsymbol{M}\left(
		\widehat{\boldsymbol{V}^{\left(1\right)}}\left(\beta_{m}-\beta_{n}\right) 
	\right)
	\boldsymbol{\Phi}_{n}\,.
\label{eq: S-matrixF}
\end{equation}

\begin{figure}[h!]
	\centering
	\includegraphics[height=30mm]{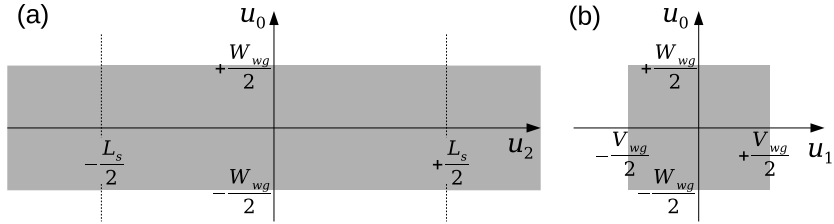}
	\caption{
		Waveguide with abrupt shape. (a) top-view. (b) cross-section.
	}
		\label{fig: waveguide}
\end{figure}

Let us apply eqs. (\ref{eq: Fgmn}) and (\ref{eq: S-matrixF}) to wave scattering by waveguide edge roughness.
The straight waveguide without roughness has a constant waveguide-width $W_{\mathrm{wg}}$ and a constant waveguide-height $V_{\mathrm{wg}}$, such as the abrupt case shown in Fig. \ref{fig: waveguide}. 
The $L_{\mathrm{s}}$ means scattering region as mentioned in section \ref{sec: notation}.
This section focuses on roughness by two edges at $u_{0}=\pm W_{\mathrm{wg}}/2$.
\noindent
Note that the unperturbed part $\boldsymbol{V}^{\left(0\right)}$ does not depend on $u_{2}$, and it is function only of $u_{0}$ and $u_{1}$.

\begin{figure}[h!]
	\centering
	\includegraphics[height=30mm]{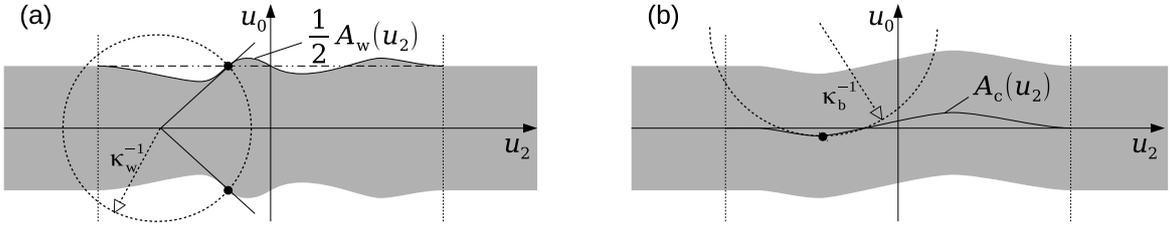}
	\caption{
		Two kinds of edge roughness: (a) line width roughness (LWR) and (b) line center roughness (LCR).
	}
		\label{fig: LWR and LCR}
\end{figure}

Here, we add two kinds of roughness functions $A_{\mathrm{w}}\left(z\right)$
 and $A_{\mathrm{c}}\left(z\right)$ as shown in Fig. \ref{fig: LWR and LCR} to the straight waveguide in Fig. \ref{fig: waveguide}.
The $A_{\mathrm{w}}$ and $A_{\mathrm{c}}$ represent
 line width roughness (LWR) and line center roughness (LCR)\cite{Constantoudis}, respectively.
We try to import the roughness functions into the perturbed part $\boldsymbol{V}^{\left(1\right)}$ of eq. \eqref{eq: S-matrixF}.
There are two approaches of importing the roughness functions.
 
\subsection{Approach I\label{subsec: approach I}}

The first approach describes the roughness as displacement of waveguide media, and then the $\left(u_{0},u_{1},u_{2}\right)$ coordinates are merely identity transformation of the  $\left(x_{0},x_{1},x_{2}\right)$ coordinates, \textit{i.e.} $h_{j} = 1$ for $j=0,1,2$.
The perturbed part $\boldsymbol{V}^{\left(1\right)}$ can be regarded as function of $u_{2}$ via $A_{\mathrm{w}}$ and $A_{\mathrm{c}}$ from Fig. \ref{fig: LWR and LCR}:
\begin{equation}
\begin{aligned}
&\quad \boldsymbol{V}^{\left(1\right)}\left(A_{\mathrm{w}}\left(u_{2}\right), A_{\mathrm{c}}\left(u_{2}\right)\right) = \boldsymbol{V}^{\left(0\right)}\left(\frac{u_{0}}{1 + A_{\mathrm{w}}/W_{\mathrm{wg}}}-A_{\mathrm{c}}, u_{1}\right)
 - \boldsymbol{V}^{\left(0\right)}\left(u_{0}, u_{1}\right)\\
 & \simeq \boldsymbol{V}^{\left(0\right)}\left(u_{0}-\frac{u_{0}A_{\mathrm{w}}}{W_{\mathrm{wg}}}-A_{\mathrm{c}}, u_{1}\right)
 - \boldsymbol{V}^{\left(0\right)}\left(u_{0}, u_{1}\right)
 \simeq - \left[\frac{u_{0}A_{\mathrm{w}}\left(u_{2}\right)}{W_{\mathrm{wg}}}+A_{\mathrm{c}}\left(u_{2}\right)\right]\frac{\partial\boldsymbol{V}^{\left(0\right)}\left(u_{0}, u_{1}\right)}{\partial u_{0}}
\end{aligned}
\label{eq: approach1 V1}
\end{equation}
when $A_{\mathrm{w}}/2,A_{\mathrm{c}} \ll W/2$.
In the framework of the above formulation (\ref{eq: approach1 V1}), the $\widehat{\boldsymbol{V}^{\left(1\right)}}\left(k\right)$ in eq. (\ref{eq: S-matrixF}) is given by 
\begin{equation}
\widehat{\boldsymbol{V}^{\left(1\right)}}\left(k\right)
	=
	- \left[\frac{u_{0}\widehat{A_{\mathrm{w}}}\left(k\right)}{W_{\mathrm{wg}}}+\widehat{A_{\mathrm{c}}}\left(k\right)\right]\frac{\partial\boldsymbol{V}^{\left(0\right)}\left(u_{0}, u_{1}\right)}{\partial u_{0}}.
	\label{eq: approach1 FM1V1}
\end{equation}
For abrupt structure of the waveguide, $\boldsymbol{V}^{\left(0\right)}$ and $\boldsymbol{V}^{\left(1\right)}$ can be set as
\begin{equation}
\left\{
\begin{aligned}
\boldsymbol{V}^{\left(0\right)} &= \boldsymbol{V}_{\mathrm{clad}}+\boldsymbol{V}_{\mathrm{pp}}\left(u_{1}\right)
 H\left(\frac{W_{\mathrm{wg}}}{2} - u_{0} \right)H\left(\frac{W_{\mathrm{wg}}}{2} + u_{0} \right)\,,
\\
\boldsymbol{V}^{\left(1\right)}  &= - \left[\frac{u_{0}A_{\mathrm{w}}\left(u_{2}\right)}{W_{\mathrm{wg}}}+A_{\mathrm{c}}\left(u_{2}\right)\right]\boldsymbol{V}_{\mathrm{pp}}\left(u_{1}\right)\left[-\delta\left(\frac{W_{\mathrm{wg}}}{2}-u_{0}\right) + \delta\left(\frac{W_{\mathrm{wg}}}{2}+u_{0}\right)\right]
\\
 &= 
	\left[\frac{1}{2}A_{\mathrm{w}}\left(u_{2}\right)
	+ \frac{u_{0}}{W_{\mathrm{wg}}/2}A_{\mathrm{c}}\left(u_{2}\right)\right]
	\boldsymbol{V}_{\mathrm{pp}}\left(u_{1}\right)
	\delta\left(\frac{W_{\mathrm{wg}}}{2}-\left|u_{0}\right|\right)
	\,,
\\
\widehat{\boldsymbol{V}^{\left(1\right)}} &= 
	\left[\frac{1}{2}\widehat{A_{\mathrm{w}}}\left(k\right)
	+ \frac{u_{0}}{W_{\mathrm{wg}}/2}\widehat{A_{\mathrm{c}}}\left(k\right)\right]
	\boldsymbol{V}_{\mathrm{pp}}\left(u_{1}\right)
	\delta\left(\frac{W_{\mathrm{wg}}}{2}-\left|u_{0}\right|\right)
	\,.
\end{aligned}
\right.\label{eq: approach1 abrupt V1}
\end{equation}
The first term $\boldsymbol{V}_{\mathrm{clad}}$ for $\boldsymbol{V}^{\left(0\right)}$ is constant. 
The $H$ in the above equation is Heaviside step function used for eq. \eqref{eq: Green operator 0}.
The $\boldsymbol{V}^{\left(1\right)}$ and $\widehat{\boldsymbol{V}^{\left(1\right)}}$ of eq. (\ref{eq: approach1 abrupt V1}) are derived by using eqs. (\ref{eq: approach1 V1}) and (\ref{eq: approach1 FM1V1}) respectively.

\subsection{Approach II\label{subsec: approach II}}

The second approach describes the roughness as space curvature.
In the $\left(u_{0},u_{1},u_{2}\right)$ coordinates, the waveguide with roughness becomes straight as shown in  Fig. \ref{fig: waveguide}.
The perturbed part $ \boldsymbol{V}^{\left(1\right)}$ can be regarded as a function of $u_{2}$ via the changes of scale factors $\mathit{\Delta}h_{0}$ and $\mathit{\Delta}h_{2}$ which are defined in eq. (\ref{eq: approx2 h2-h0}), and then it is represented as $ \boldsymbol{V}^{\left(1\right)}\left(\mathit{\Delta}h_{0}, \mathit{\Delta}h_{2}\right)$.

From Section \ref{Coordinate trans} and Fig. \ref{fig: LWR and LCR}, the two curvatures $\kappa_{w}$ and $\kappa_{b}$ are related to the two roughness $A_{\mathrm{w}}$ and $A_{\mathrm{c}}$ when $\left|\kappa_{w}\right|,\,\left|\kappa_{b}\right| \ll W_{\mathrm{wg}}^{-1}$.
The $\kappa_{w}$ and $\kappa_{b}$ are given by
\[
\kappa_{w}\left(u_{2}\right) \simeq \frac{1}{W_{\mathrm{wg}}} \frac{d A_{\mathrm{w}}\left(u_{2}\right)}{d u_{2}}
\quad \mathrm{and} \quad
\kappa_{b}\left(u_{2}\right) \simeq  \frac{d^{2} A_{\mathrm{c}}\left(u_{2}\right)}{d u_{2}^{2}}
\,.
\]
From eq. (\ref{eq: approx2 h2-h0}), the $\mathit{\Delta}h_{0}$ and $\mathit{\Delta}h_{2}$ are given by the roughness functions:
\[
\mathit{\Delta}h_{0}\left(u_{2}\right) \simeq  \frac{A_{\mathrm{w}}\left(u_{2}\right)}{W_{\mathrm{wg}}} 
\quad \mathrm{and} \quad
\mathit{\Delta}h_{2}\left(u_{0},u_{2}\right) \simeq  - u_{0} \frac{d^{2} A_{\mathrm{c}}}{d u_{2}^{2}} - \frac{u_{0}^{2}}{2W_{\mathrm{wg}}} \frac{d^{2} A_{\mathrm{w}}}{d u_{2}^{2}}\,.
\]
The $\widehat{\boldsymbol{V}^{\left(1\right)}}\left(k\right)$ of eq. \eqref{eq: S-matrixF} is given by FT of $\mathit{\Delta}h_{0}$ and $\mathit{\Delta}h_{2}$ as
\begin{equation}
\left\{
\begin{gathered}
\widehat{\boldsymbol{V}^{\left(1\right)}} \left(k\right) 
	 \simeq \boldsymbol{V}^{\left(1\right)}\left(\widehat{\mathit{\Delta}h_{0}}\left(k\right),\widehat{\mathit{\Delta}h_{2}}\left(u_{0},k\right)\right) \,,\\
\widehat{\mathit{\Delta}h_{0}}\left(k\right)  \simeq  \frac{1}{W_{\mathrm{wg}}}\widehat{A_{\mathrm{w}}}\left(k\right) \quad \mathrm{and} \quad
\widehat{\mathit{\Delta}h_{2}}\left(u_{0},k\right)  \simeq   u_{0} k^{2} \widehat{A_{\mathrm{c}}}\left(k\right)
 + \frac{u_{0}^{2}}{2W_{\mathrm{wg}}} k^{2} \widehat{A_{\mathrm{w}}}\left(k\right)\,.
\end{gathered}
\right.
\label{eq: approach2 FM1V1}
\end{equation}
The representation
 of eq. (\ref{eq: approach1 FM1V1}) is different from one of eq. (\ref{eq: approach2 FM1V1}), but two representations should give us the same results for roughness scattering.
Cross-check by eqs. (\ref{eq: approach1 FM1V1}) and (\ref{eq: approach2 FM1V1}) will show the validity of the formulation in this chapter. 

\section{Auto-correlation function of roughness\label{sec: Auto-correlation function}}

Let us consider ensemble average for roughness $A_{\nu}$ as $\nu = \mathrm{w},\,\mathrm{c}$.
This section introduces normalized roughness $a_{\nu}$ defined by
\begin{equation} 
	a_{\nu}\left(u_{2}\right) = \frac{A_{\nu}\left(u_{2}\right)}{\sqrt{L_{\mathrm{s}}}}\,,
\label{eq: roughness A 2 a}
\end{equation} 
since we presume that the integral of $A_{\nu}^{2}\left(u_{2}\right)$ is related to $L_{\mathrm{s}}$ under randomness.
Note that $ A_{\nu}\left(u_{2}\right) = a_{\nu}\left(u_{2}\right) = 0$ when $\left|u_{2}\right| \geq L_{\mathrm{s}}/2$.

We define an auto-correlation function $R_{\nu}$ for roughness $A_{\nu}$:
\begin{equation}
R_{\nu}\left(z\right) \triangleq \frac{1}{L_{\mathrm{s}}}\int_{-\infty}^{\infty}\left\langle
	A_{\nu}\left(z^{'}\right)A_{\nu}\left(z^{'}+z\right)
\right\rangle \,dz^{'}
= \int_{-\infty}^{\infty}\left\langle
	a_{\nu}\left(z^{'}\right)a_{\nu}\left(z^{'}+z\right)
\right\rangle \,dz^{'}
\,,
\label{eq: ACF}
\end{equation}
where $\left\langle\cdots\right\rangle $ means ensemble average.
A power spectral density (PSD) $G_{\nu}\left(k\right)$ can be also defined by 
\begin{equation}
G_{\nu}\left(k\right)  \triangleq \frac{1}{L_{\mathrm{s}}}\left\langle
	\left|\widehat{A_{\nu}}\left(k\right)\right|^{2}
\right\rangle 
 = \left\langle
	\left|\widehat{a_{\nu}}\left(k\right)\right|^{2}
\right\rangle
 \,.
\label{eq: PSD}
\end{equation}
Equations (\ref{eq: ACF}) and (\ref{eq: PSD}) give us the Wiener\textendash Khinchin theorem as 
\[
\begin{aligned}
G_{\nu}\left(k\right) &= \left\langle
	\int_{-\infty}^{\infty}a_{\nu}\left(z\right)\exp\left(ikz\right)\,dz
	\int_{-\infty}^{\infty}a_{\nu}\left(z^{'}\right)\exp\left(-ik z^{'}\right)\,dz^{'}
\right\rangle \\
 & =  \int_{-\infty}^{\infty}\,dz\,\int_{-\infty}^{\infty}\,du\,\left\langle
  a_{\nu}\left(z\right)a_{\nu}\left(z+u\right)
\right\rangle \exp\left(-iku\right)
  =\int_{-\infty}^{\infty}\,R_{\nu}\left(u\right)\exp\left(-iku\right)\,du
 = \widehat{R_{\nu}}\left(k\right)\,.
\end{aligned}\]
Then
\begin{equation}
\lim_{L_{\mathrm{s}} \rightarrow +\infty} \left|\widehat{a_{\nu}}\left(k\right)\right|^{2} = \left\langle \left|\widehat{a_{\nu}}\left(k_{n}\right)\right|^{2}\right\rangle = G_{\nu}\left(k\right)\,.\label{eq: Ginf1}
\end{equation}
when roughness is ergodic.

General discussion~\cite{Mack} of roughness uses a three-parameter model, and we apply it to $R_{\mathrm{w}}\left(z\right)$:
\begin{equation}
R_{\mathrm{w}}\left(z\right)=\sigma^{2}\exp\left(-\left(\frac{\left|z\right|}{L_{c}}\right)^{2\alpha}\right)\label{eq: Rinf}
\end{equation}
with standard deviation $\sigma$, roughness (or Hurst) exponent $\alpha$ and correlation length $L_{c}$. 
Three parameters are given by scanning electron microscope (SEM) measurements for waveguides. 
From eq. (\ref{eq: Rinf}),
\[
\ln\left(\ln\frac{R_{\mathrm{w}}\left(0\right)}{R_{\mathrm{w}}\left(z\right)}\right)=2\alpha\ln\left|z\right|-2\alpha\ln L_{c}\,.
\]
We can obtain the parameters from measurement data by using the above equation. 
\begin{equation}
G_{\mathrm{w}}\left(k\right)
=\sigma^{2}\int_{-\infty}^{\infty}\exp\left(-\left(\frac{\left|z\right|}{L_{c}}\right)^{2\alpha}- ik z\right)\,dz
=\sigma^{2}\int_{-\infty}^{\infty}\exp\left(-\left(\frac{\left|z\right|}{L_{c}}\right)^{2\alpha}\right)\cos\left(kz\right)\,dz\,.\label{eq: Ginf2}
\end{equation}
Let us approximate $\widehat{a_{\mathrm{w}}}\left(k\right)$ with finite $L_{\mathrm{s}}$ by using eqs. (\ref{eq: Ginf1}) and (\ref{eq: Ginf2}):
$
\left|\widehat{a_{\mathrm{w}}}\left(k\right)\right|^{2} \simeq G_{\mathrm{w}}\left(k\right)$. 
Then, we can numerically obtain 
$
\widehat{a_{\mathrm{w}}}\left(k\right) = \pm\sqrt{G_{\mathrm{w}}\left(k\right)}
\exp\left(i\phi\left(k\right)\right)
$, 
where the real function $\phi\left(k\right)$ and ``$\pm$'' are randomly given, and it also satisfies $\phi\left(-k\right) = -\phi\left(k\right)$.

The waveguide has two line edges with line edge roughness (LER) $a_{1}\left(z\right)$ and $a_{2}\left(z\right)$. 
LWR $a_{\mathrm{w}}$ and LCR $a_{\mathrm{c}}$ can be 
represented by LER $a_{1}$ and $a_{2}$ as follows:
\[\left\{
\begin{aligned}a_{\mathrm{w}}\left(z\right) & = a_{1}\left(z\right) - a_{2}\left(z\right),\\
a_{\mathrm{c}}\left(z\right) & =\frac{a_{1}\left(z\right) + a_{2}\left(z\right)}{2}.
\end{aligned}
\right.\]
If $A_{1}$ and $A_{2}$ have the same three-parameters and are not correlated,
the standard deviations for LWR, LER and LCR are $\sigma$, $\sigma/\sqrt{2}$ and $\sigma/2$, respectively~\cite{Kato}.

%% file: Chap/6_Discretization_v12.tex
\chapter{Numerical discretization for Maxwell equation\label{ch: discretization}}
This chapter shows a way of numerical discretization in order to calculate wave propagation in the waveguides. 
As in Appendix \ref{ch: optical scattering}, 
we consider Maxwell equation with scalar permittivity $\varepsilon$ and scalar magnetic-permeability $\mu$, and then
the generalized Maxwell equation (\ref{eq: Generalized Maxwell eq}) in frequency domain
 is simplified to
\begin{equation}
\left(\begin{array}{cc}
\boldsymbol{0} & i\boldsymbol{\nabla}\times\\
-i\boldsymbol{\nabla}\times & \boldsymbol{0}
\end{array}\right)\left(\begin{array}{c}
\boldsymbol{E}\\
\boldsymbol{H}
\end{array}\right)=\omega\left(\begin{array}{cc}
\varepsilon & 0\\
0 & \mu
\end{array}\right)\left(\begin{array}{c}
\boldsymbol{E}\\
\boldsymbol{H}
\end{array}\right).\label{eq: simplest Maxwell eq}
\end{equation}

We should avoid to use huge or tiny value in numerical calculation, and then it is better to normalize the above equation by using normalization constants.

\section{Normalization constants for numerical formulation}

We introduce a characteristic wave-number $k_{0}$ as normalization constant, \textit{e.\,g.} $k_{0}= 2\pi/1\,\mu\mathrm{m}$ to the cases of c-band or o-band.
Note that the $1\,\mu\mathrm{m}$ is not equal to the wavelength, but it is set as close value.
Then, length and wave-number (propagation constant) are normalized as
\[
\frac{\boldsymbol{u}}{k_{0}^{-1}}  \Rightarrow \boldsymbol{u}\,,
\quad \mathrm{and} \quad 
\frac{\beta}{k_{0}}  \Rightarrow \beta\,.
\]
The permittivity and magnetic permeability are also normalized by electric constant $\varepsilon_{0}$ and magnetic constant $\mu_{0}= 4\pi\times 10^{-7}\,\mathrm{N A^{-2}}$, respectively:
\[
\frac{\varepsilon}{\varepsilon_{0}} \Rightarrow \varepsilon\,, 
\quad \mathrm{and} \quad 
\frac{\mu}{\mu_{0}} \Rightarrow \mu\,.
\]
Then, the $\omega$ and time $t$ are normalized to 
\[
\frac{\omega}{\frac{k_{0}}{\sqrt{\varepsilon_{0}\mu_{0}}}} \Rightarrow \omega\,.
\quad \mathrm{and} \quad 
\frac{t}{\frac{\sqrt{\varepsilon_{0}\mu_{0}}}{k_{0}}}  \Rightarrow t\,.
\]
Finally, we try to normalize the electromagnetic field $\boldsymbol{E}$ and $\boldsymbol{H}$:
\[
\frac{\boldsymbol{E}}{\sqrt{\frac{\hbar k_{0}^{4}}{
\varepsilon_{0}\sqrt{\varepsilon_{0}\mu_{0}}}}}  \Rightarrow \boldsymbol{E}\,,
\quad \mathrm{and} \quad 
\frac{\boldsymbol{H}}{\sqrt{\frac{\hbar k_{0}^{4}}{
\mu_{0}\sqrt{\varepsilon_{0}\mu_{0}}}}}  \Rightarrow \boldsymbol{H}\,.
\]
Note that the density of electromagnetic energy derived from two normalization constants is equal to a photon energy with wave-number $k_{0}$ in a cube $k_{0}^{-3}$ with vacuum.

The normalization, which uses the above seven constants, does not change the representation of  eq. (\ref{eq: simplest Maxwell eq}).

\section{Two steps of transformation: $\boldsymbol{u} \Rightarrow \boldsymbol{\xi} \Rightarrow \boldsymbol{l}$
\label{sec: u to xi}}

Numerical analysis requires to discretize the dimensionless $\boldsymbol{u}$-space: $u_{j} \Rightarrow {\xi}_{j} \Rightarrow l_{j}$, where $l_{j}$ as $j=0,\,1,\,2$ is non-negative integer.
Then, we can define positive integer $L_{j}$:
\[
L_{j} \triangleq 1 + \max \left( l_{j} \right),
\quad \mathrm{then} \quad
0 \leq l_{j} < L_{j}\,.
\]
The ${\xi}_{j}$ is continuous variable, and $0\leq{\xi}_{j}<L_{j}$.
The $u_{j}$ is a monotonically increasing function of ${\xi}_{j}$, \textit{i.e.} $u_{j}^{'} \triangleq du_{j}/d{\xi}_{j}$ is always positive. 
The system length $L_{s}$ as shown in Section \ref{sec: notation} is given by $L_{s}=\int^{L_{2}}_{0} u^{'}_{2} d\xi_{2}$.
If we apply periodical (or anti-periodical) boundary condition to electromagnetic field along $u_{j}$ axis, 
periodical boundary condition has to be applied to $u_{j}^{'}$ as
\[
u_{j}^{'}\left({\xi}_{j}\right) = u_{j}^{'}\left({\xi}_{j}+L_{j}\right).
\]
Even in the case of bend waveguide as in Chapter \ref{ch: GeneralWaveguides},  the above periodical condition could be applied to the case of $j=1$ at least. 
Appendix \ref{ch: Non-uniform mesh} shows an example of non-uniform mesh, and eq. (\ref{eq: function u_j}) gives us  functions $u_{j}\left({\xi}_{j}\right)$ and $u_{j}^{'}\left({\xi}_{j}\right)$.

In the same manner as Section \ref{sec: deformed rotation}, the rotation operator $\boldsymbol{\nabla}\times$ in the coordinates $\boldsymbol{x}=\left(x_{0},x_{1},x_{2}\right)$ can be transformed to an operator in other coordinates $\boldsymbol{\xi}=\left(\xi_{0},\xi_{1},\xi_{2}\right)$ via the coordinates $\boldsymbol{u}=\left(u_{0},u_{1},u_{2}\right)$.
\begin{equation}
\boldsymbol{\nabla}\times 
 = \prod_{j=0}^{2}\frac{1}{u_{j}^{'}h_{j}}\boldsymbol{f}_{\xi}\left(\boldsymbol{\nabla}_{\xi}\times\right)\boldsymbol{f}_{\xi}\,,
\label{eq: transformed rotation}
\end{equation}
where
\[
\boldsymbol{f}_{\xi}  \triangleq 
\left(\begin{array}{ccc}
u_{0}^{'}h_{0} & 0 & 0\\
0 & u_{1}^{'}h_{1} & 0\\
0 & 0 & u_{2}^{'}h_{2}
\end{array}\right),
\quad
\boldsymbol{\nabla}_{\xi}\times  \triangleq
\left(\begin{array}{ccc}
0 & -\partial_{\xi2} & \partial_{\xi1}\\
\partial_{\xi2} & 0 & -\partial_{\xi0}\\
-\partial_{\xi1} & \partial_{\xi0} & 0
\end{array}\right)
\quad\mathrm{as}\quad
\partial_{\xi j}\triangleq\frac{\partial}{\partial\xi_{j}}\,.
\]
If we use the scheme of Chapter \ref{ch: GeneralWaveguides}, the scale factors $h_{0}$ and $h_{2}$ are given by eq. (\ref{eq: h2-h0}), and $h_{1} = 1$.

By using eq. (\ref{eq: transformed rotation}), the Maxwell equation (\ref{eq: simplest Maxwell eq}) could be deformed to
\begin{equation}
\left(\begin{array}{cc}
0 & i\boldsymbol{\nabla}_{\xi}\times\\
-i\boldsymbol{\nabla}_{\xi}\times & 0
\end{array}\right)
\left(\begin{array}{c}
\boldsymbol{f}_{\xi}\boldsymbol{E}\\
\boldsymbol{f}_{\xi}\boldsymbol{H}
\end{array}\right)
=\omega\left(\begin{array}{cc}
{\boldsymbol{\varepsilon}}_{\xi} & 0\\
0 & {\boldsymbol{\mu}}_{\xi}
\end{array}\right)
\left(\begin{array}{c}
\boldsymbol{f}_{\xi}\boldsymbol{E}\\
\boldsymbol{f}_{\xi}\boldsymbol{H}
\end{array}\right),
\label{eq: deformed Maxwell eq}
\end{equation}
where 
\begin{equation}
\left\{
\begin{gathered}
\boldsymbol{f}_{\xi}\boldsymbol{E}\left(\boldsymbol{\xi}\right) = \left(\begin{array}{c}
u_{0}^{'}h_{0}E_{0}\\
u_{1}^{'}h_{1}E_{1}\\
u_{2}^{'}h_{2}E_{2}
\end{array}\right)
,\quad
\boldsymbol{f}_{\xi}\boldsymbol{H}\left(\boldsymbol{\xi}\right) = \left(\begin{array}{c}
u_{0}^{'}h_{0}H_{0}\\
u_{1}^{'}h_{1}H_{1}\\
u_{2}^{'}h_{2}H_{2}
\end{array}\right),
\\
{\boldsymbol{\varepsilon}}_{\xi}\left(\boldsymbol{\xi}\right)  =  \prod_{n=0}^{2}u_{n}^{'}h_{n}\boldsymbol{f}_{\xi}^{-1}\varepsilon\boldsymbol{f}_{\xi}^{-1}=
\left(\begin{array}{ccc}
{\varepsilon}_{\xi\,00} & 0 & 0\\
0 & {\varepsilon}_{\xi\,11} & 0\\
0 & 0 & {\varepsilon}_{\xi\,22}
\end{array}\right),
\quad 
{\varepsilon}_{\xi\,jj}  =  \frac{u_{j+1}^{'}u_{j+2}^{'}h_{j+1}h_{j+2}}{u_{j}^{'}h_{j}}\varepsilon\left(\boldsymbol{\xi}\right),
\\
{\boldsymbol{\mu}}_{\xi}\left(\boldsymbol{\xi}\right)  =  \prod_{n=0}^{2}u_{n}^{'}h_{n}\boldsymbol{f}_{\xi}^{-1}\mu\boldsymbol{f}_{\xi}^{-1}=
\left(\begin{array}{ccc}
{\mu}_{\xi\,00} & 0 & 0\\
0 & {\mu}_{\xi\,11} & 0\\
0 & 0 & {\mu}_{\xi\,22}
\end{array}\right),
\quad 
{\mu}_{\xi\,jj}  =  \frac{u_{j+1}^{'}u_{j+2}^{'}h_{j+1}h_{j+2}}{u_{j}^{'}h_{j}}\mu\left(\boldsymbol{\xi}\right).
\end{gathered}\right.
\label{eq: f_xi E, f_xi H}
\end{equation}
Let us approximate eq. (\ref{eq: deformed Maxwell eq}) to discretized equation with transforming $\boldsymbol{\xi}=\left(\xi_{0},\xi_{1},\xi_{2}\right) \, \Rightarrow \, \boldsymbol{l}=\left(l_{0},l_{1},l_{2}\right)$ in the framework of Yee's lattice~\cite{Yee}.

\section{Discretization by Yee's lattice}

\begin{figure}
\begin{centering}
\includegraphics[width=0.5\columnwidth]{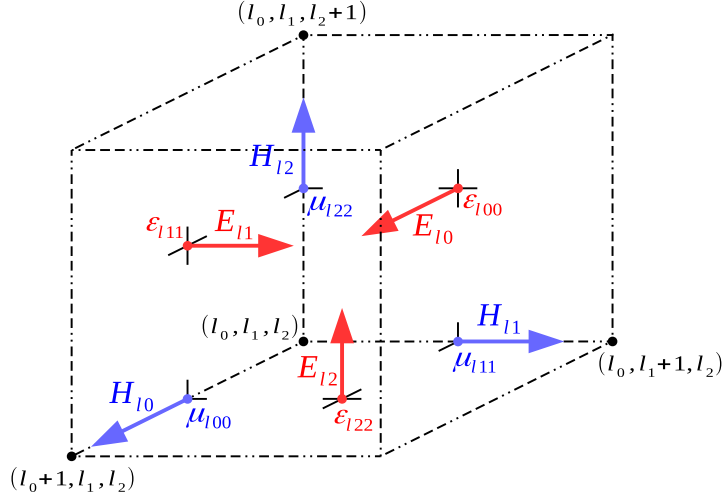}
\par\end{centering}
\caption{Yee cell~\cite{Yee}. We select the cell where $E_{l0}$ and $E_{l1}$ exist at half-integer $\xi_{2}$, because discussion about optics frequently focuses on properties of electric field. \label{fig: Yee cell}}
\end{figure}

 Figure \ref{fig: Yee cell} gives us an arrangement of discrete functions from the continuous functions $\boldsymbol{f}_{\xi}\boldsymbol{E}$, $\boldsymbol{f}_{\xi}\boldsymbol{H}$, ${\boldsymbol{\varepsilon}}_{\xi}$ and ${\boldsymbol{\mu}}_{\xi}$, where the discrete functions are allocated at a cell address $\left[\boldsymbol{l}\right]$.
Discrete electromagnetic fields $\boldsymbol{f}_{\xi}\boldsymbol{E}\left(\boldsymbol{\xi}\right) \Rightarrow \boldsymbol{E}_{l}\left[\boldsymbol{l}\right]$ and $\boldsymbol{f}_{\xi}\boldsymbol{H}\left(\boldsymbol{\xi}\right) \Rightarrow \boldsymbol{H}_{l}\left[\boldsymbol{l}\right]$ are derived by eq. (\ref{eq: f_xi E, f_xi H}) and Fig. \ref{fig: Yee cell}.
The components of $\boldsymbol{E}_{l}\left[\boldsymbol{l}\right]$ and $\boldsymbol{H}_{l}\left[\boldsymbol{l}\right]$ are defined as
\begin{equation}
\left\{\begin{aligned}
{E}_{lj}\left[\boldsymbol{l}\right] &\triangleq u_{j}^{'}h_{j}E_{j}\left(l_{0}+\frac{1-\delta_{j0}}{2},\,l_{1}+\frac{1-\delta_{j1}}{2},\,l_{2}+\frac{1-\delta_{j2}}{2}\right)\,,
\\
{H}_{lj}\left[\boldsymbol{l}\right] &\triangleq  u_{j}^{'}h_{j}H_{j}\left(l_{0}+\frac{\delta_{j0}}{2},\,l_{1}+\frac{\delta_{j1}}{2},\,l_{2}+\frac{\delta_{j2}}{2}\right)\,.
\end{aligned}\right.
\label{eq: E_l H_l}
\end{equation}
Discrete medium, which consists of ${\boldsymbol{\varepsilon}}_{\xi}\left(\boldsymbol{\xi}\right) \Rightarrow {\boldsymbol{\varepsilon}}_{l}\left[\boldsymbol{l}\right]$ and ${\boldsymbol{\mu}}_{\xi}\left(\boldsymbol{\xi}\right) \Rightarrow {\boldsymbol{\mu}}_{l}\left[\boldsymbol{l}\right]$, is also derived by eq. (\ref{eq: f_xi E, f_xi H}) and Fig. \ref{fig: Yee cell}.
The components of ${\boldsymbol{\varepsilon}}_{l}\left[\boldsymbol{l}\right]$ and ${\boldsymbol{\mu}}_{l}\left[\boldsymbol{l}\right]$ are defined with PML factor $\tilde{\sigma}_{j}\left(\xi_{j}\right)$ from eq. \eqref{eq: PML equation} and correction factor $\eta_{j}$ from eqs. \eqref{eq: disc eta} and \eqref{eq: average epsilon and mu}: 
\begin{equation}
\left\{\begin{aligned}
\varepsilon_{ljj}\left[\boldsymbol{l}\right] \triangleq \frac{u_{j+1}^{'}u_{j+2}^{'}h_{j+1}h_{j+2}}{u_{j}^{'}h_{j}} & 
\frac{\left[1+i\tilde{\sigma}_{j+1}\left(l_{j+1}+1/2\right)\right]\left[1+i\tilde{\sigma}_{j+2}\left(l_{j+2}+1/2\right)\right]}{1+i\tilde{\sigma}_{j}\left(l_{j}\right)}\\
& \quad \times{}
\frac{\varepsilon\left(l_{0}+\left({1-\delta_{j0}}\right)/{2},\,l_{1}+\left({1-\delta_{j1}}\right)/{2},\,l_{2}+\left({1-\delta_{j2}}\right)/{2}\right)}
{1+\eta_{j}\left(l_{0}+\left({1-\delta_{j0}}\right)/{2},\,l_{1}+\left({1-\delta_{j1}}\right)/{2},\,l_{2}+\left({1-\delta_{j2}}\right)/{2}\right)}\,,
\\
\mu_{ljj}\left[\boldsymbol{l}\right] \triangleq \frac{u_{j+1}^{'}u_{j+2}^{'}h_{j+1}h_{j+2}}{u_{j}^{'}h_{j}} & 
\frac{\left[1+i\tilde{\sigma}_{j+1}\left(l_{j+1}\right)\right]\left[1+i\tilde{\sigma}_{j+2}\left(l_{j+2}\right)\right]}{1+i\tilde{\sigma}_{j}\left(l_{j}+1/2\right)}\\
& \quad \times{}
\frac{\mu\left(l_{0}+{\delta_{j0}}/{2},\,l_{1}+{\delta_{j1}}/{2},\,l_{2}+{\delta_{j2}}/{2}\right)}{1+\eta_{j}\left(l_{0}+{\delta_{j0}}/{2},\,l_{1}+{\delta_{j1}}/{2},\,l_{2}+{\delta_{j2}}/{2}\right)}\,.
\end{aligned}\right.
\label{eq: epsilon_l mu_l}
\end{equation}

Numerical calculation considers finite numbers of Yee cells as $0\leq l_{j} < L_{j}$.
Function $\mathcal{G}\left[\boldsymbol{l}\right]$, which is $H_{lj}$ for example, satisfies the following boundary conditions for $l_{0}$ and $l_{1}$.
\begin{equation}
\left\{
\begin{gathered}
\mathcal{G}\left[L_{0},\,l_{1},\,l_{2}\right] = B_{0}\mathcal{G}\left[0,\,l_{1},\,l_{2}\right],
\quad 
\mathcal{G}\left[-1,\,l_{1},\,l_{2}\right]  = B_{0}^{*}\mathcal{G}\left[L_{0}-1,\,l_{1},\,l_{2}\right],
\\
\mathcal{G}\left[l_{0},\,L_{1},\,\,l_{2}\right]  =B_{1}\mathcal{G}\left[l_{0},\,0,\,l_{2}\right],
\quad 
\mathcal{G}\left[l_{0},\,-1,\,\,l_{2}\right]  =B_{1}^{*}\mathcal{G}\left[l_{0},\,L_{1}-1,\,l_{2}\right],
\end{gathered}
\right.
\label{eq: boundary condition}
\end{equation}
where $\left| B_{j} \right| = 1\,\,\mathrm{or}\,\, 0$, and $\mathcal{G}=H_{lj}$, $E_{lj}$ for example.
The two parameters $B_{0}$ and $B_{1}$ are generally complex number, and we could choose a periodic condition $B_{j} = 1$ or an anti-periodic condition $B_{j} = -1$.
We can also set $B_{0} = 0$ when $\kappa_{b}\neq 0$, since the system does not become periodic along the $u_{0}$-axis.

Let us introduce forward and backward difference operators as 
\begin{equation}
\left\{
\begin{aligned}
\bigtriangleup_{j}\mathcal{G}\left[l_{0},\,l_{1},\,l_{2}\right] &\triangleq \mathcal{G}\left[l_{0}+\delta_{j0},\,l_{1}+\delta_{j1},\,l_{2}+\delta_{j2}\right]-\mathcal{G}\left[l_{0},\,l_{1},\,l_{2}\right],
\\
\bigtriangledown_{j}\mathcal{G}\left[l_{0},\,l_{1},\,l_{2}\right] &\triangleq \mathcal{G}\left[l_{0},\,l_{1},\,l_{2}\right] - \mathcal{G}\left[l_{0}-\delta_{j0},\,l_{1}-\delta_{j1},\,l_{2}-\delta_{j2}\right].
\end{aligned}
\right.
\label{eq: difference operator}
\end{equation}
From eqs. (\ref{eq: boundary condition}) and (\ref{eq: difference operator}), the difference operators at boundaries are given by
\[
\left\{
\begin{aligned}
\bigtriangleup_{0}\mathcal{G}\left[L_{0}-1,\,l_{1},\,l_{2}\right] & = B_{0}\mathcal{G}\left[0,\,l_{1},\,l_{2}\right]-\mathcal{G}\left[L_{0}-1,\,l_{1},\,l_{2}\right],
\; 
\bigtriangledown_{0}\mathcal{G}\left[0,\,l_{1},\,l_{2}\right] = \mathcal{G}\left[0,\,l_{1},\,l_{2}\right]-B_{0}^{*}\mathcal{G}\left[L_{0}-1,\,l_{1},\,l_{2}\right],
\\
\bigtriangleup_{1}\mathcal{G}\left[l_{0},\,L_{1}-1,\,l_{2}\right] & = B_{1}\mathcal{G}\left[l_{0},\,0,\,l_{2}\right]-\mathcal{G}\left[l_{0},\,L_{1}-1,\,l_{2}\right],
\; 
\bigtriangledown_{1}\mathcal{G}\left[l_{0},\,0,\,l_{2}\right] = \mathcal{G}\left[l_{0},\,0,\,l_{2}\right]-B_{1}^{*}\mathcal{G}\left[l_{0},\,L_{1}-1,\,l_{2}\right].
\end{aligned}
\right.
\]
Therefore, discrete difference operators satisfy 
$
\bigtriangledown_{j} = -\bigtriangleup_{j}^{\dagger}
$
, but
$
\bigtriangleup_{j} \neq -\bigtriangleup_{j}^{\dagger}
$
.
This relation is different from the case of continuous difference operators $\partial_{j}=-\partial_{j}^{\dagger}$ under the boundary condition of eq. (\ref{eq: boundary condition}). 

From eqs. (\ref{eq: E_l H_l}), (\ref{eq: epsilon_l mu_l}) and Fig. \ref{fig: Yee cell}, the deformed Maxwell equation (\ref{eq: deformed Maxwell eq}) could be discretized into
\begin{equation}
\left(\begin{array}{cc}
0 & i\boldsymbol{R}\\
-i\boldsymbol{R}^{\dagger} & 0
\end{array}\right)
\left(\begin{array}{c}
\boldsymbol{E}_{l}\\
\boldsymbol{H}_{l}
\end{array}\right)
=\omega\left(\begin{array}{cc}
{\boldsymbol{\varepsilon}}_{l} & 0\\
0 & {\boldsymbol{\mu}}_{l}
\end{array}\right)
\left(\begin{array}{c}
\boldsymbol{E}_{l}\\
\boldsymbol{H}_{l}
\end{array}\right),
\label{eq: discrete Maxwell eq}
\end{equation}
where the discrete rotation operators $ \boldsymbol{R} $ and $ \boldsymbol{R}^{\dagger} $ are defined by using eqs. (\ref{eq: difference operator}):
\begin{equation}
\boldsymbol{R} \triangleq 
\left(\begin{array}{ccc}
0 & -\bigtriangleup_{2} & \bigtriangleup_{1}\\
\bigtriangleup_{2} & 0 & -\bigtriangleup_{0}\\
-\bigtriangleup_{1} & \bigtriangleup_{0} & 0
\end{array}\right)
\quad \mathrm{and} \quad
\boldsymbol{R}^{\dagger} \triangleq 
\left(\begin{array}{ccc}
0 & \bigtriangleup_{2}^{\dagger} & -\bigtriangleup_{1}^{\dagger}\\
-\bigtriangleup_{2}^{\dagger} & 0 & \bigtriangleup_{0}^{\dagger}\\
\bigtriangleup_{1}^{\dagger} & -\bigtriangleup_{0}^{\dagger} & 0
\end{array}\right)
=
\left(\begin{array}{ccc}
0 & -\bigtriangledown_{2} & \bigtriangledown_{1}\\
\bigtriangledown_{2} & 0 & -\bigtriangledown_{0}\\
-\bigtriangledown_{1} & \bigtriangledown_{0} & 0
\end{array}\right).
\label{eq: operator R}
\end{equation}
The $ \boldsymbol{R} $ and $ \boldsymbol{R}^{\dagger} $ are used in eq. \eqref{eq: Time-Maxwell eq}.

%% file: Chap/7_PropagationEqYee_v2.tex
\chapter{Propagation equation for Yee's lattice\label{ch: propagation Yee}}

This chapter shows discrete formulae derived from the discrete Maxwell-equation (\ref{eq: discrete Maxwell eq}).

\section{Discrete propagation-equation}

Discrete formulation can be represented as similar to propagation equation (\ref{eq: propagation-equation}) with eq. (\ref{eq: normal Maxwell equation}).
In order to simplify mathematical notations , we only show a discrete parameter $l_{2}$ instead of $\boldsymbol{l}$ in the following equations.
\begin{equation}
\boldsymbol{M}_{\pm}\boldsymbol{\Psi}_{\pm} = 
 -i\left(\begin{array}{cc}
\boldsymbol{0} & \bigtriangledown_{2}\\
\bigtriangleup_{2} & \boldsymbol{0}
\end{array}\right)\boldsymbol{\Psi}_{\pm} \quad \mathrm{as} \quad
\left\{\begin{aligned}
\boldsymbol{M}_{+} & \triangleq  
\left(\begin{array}{cc}
\boldsymbol{m}_{aa}\left(\boldsymbol{V}\left[ l_{2} \right]\right) & \boldsymbol{0}\\
\boldsymbol{0} & \boldsymbol{m}_{bb}\left(\boldsymbol{V}\left[ l_{2} \right]\right)
\end{array}\right),
\\
\boldsymbol{M}_{-} & \triangleq  
\left(\begin{array}{cc}
\boldsymbol{m}_{bb}\left(\boldsymbol{V}\left[ l_{2} \right]\right) & \boldsymbol{0}\\
\boldsymbol{0} & \boldsymbol{m}_{aa}\left(\boldsymbol{V}\left[ l_{2}+1 \right]\right)
\end{array}\right).
\end{aligned}\right.
\label{eq: discrete propagation-eq}
\end{equation}
The subscript $+$ ($-$) in eq. (\ref{eq: discrete propagation-eq}) means a configuration as advanced (retarded) electric fields to magnetic fields.
From eq. (\ref{eq: details of m_aa and m_bb}), the discrete $\boldsymbol{m}_{aa}$, $\boldsymbol{m}_{bb}$ and $\boldsymbol{\Psi}_{\pm}$ are defined by
\begin{equation}
\left\{
\begin{gathered}
\boldsymbol{m}_{aa}\left( \boldsymbol{V} \right) =
\left(\begin{array}{cc}
{V}_{0} + \bigtriangledown_{1}{V}_{2} \bigtriangleup_{1} &
-\bigtriangledown_{1}{V}_{2} \bigtriangleup_{0} \\
-\bigtriangledown_{0}{V}_{2} \bigtriangleup_{1} &
{V}_{1} + \bigtriangledown_{0}{V}_{2} \bigtriangleup_{0}
\end{array}\right),
\quad
\boldsymbol{m}_{bb}\left( \boldsymbol{V} \right) =
\left(\begin{array}{cc}
{V}_{3} + \bigtriangleup_{0}{V}_{5} \bigtriangledown_{0} &
\bigtriangleup_{0}{V}_{5} \bigtriangledown_{1} \\
\bigtriangleup_{1}{V}_{5} \bigtriangledown_{0} &
{V}_{4} + \bigtriangleup_{1}{V}_{5} \bigtriangledown_{1}
\end{array}\right),
\\
{V}_{0}=\omega \mu_{l00},\quad {V}_{1}=\omega \mu_{l11},\quad {V}_{2}= \frac{1}{\omega \varepsilon_{l22}},\quad {V}_{3}= \omega \varepsilon_{l11},\quad {V}_{4}= \omega \varepsilon_{l00},\quad {V}_{5}= \frac{1}{\omega \mu_{l22}},
\\
\boldsymbol{\Psi}_{+} \left[ l_{2} \right]  \triangleq 
\left(\begin{array}{c}
\boldsymbol{H}_{2D}\left[ l_{2} \right]\\
\boldsymbol{E}_{2D}\left[ l_{2} \right]
\end{array}\right),\;
\boldsymbol{\Psi}_{-} \left[ l_{2} \right]  \triangleq 
\left(\begin{array}{c}
\boldsymbol{E}_{2D}\left[ l_{2} \right]\\
\boldsymbol{H}_{2D}\left[ l_{2}+1 \right]
\end{array}\right),
\\
\boldsymbol{H}_{2D}\left[ l_{2} \right] = 
\left(\begin{array}{c} H_{l0}\left[ l_{2} \right] \\ H_{l1}\left[ l_{2} \right] \end{array}\right),\;
\boldsymbol{E}_{2D}\left[ l_{2} \right] = 
\left(\begin{array}{c} -E_{l1}\left[ l_{2} \right] \\ E_{l0}\left[ l_{2} \right] \end{array}\right).
\end{gathered}
\right.
\label{eq: discrete m_aa m_bb Psi}
\end{equation}
%
The operators $\bigtriangleup_{j}$ and $\bigtriangledown_{j}$ for $j = 0,1$ in $\boldsymbol{m}_{aa}$ and $\boldsymbol{m}_{bb}$ of eq. (\ref{eq: discrete m_aa m_bb Psi}) can be regarded as $L_{0}L_{1}\times L_{0}L_{1}$ matrices. 
The ${V}_{0},\cdots,{V}_{5}$ also become $L_{0}L_{1}\times L_{0}L_{1}$ diagonal matrices, and then $\boldsymbol{V}$ can be set as a $6L_{0}L_{1}\times 6L_{0}L_{1}$ diagonal matrix.
The $\mu_{ljj}$ and $\varepsilon_{ljj}$ in eq. (\ref{eq: discrete m_aa m_bb Psi}) were defined by eq. (\ref{eq: epsilon_l mu_l}).
The factor $\sqrt{du_{0}du_{1}}/2$ of eq. (\ref{eq: Psi for Maxwell}) is removed from the discrete $\boldsymbol{\Psi}$ in eq. (\ref{eq: discrete m_aa m_bb Psi}), since it does not affect the following discussion.
The $H_{lj}$ and $E_{lj}$ for $j = 0,1$ are $L_{0}L_{1}\times 1$ column vectors.
Note that $\boldsymbol{M}_{\pm} = \boldsymbol{M}_{\pm}^{\dagger}$ when $\varepsilon$ and $\mu$ are real number.

We define two types of power flow for eq. (\ref{eq: discrete propagation-eq}) as similar to the first of eq. (\ref{eq: power flow and the conservation}):
\[
P_{\pm}\left[ l_{2} \right] \triangleq \boldsymbol{\Psi}_{\pm}^{\dagger}\left[ l_{2} \right]\left(\begin{array}{cc}
0 & 1\\
1 & 0
\end{array}\right)\boldsymbol{\Psi}_{\pm}\left[ l_{2} \right] .
\]
From the above definition and eq. (\ref{eq: details power flow}), we show two relations for $ P_{\pm}$ as follows.
\begin{equation}\left\{\begin{aligned}
P_{+}\left[ l_{2} \right] - P_{-}\left[ l_{2}-1\right]
&=
i\boldsymbol{H}_{2D}^{\dagger}\left[ l_{2} \right]\left(\boldsymbol{m}_{aa}\left[ l_{2} \right]-\boldsymbol{m}_{aa}^{\dagger}\left[ l_{2} \right]\right)\boldsymbol{H}_{2D}\left[ l_{2} \right],
\\
P_{-}\left[ l_{2} \right] - P_{+}\left[ l_{2}\right]
&=
i\boldsymbol{E}_{2D}^{\dagger}\left[ l_{2} \right]\left(\boldsymbol{m}_{bb}\left[ l_{2} \right]-\boldsymbol{m}_{bb}^{\dagger}\left[ l_{2} \right]\right)\boldsymbol{E}_{2D}\left[ l_{2} \right] .
\end{aligned}\right.
\label{eq: P_+- relations}
\end{equation}
Note that $P_{+} = P_{-}$ when $\boldsymbol{M}_{+}^{\dagger}=\boldsymbol{M}_{+}$.
Equation (\ref{eq: P_+- relations}) gives us conservation rule of ${\Psi}_{\pm}$ as similar to the second of eq. (\ref{eq: power flow and the conservation}):
\[
\bigtriangleup_{2}P_{+}
=
i\boldsymbol{\Psi}_{-}^{\dagger}\left(\boldsymbol{M}_{-}-\boldsymbol{M}_{-}^{\dagger}\right)\boldsymbol{\Psi}_{-}^{\dagger}
\quad \mathrm{and} \quad
\bigtriangledown_{2}P_{-}
=
i\boldsymbol{\Psi}_{+}^{\dagger}\left(\boldsymbol{M}_{+}-\boldsymbol{M}_{+}^{\dagger}\right)\boldsymbol{\Psi}_{+}^{\dagger}\,.
\]
In the following discussion, we only use the configuration with subscript $+$ for eqs. (\ref{eq: discrete propagation-eq}) and (\ref{eq: discrete m_aa m_bb Psi}), since it can be directly associated with $H_{lj}$ and $E_{lj}$ of the Yee cell in Fig. \ref{fig: Yee cell}.
The subscript  $+$ will be abbreviated as $\boldsymbol{\Psi}$.

From eq. (\ref{eq: derivation for transfer matrix}), we can define transfer matrix $\boldsymbol{T}_{RL}\left[ l_{2} \right]$ as follows.
\begin{equation}
\boldsymbol{\Psi}\left[ l_{2}+1 \right] = \boldsymbol{T}_{RL}\left[ l_{2} \right] \boldsymbol{\Psi}\left[ l_{2} \right]
\quad \mathrm{as} \quad
\boldsymbol{T}_{RL}\left[ l_{2} \right] \triangleq 
\left(\begin{array}{cc}
1 & i\boldsymbol{m}_{bb}\left[ l_{2} \right]\\
i\boldsymbol{m}_{aa}\left[ l_{2}+1 \right] & 1 - \boldsymbol{m}_{aa}\left[ l_{2}+1 \right]\boldsymbol{m}_{bb}\left[ l_{2} \right]
\end{array}\right).
\label{eq: discrete transfer matrix}
\end{equation}

\section{Discrete mode-equation}

This section considers a scattering region and two outer regions.
The scattering region is in $0 \leq l_{2} < L_{2}$, and the outer regions are in $l_{2} < 0$ and $l_{2} \geq L_{2}$, \textit{i.e.} $\left|L_{2}-1-2l_{2}\right| > L_{2}-1$.
The outer region set that the $\boldsymbol{m}_{aa}$ and $\boldsymbol{m}_{bb}$ are not depend on $l_{2}$, and $\boldsymbol{m}_{aa}=\boldsymbol{m}_{aa}^{\dagger}$ and $\boldsymbol{m}_{bb}=\boldsymbol{m}_{bb}^{\dagger}$.
The definitions in eq. (\ref{eq: discrete m_aa m_bb Psi}) derive that
\begin{equation}
\boldsymbol{m}_{aa}^{\dagger} = \boldsymbol{m}_{aa}^{*} = \boldsymbol{m}_{aa}
\quad \mathrm{and} \quad
\boldsymbol{m}_{bb}^{\dagger} = \boldsymbol{m}_{bb}^{*} = \boldsymbol{m}_{bb}\,.
\label{eq: real m_aa and m_bb}
\end{equation}
By using definition of eq. \eqref{eq: propagation-equation M'}, $\boldsymbol{m}^{\left(0\right)}_{aa}$ and $\boldsymbol{m}^{\left(0\right)}_{bb}$ satisfy eq. \eqref{eq: real m_aa and m_bb} even in the scattering region.
Discrete mode-function $\boldsymbol{\Phi}_{m}$ as similar to eq. (\ref{eq: matrix equation}) could be given by solving an eigenvalue equation.
\begin{equation}
\left(\begin{array}{cc}
\boldsymbol{m}^{\left(0\right)}_{aa}\left[ l_{2} \right] & \boldsymbol{0}\\
\boldsymbol{0} & \boldsymbol{m}^{\left(0\right)}_{bb}\left[ l_{2} \right]
\end{array}\right)
\boldsymbol{\Phi}_{m}\left[ l_{2} \right] = 
 2 \sin \left(
 {\theta_{m}\left[ l_{2} \right]}/{2}
 \right)\left(\begin{array}{cc}
0 & 1\\
1 & 0
\end{array}\right)\boldsymbol{\Phi}_{m}\left[ l_{2} \right]
\quad \mathrm{as} \quad
\boldsymbol{\Phi}_{m}\left[ l_{2} \right] \triangleq 
\left(\begin{array}{c}
\boldsymbol{h}_{m}\left[ l_{2} \right] \\
\boldsymbol{e}_{m}\left[ l_{2} \right]
\end{array}\right).
\label{eq: discrete eigenvalue eq}
\end{equation}
Note that $\sin a \triangleq -i\left(e^{ia} - e^{-ia}\right)/2$, and then $\left( \sin a \right)^{*} = i\left(e^{-ia^{*}} - e^{ia^{*}}\right)/2 = \sin a^{*}$.
The cosine and the tangent are also defined in the same manner.
Orthogonality and normalization for the discrete $\boldsymbol{\Phi}_{m}$ are given by
\[
\boldsymbol{\Phi}_{n}^{\dagger}
\left(\begin{array}{cc}
0 & 1\\
1 & 0
\end{array}\right)
\boldsymbol{\Phi}_{m} 
= \boldsymbol{h}_{n}^{\dagger}\boldsymbol{e}_{m} + \boldsymbol{e}_{n}^{\dagger}\boldsymbol{h}_{m}
= \begin{cases}
\frac{m}{\left|m\right|}\frac{\delta_{n\,m}}{\cos\left(\theta_{m}/2\right)} & \mathrm{for}\quad 0 < \left| m \right| \leq n_{\max}\,,
\\
\frac{\delta_{-n\,m}}{\cos\left(\theta_{m}/2\right)} & \mathrm{for}\quad \left| m \right| > n_{\max}\,.
\end{cases}
\]
where the above numbering rule is the same as the rule of Section \ref{sec: basic relation of prop eq}.
Note that the denominator $\cos\left(\theta_{m}/2\right)$ in the normalization is caused by the discretization, and it is different from the case in Section \ref{sec: special mode eq}.
Small $\theta_{m}$ 
is related to the propagation constant $\beta_{m}$ in eq. (\ref{eq: matrix equation}) by referring to Section \ref{sec: u to xi} and eq. \eqref{eq: epsilon_l mu_l}:
\[
\beta_{m} = \lim_{h_{2} {u_{2}^{'}}\rightarrow 0}\frac{2\sin\left({\theta_{m}}/{2}\right)}{h_{2} {u_{2}^{'}}} \simeq \frac{{\theta_{m}}}{h_{2} {u_{2}^{'}}}\,. 
\]
Equation (\ref{eq: discrete eigenvalue eq}) gives us a reduced eigenvalue-equation of $\boldsymbol{h}_{m}$ ( or $\boldsymbol{e}_{m}$ ) and a relation between $\boldsymbol{e}_{m}$ and $\boldsymbol{h}_{m}$:
\begin{equation}
\left\{
\begin{aligned}
\boldsymbol{m}^{\left(0\right)}_{bb}\boldsymbol{m}^{\left(0\right)}_{aa}\boldsymbol{h}_{m} & = 4\sin^{2}\left({\theta_{m}}/{2}\right)\boldsymbol{h}_{m}\,,\\
\boldsymbol{e}_{m} & = \frac{\boldsymbol{m}^{\left(0\right)}_{aa}\boldsymbol{h}_{m}}{2\sin\left({\theta_{m}}/{2}\right)}\,.
\end{aligned}
\right.
\quad\left(
\,\mathrm{or}\quad
\left\{
\begin{aligned}
\boldsymbol{m}^{\left(0\right)}_{aa}\boldsymbol{m}^{\left(0\right)}_{bb}\boldsymbol{e}_{m} & = 4\sin^{2}\left({\theta_{m}}/{2}\right)\boldsymbol{e}_{m}\,,\\
\boldsymbol{h}_{m} & = \frac{\boldsymbol{m}^{\left(0\right)}_{bb}\boldsymbol{e}_{m}}{2\sin\left({\theta_{m}}/{2}\right)}\,.
\end{aligned}
\right.
\quad\right)
\label{eq: h_m eigenvalue eq and relation of e_m and h_m}
\end{equation}
The $\boldsymbol{h}_{m}$ and $\boldsymbol{e}_{m}$ become real vector if $\sin\left({\theta_{m}}/{2}\right)$ is real, since $\boldsymbol{m}_{aa}$ and $\boldsymbol{m}_{bb}$ are real matrix from eq. (\ref{eq: real m_aa and m_bb}).
Furthermore, $\boldsymbol{e}_{n}^{\mathrm T}\boldsymbol{h}_{m} = 0$ if $\sin^{2}\left({\theta_{m}}/{2}\right) \neq \sin^{2}\left({\theta_{n}}/{2}\right)$ for real $\sin\left({\theta_{m}}/{2}\right)$ and $\sin\left({\theta_{n}}/{2}\right)$.
However, we have to be careful in handling eigenvectors (\textit{i.e.} modes) for other case as discussed in Section \ref{sec: special mode eq}.
From the definition in eq. (\ref{eq: discrete eigenvalue eq}) and the relation of $\boldsymbol{h}_{m}$ and $\boldsymbol{e}_{m}$ in eq. (\ref{eq: h_m eigenvalue eq and relation of e_m and h_m}), the orthogonality and normalization of the $\boldsymbol{\Phi}_{m}$ are reduced to
\begin{equation}
 \begin{cases}
 \boldsymbol{h}_{n}^{\mathrm T}\boldsymbol{e}_{m} = \boldsymbol{e}_{n}^{\mathrm T}\boldsymbol{h}_{m} =
\frac{m\delta_{n\,m}}{2\left|m\right|\cos\left(\theta_{m}/2\right)} & \mathrm{for}\quad 0 < \left| m \right| \leq n_{\max}\,,
\\
\boldsymbol{h}_{n}^{\dagger}\boldsymbol{e}_{m} = \boldsymbol{e}_{n}^{\dagger}\boldsymbol{h}_{m} =
\frac{\delta_{-n\,m}}{2\cos\left(\theta_{m}/2\right)} & \mathrm{for}\quad\left| m \right| > n_{\max}\,.
\end{cases}
\label{eq: h_n e_m orthogonality}
\end{equation}
Furthermore, we can add special relations
\begin{equation}
\boldsymbol{h}_{-m} = \boldsymbol{h}_{m}
\quad \mathrm{and} \quad
\boldsymbol{e}_{-m} = -\boldsymbol{e}_{m}
\quad \mathrm{for} \quad
0 < \left| m \right| \leq n_{\max}
\label{eq: h_m e_m special relations}
\end{equation}
to eq. (\ref{eq: h_n e_m orthogonality}) without loss of generality.
Discrete electromagnetic field $\boldsymbol{\Psi}_{+}\left[ l_{2} \right]$ in eq. (\ref{eq: discrete m_aa m_bb Psi}) for $\left| 2l_{2}+1-L_{2}\right|>L_{2} $ can be expanded into series of modified mode function $\boldsymbol{\Xi}_{m}$:
\begin{equation}
\boldsymbol{\Psi}_{+}\left[ l_{2} \right] = \sum_{m\neq 0} c_{m} \exp\left(i\theta_{m}l_{2}\right)\boldsymbol{\Xi}_{m}
\quad \mathrm{as} \quad
\boldsymbol{\Xi}_{m} \triangleq
\left(\begin{array}{cc}
\exp\left(-i\theta_{m}/2\right) & 0\\
0 & 1
\end{array}\right)
\left(\begin{array}{c}
\boldsymbol{h}_{m} \\
\boldsymbol{e}_{m}
\end{array}\right).
\label{eq: Xi definition}
\end{equation}
By checking eq. (\ref{eq: discrete propagation eq by Xi}), we can confirm that the above expansion has consistency with the discrete propagation equation (\ref{eq: discrete propagation-eq}).
The $\boldsymbol{\Xi}_{m}$ maintains the orthogonality as
\begin{equation}
\boldsymbol{\Xi}_{n}^{\dagger}
\left(\begin{array}{cc}
0 & 1\\
1 & 0
\end{array}\right)
\boldsymbol{\Xi}_{m} 
= 
\boldsymbol{h}_{n}^{\dagger} \exp\left(i\frac{\theta_{n}^{*}}{2}\right)\boldsymbol{e}_{m}
+\boldsymbol{e}_{n}^{\dagger} \exp\left(-i\frac{\theta_{m}}{2}\right)\boldsymbol{h}_{m}
=
\begin{cases}
\frac{m}{\left|m\right|}\delta_{n\,m} & \mathrm{for}\quad 0 < \left| m \right| \leq n_{\max}\,,
\\
\delta_{-n\,m} & \mathrm{for}\quad \left| m \right| > n_{\max}\,.
\end{cases}
\label{eq: Xi orthogoanlity}
\end{equation}
The power flow $P_{+}$ is represented by $ c_{m} $ from eqs. (\ref{eq: Xi definition}) and (\ref{eq: Xi orthogoanlity}):
\[
P_{+}\left[ l_{2} \right] =
\boldsymbol{\Psi}_{+}^{\dagger} \left[ l_{2} \right]
\left(\begin{array}{cc}
0 & 1\\
1 & 0
\end{array}\right)
\boldsymbol{\Psi}_{+} \left[ l_{2} \right] = 
\sum_{0 < \left| m \right| \leq n_{\max}} \frac{m}{\left| m \right|}\left| c_{m} \right|^{2}
+ \sum_{\left| m \right| > n_{\max}} c_{-m}^{*} c_{m}\,.
\]

\section{Discrete scattering-matrix}

Scattering matrix of eq. (\ref{eq: S-matrix}) can be discretized by the discrete propagation-equation (\ref{eq: discrete propagation-eq}) and the modified mode function (\ref{eq: Xi definition}).
\begin{equation}
S_{mn} \triangleq \exp\left(-i \Theta_{n}\left[ b_{m} \right] \right) \boldsymbol{\Xi}_{m}^{\dagger}\left[ b_{m} \right]\frac{m}{\left| m \right|}
\left(\begin{array}{cc}
0&1 \\
1&0
\end{array}\right)
\boldsymbol{\Psi}_{n}\left[ b_{m} \right]
\label{eq: discrete S-matrix}
\end{equation}
for $0 < \left| m \right| \leq n_{\max}$ and $0 < \left| n \right| \leq n_{\max}$.
The $b_{m}$ in the right-hand side of eq. (\ref{eq: discrete S-matrix}) is redefined as
\[
b_{m} = \frac{\left| m \right| + m}{2\left| m \right|} \left( L_{2} - 1 \right)
= 
\begin{cases}
L_{2} - 1 & \mathrm{for}\quad m > 0\,,
\\
0 & \mathrm{for}\quad m < 0\,,
\end{cases}
\]
and $\Theta_{n}\left[ b_{m} \right]$ is given by eq. (\ref{eq: Theta(l_2)}):
\begin{equation}
\begin{split}
\frac{m}{\left| m \right|} \Theta_{n}\left[ b_{m} \right] & \triangleq 
\sum_{\frac{b_{m}}{2} \leq l_{2} < \frac{L_{2}-1+b_{m}}{2}}\frac{\theta_{n}\left[ l_{2} \right]}{2} + \sum_{\frac{b_{m}}{2} < l_{2}\leq \frac{L_{2}-1+b_{m}}{2}}\frac{\theta_{n}\left[ l_{2} \right]}{2}
\\
& =
\begin{cases}
\frac{\theta_{n}\left[\frac{L_{2}-1}{2}\right]}{2} + \sum_{\frac{L_{2}-1}{2}<l_{2}<L_{2}-1}\theta_{n}\left[ l_{2} \right] +\frac{\theta_{n}\left[L_{2}-1\right]}{2}  & \mathrm{for} \quad m > 0 \quad \mathrm{and} \quad L_{2}=\mathrm{odd}\,,
\\
\qquad \qquad \quad \sum_{\frac{L_{2}-1}{2}<l_{2}<L_{2}-1}\theta_{n}\left[ l_{2} \right] + \frac{\theta_{n}\left[L_{2}-1\right]}{2} & \mathrm{for} \quad m > 0 \quad \mathrm{and} \quad L_{2}=\mathrm{even}\,,
\\
\frac{\theta_{n}\left[0\right]}{2} + \sum_{0<l_{2}<\frac{L_{2}-1}{2}}\theta_{n}\left[ l_{2} \right] + \frac{\theta_{n}\left[\frac{L_{2}-1}{2}\right]}{2} & \mathrm{for} \quad m < 0 \quad \mathrm{and} \quad L_{2}=\mathrm{odd}\,,
\\
\frac{\theta_{n}\left[0\right]}{2} + \sum_{0<l_{2}<\frac{L_{2}-1}{2}}\theta_{n}\left[ l_{2} \right] & \mathrm{for} \quad m < 0 \quad \mathrm{and} \quad L_{2}=\mathrm{even}\,.
\end{cases}
\end{split}
\label{eq: Theta(b_m)}
\end{equation}
The $\boldsymbol{\Psi}_{n}$ in eq. (\ref{eq: discrete S-matrix}) is redefined as a particular solution of $\boldsymbol{\Psi}_{+}$ in eq. (\ref{eq: discrete m_aa m_bb Psi}), which satisfies boundary conditions at outer regions:
\begin{equation}
\boldsymbol{\Psi}_{n} \left[ b_{m} \right] = \exp\left(i \Theta_{n}\left[ b_{m} \right] \right)\left(\frac{\left| mn \right| - mn}{2\left| mn \right|}\boldsymbol{\Xi}_{n}\left[ b_{m} \right]
+ \sum_{ 0 < \frac{m^{'}m}{\left| m \right|} \leq n_{\max}} S_{m^{'}n} \boldsymbol{\Xi}_{m^{'}}\left[ b_{m} \right]\right).
\label{eq: discrete Psi b_m}
\end{equation}
The following discussion is based on the framework of Section \ref{sec: discrete Green function}.
Equation (\ref{eq: discrete Green func}) at $l_{2}= b_{m}$ shows that
\[
\begin{split}
\boldsymbol{\Psi}_{n}\left[ b_{m} \right]  &= \exp\left( i\Theta_{n}\left[ b_{m} \right] \right)\left( \boldsymbol{\Xi}_{n}\left[ b_{m} \right] + \sum_{1 \leq j < L_{2}-1} \boldsymbol{G}\left[b_{m},j \right]\right),
\\
\boldsymbol{G}\left[b_{m},j \right] & = \sum_{ 0 < \frac{m^{'}m}{\left| m \right|} \leq n_{\max}} g_{m^{'} n}\left[j \right]
\exp\left\{ i\left(\Theta_{m^{'}}\left[ b_{m} \right]-\Theta_{n}\left[ b_{m} \right]\right) -i\left(\Theta_{m^{'}}\left[ j \right] - \Theta_{n}\left[ j \right]\right)\right\}
\boldsymbol{\Xi}_{m^{'}}\left[ b_{m} \right]\,.
\end{split}
\]
By comparing the above equations and eq. (\ref{eq: discrete Psi b_m}), the $S_{mn}$ is related to the $g_{m n}$:
\begin{equation}
S_{mn} - \delta_{mn} = \sum_{1 \leq j < L_{2}-1} g_{mn}\left[j \right]
\exp\left\{ i\left(\Theta_{m}\left[ b_{m} \right]-\Theta_{n}\left[ b_{m} \right]\right) -i\left(\Theta_{m}\left[j \right]-\Theta_{n}\left[j \right]\right)\right\}\,.
\label{eq: discrete S_mn and g_mn relation}
\end{equation}
When ${\boldsymbol{M}}_{+}^{\left(0\right)}$ in eq. (\ref{eq: discrete M^0 M^1 Psi_n}) is independent of $l_{2}$, the $S_{mn}$ of eq.(\ref{eq: discrete S_mn and g_mn relation}) is simplified to
\[
S_{mn} - \delta_{mn} = \sum_{0 \leq j < L_{2}} g_{mn}\left[j \right]
\exp\left\{ i\left(\theta_{m}\left[b_{m} - j \right]-\theta_{n}\left[b_{m} - j \right]\right) \right\}
 = \exp\left( i\frac{m}{\left|m\right|}\left(\theta_{m}-\theta_{n}\right)\frac{L_{2}-1}{2} \right) \widehat{g_{mn}}_{\mathrm{DFT}}\left(\theta_{m} -\theta_{n} \right),
\]
where $g_{mn}\left[0 \right]=g_{mn}\left[L_{2}-1 \right]=0$, and ``$\,\widehat{\quad}_{\mathrm{DFT}}$'' is defined by eq. (\ref{eq: DFT definition}). 
From detail of the $\widehat{g_{mn}}_{\mathrm{DFT}}$ in eq. (\ref{eq: DFT g_mn when M^(0) is independent of l_2}), the above equation is represented as
\begin{equation}
\begin{split}
\left| S_{mn} - \delta_{mn} \right|^{2} &= \left| \widehat{g_{mn}}_{\mathrm{DFT}}\left(\theta_{m} -\theta_{n} \right) \right|^{2} = \left| \boldsymbol{\Xi}_{m}^{\dagger}
\left(\begin{array}{cc}
 e^{i\theta_{m}}\widehat{\boldsymbol{m}_{aa}^{\left(1\right)}}_{\mathrm{DFT}}\left(\theta_{m} -\theta_{n} \right) & \boldsymbol{0} \\
\boldsymbol{0} &  e^{-i\theta_{m}}\widehat{\boldsymbol{m}_{bb}^{\left(1\right)}}_{\mathrm{DFT}}\left(\theta_{m} -\theta_{n} \right)
\end{array}\right)\boldsymbol{\Xi}_{n} \right|^{2}
\\
&=\left|
 e^{i\theta_{m}}e^{i\left(\theta_{m} -\theta_{n} \right)/2}\boldsymbol{h}_{m}^{\mathrm {T}}\left[\widehat{\boldsymbol{m}_{aa}^{\left(1\right)}}_{\mathrm{DFT}}\left(\theta_{m} -\theta_{n} \right)\right]\boldsymbol{h}_{n}
+
 e^{-i\theta_{m}}\boldsymbol{e}_{m}^{\mathrm {T}}\left[\widehat{\boldsymbol{m}_{bb}^{\left(1\right)}}_{\mathrm{DFT}}\left(\theta_{m} -\theta_{n} \right)\right]\boldsymbol{e}_{n}
\right|^{2}.
\end{split}
\label{eq: discrete S_mn and DFT g_mn}
\end{equation}

\section{Discrete edge-roughness scattering of Section \ref{sec: edge-roughness for optical WG}}
This section shows $\widehat{\boldsymbol{m}_{aa}^{\left(1\right)}}_{\mathrm{DFT}}$ and $\widehat{\boldsymbol{m}_{bb}^{\left(1\right)}}_{\mathrm{DFT}}$ in eq. (\ref{eq: discrete S_mn and DFT g_mn}) by separating unperturbed part $\boldsymbol{V}^{\left(0\right)}$ and perturbed part $\boldsymbol{V}^{\left(1\right)}$ from the discrete $\boldsymbol{V}$ in eq. (\ref{eq: discrete m_aa m_bb Psi}).
Elements of $\boldsymbol{V}^{\left(0\right)}$ are given by
\[
{V}_{0}^{\left(0\right)}=\omega \mu_{l00},\quad {V}_{1}^{\left(0\right)}=\omega \mu_{l11},\quad {V}_{2}^{\left(0\right)}= \frac{1}{\omega \varepsilon_{l22}},\quad {V}_{3}^{\left(0\right)}= \omega \varepsilon_{l11},\quad {V}_{4}^{\left(0\right)}= \omega \varepsilon_{l00},\quad {V}_{5}^{\left(0\right)}= \frac{1}{\omega \mu_{l22}},
\]
where $\boldsymbol{V}^{\left(0\right)}$ is not depend on $u_{2}$.
The DFT of $\boldsymbol{V}^{\left(1\right)}$ can be approximated to the FT with care of phase shift caused by half shift in eq. (\ref{eq: epsilon_l mu_l}).
\[
\widehat{{V}^{\left(1\right)}_{j+3J}}_{\mathrm{DFT}}\left( \phi \right)
\simeq \frac{e^{i\phi J/2}}{u^{'}_{2}}\widehat{{V}^{\left(1\right)}_{j+3J}}\left(\frac{\phi}{u^{'}_{2}}\right)
\quad \mathrm{for} \quad
\left\{ \begin{aligned}
j&=0,1,2,\\J&=0,1,
\end{aligned} \right.
\]
when $u^{'}_{2}/L_{\mathrm{s}}$ is small, \textit{i.e.} $L_{2} \gg 1$ from eq. (\ref{eq: DFT FT relations}).
Then,
\[
\widehat{\boldsymbol{m}_{aa}^{\left(1\right)}}_{\mathrm{DFT}}\left(\phi\right)
= \frac{\sqrt{L_{\mathrm{s}}}}{u^{'}_{2}} \boldsymbol{m}_{aa}\left(
 \frac{1}{\sqrt{L_{\mathrm{s}}}}\widehat{\boldsymbol{V}^{\left(1\right)}}\left(\frac{\phi}{u^{'}_{2}}\right)\right),
\quad
\widehat{\boldsymbol{m}_{bb}^{\left(1\right)}}_{\mathrm{DFT}}\left(\phi\right)
= \frac{e^{i\phi/2}\sqrt{L_{\mathrm{s}}}}{u^{'}_{2}} \boldsymbol{m}_{bb}\left( 
\frac{1}{\sqrt{L_{\mathrm{s}}}}\widehat{\boldsymbol{V}^{\left(1\right)}}\left(\frac{\phi}{u^{'}_{2}}\right)\right).
\]
Note that $\boldsymbol{m}_{aa}$ ($\boldsymbol{m}_{bb}$) is a function of only $V_{0}$, $V_{1}$ and $V_{2}$ ($V_{3}$, $V_{4}$ and $V_{5}$).
Equation (\ref{eq: discrete S_mn and DFT g_mn}) is represented as
\begin{equation}
\left\{ \begin{aligned}
\left| S_{mn} - \delta_{mn} \right|^{2} &=\frac{L_{\mathrm{s}}}{{u^{'}_{2}}^{2}}\left|
 e^{i\theta_{m}}\boldsymbol{h}_{m}^{\mathrm {T}}\left[\boldsymbol{m}_{aa}\left( \frac{1}{\sqrt{L_{\mathrm{s}}}}\widehat{\boldsymbol{V}^{\left(1\right)}}\left(k_{mn}\right)\right)
 \right]\boldsymbol{h}_{n}
+ 
 e^{-i\theta_{m}}\boldsymbol{e}_{m}^{\mathrm {T}}\left[\boldsymbol{m}_{bb}\left( \frac{1}{\sqrt{L_{\mathrm{s}}}}\widehat{\boldsymbol{V}^{\left(1\right)}}\left(k_{mn}\right)\right)
 \right]\boldsymbol{e}_{n}
\right|^{2},
\\
k_{mn} &= \frac{\theta_{m} -\theta_{n}}{u^{'}_{2}}.
\end{aligned} \right.
\label{eq: discrete S_mn and FT V1}
\end{equation}
We can estimate roughness scattering by using eq. (\ref{eq: discrete S_mn and FT V1}) with the following eq. (\ref{eq: Approach I for Yee lattice}) or (\ref{eq: Approach II for Yee lattice}).
Therefore, we should crosscheck numerical results against another approach.

\subsection{Discrete representation of Approach I for \ref{subsec: optical approach I}}

This subsection considers that the $\boldsymbol{V}$ is a function of $A_{\mathrm{w}}$ and $A_{\mathrm{c}}$ as shown in Fig. \ref{fig: LWR and LCR}.
Here, we use eq. (\ref{eq: epsilon_l mu_l}) for $h_{0}=h_{2}=1$ and constant $u^{'}_{2}$.
By using the framework of eq. (\ref{eq: approach1 V1}), elements of $\boldsymbol{V}^{\left(1\right)}\left(A_{\mathrm{w}},A_{\mathrm{c}}\right)
$ are given as
\begin{equation}
\begin{split}
{V}_{j+3J}^{\left(1\right)}\left(A_{\mathrm{w}},A_{\mathrm{c}}\right)  &= \frac{{V}_{j+3J}\left(\frac{\varDelta A_{\mathrm{w}}}{2},0\right)-{V}_{j+3J}\left(-\frac{\varDelta A_{\mathrm{w}}}{2},0\right)}{\varDelta A_{\mathrm{w}}}A_{\mathrm{w}}\left(u_{2}\left(l_{2}+J/2\right)\right)
\\
  &\quad + \frac{{V}_{j+3J}\left(0,\frac{\varDelta A_{\mathrm{c}}}{2}\right)-{V}_{j+3J}\left(0,-\frac{\varDelta A_{\mathrm{c}}}{2}\right)}{\varDelta A_{\mathrm{c}}}A_{\mathrm{c}}\left(u_{2}\left(l_{2}+J/2\right)\right) 
\end{split}
\quad \mathrm{for} \quad
\left\{ \begin{aligned}
j&=0,1,2,\\J&=0,1,
\end{aligned} \right.
\label{eq: Discrete optical approach1  V0V1}
\end{equation}
where $\varDelta A_{\mathrm{w}}$ and $\varDelta A_{\mathrm{c}}$ are parameters of finite difference, and we usually set $u^{'}_{0}\lesssim \varDelta A_{\mathrm{w}}/2=\varDelta A_{\mathrm{c}} \lesssim 2u^{'}_{0}$.
By using eq. (\ref{eq: Discrete optical approach1  V0V1}), the $\widehat{\boldsymbol{V}^{\left(1\right)}}/\sqrt{L_{\mathrm{s}}}$ in eq. (\ref{eq: discrete S_mn and FT V1}) can be given as
\begin{equation}
\frac{1}{\sqrt{L_{\mathrm{s}}}}\widehat{\boldsymbol{V}^{\left(1\right)}}\left(k_{mn}\right) 
= \frac{\boldsymbol{V}\left(\frac{\varDelta A_{\mathrm{w}}}{2},0\right)-\boldsymbol{V}\left(-\frac{\varDelta A_{\mathrm{w}}}{2},0\right)}{\varDelta A_{\mathrm{w}}}\widehat{a_{\mathrm{w}}}\left(k_{mn}\right)
+ \frac{\boldsymbol{V}\left(0,\frac{\varDelta A_{\mathrm{c}}}{2}\right)-\boldsymbol{V}\left(0,-\frac{\varDelta A_{\mathrm{c}}}{2}\right)}{\varDelta A_{\mathrm{c}}}\widehat{a_{\mathrm{c}}}\left(k_{mn}\right)\, .
\label{eq: Approach I for Yee lattice}
\end{equation}

\subsection{Discrete representation of Approach II for \ref{subsec: optical approach II}}

This subsection uses the parameter $s_{j+3J}\left(u_{0},u_{2}\right)$ of eq. (\ref{eq: V^1 for opitcal roughness 2}) as
\begin{equation}
u_{0}\left( l_{0} + \frac{\left[\left(j+1\right)\%3\right]\%2 + \left\{\left[\left(j+2\right)\%3\right]\%2\right\}\left(1-J\right)}{2}\right)
,\quad
u_{2}\left(l_{2}+\frac{J}{2}\right) = \left[l_{2}+\frac{J}{2}-\frac{L_{2}}{2}\right]u^{'}_{2}
\quad \mathrm{for} \quad
\left\{ \begin{aligned}
j&=0,1,2,\\J&=0,1.
\end{aligned} \right.
\label{eq: u0 u2 Yee lattice}
\end{equation}
By using eq. (\ref{eq: opitcal roughness 2}) and the $u_{0}$ in eq. (\ref{eq: u0 u2 Yee lattice}), elements of the $\widehat{\boldsymbol{V}^{\left(1\right)}}/\sqrt{L_{\mathrm{s}}}$ in eq. (\ref{eq: discrete S_mn and FT V1}) can be given as
\begin{equation}
\frac{1}{\sqrt{L_{\mathrm{s}}}}\widehat{{V}^{\left(1\right)}_{j+3J}}\left(k_{mn}\right) 
= {V}^{\left(0\right)}_{j+3J}
\left[
 u_{0} k_{mn}^{2} \widehat{a_{\mathrm{c}}}\left(k_{mn}\right)
 + \frac{1}{W_{\mathrm{wg}}} \left( \frac{1}{2} u_{0}^{2} k_{mn}^{2} - (-1)^{\left[\left(j+J\right)\%3\right]\%2}  \right) \widehat{a_{\mathrm{w}}}\left(k_{mn}\right)
 \right] 
 \quad \mathrm{for} \quad
\left\{ \begin{aligned}
j&=0,1,2,\\J&=0,1.
\end{aligned} \right.
\label{eq: Approach II for Yee lattice}
\end{equation}


%% file: Chap/8_FDTD_v2.tex
\chapter{Finite Difference Time Domain\label{ch: FDTD}}
This chapter shows a way of Finite Difference Time Domain (FDTD). 
From eq. \eqref{eq: discrete Maxwell eq}, we consider Maxwell equation in time domain, which has already been discretized for the 3D-space:
\[
\left(\begin{array}{cc}
0 & i\boldsymbol{R}\\
-i\boldsymbol{R}^{\dagger} & 0
\end{array}\right)
\left(\begin{array}{c}
\boldsymbol{E}_{l}\\
\boldsymbol{H}_{l}
\end{array}\right)
= i\frac{\partial}{\partial t}\left(\begin{array}{cc}
{\boldsymbol{\varepsilon}}_{l} & 0\\
0 & {\boldsymbol{\mu}}_{l}
\end{array}\right)
\left(\begin{array}{c}
\boldsymbol{E}_{l}\\
\boldsymbol{H}_{l}
\end{array}\right)\, ,
\]
where the discrete rotation operators $ \boldsymbol{R} $ and $ \boldsymbol{R}^{\dagger} $ are defined by using eq. \eqref{eq: operator R}.
In order to consider $\boldsymbol{E}_{l},\,\boldsymbol{H}_{l}\in \Re$, the above equation can be approximated to
\begin{equation}
\left(\begin{array}{c}
\boldsymbol{R} \boldsymbol{H}_{l}\\
-\boldsymbol{R}^{\dagger}\boldsymbol{E}_{l}
\end{array}\right)
\simeq \frac{\partial}{\partial t}
\left(\begin{array}{c}
\left({\mathrm{Re}\,\boldsymbol{\varepsilon}_{l}}\right)\boldsymbol{E}_{l}\\
\left({\mathrm{Re}\,\boldsymbol{\mu}_{l}}\right)\boldsymbol{H}_{l}
\end{array}\right){}
+\omega_{c}
\left(\begin{array}{c}
\left({\mathrm{Im}\,\boldsymbol{\varepsilon}_{l}}\right)\boldsymbol{E}_{l}\\
\left({\mathrm{Im}\,\boldsymbol{\mu}_{l}}\right)\boldsymbol{H}_{l}
\end{array}\right)
\label{eq: Time-Maxwell eq}
\end{equation}
with introducing the center frequency $\omega_{c}$.

\section{Discretization for time domain}

\begin{figure}[h]
\begin{centering}
\includegraphics[width=0.6\columnwidth]{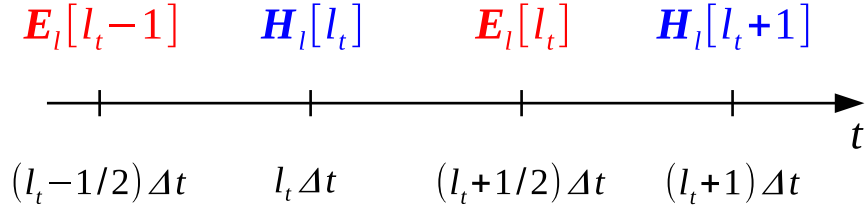}
\par\end{centering}
\caption{We consider discretized cells of $t$. \label{fig: FDTD}}
\end{figure}

Figure \ref{fig: FDTD} shows the discretized $\boldsymbol{E}_{l}\left[\boldsymbol{l},l_{t}\right]$ and $\boldsymbol{H}_{l}\left[\boldsymbol{l},l_{t}\right]$ for time $t$ using the same rule as shown in Fig. \ref{fig: Yee cell}.
We simplify the cell address $\left[\boldsymbol{l},l_{t}\right]$ into $\left[l_{t}\right]$ or $\left[l_{2}, l_{t}\right]$ in the following equations.
Time step $\Delta t$ discretizes the $t$ to time cells.

Equation \eqref{eq: Time-Maxwell eq} is discretized as follows.
\begin{equation}
\begin{split}
\left(\frac{\mathrm{Re}\,\boldsymbol{\varepsilon}_{l}\left[l_{t}\right]}{\tau_{R}} + \frac{\mathrm{Im}\,\boldsymbol{\varepsilon}_{l}\left[l_{t}\right]}{\tau_{I}}\right)\boldsymbol{E}_{l}\left[l_{t}\right]
&= \boldsymbol{R} \boldsymbol{H}_{l}\left[l_{t}\right] +
\left(\frac{\mathrm{Re}\,\boldsymbol{\varepsilon}_{l}\left[l_{t}-1\right]}{\tau_{R}} - \frac{\mathrm{Im}\,\boldsymbol{\varepsilon}_{l}\left[l_{t}-1\right]}{\tau_{I}}\right)\boldsymbol{E}_{l}\left[l_{t}-1\right]
\, ,\\
\left(\frac{\mathrm{Re}\,\boldsymbol{\mu}_{l}\left[l_{t}+1\right]}{\tau_{R}} + \frac{\mathrm{Im}\,\boldsymbol{\mu}_{l}\left[l_{t}+1\right]}{\tau_{I}}\right)\boldsymbol{H}_{l}\left[l_{t}+1\right]
&= -\boldsymbol{R}^{\dagger}\boldsymbol{E}_{l}\left[l_{t}\right] + 
\left(\frac{\mathrm{Re}\,\boldsymbol{\mu}_{l}\left[l_{t}\right]}{\tau_{R}} - \frac{\mathrm{Im}\,\boldsymbol{\mu}_{l}\left[l_{t}\right]}{\tau_{I}}\right)\boldsymbol{H}_{l}\left[l_{t}\right]
\, ,
\end{split}
\label{eq: Disc Time-Maxwell eq}
\end{equation}
where $\tau_{R}^{-1}$ and $\tau_{I}^{-1}$ are modified from ${\Delta t}^{-1}$ and ${\omega_{c}}/{2}$ for correcting discretization errors:
\[ 
\frac{1}{\tau_{R}}=\frac{1}{\Delta t}\frac{\omega_{c} \Delta t /2}{\sin\left(\omega_{c} \Delta t /2\right)}
\quad \mathrm{and} \quad 
\frac{1}{\tau_{I}}=\frac{\omega_{c}}{2}\frac{1}{\cos\left(\omega_{c} \Delta t /2\right)}\, .
\]

The time step $\Delta t$ has to satisfy the Courant-Friedrichs-Lewy (CFL) condition, and eqs. \eqref{eq: Dx and kr} and \eqref{eq: epsilon_l mu_l} without $\tilde{\sigma}_{j}$ are applied to the CFL condition:
\begin{equation}
\tau_{R}^{2} < \left(\Delta t\right)^{2} \leq \min_{\boldsymbol{\xi}}
{\frac{\varepsilon\left(\boldsymbol{\xi}\right)\,\mu\left(\boldsymbol{\xi}\right)}{\sum_{j=0}^{2}\left(u_{j}^{'}\left({\xi}_{j}\right)\, h_{j}\left(\boldsymbol{u}\left(\boldsymbol{\xi}\right)\right)\right)^{-2}}}
\simeq
\sqrt{
\min_{\boldsymbol{l}}
\frac{
	{\prod_{j=0}^{2} {\varepsilon}_{ljj}\left[\boldsymbol{l}\right]}\,
	{ {\mu}_{ljj}\left[\boldsymbol{l}\right]}
}{
	\left({\sum_{j=0}^{2} {\varepsilon}_{ljj}\left[\boldsymbol{l}\right]}\right)
	\left({\sum_{j=0}^{2} {\mu}_{ljj}\left[\boldsymbol{l}\right]}\right)
}}
\, . 
\label{eq: CFL condition}
\end{equation}

\section{Mode source}

Equations \eqref{eq: discrete m_aa m_bb Psi} and \eqref{eq: Xi definition} give us the $m$-th mode fields in frequency domain:
\[{}
\left(
	\begin{array}{cccc}
		H_{l0}^{\left( m \right)}\left( \omega_{c} \right) & H_{l1}^{\left( m \right)}\left( \omega_{c} \right) & -E_{l1}^{\left( m \right)}\left( \omega_{c} \right) & E_{l0}^{\left( m \right)}\left( \omega_{c} \right)
	\end{array}
\right)^{\mathrm{T}}
= \boldsymbol{\Xi}_{m}\left( \omega_{c} \right)\, .
\]
The $m$-th mode fields in time domain can be defined by using the above mode fields.
\begin{equation}
\left\{ 
\begin{split}
H_{lj}^{\left( m \right)}\left[ l_{t}\right]  &\triangleq f_{\mathrm{env}}\left(l_{t} \Delta t\right)\mathrm{Re}\left(e^{-i\omega_{c}l_{t} \Delta t}
		H_{lj}^{\left( m \right)}\left( \omega_{c} \right)
\right),\\
{E}_{lj}^{\left( m \right)}\left[ l_{t}\right]  &\triangleq f_{\mathrm{env}}\left(\left(l_{t}+
{1}/{2}\right) \Delta t\right)\mathrm{Re}\left(e^{-i\omega_{c}\left(l_{t}+{1}/{2}\right)\Delta t}
		E_{lj}^{\left( m \right)}\left( \omega_{c} \right)
\right),
\end{split}
\right.
\quad \mathrm{as} \quad j= 0,1\, ,
\label{eq: mode in time domain}
\end{equation}
where $f_{\mathrm{env}}\left(t\right)$ is non-negative function, and $f_{\mathrm{env}}\left(0\right)=0$ and $\left| \mathrm{d} f_{\mathrm{env}}/\mathrm{d}t\right|\ll\omega_{c}$.

\begin{figure}[h]
\begin{centering}
\includegraphics[width=0.4\columnwidth]{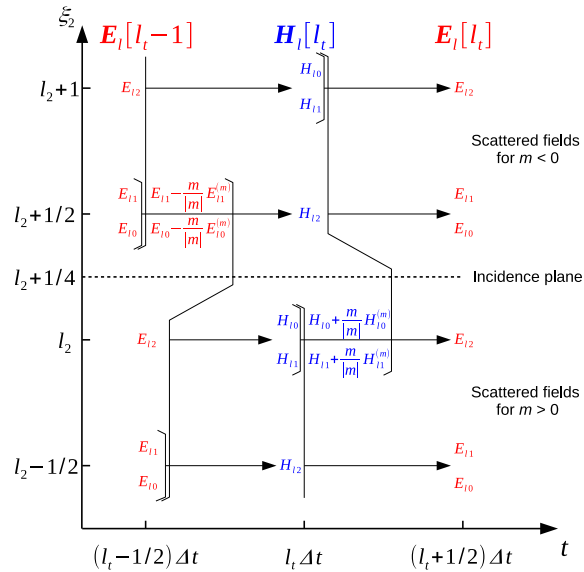}
\par\end{centering}
\caption{Calculation manner of eqs. \eqref{eq: Disc Time-Maxwell eq} and \eqref{eq: incidence plane} near the incidence plane for the $m$-th mode.
See also Fig. 2(b) in \cite{Umashankar}.
 \label{fig: mode source}}
\end{figure}

Let us induce the $m$-th mode fields of eq. \eqref{eq: mode in time domain} from an incidence plane at $\xi_{2} = l_{2} + 1/4$ into the discretized $\boldsymbol{H}_{l}$ and $\boldsymbol{E}_{l}$.
The first terms in the right hand sides of eqs. \eqref{eq: Disc Time-Maxwell eq}, which are $\boldsymbol{R} \boldsymbol{H}_{l}\left[l_{2},l_{t}\right]$ and $\boldsymbol{R}^{\dagger}\boldsymbol{E}_{l}\left[l_{2},l_{t}\right]$, are modified into
\begin{equation}
\begin{split}
\boldsymbol{R} \boldsymbol{H}_{l}\left[l_{2},l_{t}\right] &= \left(\begin{array}{ccc}
0 & -\bigtriangleup_{2} & \bigtriangleup_{1}\\
\bigtriangleup_{2} & 0 & -\bigtriangleup_{0}\\
-\bigtriangleup_{1} & \bigtriangleup_{0} & 0
\end{array}\right)
\left(\begin{array}{c}
H_{l0}\left[l_{2},l_{t}\right]\\H_{l1}\left[l_{2},l_{t}\right]\\H_{l2}\left[l_{2},l_{t}\right]
\end{array}\right)
\\ &
\Rightarrow
\left(\begin{array}{c}
-H_{l1}\left[l_{2}+1,l_{t}\right] + H_{l1}\left[l_{2},l_{t}\right] +\frac{m}{\left| m \right|}H_{l1}^{\left( m \right)}\left[l_{t}\right] + \bigtriangleup_{1}H_{l2}\left[l_{2},l_{t}\right]\\
 H_{l0}\left[l_{2}+1,l_{t}\right] - H_{l0}\left[l_{2},l_{t}\right] -\frac{m}{\left| m \right|}H_{l0}^{\left( m \right)}\left[l_{t}\right] - \bigtriangleup_{0}H_{l2}\left[l_{2},l_{t}\right]\\
-\bigtriangleup_{1}H_{l0}\left[l_{2},l_{t}\right]+\bigtriangleup_{0}H_{l1}\left[l_{2},l_{t}\right]
\end{array}\right)\, ,
\\
\boldsymbol{R}^{\dagger}\boldsymbol{E}_{l}\left[l_{2},l_{t}\right] &= \left(\begin{array}{ccc}
0 & -\bigtriangledown_{2} & \bigtriangledown_{1}\\
\bigtriangledown_{2} & 0 & -\bigtriangledown_{0}\\
-\bigtriangledown_{1} & \bigtriangledown_{0} & 0
\end{array}\right)
\left(\begin{array}{c}
E_{l0}\left[l_{2},l_{t}\right]\\E_{l1}\left[l_{2},l_{t}\right]\\E_{l2}\left[l_{2},l_{t}\right]
\end{array}\right)
\\ &
\Rightarrow
\left(\begin{array}{c}
-E_{l1}\left[l_{2},l_{t}\right] +\frac{m}{\left| m \right|}E_{l1}^{\left( m \right)}\left[l_{t}\right] + E_{l1}\left[l_{2}-1,l_{t}\right] + \bigtriangledown_{1}E_{l2}\left[l_{2},l_{t}\right]\\
 E_{l0}\left[l_{2},l_{t}\right] -\frac{m}{\left| m \right|}E_{l0}^{\left( m \right)}\left[l_{t}\right] - E_{l0}\left[l_{2}-1,l_{t}\right] - \bigtriangledown_{0}E_{l2}\left[l_{2},l_{t}\right]\\
-\bigtriangledown_{1}E_{l0}+\bigtriangledown_{0}E_{l1}\left[l_{2},l_{t}\right]
\end{array}\right)\, .
\end{split}
\label{eq: incidence plane}
\end{equation}
Figure \ref{fig: mode source} shows calculation manner of eq. \eqref{eq: incidence plane} with the discretized $\boldsymbol{H}_{l}$ and $\boldsymbol{E}_{l}$ near the incidence plane.

%% file: App/A_GeneralWaveguides_v7.tex
\chapter{Two steps of coordinate transformation\label{ch: Details of two steps}}
This chapter shows details of the transformation in Chapter \ref{ch: GeneralWaveguides}.

\section{$\left(x_{0},\,x_{2}\right)\Rightarrow\left(r_{0},\,r_{2}\right)$\label{sec: details of x to r}}

From eq. (\ref{eq: 1st-step trans}), partial derivatives of $x_{0}$ and $x_{2}$ by $r_{0}$ and $r_{2}$ are that 
\begin{gather}
\frac{\partial x_{0}}{\partial r_{0}}  = f\left(r_{2}\right),  \quad 
\frac{\partial x_{2}}{\partial r_{0}}  = -g\left(r_{2}\right),\nonumber\\
\frac{\partial x_{0}}{\partial r_{2}} = \left(1-r_{0}\kappa_{b}\left(r_{2}\right)\right)g\left(r_{2}\right),  \quad  
\frac{\partial x_{2}}{\partial r_{2}} = \left(1-r_{0}\kappa_{b}\left(r_{2}\right)\right)f\left(r_{2}\right).\nonumber
\end{gather}
Obviously, 
\begin{align}
\frac{\partial x_{0}}{\partial r_{2}}\frac{\partial x_{0}}{\partial r_{0}}+\frac{\partial x_{2}}{\partial r_{2}}\frac{\partial x_{2}}{\partial r_{0}} &= 0,\label{eq: x-r ortho}\\
\left(\frac{\partial x_{0}}{\partial r_{0}}\right)^{2}+\left(\frac{\partial x_{2}}{\partial r_{0}}\right)^{2} &= 1,\label{eq: x-r scale0}\\
\left(\frac{\partial x_{0}}{\partial r_{2}}\right)^{2}+\left(\frac{\partial x_{2}}{\partial r_{2}}\right)^{2} &=  \left(1-r_{0}\kappa_{b}\left(r_{2}\right)\right)^{2}.\label{eq: x-r scale2}
\end{align}
For $r_{0}=0$, the partial derivatives are that
\begin{gather}
\frac{\partial x_{0}}{\partial r_{2}}=g,  \quad \frac{\partial x_{2}}{\partial r_{2}}=f,\nonumber\\
\frac{\partial^{2}x_{0}}{\partial r_{2}^{2}}=\kappa_{b}f,  \quad  \frac{\partial^{2}x_{2}}{\partial r_{2}^{2}}=-\kappa_{b}g.\nonumber
\end{gather}
Then
\begin{equation}
\left\{
\begin{aligned}
\frac{\partial x_{2}}{\partial r_{2}}\frac{\partial^{2}x_{0}}{\partial r_{2}^{2}}-\frac{\partial^{2}x_{2}}{\partial r_{2}^{2}}\frac{\partial x_{0}}{\partial r_{2}} &=
\kappa_{b} f^{2} + \kappa_{b} g^{2} = \kappa_{b}\,,\\
\left(\frac{\partial x_{0}}{\partial r_{2}}\right)^{2}+\left(\frac{\partial x_{2}}{\partial r_{2}}\right)^{2}  &=
f^{2} + g^{2} = 1
\end{aligned}
\right.
\label{eq: bending kappa}
\end{equation}
at $r_{0}=0$. 

\section{$\left(r_{0},\,r_{2}\right)\Rightarrow\left(u_{0},\,u_{2}\right)$\label{sec: details of r to u}}

We focus on orthogonal condition for the second step $\left(r_{0},\,r_{2}\right)\Rightarrow\left(u_{0},\,u_{2}\right)$. 
Orthogonal condition $\left(x_{0},\,x_{2}\right)\Rightarrow\left(u_{0},\,u_{2}\right)$ is that
\[
\frac{\partial x_{0}}{\partial u_{0}}\frac{\partial x_{0}}{\partial u_{2}}+\frac{\partial x_{2}}{\partial u_{0}}\frac{\partial x_{2}}{\partial u_{2}}=0\,.
\]
Here,
\begin{align}
\frac{\partial x_{0}}{\partial u_{0}}\frac{\partial x_{0}}{\partial u_{2}} & = \left(\frac{\partial x_{0}}{\partial r_{0}}\frac{\partial r_{0}}{\partial u_{0}}+\frac{\partial x_{0}}{\partial r_{2}}\frac{\partial r_{2}}{\partial u_{0}}\right)\left(\frac{\partial x_{0}}{\partial r_{0}}\frac{\partial r_{0}}{\partial u_{2}}+\frac{\partial x_{0}}{\partial r_{2}}\frac{\partial r_{2}}{\partial u_{2}}\right)\nonumber\\
 & = \left(\frac{\partial x_{0}}{\partial r_{0}}\right)^{2}\frac{\partial r_{0}}{\partial u_{0}}\frac{\partial r_{0}}{\partial u_{2}}+\left(\frac{\partial x_{0}}{\partial r_{2}}\right)^{2}\frac{\partial r_{2}}{\partial u_{0}}\frac{\partial r_{2}}{\partial u_{2}}+\frac{\partial x_{0}}{\partial r_{0}}\frac{\partial x_{0}}{\partial r_{2}}\left(\frac{\partial r_{2}}{\partial u_{0}}\frac{\partial r_{0}}{\partial u_{2}}+\frac{\partial r_{0}}{\partial u_{0}}\frac{\partial r_{2}}{\partial u_{2}}\right),\nonumber
\end{align}
and
\begin{align}
\frac{\partial x_{2}}{\partial u_{0}}\frac{\partial x_{2}}{\partial u_{2}} & = \left(\frac{\partial x_{2}}{\partial r_{0}}\frac{\partial r_{0}}{\partial u_{0}}+\frac{\partial x_{2}}{\partial r_{2}}\frac{\partial r_{2}}{\partial u_{0}}\right)\left(\frac{\partial x_{2}}{\partial r_{0}}\frac{\partial r_{0}}{\partial u_{2}}+\frac{\partial x_{2}}{\partial r_{2}}\frac{\partial r_{2}}{\partial u_{2}}\right)\nonumber\\
 & = \left(\frac{\partial x_{2}}{\partial r_{0}}\right)^{2}\frac{\partial r_{0}}{\partial u_{0}}\frac{\partial r_{0}}{\partial u_{2}}+\left(\frac{\partial x_{2}}{\partial r_{2}}\right)^{2}\frac{\partial r_{2}}{\partial u_{0}}\frac{\partial r_{2}}{\partial u_{2}}+\frac{\partial x_{2}}{\partial r_{0}}\frac{\partial x_{2}}{\partial r_{2}}\left(\frac{\partial r_{2}}{\partial u_{0}}\frac{\partial r_{0}}{\partial u_{2}}+\frac{\partial r_{0}}{\partial u_{0}}\frac{\partial r_{2}}{\partial u_{2}}\right)\,.\nonumber
\end{align}
From equation (\ref{eq: x-r ortho}),
\[
\frac{\partial x_{0}}{\partial u_{0}}\frac{\partial x_{0}}{\partial u_{2}}+\frac{\partial x_{2}}{\partial u_{0}}\frac{\partial x_{2}}{\partial u_{2}}=\left(\left(\frac{\partial x_{0}}{\partial r_{0}}\right)^{2}+\left(\frac{\partial x_{2}}{\partial r_{0}}\right)^{2}\right)\frac{\partial r_{0}}{\partial u_{0}}\frac{\partial r_{0}}{\partial u_{2}}+\left(\left(\frac{\partial x_{0}}{\partial r_{2}}\right)^{2}+\left(\frac{\partial x_{2}}{\partial r_{2}}\right)^{2}\right)\frac{\partial r_{2}}{\partial u_{0}}\frac{\partial r_{2}}{\partial u_{2}}\,.
\]
From equations (\ref{eq: x-r scale0}) and (\ref{eq: x-r scale2}), the orthogonal condition can be modified as
\begin{equation}
\frac{\partial r_{0}}{\partial u_{0}}\frac{\partial r_{0}}{\partial u_{2}}+\left(1-r_{0}\kappa_{b}\left(r_{2}\right)\right)^{2}\frac{\partial r_{2}}{\partial u_{0}}\frac{\partial r_{2}}{\partial u_{2}}=0\,.\label{eq: mod. ortho}
\end{equation}

The partial derivatives for eq. (\ref{eq: 2nd-step trans}) are that 
\begin{equation}
\left\{\begin{aligned}
\frac{\partial r_{2}}{\partial u_{0}} 
& = -F_{\mathrm{2D}}\left(r_{0}\left(u_{0},\,u_{2}\right),\,r_{2}\left(u_{0},\,u_{2}\right)\right),\\
\frac{\partial r_{0}}{\partial u_{0}} 
& = \zeta\left(r_{2}\right) + u_{0}\zeta^{'}\left(r_{2}\right)\frac{\partial r_{2}}{\partial u_{0}}
= \zeta\left(r_{2}\right) -  r_{0}\kappa_{w}\left(r_{2}\right)F_{\mathrm{2D}}\left(r_{0},\,r_{2}\right)\,,\\
\frac{\partial r_{0}}{\partial u_{2}} 
& = u_{0}\zeta^{'}\left(r_{2}\right)\frac{\partial r_{2}}{\partial u_{2}}
= r_{0}\kappa_{w}\left(r_{2}\right)\frac{\partial r_{2}}{\partial u_{2}}\,.
\end{aligned}\right.\label{eq: partial r2 r0 by u0 u2}
\end{equation}
where $\zeta^{'}\left(r_{2}\right) = \kappa_{w}\left(r_{2}\right)\zeta\left(r_{2}\right)$.
From eqs. (\ref{eq: mod. ortho}) and (\ref{eq: partial r2 r0 by u0 u2}),
\begin{equation}
\begin{aligned}
 0 &= \frac{\partial r_{0}}{\partial u_{0}}\frac{\partial r_{0}}{\partial u_{2}}+\left(1-r_{0}\kappa_{b}\left(r_{2}\right)\right)^{2}\frac{\partial r_{2}}{\partial u_{0}}\frac{\partial r_{2}}{\partial u_{2}}\\
& = \left\{	r_{0}\kappa_{w}\left(r_{2}\right)\zeta\left(r_{2}\right) 
-\left[\left(1-r_{0}\kappa_{b}\left(r_{2}\right)\right)^{2}+\left(r_{0}\kappa_{w}\left(r_{2}\right)\right)^{2}\right]F_{\mathrm{2D}}\left(r_{0},\,r_{2}\right)\right\} \frac{\partial r_{2}}{\partial u_{2}}
\end{aligned}\label{eq: F definition}
\end{equation}
The definition of $F_{\mathrm{2D}}$ in eq.  (\ref{eq: 2nd-step trans}) always satisfies the above eq. (\ref{eq: F definition}).
For $u_{0}=0$, 
\begin{gather}
\left.\frac{\partial r_{0}}{\partial u_{0}}\right|_{u_{0}=0} = \zeta, \quad 
\left.\frac{\partial r_{2}}{\partial u_{0}}\right|_{u_{0}=0} = 0,\nonumber\\
\left.\frac{\partial^{2}r_{2}}{\partial u_{0}^{2}}\right|_{u_{0}=0}
 = -\left[
  \frac{\partial r_{0}}{\partial u_{0}}\frac{\partial}{\partial r_{0}}
+ \frac{\partial r_{2}}{\partial u_{0}}\frac{\partial}{\partial r_{2}}
\right] F_{\mathrm{2D}}\left(r_{0},\,r_{2}\right)
= -\zeta\frac{\partial F_{\mathrm{2D}}\left(r_{0},\,r_{2}\right)}{\partial r_{0}}
= -\kappa_{w}\zeta^{2}.\nonumber
\end{gather}
Then
\begin{equation}
\left\{
\begin{aligned}
\frac{\partial r_{2}}{\partial u_{0}}\frac{\partial^{2}r_{0}}{\partial u_{0}^{2}}-\frac{\partial^{2}r_{2}}{\partial u_{0}^{2}}\frac{\partial r_{0}}{\partial u_{0}} &=
\kappa_{w}\zeta^{3}\,,\\
\left(\frac{\partial r_{0}}{\partial u_{0}}\right)^{2}+\left(\frac{\partial r_{2}}{\partial u_{0}}\right)^{2} &=
\zeta^{2}
\end{aligned}
\right.
\label{eq: width kappa}
\end{equation}
at $u_{0}=0$. 

\section{Partial derivatives and scale factors\label{sec: details of scale factor}}

Let us obtain scale factor $h_{0}$ and $h_{2}$. 
The partial derivatives by $u_{0}$ are that 
\begin{align}
\frac{\partial x_{0}}{\partial u_{0}} & = \frac{\partial x_{0}}{\partial r_{0}}\frac{\partial r_{0}}{\partial u_{0}}+\frac{\partial x_{0}}{\partial r_{2}}\frac{\partial r_{2}}{\partial u_{0}}
= \frac{f\left(1-u_{0}\zeta\kappa_{b}\right)^{2}\zeta}{\left(1-u_{0}\zeta\kappa_{b}\right)^{2}+\left(u_{0}\zeta^{'}\right)^{2}}-\frac{\left(1-u_{0}\zeta\kappa_{b}\right)gu_{0}\zeta^{'}\zeta}{\left(1-u_{0}\zeta\kappa_{b}\right)^{2}+\left(u_{0}\zeta^{'}\right)^{2}}\nonumber\\
 & = \frac{\left(1-u_{0}\zeta\kappa_{b}\right)\zeta\left(f\left(1-u_{0}\zeta\kappa_{b}\right)-gu_{0}\zeta^{'}\right)}{\left(1-u_{0}\zeta\kappa_{b}\right)^{2}+\left(u_{0}\zeta^{'}\right)^{2}}\nonumber
\end{align}
and
\begin{align}
\frac{\partial x_{2}}{\partial u_{0}} & = \frac{\partial x_{2}}{\partial r_{0}}\frac{\partial r_{0}}{\partial u_{0}}+\frac{\partial x_{2}}{\partial r_{2}}\frac{\partial r_{2}}{\partial u_{0}}
= -\frac{g\left(1-u_{0}\zeta\kappa_{b}\right)^{2}\zeta}{\left(1-u_{0}\zeta\kappa_{b}\right)^{2}+\left(u_{0}\zeta^{'}\right)^{2}}-\frac{\left(1-u_{0}\zeta\kappa_{b}\right)fu_{0}\zeta^{'}\zeta}{\left(1-u_{0}\zeta\kappa_{b}\right)^{2}+\left(u_{0}\zeta^{'}\right)^{2}}\nonumber\\
 & = -\frac{\left(1-u_{0}\zeta\kappa_{b}\right)\zeta\left(g\left(1-u_{0}\zeta\kappa_{b}\right)+fu_{0}\zeta^{'}\right)}{\left(1-u_{0}\zeta\kappa_{b}\right)^{2}+\left(u_{0}\zeta^{'}\right)^{2}}.\nonumber
\end{align}
Then,
\begin{equation}\begin{aligned}
h_{0}  
= \sqrt{
	\left(\frac{\partial x_{0}}{\partial u_{0}}\right)^{2}
	+\left(\frac{\partial x_{2}}{\partial u_{0}}\right)^{2}
	}
 & = \frac{\left(1-u_{0}\zeta\kappa_{b}\right)\zeta}{\left(1-u_{0}\zeta\kappa_{b}\right)^{2}+\left(u_{0}\zeta^{'}\right)^{2}}\sqrt{\left(f\left(1-u_{0}\zeta\kappa_{b}\right)-gu_{0}\zeta^{'}\right)^{2}+\left(g\left(1-u_{0}\zeta\kappa_{b}\right)+fu_{0}\zeta^{'}\right)^{2}} \\
 & = \frac{\left(1-u_{0}\zeta\kappa_{b}\right)\zeta}{\sqrt{\left(1-u_{0}\zeta\kappa_{b}\right)^{2}+\left(u_{0}\zeta^{'}\right)^{2}}}\,.
\end{aligned}\label{eq: h0 with curvatures}\end{equation}
The partial derivatives by $u_{2}$ are that
\[\begin{aligned}
\frac{\partial x_{0}}{\partial u_{2}}&=\frac{\partial x_{0}}{\partial r_{0}}\frac{\partial r_{0}}{\partial u_{2}}+\frac{\partial x_{0}}{\partial r_{2}}\frac{\partial r_{2}}{\partial u_{2}}=\left(fu_{0}\zeta^{'}+\left(1-u_{0}\zeta\kappa_{b}\right)g\right)\frac{\partial r_{2}}{\partial u_{2}}\,,\\
\frac{\partial x_{2}}{\partial u_{2}}&=\frac{\partial x_{2}}{\partial r_{0}}\frac{\partial r_{0}}{\partial u_{2}}+\frac{\partial x_{2}}{\partial r_{2}}\frac{\partial r_{2}}{\partial u_{2}}=\left(-gu_{0}\zeta^{'}+\left(1-u_{0}\zeta\kappa_{b}\right)f\right)\frac{\partial r_{2}}{\partial u_{2}}\,.
\end{aligned}\]
Then,
\begin{equation}\begin{aligned}
h_{2} 
= \sqrt{
	\left(\frac{\partial x_{0}}{\partial u_{2}}\right)^{2}
	+\left(\frac{\partial x_{2}}{\partial u_{2}}\right)^{2}
	}
 & = \frac{\partial r_{2}}{\partial u_{2}}\sqrt{\left(fu_{0}\zeta^{'}+\left(1-u_{0}\zeta\kappa_{b}\right)g\right)^{2}+\left(-gu_{0}\zeta^{'}+\left(1-u_{0}\zeta\kappa_{b}\right)f\right)^{2}}\\
 & = \frac{\partial r_{2}}{\partial u_{2}}\sqrt{\left(1-u_{0}\zeta\kappa_{b}\right)^{2}+\left(u_{0}\zeta^{'}\right)^{2}}\,.
\end{aligned}\label{eq: h2 with curvatures}\end{equation}
The $\partial r_{2}/\partial u_{2}$ in the right side of eq. (\ref{eq: h2 with curvatures}) can be numerically solved by using eq. (\ref{eq: r2u2}). 
This section shows details of $\partial F_{\mathrm{2D}}/\partial u_{2}$.
From eq. (\ref{eq: partial r2 r0 by u0 u2}),
\[
\frac{\partial F_{\mathrm{2D}}}{\partial u_{2}} = 
	\frac{\partial r_{0}}{\partial u_{2}}\frac{\partial F_{\mathrm{2D}}}{\partial r_{0}}
+ \frac{\partial r_{2}}{\partial u_{2}}\frac{\partial F_{\mathrm{2D}}}{\partial r_{2}}
= \frac{\partial r_{2}}{\partial u_{2}}\left(
	r_{0}\kappa_{w}\frac{\partial F_{\mathrm{2D}}}{\partial r_{0}}
	+ \frac{\partial F_{\mathrm{2D}}}{\partial r_{2}}
\right)\,.
\]
From eq. (\ref{eq: 2nd-step trans}),
\begin{equation}\begin{aligned}
\frac{\partial F_{\mathrm{2D}}}{\partial r_{0}} &= 
\frac{\kappa_{w}\zeta}{\left(1-r_{0}\kappa_{b}\right)^{2}+\left(r_{0}\kappa_{w}\right)^{2}}
- \frac{
	r_{0}\kappa_{w}\zeta\left[2\left(1-r_{0}\kappa_{b}\right)\left(-\kappa_{b}\right)
	+2r_{0}\kappa_{w}^{2}\right]
}{\left[
\left( 1 - r_{0}\kappa_{b} \right)^{2} + \left( r_{0} \kappa_{w} \right)^{2}
\right]^{2}}\,,\\
\frac{\partial F_{\mathrm{2D}}}{\partial r_{2}} &=
\frac{r_{0}\left(\kappa_{w}^{'}+\kappa_{w}^{2}\right)\zeta}
{\left(1-r_{0}\kappa_{b}\right)^{2}+\left(r_{0}\kappa_{w}\right)^{2}}
- \frac{
	r_{0}\kappa_{w}\zeta\left[2\left(1-r_{0}\kappa_{b}\right)\left(-r_{0}\kappa_{b}^{'}\right)
	+2r_{0}^{2}\kappa_{w}\kappa_{w}^{'}\right]
}{\left[
\left( 1 - r_{0}\kappa_{b} \right)^{2} + \left( r_{0} \kappa_{w} \right)^{2}
\right]^{2}}\,.
\end{aligned}\nonumber\end{equation}
Then
\begin{equation}\begin{aligned}
\frac{\partial F_{\mathrm{2D}}}{\partial u_{2}} &= \frac{\partial r_{2}}{\partial u_{2}}\left\{
	\frac{r_{0}\left(\kappa_{w}^{'}+2\kappa_{w}^{2}\right)\zeta}
	{\left(1-r_{0}\kappa_{b}\right)^{2}+\left(r_{0}\kappa_{w}\right)^{2}}
	- \frac{
		r_{0}^{2}\kappa_{w}\zeta\left[
			2\left(1-r_{0}\kappa_{b}\right)\left(-\kappa_{b}\kappa_{w}-\kappa_{b}^{'}\right)
		+2r_{0}\kappa_{w}\left(\kappa_{w}^{2} + \kappa_{w}^{'}\right)\right]
	}{\left[
		\left( 1 - r_{0}\kappa_{b} \right)^{2} + \left( r_{0} \kappa_{w} \right)^{2}
	\right]^{2}}
\right\}\\
	&= \frac{\partial r_{2}}{\partial u_{2}} r_{0}\zeta \left\{
	\frac{\kappa_{w}^{'}+2\kappa_{w}^{2}}
	{\left(1-r_{0}\kappa_{b}\right)^{2}+\left(r_{0}\kappa_{w}\right)^{2}}
	- \frac{
		r_{0}\kappa_{w}\left[
			2\left(1-r_{0}\kappa_{b}\right)\left(-\kappa_{b}\kappa_{w}-\kappa_{b}^{'}\right)
		+2r_{0}\kappa_{w}\left(\kappa_{w}^{2} + \kappa_{w}^{'}\right)\right]
	}{\left[
		\left( 1 - r_{0}\kappa_{b} \right)^{2} + \left( r_{0} \kappa_{w} \right)^{2}
	\right]^{2}}
\right\}\\
	&= \frac{\partial r_{2}}{\partial u_{2}} r_{0}\zeta 
	\frac{
		\kappa_{w}^{'}\left(1 - r_{0}\kappa_{b}\right)^{2}
		- \kappa_{w}^{'}\left(r_{0}\kappa_{w}\right)^{2}
		+ 2\kappa_{w}\left(1 - r_{0}\kappa_{b}\right)\left(\kappa_{w} + r_{0}\kappa_{b}^{'}\right)
	}{\left[
		\left( 1 - r_{0}\kappa_{b} \right)^{2} + \left( r_{0} \kappa_{w} \right)^{2}
	\right]^{2}}
\,.
\end{aligned}\label{eq: partial F partial u2}\end{equation}

%% file: App/B_PeriodicWG_v2.tex
\chapter{S-matrix, Transfer matrix and Periodic system\label{ch: Periodic WG}}

This chapter shows S-matrix character, derivation of Transfer matrix and mode equation of periodic system.

\section{S-matrix\label{sec: scattering matrix}}
The S-matrix was first introduced by J. A. Wheeler in the 1937 paper~\cite{Wheeler} for nuclear physics.
It described the scattering between quantum states indexed by spin and angular momentum of nuclei, and its unitarity had been already discussed.
In the framework of circuit theory, the concept of scattering matrix was introduced by V. Belevitch in the 1945 thesis~\cite{Vandewalle}.
The S-matrix ``$\boldsymbol{S}$'' was independently introduced by R. H. Dicke~\cite{Dicke} as the work of The Radiation Laboratory~\cite{Montgomery}.
The ``$\boldsymbol{S}$'' is defined between multi-terminals which are connected to a waveguide junction, and then it can also be represented as eq. \eqref{eq: S-matrix}.
The S-matrix symmetry (reciprocity) was discussed in Sec. II of Ref.~\cite{Belevitch} and for Eq. (92) in Chap. 5 of Ref.~\cite{Montgomery}.
Its unitarity was also given by Eq. (101) in Chap. 5 of Ref.~\cite{Montgomery}.
The S-matrix of multichannel system was also studied for quantum transport, its unitarity and symmetry were shown by eqs. (3.1) and (3.2) of Ref.~\cite{Buttiker}.

This section will try to show the unitarity for no-loss system and the symmetry for no-loss and time-reversal invariant system without using any specific model.
Unitarity of S-matrix directly derives from flow conservation of the no-loss system.
Flow conservation in case of mode ``$m$'' incident is represented by $\sum_{j}\left|S_{jm}\right|^{2}=1$.
 In case of incident for two modes ``$m$'' and ``$n$'', flow conservation of a linear system gives us
\[
 \sum_{j}\left|S_{jm}+S_{jn}\right|^{2}=\sum_{j}\left|S_{jm}+iS_{jn}\right|^{2}=2 \quad \mathrm{for} \quad m \neq n\,.
\]
Then,
\[
\sum_{j}S_{jm}^{*}S_{jn}+S_{jn}^{*}S_{jm}=i\sum_{j}S_{jm}^{*}S_{jn}-S_{jn}^{*}S_{jm}=0 \quad \mathrm{for} \quad m \neq n\,.
\]
Therefore $\sum_{j}S_{jm}^{*}S_{jn}=\sum_{j}S_{mj}^{\dagger}S_{jn}=\delta_{mn}$, that is
\begin{equation}
\boldsymbol{S}^{\dagger}\boldsymbol{S}=\boldsymbol{S}\boldsymbol{S}^{\dagger}=1.
\label{eq: unitaryS PeriodicWG}
\end{equation}
Additional symmetry of S-matrix derives from time-reversal invariance.
Reversed propagation for mode ``$m$'' incident (exit) is equivalent to complex conjugate of propagation for mode ``$m$'' exit (incident) only if the system remains time-reversal invariance: $\sum_{j}S_{nj}S_{jm}^{*}=\delta_{mn}$, that is
\[
\boldsymbol{S}^{*}\boldsymbol{S}=\boldsymbol{S}\boldsymbol{S}^{*}=1.
\]
 When the linear system satisfies both of flow conservation and time-reversal invariance, S-matrix satisfies the symmetry (reciprocity) $\boldsymbol{S}^{\mathrm{T}}=\boldsymbol{S}$
  from Eqs. \eqref{eq: unitaryS PeriodicWG} and the above relation.

\section{Transfer Matrix\label{sec: transfer matrix}}

Let us consider a scattering matrix which consists of 4 submatrices for the left-hand and right-hand sides.
\begin{equation}
\boldsymbol{S}=\left(\begin{array}{cc}
	\boldsymbol{s}_{LL} & \boldsymbol{s}_{LR}\\
	\boldsymbol{s}_{RL} & \boldsymbol{s}_{RR}
\end{array}\right),
\label{eq: 4subM PeriodicWG}
\end{equation}
where we set $\boldsymbol{s}_{LR}$ and $\boldsymbol{s}_{RL}$ as regular matrices.
We also consider 4 column vectors for forward and backward modes in the left-hand and right-hand sides as $\boldsymbol{f}_{L}$, $\boldsymbol{b}_{L}$, $\boldsymbol{f}_{R}$ and $\boldsymbol{b}_{R}$.
They are related to each other by Eq. \eqref{eq: 4subM PeriodicWG}:
\begin{equation}
\left\{
	\begin{split}
		\boldsymbol{f}_{R} & =\boldsymbol{s}_{RL}\boldsymbol{f}_{L}+\boldsymbol{s}_{RR}\boldsymbol{b}_{R},\\
		\boldsymbol{b}_{L} & =\boldsymbol{s}_{LL}\boldsymbol{f}_{L}+\boldsymbol{s}_{LR}\boldsymbol{b}_{R}.
	\end{split}
\right.
\label{eq: frbl PeriodicWG}
\end{equation}
A transfer matrix $\boldsymbol{T}_{RL}$ is defined as
\begin{equation}
\left(\begin{array}{c}
\boldsymbol{f}_{R}\\
\boldsymbol{b}_{R}
\end{array}\right)=\boldsymbol{T}_{RL}\left(\begin{array}{c}
\boldsymbol{f}_{L}\\
\boldsymbol{b}_{L}
\end{array}\right).
\label{eq: T-matrix PeriodicWG}
\end{equation}
The above and Eq. \eqref{eq: frbl PeriodicWG} give us the following relation.
\begin{equation}
\boldsymbol{T}_{RL}=\left(\begin{array}{cc}
\boldsymbol{t}_{ff} & \boldsymbol{t}_{fb}\\
\boldsymbol{t}_{bf} & \boldsymbol{t}_{bb}
\end{array}\right)
=
\left(
\begin{array}{cc}
	\boldsymbol{s}_{RL}-\boldsymbol{s}_{RR}\frac{1}{\boldsymbol{s}_{LR}}\boldsymbol{s}_{LL} & \boldsymbol{s}_{RR}\frac{1}{\boldsymbol{s}_{LR}}\\
	-\frac{1}{\boldsymbol{s}_{LR}}\boldsymbol{s}_{LL} & \frac{1}{\boldsymbol{s}_{LR}}
\end{array}
\right)
=
\left(
\begin{array}{cc}
	\frac{1}{\boldsymbol{s}_{RL}} & -\frac{1}{\boldsymbol{s}_{RL}}\boldsymbol{s}_{RR}\\
	\boldsymbol{s}_{LL}\frac{1}{\boldsymbol{s}_{RL}} & \boldsymbol{s}_{LR}-\boldsymbol{s}_{LL}\frac{1}{\boldsymbol{s}_{RL}}\boldsymbol{s}_{RR}
\end{array}
\right)^{-1}.
\label{eq: 4matricesT PeriodicWG}
\end{equation}
The S-matrix can also be represented by the 4 submatrices of the $\boldsymbol{T}_{RL}$.
\[
\boldsymbol{S}=
\left(\begin{array}{cc}
	\boldsymbol{s}_{LL} & \boldsymbol{s}_{LR}\\
	\boldsymbol{s}_{RL} & \boldsymbol{s}_{RR}
\end{array}\right)
=
\left(
\begin{array}{cc}
	-\frac{1}{\boldsymbol{t}_{bb}}\boldsymbol{t}_{bf} & \frac{1}{\boldsymbol{t}_{bb}}\\
	\boldsymbol{t}_{ff}-\boldsymbol{t}_{fb}\frac{1}{\boldsymbol{t}_{bb}}\boldsymbol{t}_{bf} & \boldsymbol{t}_{fb}\frac{1}{\boldsymbol{t}_{bb}}
\end{array}
\right)=
\left(
\begin{array}{cc}
	-\frac{1}{\boldsymbol{t}_{ff}}\boldsymbol{t}_{fb} & \frac{1}{\boldsymbol{t}_{ff}}\\
	\boldsymbol{t}_{bb}-\boldsymbol{t}_{bf}\frac{1}{\boldsymbol{t}_{ff}}\boldsymbol{t}_{fb} & \boldsymbol{t}_{bf}\frac{1}{\boldsymbol{t}_{ff}}
\end{array}
\right)^{-1}.
\]
Note that the transfer matrix $\boldsymbol{T}_{RL}$ of eq. \eqref{eq: T-matrix PeriodicWG} is different from the T-matrix $\boldsymbol{T}^{\left(\pm\right)}$ of eq. \eqref{eq: T-matrix equation}.

\section{Periodic waveguide}

Both sides of a block of periodic system are connected to two hypothetical
waveguides which have the same structure. First, let us redefine evanescent
modes $\boldsymbol{\Phi}_{\pm m}$ ($m>n_{\max}$) for the hypothetical
waveguide. Redefined mode $\boldsymbol{\Phi}_{\pm m}^{'}$ is given
by
\begin{equation}
\boldsymbol{\Phi}_{\pm m}^{'}=\frac{\boldsymbol{\Phi}_{m}\pm\boldsymbol{\Phi}_{-m}}{\sqrt{2}}\quad\mathrm{for}\quad m>n_{\max},
\label{eq: redefined evanescent mode PeriodicWG}
\end{equation}
where original evanescent and divergent modes satisfy the normalization rule of Eq. \eqref{eq: evanescent wave normalization}.
Then, the $\boldsymbol{\Phi}_{n}^{'}$ is applied to the same orthogonality as Eq. \eqref{eq: propagating wave normalization}: 
\[
\boldsymbol{\Phi}_{n}^{'\dagger}\left(\begin{array}{cc}
0 & 1\\
1 & 0
\end{array}\right)\boldsymbol{\Phi}_{m}^{'}=\frac{n}{\left|n\right|}\delta_{nm}.
\]
Accordingly, we will use the redefined mode $\boldsymbol{\Phi}_{n}^{'}$ as one of propagating modes in the following discussion.
Note that the $\boldsymbol{\Phi}_{n}^{'}$ is not eigenvector of eq. \eqref{eq: matrix equation} for the hypothetical waveguide, but it is still a solution of eq.\eqref{eq: propagation-equation}.

Column vector in forward-mode ( backward-mode ) space of the hypothetical waveguides is shown as $\boldsymbol{f}$ ( $\boldsymbol{b}$ ).
Bloch function of periodic system is given by Eqs. \eqref{eq: T-matrix PeriodicWG}:
\[
\boldsymbol{T}_{RL}\left(\begin{array}{c} \boldsymbol{f} \\ \boldsymbol{b} \end{array}\right)
 = e^{i\theta}\left(\begin{array}{c} \boldsymbol{f} \\ \boldsymbol{b} \end{array}\right),\quad\mathrm{and}\quad
\boldsymbol{T}_{RL}^{-1}\left(\begin{array}{c} \boldsymbol{f} \\ \boldsymbol{b} \end{array}\right)
 = e^{-i\theta}\left(\begin{array}{c} \boldsymbol{f} \\ \boldsymbol{b} \end{array}\right).
\]
Then we obtain an eigenvalue equation:
\begin{equation}
\frac{1}{2i}\left(\boldsymbol{T}_{RL}-\boldsymbol{T}_{RL}^{-1}\right)\left(\begin{array}{c}
\boldsymbol{f}\\
\boldsymbol{b}
\end{array}\right)=\sin\theta\left(\begin{array}{c}
\boldsymbol{f}\\
\boldsymbol{b}
\end{array}\right).
\label{eq: eigen1 PeriodicWG}
\end{equation}

From Eqs. \eqref{eq: unitaryS PeriodicWG} and \eqref{eq: 4subM PeriodicWG}, no-loss system satisfies that
\[
-\boldsymbol{s}_{LL}\frac{1}{\boldsymbol{s}_{RL}} = \frac{1}{\boldsymbol{s}_{LR}^{\dagger}}\boldsymbol{s}_{RR}^{\dagger},\quad
-\frac{1}{\boldsymbol{s}_{LR}}\boldsymbol{s}_{LL} = \boldsymbol{s}_{RR}^{\dagger}\frac{1}{\boldsymbol{s}_{RL}^{\dagger}},\quad
\boldsymbol{s}_{RL}-\boldsymbol{s}_{RR}\frac{1}{\boldsymbol{s}_{LR}}\boldsymbol{s}_{LL} = \frac{1}{\boldsymbol{s}_{RL}^{\dagger}},\; \mathrm{and}\; 
\boldsymbol{s}_{LR}-\boldsymbol{s}_{LL}\frac{1}{\boldsymbol{s}_{RL}}\boldsymbol{s}_{RR} = \frac{1}{\boldsymbol{s}_{LR}^{\dagger}}.
\]
From Eq. \eqref{eq: 4matricesT PeriodicWG} and the above,
$\left(\boldsymbol{T}_{RL}-\boldsymbol{T}_{RL}^{-1}\right)/\left(2i\right)$
can be deformed to
\begin{eqnarray*}
\frac{1}{2i}\left(\boldsymbol{T}_{RL}-\boldsymbol{T}_{RL}^{-1}\right) & = & \frac{1}{2i}\left(\begin{array}{cc}
\boldsymbol{s}_{RL}-\boldsymbol{s}_{RR}\frac{1}{\boldsymbol{s}_{LR}}\boldsymbol{s}_{LL}-\frac{1}{\boldsymbol{s}_{RL}} & \boldsymbol{s}_{RR}\frac{1}{\boldsymbol{s}_{LR}}+\frac{1}{\boldsymbol{s}_{RL}}\boldsymbol{s}_{RR}\\
-\frac{1}{\boldsymbol{s}_{LR}}\boldsymbol{s}_{LL}-\boldsymbol{s}_{LL}\frac{1}{\boldsymbol{s}_{RL}} & \frac{1}{\boldsymbol{s}_{LR}}-\left(\boldsymbol{s}_{LR}-\boldsymbol{s}_{LL}\frac{1}{\boldsymbol{s}_{RL}}\boldsymbol{s}_{RR}\right)
\end{array}\right)\\
 & = & \frac{1}{2i}\left(\begin{array}{cc}
\frac{1}{\boldsymbol{s}_{RL}^{\dagger}}-\frac{1}{\boldsymbol{s}_{RL}} & \boldsymbol{s}_{RR}\frac{1}{\boldsymbol{s}_{LR}}+\frac{1}{\boldsymbol{s}_{RL}}\boldsymbol{s}_{RR}\\
\boldsymbol{s}_{RR}^{\dagger}\frac{1}{\boldsymbol{s}_{RL}^{\dagger}}+\frac{1}{\boldsymbol{s}_{LR}^{\dagger}}\boldsymbol{s}_{RR}^{\dagger} & \frac{1}{\boldsymbol{s}_{LR}}-\frac{1}{\boldsymbol{s}_{LR}^{\dagger}}
\end{array}\right).
\end{eqnarray*}
Then, Eq. \eqref{eq: eigen1 PeriodicWG} is deformed to 
\[
\frac{1}{2i}\left(\begin{array}{cc}
\frac{1}{\boldsymbol{s}_{RL}^{\dagger}}-\frac{1}{\boldsymbol{s}_{RL}} & \boldsymbol{s}_{RR}\frac{1}{\boldsymbol{s}_{LR}}+\frac{1}{\boldsymbol{s}_{RL}}\boldsymbol{s}_{RR}\\
-\boldsymbol{s}_{RR}^{\dagger}\frac{1}{\boldsymbol{s}_{RL}^{\dagger}}-\frac{1}{\boldsymbol{s}_{LR}^{\dagger}}\boldsymbol{s}_{RR}^{\dagger} & \frac{1}{\boldsymbol{s}_{LR}^{\dagger}}-\frac{1}{\boldsymbol{s}_{LR}}
\end{array}\right)\left(\begin{array}{c}
\boldsymbol{f}\\
\boldsymbol{b}
\end{array}\right)=\sin\theta\left(\begin{array}{cc}
1 & 0\\
0 & -1
\end{array}\right)\left(\begin{array}{c}
\boldsymbol{f}\\
\boldsymbol{b}
\end{array}\right).
\]
Finally, we can obtain the mode equation \eqref{eq: matrix equation} for the periodic system as
\begin{equation}
\boldsymbol{M}_{p}\left(\begin{array}{c}
\frac{\boldsymbol{f}+\boldsymbol{b}}{\sqrt{2}}\\
\frac{\boldsymbol{f}-\boldsymbol{b}}{\sqrt{2}}
\end{array}\right)=\sin\theta\left(\begin{array}{cc}
0 & 1\\
1 & 0
\end{array}\right)\left(\begin{array}{c}
\frac{\boldsymbol{f}+\boldsymbol{b}}{\sqrt{2}}\\
\frac{\boldsymbol{f}-\boldsymbol{b}}{\sqrt{2}}
\end{array}\right),{}
\label{eq: periodic mode eq}
\end{equation}
where $\boldsymbol{M}_{p}=\boldsymbol{M}_{p}^{\dagger}$, because
\[
\boldsymbol{M}_{p} = \left(\begin{array}{cc}
\frac{1}{\sqrt{2}} & \frac{1}{\sqrt{2}}\\
\frac{1}{\sqrt{2}} & -\frac{1}{\sqrt{2}}
\end{array}\right)\frac{1}{2i}\left(\begin{array}{cc}
\frac{1}{\boldsymbol{s}_{RL}^{\dagger}}-\frac{1}{\boldsymbol{s}_{RL}} & \boldsymbol{s}_{RR}\frac{1}{\boldsymbol{s}_{LR}}+\frac{1}{\boldsymbol{s}_{RL}}\boldsymbol{s}_{RR}\\
-\boldsymbol{s}_{RR}^{\dagger}\frac{1}{\boldsymbol{s}_{RL}^{\dagger}}-\frac{1}{\boldsymbol{s}_{LR}^{\dagger}}\boldsymbol{s}_{RR}^{\dagger} & \frac{1}{\boldsymbol{s}_{LR}^{\dagger}}-\frac{1}{\boldsymbol{s}_{LR}}
\end{array}\right)\left(\begin{array}{cc}
\frac{1}{\sqrt{2}} & \frac{1}{\sqrt{2}}\\
\frac{1}{\sqrt{2}} & -\frac{1}{\sqrt{2}}
\end{array}\right).
\]
The power flow of Eq. \eqref{eq: periodic mode eq} is given by
\[
\left(\begin{array}{cc}
\frac{\boldsymbol{f}^{\dagger}+\boldsymbol{b}^{\dagger}}{\sqrt{2}} & \frac{\boldsymbol{f}^{\dagger}-\boldsymbol{b}^{\dagger}}{\sqrt{2}}\end{array}\right)\left(\begin{array}{cc}
0 & 1\\
1 & 0
\end{array}\right)\left(\begin{array}{c}
\frac{\boldsymbol{f}+\boldsymbol{b}}{\sqrt{2}}\\
\frac{\boldsymbol{f}-\boldsymbol{b}}{\sqrt{2}}
\end{array}\right)=\left(\begin{array}{cc}
\boldsymbol{f}^{\dagger} & \boldsymbol{b}^{\dagger}\end{array}\right)\left(\begin{array}{cc}
1 & 0\\
0 & -1
\end{array}\right)\left(\begin{array}{c}
\boldsymbol{f}\\
\boldsymbol{b}
\end{array}\right)=\boldsymbol{f}^{\dagger}\boldsymbol{f}-\boldsymbol{b}^{\dagger}\boldsymbol{b}\,.
\]
Note that the redefined mode $\boldsymbol{\Phi}_{\left|m\right|}^{'}$
( $\boldsymbol{\Phi}_{-\left|m\right|}^{'}$ ) of Eq. \eqref{eq: redefined evanescent mode PeriodicWG} are included in $\boldsymbol{f}$
( $\boldsymbol{b}$ ).

%% file: App/C_ShroedingerEq_v1.tex
\chapter{Shr{\"o}dinger equation \label{ch: Shroedinger}}

We will show two cases of quantum mechanics.

\section{Equation for an electron in static electromagnetic field}

Within Pauli approximation, the Shr{\"o}dinger equation of an electron
shows that
\begin{equation}
E\boldsymbol{\psi}=\left[
	\frac{1}{2m_{\mathrm{e}}}\left(
		\frac{\hbar}{i}\boldsymbol{\nabla} + e\boldsymbol{A}
	\right)^{2}
	+ V +
	\frac{e\hbar}{2m_{\mathrm{e}}}\boldsymbol{\sigma}\cdot\boldsymbol{B}
\right]\boldsymbol{\psi},\label{eq: Shroedinger eq}
\end{equation}
where $\boldsymbol{B}=\boldsymbol{\nabla}\times\boldsymbol{A}$ and
wave function
\[
\boldsymbol{\psi}=
\left(\begin{array}{c}
\psi_{\uparrow}\left(\boldsymbol{r}\right)\\
\psi_{\downarrow}\left(\boldsymbol{r}\right)
\end{array}\right).
\]
We will introduce $\mathcal{H}_{xy}$ as $x$ and $y$ components of eq.  (\ref{eq: Shroedinger eq}) as follows.
\[
\mathcal{H}_{xy}  =  \frac{1}{2m_{\mathrm{e}}}\left(
	\frac{\hbar}{i}\frac{\partial}{\partial x}+eA_{x}
\right)^{2}
+ \frac{1}{2m_{\mathrm{e}}}\left(
	\frac{\hbar}{i}\frac{\partial}{\partial y}+eA_{y}
\right)^{2}
+ V +
\frac{e\hbar}{2m_{\mathrm{e}}}\boldsymbol{\sigma}\cdot\boldsymbol{B}\,.
\]
Note that $z \mathcal{H}_{xy} = \mathcal{H}_{xy} z$.
Equation (\ref{eq: Shroedinger eq}) can be modified to
\[
\left(E-\mathcal{H}_{xy}\right)\boldsymbol{\psi}=\frac{m_{\mathrm{e}}}{2}\left(\frac{\hbar}{im_{\mathrm{e}}}\frac{\partial}{\partial z}
+\frac{e}{m_{\mathrm{e}}}A_{z}\right)^{2}\boldsymbol{\psi}\,.
\]
Then, we obtain a propagation equation:
\begin{equation}
\left(\begin{array}{cc}
	\frac{2}{\hbar}\left(E - \mathcal{H}_{xy}\right) & - \frac{e}{\hbar}A_{z}\\
	-\frac{e}{\hbar}A_{z} & \frac{m_{\mathrm{e}}}{\hbar}
\end{array}\right) \boldsymbol{\Psi} =
-i\frac{\partial}{\partial z}
\left(\begin{array}{cc}
	\boldsymbol{0} & \boldsymbol{1}\\
	\boldsymbol{1} & \boldsymbol{0}
\end{array}\right) \boldsymbol{\Psi}\,,\label{eq: case1}
\end{equation}
and $\boldsymbol{\Psi}$ is given by
\[
\boldsymbol{\Psi}=
\sqrt{\frac{dx\,dy}{2}}
\left(\begin{array}{c}
\boldsymbol{\psi}\\
\left(\frac{\hbar}{im_{\mathrm{e}}}\frac{\partial}{\partial z}+\frac{e}{m_{\mathrm{e}}}A_{z}\right)\boldsymbol{\psi}
\end{array}\right).
\]
The $\sqrt{dx\,dy}$ is formally added to the $\boldsymbol{\Psi}$ for integral of the cross section,
 and then numerical discrete formulation does not have it.
The author is grateful to Dr. Motomu Takatsu for his suggestions to eq. (\ref{eq: case1}).

When $A_{z} = 0$, eq. (\ref{eq: case1}) can be reduced to 
\begin{equation}
\left(\begin{array}{cc}
	\frac{2}{\hbar}\left(E - \mathcal{H}_{xy}\right) & 0\\
	0 & \frac{m_{\mathrm{e}}}{\hbar}
\end{array}\right) \boldsymbol{\Psi} =
-i\frac{\partial}{\partial z}
\left(\begin{array}{cc}
	\boldsymbol{0} & \boldsymbol{1}\\
	\boldsymbol{1} & \boldsymbol{0}
\end{array}\right) \boldsymbol{\Psi}\,.\label{eq: Shroedinger eq. Az=0}
\end{equation}

\section{Ando model for 2D system}

Equation (2.6) in \cite{Ando} shows a vector $\boldsymbol{C}_{j}$ that satisfies 
\begin{equation}
\left(E-\mathcal{H}_{0}\right)\boldsymbol{C}_{j}+t\boldsymbol{P}\boldsymbol{C}_{j-1}+t\boldsymbol{P}^{*}\boldsymbol{C}_{j+1}=\boldsymbol{0}\,,\label{eq: Ando model}
\end{equation}
where $\boldsymbol{P}$ is a diagonal matrix and $\boldsymbol{P}\boldsymbol{P}^{*}=1$.
Note that $\mathcal{H}_{j}=\mathcal{H}_{0}$, and we can set that $ \boldsymbol{C}_{j}  =  e^{i\theta j}\boldsymbol{C}_{0} $.
Equation (\ref{eq: Ando model}) can be modified to
\[
e^{i\theta}\boldsymbol{C}_{j}  =  \boldsymbol{P}\frac{\mathcal{H}_{0}-E}{t}\boldsymbol{C}_{j}-\boldsymbol{P}^{2}\boldsymbol{C}_{j-1}
\]
with $ e^{i\theta}\boldsymbol{C}_{j-1}  =  \boldsymbol{C}_{j} $.
The above leads to the following eigenvalue problem:
\begin{eqnarray*}
e^{i\theta}\left(\begin{array}{c}
\boldsymbol{C}_{j}\\
\boldsymbol{C}_{j-1}
\end{array}\right) & = & \left(\begin{array}{cc}
\boldsymbol{P}\frac{\mathcal{H}_{0}-E}{t} & -\boldsymbol{P}^{2}\\
1 & 0
\end{array}\right)\left(\begin{array}{c}
\boldsymbol{C}_{j}\\
\boldsymbol{C}_{j-1}
\end{array}\right).
\end{eqnarray*}
Furthermore, we can show mode equation for $\sin\theta$.
Equation (\ref{eq: Ando model}) can also be modified to
\[
e^{-i\theta}\boldsymbol{C}_{j}=\boldsymbol{P}^{*}\frac{\mathcal{H}_{0}-E}{t}\boldsymbol{C}_{j}-\left(\boldsymbol{P}^{*}\right)^{2}\boldsymbol{C}_{j+1}\,
\]
with $e^{-i\theta}\boldsymbol{C}_{j+1}  =  \boldsymbol{C}_{j}$. 
Then the above leads to another eigenvalue problem:
\begin{eqnarray*}
e^{-i\theta}\left(\begin{array}{c}
\boldsymbol{C}_{j}\\
\boldsymbol{C}_{j-1}
\end{array}\right) & = & \left(\begin{array}{cc}
0 & 1\\
-\left(\boldsymbol{P}^{*}\right)^{2} & \boldsymbol{P}^{*}\frac{\mathcal{H}_{0}-E}{t}
\end{array}\right)\left(\begin{array}{c}
\boldsymbol{C}_{j}\\
\boldsymbol{C}_{j-1}
\end{array}\right).
\end{eqnarray*}
From both eigenvalue problems,
\[  \left(e^{i\theta}-e^{-i\theta}\right)\left(\begin{array}{c}
\boldsymbol{C}_{j}\\
\boldsymbol{C}_{j-1}
\end{array}\right)
  =  \left(\begin{array}{cc}
\boldsymbol{P}\frac{\mathcal{H}_{0}-E}{t} & -1-\boldsymbol{P}^{2}\\
1+\left(\boldsymbol{P}^{*}\right)^{2} & -\boldsymbol{P}^{*}\frac{\mathcal{H}_{0}-E}{t}
\end{array}\right)\left(\begin{array}{c}
\boldsymbol{C}_{j}\\
\boldsymbol{C}_{j-1}
\end{array}\right).
\]
Let us introduce a diagonal matrix $\boldsymbol{Q}$: $i\boldsymbol{P}=\boldsymbol{Q}^{2}$
and $\boldsymbol{Q}\boldsymbol{Q}^{*}=1$. %
 We obtain a mode equation of the Ando model:
\begin{equation}
\left(\begin{array}{cc}
-\mathrm{Re}\boldsymbol{P} & \boldsymbol{Q}^{*}\frac{E-\mathcal{H}_{0}}{2t}\boldsymbol{Q}^{*}\\
\boldsymbol{Q}\frac{E-\mathcal{H}_{0}}{2t}\boldsymbol{Q} & -\mathrm{Re}\boldsymbol{P}
\end{array}\right)\boldsymbol{\Phi}_{n}  =\sin\theta_{n}\left(\begin{array}{cc}
\boldsymbol{0} & \boldsymbol{1}\\
\boldsymbol{1} & \boldsymbol{0}
\end{array}\right)\boldsymbol{\Phi}_{n}\,,
\label{eq: case2}
\end{equation}
where
\[
\boldsymbol{\Phi}_{n}=
\left(\begin{array}{c}
\boldsymbol{\phi}_{an}\\
\boldsymbol{\phi}_{bn}
\end{array}\right)
=\left(\begin{array}{c}
\boldsymbol{Q}^{*}\boldsymbol{C}_{j}\left(n\right)\\
-\boldsymbol{Q}\boldsymbol{C}_{j-1}\left(n\right)
\end{array}\right).
\]

Let us consider a special case that $\boldsymbol{P}=1$, \textit{i.e.} $\boldsymbol{Q}=\exp\left(i\pi/4\right)$.
Equation (\ref{eq: case2}) becomes that
\[
\left(\begin{array}{cc}
-\boldsymbol{1} & -i\boldsymbol{m}\\
i\boldsymbol{m} & -\boldsymbol{1}
\end{array}\right)\boldsymbol{\Phi}_{n}=\sin\theta_{n}\left(\begin{array}{cc}
\boldsymbol{0} & \boldsymbol{1}\\
\boldsymbol{1} & \boldsymbol{0}
\end{array}\right)\boldsymbol{\Phi}_{n}\,,
\]
where $\boldsymbol{m}=\left(E-\mathcal{H}_{0}\right)/\left(2t\right)$. %
Furthermore,
\[
\left(\begin{array}{cc}
\boldsymbol{1}-\boldsymbol{m}\boldsymbol{m} & \boldsymbol{0}\\
\boldsymbol{0} & \boldsymbol{1}-\boldsymbol{m}\boldsymbol{m}
\end{array}\right)\boldsymbol{\Phi}_{n}=\sin^{2}\theta_{n}\boldsymbol{\Phi}_{n}\,.
\]
Then, we can obtain $\boldsymbol{\Phi}_{n}$ by solving
\begin{equation}
\left(\boldsymbol{1}-\boldsymbol{m}\boldsymbol{m}\right)\boldsymbol{\phi}_{an}  
= \sin^{2}\theta_{n}\boldsymbol{\phi}_{an}\,,
\quad
\boldsymbol{\phi}_{bn}  
= \left(i\boldsymbol{m}-\sin\theta_{n}\right)\boldsymbol{\phi}_{an}\,
\label{eq: case2 eigenvalue eq}
\end{equation}
or
\[
\left(\boldsymbol{1}-\boldsymbol{m}\boldsymbol{m}\right)\boldsymbol{\phi}_{bn}  
= \sin^{2}\theta_{n}\boldsymbol{\phi}_{bn}\,,
\quad
\boldsymbol{\phi}_{an}
= -\left(i\boldsymbol{m}+\sin\theta_{n}\right)\boldsymbol{\phi}_{bn}\,.
\]
The eigenvalue $\sin^{2}\theta_{n}$ is always real, since
 $\boldsymbol{m}=\boldsymbol{m}^{\dagger}$. 

%% file: App/D_GeneralMaxwellEq_v2.tex
\chapter{Generalized Maxwell equation\label{ch: Generalized Maxwell}}

This chapter derives propagation equation (\ref{eq: propagation-equation}) for Maxwell equation. 
The Maxwell equation in frequency domain is generalized into a matrix representation:
\begin{equation}
\left(\begin{array}{cc}
\boldsymbol{0} & i\boldsymbol{\nabla}\times\\
-i\boldsymbol{\nabla}\times & \boldsymbol{0}
\end{array}\right)\left(\begin{array}{c}
\boldsymbol{E}\\
\boldsymbol{H}
\end{array}\right)=\omega\left(\begin{array}{cc}
\boldsymbol{\varepsilon} & \boldsymbol{\alpha}\\
\boldsymbol{\gamma} & \boldsymbol{\mu}
\end{array}\right)\left(\begin{array}{c}
\boldsymbol{E}\\
\boldsymbol{H}
\end{array}\right).\label{eq: Generalized Maxwell eq}
\end{equation}
We added small matrices $\boldsymbol{\alpha}$ and $\boldsymbol{\gamma}$ for multiferroics to the right hand of eq. (\ref{eq: Generalized Maxwell eq}). 
The following section will focus on coordinate transformations for rotation operator in the left hand of eq. (\ref{eq: Generalized Maxwell eq}).

\section{Rotation for arbitrary orthogonal curvilinear coordinates\label{sec: deformed rotation}}

\begin{figure}
\begin{centering}
\includegraphics[bb=0bp 0bp 317bp 372bp,clip,width=0.25\columnwidth]{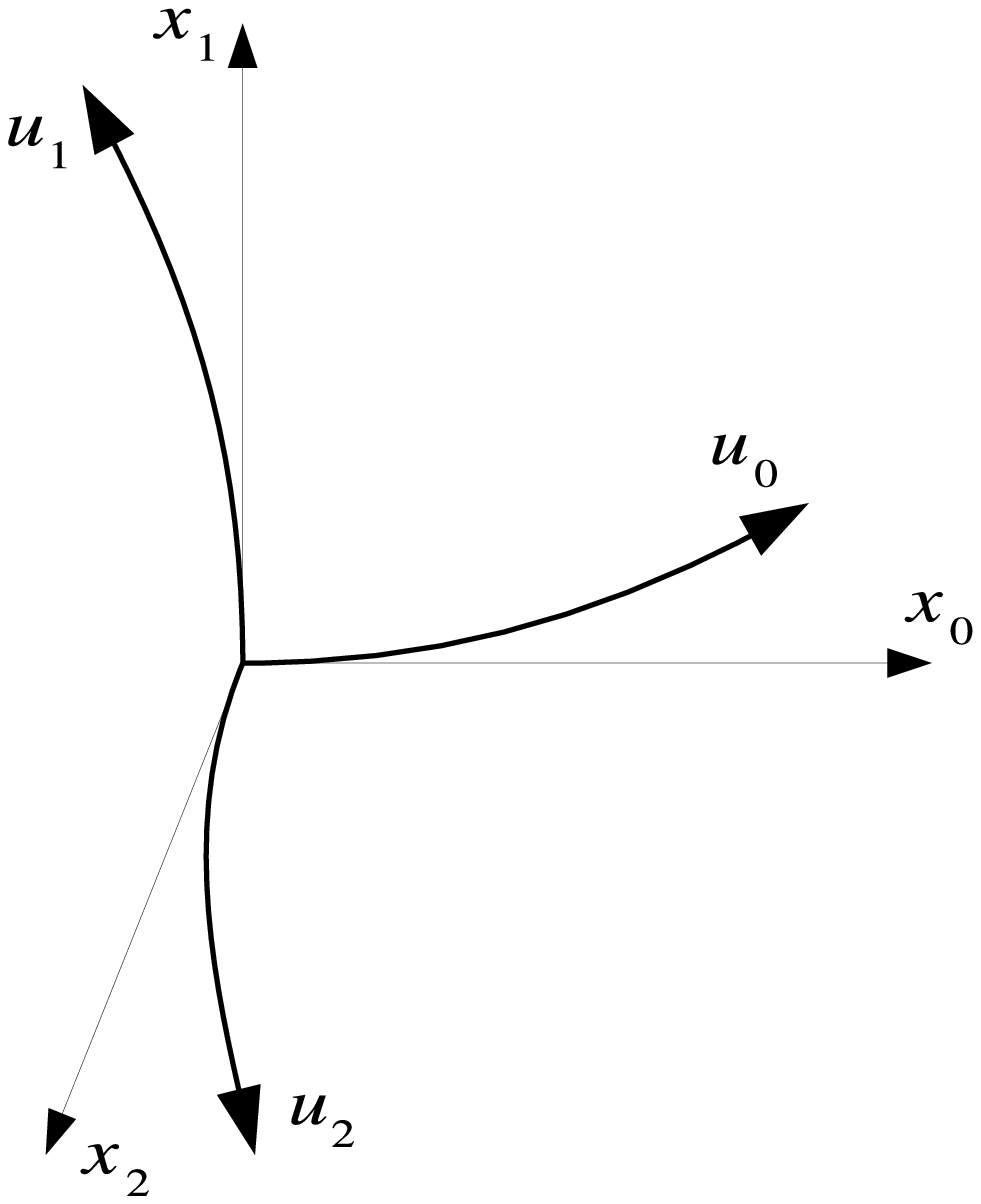}
\par\end{centering}

\caption{Cartesian coordinate $\left(x_{0},\,x_{1},\,x_{2}\right)$ and orthogonal curvilinear coordinate $\left(u_{0},\,u_{1},\,u_{2}\right)$.\label{fig: curvilinear coordinates}}
\end{figure}

We consider Cartesian coordinate $\left(x_{0},\,x_{1},\,x_{2}\right)$ and orthogonal curvilinear coordinate $\left(u_{0},\,u_{1},\,u_{2}\right)$ as shown in Fig. \ref{fig: curvilinear coordinates}. 
 Note that we use non-negative integers $0$, $1$ and $2$ for the coordinate numbers, because modulo operation can be directly applicable to the numbers.

Rotation $\boldsymbol{\nabla}\times$ in an orthogonal curvilinear coordinates is defined by %
\[
\boldsymbol{\nabla}\times\boldsymbol{X}=\sum_{j=0}^{2}\frac{1}{h_{j}h_{k}h_{l}}\left(h_{j}\frac{\partial}{\partial u_{k}}\left(h_{l}X_{l}\right)-h_{j}\frac{\partial}{\partial u_{l}}\left(h_{k}X_{k}\right)\right)\boldsymbol{u}_{j}\,,
\]
where $\boldsymbol{u}_{j}$ is a unit vector in the $\left(u_{0},\,u_{1},\,u_{2}\right)$ space, $k=j+1\:\mathrm{mod\:3}$ and $l=j+2\:\mathrm{mod\:3}$, and $h_{k}$ are scale factors:%
\begin{equation}
h_{k}=\sqrt{\sum_{j=0}^{2}\left(\frac{\partial x_{j}}{\partial u_{k}}\right)^{2}}\quad\mathrm{for}\quad k=0,1,2.\label{eq: scale factor}
\end{equation}
Matrix representation of $\boldsymbol{\nabla}\times$ as in eq. (4) of \cite{Chen} 
 is that
\begin{equation}
\boldsymbol{\nabla}\times = \frac{1}{\prod_{j=0}^{2}h_{j}}\boldsymbol{f}_{u}\left(\boldsymbol{\nabla}_{u}\times\right)\boldsymbol{f}_{u}\,,
\label{eq: modified rotation}
\end{equation}
where
\[
\boldsymbol{f}_{u}  \triangleq 
\left(\begin{array}{ccc}
h_{0} & 0 & 0\\
0 & h_{1} & 0\\
0 & 0 & h_{2}
\end{array}\right),
\quad
\boldsymbol{\nabla}_{u}\times  \triangleq
\left(\begin{array}{ccc}
0 & -\partial_{u2} & \partial_{u1}\\
\partial_{u2} & 0 & -\partial_{u0}\\
-\partial_{u1} & \partial_{u0} & 0
\end{array}\right)
\quad \mathrm{as} \quad
\partial_{u j} = \frac{\partial}{\partial u_{j}}\,.
\]
Maxwell equation of eq. (\ref{eq: Generalized Maxwell eq}) is deformed by eq. (\ref{eq: modified rotation}) into 
\begin{equation}
\left(\begin{array}{cc}
0 & i\boldsymbol{\nabla}_{u}\times\\
-i\boldsymbol{\nabla}_{u}\times & 0
\end{array}\right)
\left(\begin{array}{c}
\boldsymbol{f}_{u}\boldsymbol{E}\\
\boldsymbol{f}_{u}\boldsymbol{H}
\end{array}\right)
=\omega\left(\begin{array}{cc}
\tilde{\boldsymbol{\varepsilon}} & \tilde{\boldsymbol{\alpha}}\\
\tilde{\boldsymbol{\gamma}} & \tilde{\boldsymbol{\mu}}
\end{array}\right)
\left(\begin{array}{c}
\boldsymbol{f}_{u}\boldsymbol{E}\\
\boldsymbol{f}_{u}\boldsymbol{H}
\end{array}\right),
\label{eq: Maxwell_tilde}
\end{equation}
where
\[
\tilde{\varepsilon}_{jk} \triangleq h_{0}h_{1}h_{2}\frac{\varepsilon_{jk}}{h_{j}h_{k}}
\,,\quad
\tilde{\alpha}_{jk} \triangleq h_{0}h_{1}h_{2}\frac{\alpha_{jk}}{h_{j}h_{k}}
\,,\quad
\tilde{\gamma}_{jk} \triangleq h_{0}h_{1}h_{2}\frac{\gamma_{jk}}{h_{j}h_{k}}
\,,\quad
\tilde{\mu}_{jk} \triangleq h_{0}h_{1}h_{2}\frac{\mu_{jk}}{h_{j}h_{k}}
\,.
\]

\section{Details of deformed Maxwell equation }

We will focus on a case: $\tilde\varepsilon_{22}\tilde\mu_{22}\neq\tilde\alpha_{22}\tilde\gamma_{22}$.
As shown in Section \ref{sec: check l u}, transformed $\hat{\boldsymbol{\varepsilon}}$, $\hat{\boldsymbol{\alpha}}$,
$\hat{\boldsymbol{\gamma}}$ and $\hat{\boldsymbol{\mu}}$ can be
defined as 
\[
\left(\begin{array}{cc}
\hat{\boldsymbol{\varepsilon}} & \hat{\boldsymbol{\alpha}}\\
\hat{\boldsymbol{\gamma}} & \hat{\boldsymbol{\mu}}
\end{array}\right)=\left(1-\boldsymbol{u}_{\varepsilon}-\boldsymbol{u}_{\mu}\right)\left(\begin{array}{cc}
\tilde{\boldsymbol{\varepsilon}} & \tilde{\boldsymbol{\alpha}}\\
\tilde{\boldsymbol{\gamma}} & \tilde{\boldsymbol{\mu}}
\end{array}\right)\left(1-\boldsymbol{l}_{\varepsilon}-\boldsymbol{l}_{\mu}\right),
\]
where %
\begin{equation}
1-\boldsymbol{l}_{\varepsilon}-\boldsymbol{l}_{\mu}
=
\left(\begin{array}{cccccc}
1 & 0 & 0 & 0 & 0 & 0\\
0 & 1 & 0 & 0 & 0 & 0\\
-l_{\varepsilon0} & -l_{\varepsilon1} & 1 & -l_{\varepsilon3} & -l_{\varepsilon4} & 0\\
0 & 0 & 0 & 1 & 0 & 0\\
0 & 0 & 0 & 0 & 1 & 0\\
-l_{\mu0} & -l_{\mu1} & 0 & -l_{\mu3} & -l_{\mu4} & 1
\end{array}\right) \quad \mathrm{as} \quad
\left\{
\begin{aligned}l_{\varepsilon n} & =\frac{\tilde\mu_{22}\tilde\varepsilon_{2n}-\tilde\alpha_{22}\tilde\gamma_{2n}}{\tilde\varepsilon_{22}\tilde\mu_{22}-\tilde\alpha_{22}\tilde\gamma_{22}}\,,\\
l_{\mu n} & =\frac{\tilde\varepsilon_{22}\tilde\gamma_{2n}-\tilde\gamma_{22}\tilde\varepsilon_{2n}}{\tilde\varepsilon_{22}\tilde\mu_{22}-\tilde\alpha_{22}\tilde\gamma_{22}}\,,\\
l_{\varepsilon n+3} & =\frac{\tilde\mu_{22}\tilde\alpha_{2n}-\tilde\alpha_{22}\tilde\mu_{2n}}{\tilde\varepsilon_{22}\tilde\mu_{22}-\tilde\alpha_{22}\tilde\gamma_{22}}\,,\\
l_{\mu n+3} & =\frac{\tilde\varepsilon_{22}\tilde\mu_{2n}-\tilde\gamma_{22}\tilde\alpha_{2n}}{\tilde\varepsilon_{22}\tilde\mu_{22}-\tilde\alpha_{22}\tilde\gamma_{22}}\,,
\end{aligned}
\right.\label{eq: l-elements}
\end{equation}
and
\begin{equation}
1-\boldsymbol{u}_{\varepsilon}-\boldsymbol{u}_{\mu}=\left(\begin{array}{cccccc}
1 & 0 & -u_{\varepsilon0} & 0 & 0 & -u_{\mu0}\\
0 & 1 & -u_{\varepsilon1} & 0 & 0 & -u_{\mu1}\\
0 & 0 & 1 & 0 & 0 & 0\\
0 & 0 & -u_{\varepsilon3} & 1 & 0 & -u_{\mu3}\\
0 & 0 & -u_{\varepsilon4} & 0 & 1 & -u_{\mu4}\\
0 & 0 & 0 & 0 & 0 & 1
\end{array}\right) \quad \mathrm{as} \quad
\left\{
\begin{aligned}u_{\varepsilon n} & =\frac{\tilde\mu_{22}\tilde\varepsilon_{n2}-\tilde\gamma_{22}\tilde\alpha_{n2}}{\tilde\varepsilon_{22}\tilde\mu_{22}-\tilde\alpha_{22}\tilde\gamma_{22}}\,,\\
u_{\mu n} & =\frac{\tilde\varepsilon_{22}\tilde\alpha_{n2}-\tilde\alpha_{22}\tilde\varepsilon_{n2}}{\tilde\varepsilon_{22}\tilde\mu_{22}-\tilde\alpha_{22}\tilde\gamma_{22}}\,,\\
u_{\varepsilon n+3} & =\frac{\tilde\mu_{22}\tilde\gamma_{n2}-\tilde\gamma_{22}\tilde\mu_{n2}}{\tilde\varepsilon_{22}\tilde\mu_{22}-\tilde\alpha_{22}\tilde\gamma_{22}}\,,\\
u_{\mu n+3} & =\frac{\tilde\varepsilon_{22}\tilde\mu_{n2}-\tilde\alpha_{22}\tilde\gamma_{n2}}{\tilde\varepsilon_{22}\tilde\mu_{22}-\tilde\alpha_{22}\tilde\gamma_{22}}\,.
\end{aligned}
\right.\label{eq: u-elements}
\end{equation}
The $\hat{\boldsymbol{\varepsilon}}$, $\hat{\boldsymbol{\alpha}}$,
$\hat{\boldsymbol{\gamma}}$ and $\hat{\boldsymbol{\mu}}$ are that
\[
\hat{\boldsymbol{\varepsilon}}=\left(\begin{array}{ccc}
\hat{\varepsilon}_{00} & \hat{\varepsilon}_{01} & 0\\
\hat{\varepsilon}_{10} & \hat{\varepsilon}_{11} & 0\\
0 & 0 & \tilde\varepsilon_{22}
\end{array}\right),\:\hat{\boldsymbol{\alpha}}=\left(\begin{array}{ccc}
\hat{\alpha}_{00} & \hat{\alpha}_{01} & 0\\
\hat{\alpha}_{10} & \hat{\alpha}_{11} & 0\\
0 & 0 & \tilde\alpha_{22}
\end{array}\right),\:\hat{\boldsymbol{\gamma}}=\left(\begin{array}{ccc}
\hat{\gamma}_{00} & \hat{\gamma}_{01} & 0\\
\hat{\gamma}_{10} & \hat{\gamma}_{11} & 0\\
0 & 0 & \tilde\gamma_{22}
\end{array}\right),\:\hat{\boldsymbol{\mu}}=\left(\begin{array}{ccc}
\hat{\mu}_{00} & \hat{\mu}_{01} & 0\\
\hat{\mu}_{10} & \hat{\mu}_{11} & 0\\
0 & 0 & \tilde\mu_{22}
\end{array}\right),
\]
where
\begin{equation}\left\{
\begin{aligned}\hat{\varepsilon}_{mn} & =\tilde\varepsilon_{mn}-\tilde\varepsilon_{m2}l_{\varepsilon n}-\tilde\alpha_{m2}l_{\mu n}\,,\\
\hat{\alpha}_{mn} & =\tilde\alpha_{mn}-\tilde\varepsilon_{m2}l_{\varepsilon n+3}-\tilde\alpha_{m2}l_{\mu n+3}\,,\\
\hat{\gamma}_{mn} & =\tilde\gamma_{mn}-\tilde\gamma_{m2}l_{\varepsilon n}-\tilde\mu_{m2}l_{\mu n}\,,\\
\hat{\mu}_{mn} & =\tilde\mu_{mn}-\tilde\gamma_{m2}l_{\varepsilon n+3}-\tilde\mu_{m2}l_{\mu n+3}\,,
\end{aligned}
\right.\label{eq: trans. epsilon}
\end{equation}
for $m,\,n=0,\,1$. The $1-\boldsymbol{l}_{\varepsilon}-\boldsymbol{l}_{\mu}$
and $1+\boldsymbol{l}_{\varepsilon}+\boldsymbol{l}_{\mu}$ have the following
relation by eq. (\ref{eq: square of l_e + l_mu}):%
\begin{equation}
\left(\boldsymbol{1}-\boldsymbol{l}_{\varepsilon}-\boldsymbol{l}_{\mu}\right)\left(\boldsymbol{1}+\boldsymbol{l}_{\varepsilon}+\boldsymbol{l}_{\mu}\right)= 1\,.
\label{eq: 1-l-l 1+l+l}
\end{equation}
Then, equation (\ref{eq: Maxwell_tilde}) is transformed to 
\begin{equation}
\left(1-\boldsymbol{u}_{\varepsilon}-\boldsymbol{u}_{\mu}\right)\left(\begin{array}{cc}
0 & i\boldsymbol{\nabla}_{u}\times\\
-i\boldsymbol{\nabla}_{u}\times & 0
\end{array}\right)\left(1-\boldsymbol{l}_{\varepsilon}-\boldsymbol{l}_{\mu}\right)\left(\begin{array}{c}
\hat{\boldsymbol{E}}\\
\hat{\boldsymbol{H}}
\end{array}\right)=\omega\left(\begin{array}{cc}
\hat{\boldsymbol{\varepsilon}} & \hat{\boldsymbol{\alpha}}\\
\hat{\boldsymbol{\gamma}} & \hat{\boldsymbol{\mu}}
\end{array}\right)\left(\begin{array}{c}
\hat{\boldsymbol{E}}\\
\hat{\boldsymbol{H}}
\end{array}\right),
\label{eq: Maxwell_tilde2}
\end{equation}
where %
\[\left\{
\begin{aligned}
\left(\begin{array}{c}
\hat{\boldsymbol{E}}\\
\hat{\boldsymbol{H}}
\end{array}\right) & \triangleq \left(\boldsymbol{1}+\boldsymbol{l}_{\varepsilon}+\boldsymbol{l}_{\mu}\right)\left(\begin{array}{c}
\boldsymbol{f}_{u}\boldsymbol{E}\\
\boldsymbol{f}_{u}\boldsymbol{H}
\end{array}\right)=\left(\begin{array}{cccccc}
h_{0}{E}_{0} & h_{1}{E}_{1} & \hat{E}_{2} & h_{0}{H}_{0} & h_{1}{H}_{1} & \hat{H}_{2}\end{array}\right)^{\mathrm{T}},\\
\hat{E}_{2} & \triangleq  h_{2}{E}_{2}+\sum_{n=0}^{1}\left(l_{\varepsilon n}h_{n}{E}_{n}+l_{\varepsilon n+3}h_{n}{H}_{n}\right),\\
\hat{H}_{2} & \triangleq  h_{2}{H}_{2}+\sum_{n=0}^{1}\left(l_{\mu n}h_{n}{E}_{n}+l_{\mu n+3}h_{n}{H}_{n}\right).
\end{aligned}
\right.\]
From eq. (\ref{eq: hat E2 hat H2}) and the above definitions of $\hat{E}_{2}$ and $\hat{H}_{2}$, the component of $u_{2}$ can be represented by other components:
\begin{equation}
\begin{split}\left(\begin{array}{c}
h_{2}{E}_{2}\\
h_{2}{H}_{2}
\end{array}\right)= & \left[\frac{i c^{2}_{22}}{\omega}\left(\begin{array}{cc}
-\tilde\alpha_{22}\partial_{u1} & \tilde\alpha_{22}\partial_{u0}\\
\tilde\varepsilon_{22}\partial_{u1} & -\tilde\varepsilon_{22}\partial_{u0}
\end{array}\right)-\left(\begin{array}{cc}
l_{\varepsilon0} & l_{\varepsilon1}\\
l_{\mu0} & l_{\mu1}
\end{array}\right)\right]\left(\begin{array}{c}
h_{0}{E}_{0}\\
h_{1}{E}_{1}
\end{array}\right)\\
 & +\left[\frac{i c^{2}_{22}}{\omega}\left(\begin{array}{cc}
-\tilde\mu_{22}\partial_{u1} & \tilde\mu_{22}\partial_{u0}\\
\tilde\gamma_{22}\partial_{u1} & -\tilde\gamma_{22}\partial_{u0}
\end{array}\right)-\left(\begin{array}{cc}
l_{\varepsilon3} & l_{\varepsilon4}\\
l_{\mu3} & l_{\mu4}
\end{array}\right)\right]\left(\begin{array}{c}
h_{0}{H}_{0}\\
h_{1}{H}_{1}
\end{array}\right)
\end{split}
\label{eq: tilde z elements}
\end{equation}
with
$
c^{-2}_{22}\triangleq\tilde\varepsilon_{22}\tilde\mu_{22}-\tilde\alpha_{22}\tilde\gamma_{22}
$.
Let us introduce column vector $\boldsymbol{\Psi}$ of four components: %
\begin{equation}
\boldsymbol{\Psi} =
\frac{\sqrt{du_{0}du_{1}}}{2}
\left(\begin{array}{cccc}
h_{0}H_{0} & h_{1}H_{1} & -h_{1}E_{1} & h_{0}E_{0}\end{array}\right)^{\mathrm{T}}.
\label{eq: Psi for Maxwell}
\end{equation}
The $\sqrt{du_{0}du_{1}}/2$ is formally added to the $\boldsymbol{\Psi}$,
 and then numerical discrete formulation in eq. (\ref{eq: discrete m_aa m_bb Psi}) does not have it.
By using the $\boldsymbol{\Psi}$ of eq. (\ref{eq: Psi for Maxwell}), we can simplify the generalized Maxwell equation (\ref{eq: Generalized Maxwell eq}) into the propagation equation (\ref{eq: propagation-equation}).
As shown in eqs. (\ref{eq: check m_bb m_ba}) and (\ref{eq: check m_ab m_aa}), the $\boldsymbol{M}$ of eq. (\ref{eq: propagation-equation}) is given by
\begin{equation}\left\{
\begin{aligned}
\boldsymbol{M} & =
\left(\begin{array}{cc} \boldsymbol{m}_{aa} & \boldsymbol{m}_{ab}\\ \boldsymbol{m}_{ba} & \boldsymbol{m}_{bb} \end{array}\right)\,,
\\
{\boldsymbol{m}}_{aa} & = \left(\begin{array}{cc}
\omega\hat{\mu}_{00} + \partial_{u1}\frac{\tilde\mu_{22} c^{2}_{22}}{\omega}\partial_{u1} - i\partial_{u1}l_{\varepsilon3} - iu_{\varepsilon3}\partial_{u1} & \omega\hat{\mu}_{01}-\partial_{u1}\frac{\tilde\mu_{22} c^{2}_{22}}{\omega}\partial_{u0}-i\partial_{u1}l_{\varepsilon4}+iu_{\varepsilon3}\partial_{u0}\\
\omega\hat{\mu}_{10} - \partial_{u0}\frac{\tilde\mu_{22} c^{2}_{22}}{\omega}\partial_{u1} + i\partial_{u0}l_{\varepsilon3} - iu_{\varepsilon4}\partial_{u1} & \omega\hat{\mu}_{11} + \partial_{u0}\frac{\tilde\mu_{22} c^{2}_{22}}{\omega}\partial_{u0} + i\partial_{u0}l_{\varepsilon4} + iu_{\varepsilon4}\partial_{u0}
\end{array}\right),
\\
{\boldsymbol{m}}_{ab} & = \left(\begin{array}{cc}
\partial_{u1}\frac{\tilde\alpha_{22} c^{2}_{22}}{\omega}\partial_{u0} - \omega\hat{\gamma}_{01} + i\partial_{u1}l_{\varepsilon1} +i u_{\mu3}\partial_{u0} & \omega\hat{\gamma}_{00} + \partial_{u1}\frac{\tilde\alpha_{22} c^{2}_{22}}{\omega}\partial_{u1} - i\partial_{u1}l_{\varepsilon0} + iu_{\mu3}\partial_{u1}\\
-\partial_{u0}\frac{\tilde\alpha_{22} c^{2}_{22}}{\omega}\partial_{u0} - \omega\hat{\gamma}_{11} - i\partial_{u0}l_{\varepsilon1} + iu_{\mu4}\partial_{u0} & \omega\hat{\gamma}_{10} - \partial_{u0}\frac{\tilde\alpha_{22} c^{2}_{22}}{\omega}\partial_{u1} + i\partial_{u0}l_{\varepsilon0} + iu_{\mu4}\partial_{u1}
\end{array}\right),
\\
{\boldsymbol{m}}_{ba} & = \left(\begin{array}{cc}
\partial_{u0}\frac{\tilde\gamma_{22} c^{2}_{22}}{\omega}\partial_{u1} - \omega\hat{\alpha}_{10} + i\partial_{u0}l_{\mu3} + iu_{\varepsilon1}\partial_{u1} & -\partial_{u0}\frac{\tilde\gamma_{22} c^{2}_{22}}{\omega}\partial_{u0} - \omega\hat{\alpha}_{11} + i\partial_{u0}l_{\mu4} - iu_{\varepsilon1}\partial_{u0}\\
\omega\hat{\alpha}_{00} + \partial_{u1}\frac{\tilde\gamma_{22} c^{2}_{22}}{\omega}\partial_{u1} + i\partial_{u1}l_{\mu3} - iu_{\varepsilon0}\partial_{u1} & \omega\hat{\alpha}_{01}-\partial_{u1}\frac{\tilde\gamma_{22} c^{2}_{22}}{\omega}\partial_{u0}+i\partial_{u1}l_{\mu4}+iu_{\varepsilon0}\partial_{u0}
\end{array}\right),
\\
{\boldsymbol{m}}_{bb} & = \left(\begin{array}{cc}
\omega\hat{\varepsilon}_{11} + \partial_{u0}\frac{\tilde\varepsilon_{22} c^{2}_{22}}{\omega}\partial_{u0} - i\partial_{u0}l_{\mu1} - iu_{\mu1}\partial_{u0} & \partial_{u0}\frac{\tilde\varepsilon_{22} c^{2}_{22}}{\omega}\partial_{u1} - \omega\hat{\varepsilon}_{10} + i\partial_{u0}l_{\mu0} - iu_{\mu1}\partial_{u1}\\
\partial_{u1}\frac{\tilde\varepsilon_{22} c^{2}_{22}}{\omega}\partial_{u0}-\omega\hat{\varepsilon}_{01}-i\partial_{u1}l_{\mu1}+iu_{\mu0}\partial_{u0} & \omega\hat{\varepsilon}_{00} + \partial_{u1}\frac{\tilde\varepsilon_{22} c^{2}_{22}}{\omega}\partial_{u1} + i\partial_{u1}l_{\mu0} + iu_{\mu0}\partial_{u1}
\end{array}\right)
\end{aligned} 
\right.\label{eq: inv_h m inv_h}
\end{equation}
with
$
c^{2}_{22} = \left(\tilde\varepsilon_{22}\tilde\mu_{22}-\tilde\alpha_{22}\tilde\gamma_{22}\right)^{-1}
$.
Note that $\boldsymbol{M}=\boldsymbol{M}^{\dagger}$ when $\boldsymbol{\varepsilon} = \boldsymbol{\varepsilon}^{\dagger}$, $\boldsymbol{\mu} = \boldsymbol{\mu}^{\dagger}$, $\boldsymbol{\gamma} = \boldsymbol{\alpha}^{\dagger}$ and $\partial_{uj}=-\partial_{uj}^{\dagger}$ as the system satisfies boundary condition as eq. (\ref{eq: boundary condition}).

\section{Special case}

This section consider a case that $\boldsymbol{\varepsilon}$ and $\boldsymbol{\mu}$ are diagonal hermitian, and $\boldsymbol{\alpha}=\boldsymbol{\gamma}=0$.
Note that $l_{\mu n} = u_{\mu n} = l_{\varepsilon n} = u_{\varepsilon n} = 0$ as $n=0,1,3,4$ from eqs. (\ref{eq: l-elements}) and (\ref{eq: u-elements}).
Therefore $\boldsymbol{m}_{ab}=\boldsymbol{m}_{ba}=0$ in eq. (\ref{eq: inv_h m inv_h}), and the $\boldsymbol{M}$ in eq. (\ref{eq: propagation-equation}) can be reduced to
\begin{equation}
\boldsymbol{M}  = 
\left(\begin{array}{cc}
\boldsymbol{m}_{aa} & \boldsymbol{0}\\
\boldsymbol{0} & \boldsymbol{m}_{bb}
\end{array}\right)\quad \mathrm{as} \quad
\left\{\begin{aligned}
\boldsymbol{m}_{aa} & = \left(\begin{array}{cc}
\omega\hat{\mu}_{00} + \partial_{u1}\frac{1}{\omega\tilde\varepsilon_{22}}\partial_{u1} & -\partial_{u1}\frac{1}{\omega\tilde\varepsilon_{22}}\partial_{u0}\\
 - \partial_{u0}\frac{1}{\omega\tilde\varepsilon_{22}}\partial_{u1} & \omega\hat{\mu}_{11} + \partial_{u0}\frac{1}{\omega\tilde\varepsilon_{22}}\partial_{u0}
\end{array}\right)
\\ &
 =\left(\begin{array}{cc}
\frac{h_{1}h_{2}\omega \mu_{00}}{h_{0}} + \partial_{u1}\frac{h_{2}}{h_{0}h_{1} \omega \varepsilon_{22}}\partial_{u1} &
 -\partial_{u1}\frac{h_{2}}{h_{0}h_{1} \omega \varepsilon_{22}}\partial_{u0}\\
 -\partial_{u0}\frac{h_{2}}{h_{0}h_{1} \omega \varepsilon_{22}}\partial_{u1} &
 \frac{h_{0}h_{2}\omega \mu_{11}}{h_{1}} + \partial_{u0}\frac{h_{2}}{h_{0}h_{1} \omega \varepsilon_{22}}\partial_{u0}
\end{array}\right),
\\
\boldsymbol{m}_{bb} & =\left(\begin{array}{cc}
\omega\hat{\varepsilon}_{11} + \partial_{u0}\frac{1}{\omega\tilde\mu_{22}}\partial_{u0} & \partial_{u0}\frac{1}{\omega\tilde\mu_{22}}\partial_{u1}\\
\partial_{u1}\frac{1}{\omega\tilde\mu_{22}}\partial_{u0} & \omega\hat{\varepsilon}_{00} + \partial_{u1}\frac{1}{\omega\tilde\mu_{22}}\partial_{u1}
\end{array}\right)
\\
& = \left(\begin{array}{cc}
\frac{h_{0}h_{2}\omega \varepsilon_{11}}{h_{1}} + \partial_{0}\frac{h_{2}}{h_{0}h_{1} \omega \mu_{22}}\partial_{0} &
 \partial_{0}\frac{h_{2}}{h_{0}h_{1} \omega \mu_{22}}\partial_{1}\\
\partial_{1}\frac{h_{2}}{h_{0}h_{1} \omega \mu_{22}}\partial_{0} &
 \frac{h_{1}h_{2}\omega \varepsilon_{00}}{h_{0}} + \partial_{1}\frac{h_{2}}{h_{0}h_{1}\omega \mu_{22}}\partial_{1}
\end{array}\right)\,.
\end{aligned}\right.\label{eq: normal Maxwell equation}
\end{equation}
Equation (\ref{eq: normal Maxwell equation}) becomes the same as eqs. (3.2-9) and (3.2-10) of \cite{Marcuse} when $h_{0}=h_{1}=h_{2}=1$, $\boldsymbol{\varepsilon}=\varepsilon_{0}$ and $\boldsymbol{\mu}=\mu_{0}$. 

\section{Permittivity with damping}

Permittivity with damping becomes complex number.
Then we will check it by considering response of polarization.

Orientation polarization is represented by
$P_{o}\left(t\right)=\chi_{o}E\left(t\right)-\tau\frac{dP_{o}}{dt}$.
Then,
\[
\begin{split}P_{o}\left(\omega\right) & =\frac{\chi_{o}E\left(\omega\right)}{1-i\omega\tau}=\frac{1+i\omega\tau}{1+\omega^{2}\tau^{2}}\chi_{o}E\left(\omega\right).\end{split}
\]

Displacement polarization: $m\frac{d^{2}x}{dt^{2}}=-m\omega_{0}^{2}x\left(t\right)-m\omega_{1}\frac{dx}{dt}+qE\left(t\right)$.
Then, 
\[
x\left(\omega\right)=\frac{q}{m}\frac{\omega_{0}^{2}-\omega^{2}+i\omega_{1}\omega}{\left(\omega_{0}^{2}-\omega^{2}\right)^{2}+\omega_{1}^{2}\omega^{2}}E\left(\omega\right).
\]
Electron polarization is a special case of the displacement polarization:
$\omega_{0}^{2}=0$.

Therefore, imaginary part of permittivity becomes positive for damped case.
Note that time dependence $\exp\left(-i\omega t\right)$ is assumed.

%% file: App/Dsub_CheckGenMaxEq_v1.tex
\section{ Check of eqs. (\ref{eq: l-elements}), (\ref{eq: u-elements}), (\ref{eq: trans. epsilon}) and (\ref{eq: 1-l-l 1+l+l}) \label{sec: check l u}}

\[
\begin{split}
&\quad\left(\begin{array}{cc}
\hat{\boldsymbol{\varepsilon}} & \hat{\boldsymbol{\alpha}}\\
\hat{\boldsymbol{\gamma}} & \hat{\boldsymbol{\mu}}
\end{array}\right)
=
\left(1-\boldsymbol{u}_{\varepsilon}-\boldsymbol{u}_{\mu}\right)\left(\begin{array}{cc}
\tilde{\boldsymbol{\varepsilon}} & \tilde{\boldsymbol{\alpha}}\\
\tilde{\boldsymbol{\gamma}} & \tilde{\boldsymbol{\mu}}
\end{array}\right)\left(1-\boldsymbol{l}_{\varepsilon}-\boldsymbol{l}_{\mu}\right)
\\
&=\left(1-\boldsymbol{u}_{\varepsilon}-\boldsymbol{u}_{\mu}\right)
\left(\begin{array}{cccccc}
\tilde\varepsilon_{00} & \tilde\varepsilon_{01} & \tilde\varepsilon_{02} & \tilde\alpha_{00} & \tilde\alpha_{01} & \tilde\alpha_{02}\\
\tilde\varepsilon_{10} & \tilde\varepsilon_{11} & \tilde\varepsilon_{12} & \tilde\alpha_{10} & \tilde\alpha_{11} & \tilde\alpha_{12}\\
\tilde\varepsilon_{20} & \tilde\varepsilon_{21} & \tilde\varepsilon_{22} & \tilde\alpha_{20} & \tilde\alpha_{21} & \tilde\alpha_{22}\\
\tilde\gamma_{00} & \tilde\gamma_{01} & \tilde\gamma_{02} & \tilde\mu_{00} & \tilde\mu_{01} & \tilde\mu_{02}\\
\tilde\gamma_{10} & \tilde\gamma_{11} & \tilde\gamma_{12} & \tilde\mu_{10} & \tilde\mu_{11} & \tilde\mu_{12}\\
\tilde\gamma_{20} & \tilde\gamma_{21} & \tilde\gamma_{22} & \tilde\mu_{20} & \tilde\mu_{21} & \tilde\mu_{22}
\end{array}\right)
\left(\begin{array}{cccccc}
1 & 0 & 0 & 0 & 0 & 0\\
0 & 1 & 0 & 0 & 0 & 0\\
-l_{\varepsilon0} & -l_{\varepsilon1} & 1 & -l_{\varepsilon3} & -l_{\varepsilon4} & 0\\
0 & 0 & 0 & 1 & 0 & 0\\
0 & 0 & 0 & 0 & 1 & 0\\
-l_{\mu0} & -l_{\mu1} & 0 & -l_{\mu3} & -l_{\mu4} & 1
\end{array}\right)
\\
&=\left(1-\boldsymbol{u}_{\varepsilon}-\boldsymbol{u}_{\mu}\right)
\\
&\quad\times
\left(\begin{array}{cccccc}
\tilde\varepsilon_{00} - \tilde\varepsilon_{02}l_{\varepsilon0} - \tilde\alpha_{02}l_{\mu0} & 
\tilde\varepsilon_{01} - \tilde\varepsilon_{02}l_{\varepsilon1} - \tilde\alpha_{02}l_{\mu1} & 
\tilde\varepsilon_{02} & 
\tilde\alpha_{00} - \tilde\varepsilon_{02}l_{\varepsilon3} - \tilde\alpha_{02}l_{\mu3} & 
\tilde\alpha_{01} - \tilde\varepsilon_{02}l_{\varepsilon4} - \tilde\alpha_{02}l_{\mu4} & 
\tilde\alpha_{02}
\\
\tilde\varepsilon_{10} - \tilde\varepsilon_{12}l_{\varepsilon0} - \tilde\alpha_{12}l_{\mu0} & 
\tilde\varepsilon_{11} - \tilde\varepsilon_{12}l_{\varepsilon1} - \tilde\alpha_{12}l_{\mu1} & 
\tilde\varepsilon_{12} & 
\tilde\alpha_{10} - \tilde\varepsilon_{12}l_{\varepsilon3} - \tilde\alpha_{12}l_{\mu3} & 
\tilde\alpha_{11} - \tilde\varepsilon_{12}l_{\varepsilon4} - \tilde\alpha_{12}l_{\mu4} & 
\tilde\alpha_{12}
\\
\tilde\varepsilon_{20} - \tilde\varepsilon_{22}l_{\varepsilon0} - \tilde\alpha_{22}l_{\mu0} & 
\tilde\varepsilon_{21} - \tilde\varepsilon_{22}l_{\varepsilon1} - \tilde\alpha_{22}l_{\mu1} & 
\tilde\varepsilon_{22} & 
\tilde\alpha_{20} - \tilde\varepsilon_{22}l_{\varepsilon3} - \tilde\alpha_{22}l_{\mu3} & 
\tilde\alpha_{21} - \tilde\varepsilon_{22}l_{\varepsilon4} - \tilde\alpha_{22}l_{\mu4} & 
\tilde\alpha_{22}
\\
\tilde\gamma_{00} - \tilde\gamma_{02}l_{\varepsilon0} - \tilde\mu_{02}l_{\mu0} & 
\tilde\gamma_{01} - \tilde\gamma_{02}l_{\varepsilon1} - \tilde\mu_{02}l_{\mu1} & 
\tilde\gamma_{02} & 
\tilde\mu_{00} - \tilde\gamma_{02}l_{\varepsilon3} - \tilde\mu_{02}l_{\mu3} & 
\tilde\mu_{01} - \tilde\gamma_{02}l_{\varepsilon4} - \tilde\mu_{02}l_{\mu4} & 
\tilde\mu_{02}
\\
\tilde\gamma_{10} - \tilde\gamma_{12}l_{\varepsilon0} - \tilde\mu_{12}l_{\mu0} & 
\tilde\gamma_{11} - \tilde\gamma_{12}l_{\varepsilon1} - \tilde\mu_{12}l_{\mu1} & 
\tilde\gamma_{12} & 
\tilde\mu_{10} - \tilde\gamma_{12}l_{\varepsilon3} - \tilde\mu_{12}l_{\mu3} & 
\tilde\mu_{11} - \tilde\gamma_{12}l_{\varepsilon4} - \tilde\mu_{12}l_{\mu4} & 
\tilde\mu_{12}
\\
\tilde\gamma_{20} - \tilde\gamma_{22}l_{\varepsilon0} - \tilde\mu_{22}l_{\mu0} & 
\tilde\gamma_{21} - \tilde\gamma_{22}l_{\varepsilon1} - \tilde\mu_{22}l_{\mu1} & 
\tilde\gamma_{22} & 
\tilde\mu_{20} - \tilde\gamma_{22}l_{\varepsilon3} - \tilde\mu_{22}l_{\mu3} & 
\tilde\mu_{21} - \tilde\gamma_{22}l_{\varepsilon4} - \tilde\mu_{22}l_{\mu4} & 
\tilde\mu_{22}
\end{array}\right).
\end{split}
\]
From eq. (\ref{eq: l-elements}),
\[\left\{\begin{aligned}
\tilde\varepsilon_{2n} - \tilde\varepsilon_{22}l_{\varepsilon n} - \tilde\alpha_{22}l_{\mu n}
& =
\tilde\varepsilon_{2n} - \tilde\varepsilon_{22}
\frac{\tilde\mu_{22}\tilde\varepsilon_{2n}-\tilde\alpha_{22}\tilde\gamma_{2n}}{\tilde\varepsilon_{22}\tilde\mu_{22}-\tilde\alpha_{22}\tilde\gamma_{22}}
 - \tilde\alpha_{22}
\frac{\tilde\varepsilon_{22}\tilde\gamma_{2n}-\tilde\gamma_{22}\tilde\varepsilon_{2n}}{\tilde\varepsilon_{22}\tilde\mu_{22}-\tilde\alpha_{22}\tilde\gamma_{22}}
\\
& =
\tilde\varepsilon_{2n} - 
\frac{\tilde\varepsilon_{22}\tilde\mu_{22}\tilde\varepsilon_{2n}-\tilde\varepsilon_{22}\tilde\alpha_{22}\tilde\gamma_{2n}}{\tilde\varepsilon_{22}\tilde\mu_{22}-\tilde\alpha_{22}\tilde\gamma_{22}}
 - 
\frac{\tilde\alpha_{22}\tilde\varepsilon_{22}\tilde\gamma_{2n}-\tilde\alpha_{22}\tilde\gamma_{22}\tilde\varepsilon_{2n}}{\tilde\varepsilon_{22}\tilde\mu_{22}-\tilde\alpha_{22}\tilde\gamma_{22}}
= 0\,,
\\
\tilde\gamma_{2n} - \tilde\gamma_{22}l_{\varepsilon n} - \tilde\mu_{22}l_{\mu n}
& =
\tilde\gamma_{2n} - \tilde\gamma_{22}
\frac{\tilde\mu_{22}\tilde\varepsilon_{2n}-\tilde\alpha_{22}\tilde\gamma_{2n}}{\tilde\varepsilon_{22}\tilde\mu_{22}-\tilde\alpha_{22}\tilde\gamma_{22}}
 - \tilde\mu_{22}
\frac{\tilde\varepsilon_{22}\tilde\gamma_{2n}-\tilde\gamma_{22}\tilde\varepsilon_{2n}}{\tilde\varepsilon_{22}\tilde\mu_{22}-\tilde\alpha_{22}\tilde\gamma_{22}}
\\
& =
\tilde\gamma_{2n} - 
\frac{\tilde\gamma_{22}\tilde\mu_{22}\tilde\varepsilon_{2n}-\tilde\gamma_{22}\tilde\alpha_{22}\tilde\gamma_{2n}}{\tilde\varepsilon_{22}\tilde\mu_{22}-\tilde\alpha_{22}\tilde\gamma_{22}}
 - 
\frac{\tilde\mu_{22}\tilde\varepsilon_{22}\tilde\gamma_{2n}-\tilde\mu_{22}\tilde\gamma_{22}\tilde\varepsilon_{2n}}{\tilde\varepsilon_{22}\tilde\mu_{22}-\tilde\alpha_{22}\tilde\gamma_{22}}
= 0\,,
\\
\tilde\alpha_{2n} - \tilde\varepsilon_{22}l_{\varepsilon3+n} - \tilde\alpha_{22}l_{\mu3+n}
& =
\tilde\alpha_{2n} - \tilde\varepsilon_{22}
\frac{\tilde\mu_{22}\tilde\alpha_{2n}-\tilde\alpha_{22}\tilde\mu_{2n}}{\tilde\varepsilon_{22}\tilde\mu_{22}-\tilde\alpha_{22}\tilde\gamma_{22}}
 - \tilde\alpha_{22}
\frac{\tilde\varepsilon_{22}\tilde\mu_{2n}-\tilde\gamma_{22}\tilde\alpha_{2n}}{\tilde\varepsilon_{22}\tilde\mu_{22}-\tilde\alpha_{22}\tilde\gamma_{22}}
\\
& =
\tilde\alpha_{2n} - 
\frac{\tilde\varepsilon_{22}\tilde\mu_{22}\tilde\alpha_{2n}-\tilde\varepsilon_{22}\tilde\alpha_{22}\tilde\mu_{2n}}{\tilde\varepsilon_{22}\tilde\mu_{22}-\tilde\alpha_{22}\tilde\gamma_{22}}
 - 
\frac{\tilde\alpha_{22}\tilde\varepsilon_{22}\tilde\mu_{2n}-\tilde\alpha_{22}\tilde\gamma_{22}\tilde\alpha_{2n}}{\tilde\varepsilon_{22}\tilde\mu_{22}-\tilde\alpha_{22}\tilde\gamma_{22}}
= 0\,,
\\
\tilde\mu_{2n} - \tilde\gamma_{22}l_{\varepsilon3+n} - \tilde\mu_{22}l_{\mu3+n}
& =
\tilde\mu_{2n} - \tilde\gamma_{22}
\frac{\tilde\mu_{22}\tilde\alpha_{2n}-\tilde\alpha_{22}\tilde\mu_{2n}}{\tilde\varepsilon_{22}\tilde\mu_{22}-\tilde\alpha_{22}\tilde\gamma_{22}}
 - \tilde\mu_{22}
\frac{\tilde\varepsilon_{22}\tilde\mu_{2n}-\tilde\gamma_{22}\tilde\alpha_{2n}}{\tilde\varepsilon_{22}\tilde\mu_{22}-\tilde\alpha_{22}\tilde\gamma_{22}}
\\
& =
\tilde\mu_{2n} - 
\frac{\tilde\gamma_{22}\tilde\mu_{22}\tilde\alpha_{2n}-\tilde\gamma_{22}\tilde\alpha_{22}\tilde\mu_{2n}}{\tilde\varepsilon_{22}\tilde\mu_{22}-\tilde\alpha_{22}\tilde\gamma_{22}}
 - 
\frac{\tilde\mu_{22}\tilde\varepsilon_{22}\tilde\mu_{2n}-\tilde\mu_{22}\tilde\gamma_{22}\tilde\alpha_{2n}}{\tilde\varepsilon_{22}\tilde\mu_{22}-\tilde\alpha_{22}\tilde\gamma_{22}}
= 0\,.
\end{aligned}\right.\]
From the above relations and eq. (\ref{eq: trans. epsilon}),
\[
\begin{split}
\left(\begin{array}{cc}
\hat{\boldsymbol{\varepsilon}} & \hat{\boldsymbol{\alpha}}\\
\hat{\boldsymbol{\gamma}} & \hat{\boldsymbol{\mu}}
\end{array}\right)
&=
\left(\begin{array}{cccccc}
1 & 0 & -u_{\varepsilon0} & 0 & 0 & -u_{\mu0}\\
0 & 1 & -u_{\varepsilon1} & 0 & 0 & -u_{\mu1}\\
0 & 0 & 1 & 0 & 0 & 0\\
0 & 0 & -u_{\varepsilon3} & 1 & 0 & -u_{\mu3}\\
0 & 0 & -u_{\varepsilon4} & 0 & 1 & -u_{\mu4}\\
0 & 0 & 0 & 0 & 0 & 1
\end{array}\right)
\left(\begin{array}{cccccc}
\hat\varepsilon_{00} & \hat\varepsilon_{01} & \tilde\varepsilon_{02} & \hat\alpha_{00} & \hat\alpha_{01} & \tilde\alpha_{02}\\
\hat\varepsilon_{10} & \hat\varepsilon_{11} & \tilde\varepsilon_{12} & \hat\alpha_{10} & \hat\alpha_{11} & \tilde\alpha_{12}\\
0 & 0 & \tilde\varepsilon_{22} & 0 & 0 & \tilde\alpha_{22}\\
\hat\gamma_{00} & \hat\gamma_{01} & \tilde\gamma_{02} & \hat\mu_{00} & \hat\mu_{01} & \tilde\mu_{02}\\
\hat\gamma_{10} & \hat\gamma_{11} & \tilde\gamma_{12} & \hat\mu_{10} & \hat\mu_{11} & \tilde\mu_{12}\\
0 & 0 & \tilde\gamma_{22} & 0 & 0 & \tilde\mu_{22}
\end{array}\right)
\\
&=
\left(\begin{array}{cccccc}
\hat\varepsilon_{00} & \hat\varepsilon_{01} & 
\tilde\varepsilon_{02} -u_{\varepsilon0}\tilde\varepsilon_{22}  -u_{\mu0}\tilde\gamma_{22}& 
\hat\alpha_{00} & \hat\alpha_{01} & 
\tilde\alpha_{02} - u_{\varepsilon0}\tilde\alpha_{22} - u_{\mu0}\tilde\mu_{22}\\
\hat\varepsilon_{10} & \hat\varepsilon_{11} & 
\tilde\varepsilon_{12} -u_{\varepsilon1}\tilde\varepsilon_{22}  -u_{\mu1}\tilde\gamma_{22}& 
\hat\alpha_{10} & \hat\alpha_{11} & 
\tilde\alpha_{12} - u_{\varepsilon1}\tilde\alpha_{22} - u_{\mu1}\tilde\mu_{22}\\
0 & 0 & \tilde\varepsilon_{22} & 0 & 0 & \tilde\alpha_{22}\\
\hat\gamma_{00} & \hat\gamma_{01} & 
\tilde\gamma_{02} -u_{\varepsilon3}\tilde\varepsilon_{22}  -u_{\mu3}\tilde\gamma_{22}& 
\hat\mu_{00} & \hat\mu_{01} & 
\tilde\mu_{02} - u_{\varepsilon3}\tilde\alpha_{22} - u_{\mu3}\tilde\mu_{22} \\
\hat\gamma_{10} & \hat\gamma_{11} & 
\tilde\gamma_{12} -u_{\varepsilon4}\tilde\varepsilon_{22}  -u_{\mu4}\tilde\gamma_{22}& 
\hat\mu_{10} & \hat\mu_{11} & 
\tilde\mu_{12} - u_{\varepsilon4}\tilde\alpha_{22} - u_{\mu4}\tilde\mu_{22} \\
0 & 0 & \tilde\gamma_{22} & 0 & 0 & \tilde\mu_{22}
\end{array}\right).
\end{split}
\]
From eq. (\ref{eq: u-elements}),
\[\left\{
\begin{aligned}
\tilde\varepsilon_{n2} -u_{\varepsilon n}\tilde\varepsilon_{22}  -u_{\mu n}\tilde\gamma_{22}
&= \tilde\varepsilon_{n2} -
\frac{\tilde\mu_{22}\tilde\varepsilon_{n2}-\tilde\gamma_{22}\tilde\alpha_{n2}}{\tilde\varepsilon_{22}\tilde\mu_{22}-\tilde\alpha_{22}\tilde\gamma_{22}}
\tilde\varepsilon_{22}  -
\frac{\tilde\varepsilon_{22}\tilde\alpha_{n2}-\tilde\alpha_{22}\tilde\varepsilon_{n2}}{\tilde\varepsilon_{22}\tilde\mu_{22}-\tilde\alpha_{22}\tilde\gamma_{22}}
\tilde\gamma_{22}
\\
&= \tilde\varepsilon_{n2} -
\frac{\tilde\varepsilon_{22}\tilde\mu_{22}\tilde\varepsilon_{n2}-\tilde\varepsilon_{22}\tilde\gamma_{22}\tilde\alpha_{n2}}{\tilde\varepsilon_{22}\tilde\mu_{22}-\tilde\alpha_{22}\tilde\gamma_{22}}
  -
\frac{\tilde\varepsilon_{22}\tilde\gamma_{22}\tilde\alpha_{n2}-\tilde\alpha_{22}\tilde\gamma_{22}\tilde\varepsilon_{n2}}{\tilde\varepsilon_{22}\tilde\mu_{22}-\tilde\alpha_{22}\tilde\gamma_{22}}
= 0\,,
\\
\tilde\alpha_{n2} - u_{\varepsilon n}\tilde\alpha_{22} - u_{\mu n}\tilde\mu_{22}
&= \tilde\alpha_{n2} - 
\frac{\tilde\mu_{22}\tilde\varepsilon_{n2}-\tilde\gamma_{22}\tilde\alpha_{n2}}{\tilde\varepsilon_{22}\tilde\mu_{22}-\tilde\alpha_{22}\tilde\gamma_{22}}
\tilde\alpha_{22} - 
\frac{\tilde\varepsilon_{22}\tilde\alpha_{n2}-\tilde\alpha_{22}\tilde\varepsilon_{n2}}{\tilde\varepsilon_{22}\tilde\mu_{22}-\tilde\alpha_{22}\tilde\gamma_{22}}
\tilde\mu_{22}
\\
&= \tilde\alpha_{n2} - 
\frac{\tilde\mu_{22}\tilde\alpha_{22}\tilde\varepsilon_{n2}-\tilde\alpha_{22}\tilde\gamma_{22}\tilde\alpha_{n2}}{\tilde\varepsilon_{22}\tilde\mu_{22}-\tilde\alpha_{22}\tilde\gamma_{22}}
 - 
\frac{\tilde\varepsilon_{22}\tilde\mu_{22}\tilde\alpha_{n2}-\tilde\mu_{22}\tilde\alpha_{22}\tilde\varepsilon_{n2}}{\tilde\varepsilon_{22}\tilde\mu_{22}-\tilde\alpha_{22}\tilde\gamma_{22}}
= 0\,,
\\
\tilde\gamma_{n2} -u_{\varepsilon3+n}\tilde\varepsilon_{22}  -u_{\mu3+n}\tilde\gamma_{22}
&=\tilde\gamma_{n2} -
\frac{\tilde\mu_{22}\tilde\gamma_{n2}-\tilde\gamma_{22}\tilde\mu_{n2}}{\tilde\varepsilon_{22}\tilde\mu_{22}-\tilde\alpha_{22}\tilde\gamma_{22}}
\tilde\varepsilon_{22}  -
\frac{\tilde\varepsilon_{22}\tilde\mu_{n2}-\tilde\alpha_{22}\tilde\gamma_{n2}}{\tilde\varepsilon_{22}\tilde\mu_{22}-\tilde\alpha_{22}\tilde\gamma_{22}}
\tilde\gamma_{22}
\\
&=\tilde\gamma_{n2} -
\frac{\tilde\varepsilon_{22}\tilde\mu_{22}\tilde\gamma_{n2}-\tilde\varepsilon_{22}\tilde\gamma_{22}\tilde\mu_{n2}}{\tilde\varepsilon_{22}\tilde\mu_{22}-\tilde\alpha_{22}\tilde\gamma_{22}}
  -
\frac{\tilde\varepsilon_{22}\tilde\gamma_{22}\tilde\mu_{n2}-\tilde\alpha_{22}\tilde\gamma_{22}\tilde\gamma_{n2}}{\tilde\varepsilon_{22}\tilde\mu_{22}-\tilde\alpha_{22}\tilde\gamma_{22}}
= 0\,,
\\
\tilde\mu_{n2} - u_{\varepsilon3+n}\tilde\alpha_{22} - u_{\mu3+n}\tilde\mu_{22}
&=\tilde\mu_{n2} - 
\frac{\tilde\mu_{22}\tilde\gamma_{n2}-\tilde\gamma_{22}\tilde\mu_{n2}}{\tilde\varepsilon_{22}\tilde\mu_{22}-\tilde\alpha_{22}\tilde\gamma_{22}}
\tilde\alpha_{22} - 
\frac{\tilde\varepsilon_{22}\tilde\mu_{n2}-\tilde\alpha_{22}\tilde\gamma_{n2}}{\tilde\varepsilon_{22}\tilde\mu_{22}-\tilde\alpha_{22}\tilde\gamma_{22}}
\tilde\mu_{22}
\\
&=\tilde\mu_{n2} - 
\frac{\tilde\mu_{22}\tilde\alpha_{22}\tilde\gamma_{n2}-\tilde\alpha_{22}\tilde\gamma_{22}\tilde\mu_{n2}}{\tilde\varepsilon_{22}\tilde\mu_{22}-\tilde\alpha_{22}\tilde\gamma_{22}}
 - 
\frac{\tilde\varepsilon_{22}\tilde\mu_{22}\tilde\mu_{n2}-\tilde\mu_{22}\tilde\alpha_{22}\tilde\gamma_{n2}}{\tilde\varepsilon_{22}\tilde\mu_{22}-\tilde\alpha_{22}\tilde\gamma_{22}}
= 0 \,.
\end{aligned}
\right.\]
Then,
\[
\left(\begin{array}{cc}
\hat{\boldsymbol{\varepsilon}} & \hat{\boldsymbol{\alpha}}\\
\hat{\boldsymbol{\gamma}} & \hat{\boldsymbol{\mu}}
\end{array}\right)
=
\left(\begin{array}{cccccc}
\hat\varepsilon_{00} & \hat\varepsilon_{01} & 
0 & 
\hat\alpha_{00} & \hat\alpha_{01} & 
0 \\
\hat\varepsilon_{10} & \hat\varepsilon_{11} & 
0 & 
\hat\alpha_{10} & \hat\alpha_{11} & 
0 \\
0 & 0 & \tilde\varepsilon_{22} & 0 & 0 & \tilde\alpha_{22}\\
\hat\gamma_{00} & \hat\gamma_{01} & 
0 & 
\hat\mu_{00} & \hat\mu_{01} & 
0 \\
\hat\gamma_{10} & \hat\gamma_{11} & 
0 & 
\hat\mu_{10} & \hat\mu_{11} & 
0 \\
0 & 0 & \tilde\gamma_{22} & 0 & 0 & \tilde\mu_{22}
\end{array}\right).
\]
We checked the consistency of eqs. (\ref{eq: l-elements}), (\ref{eq: u-elements}) and (\ref{eq: trans. epsilon}).
Furthermore, we can confirm eq. (\ref{eq: 1-l-l 1+l+l}), since square of $\boldsymbol{l}_{\varepsilon}+\boldsymbol{l}_{\mu}$ is equal to zero:
\begin{equation}
\left(\boldsymbol{l}_{\varepsilon}+\boldsymbol{l}_{\mu}\right)^{2}
=
\left(\begin{array}{cccccc}
0 & 0 & 0 & 0 & 0 & 0\\
0 & 0 & 0 & 0 & 0 & 0\\
l_{\varepsilon0} & l_{\varepsilon1} & 0 & l_{\varepsilon3} & l_{\varepsilon4} & 0\\
0 & 0 & 0 & 0 & 0 & 0\\
0 & 0 & 0 & 0 & 0 & 0\\
l_{\mu0} & l_{\mu1} & 0 & l_{\mu3} & l_{\mu4} & 0
\end{array}\right)
\left(\begin{array}{cccccc}
0 & 0 & 0 & 0 & 0 & 0\\
0 & 0 & 0 & 0 & 0 & 0\\
l_{\varepsilon0} & l_{\varepsilon1} & 0 & l_{\varepsilon3} & l_{\varepsilon4} & 0\\
0 & 0 & 0 & 0 & 0 & 0\\
0 & 0 & 0 & 0 & 0 & 0\\
l_{\mu0} & l_{\mu1} & 0 & l_{\mu3} & l_{\mu4} & 0
\end{array}\right)=
0\,.
\label{eq: square of l_e + l_mu}
\end{equation}

\section{Check of eqs. (\ref{eq: tilde z elements}) and (\ref{eq: inv_h m inv_h})}

Details of eq. (\ref{eq: Maxwell_tilde2}) are represented as
\begin{equation}
\begin{split}
&\left(1-\boldsymbol{u}_{\varepsilon}-\boldsymbol{u}_{\mu}\right)
\left(\begin{array}{cc}
0 & i\boldsymbol{\nabla}_{u}\times\\
-i\boldsymbol{\nabla}_{u}\times & 0
\end{array}\right)
\left(1-\boldsymbol{l}_{\varepsilon}-\boldsymbol{l}_{\mu}\right)
\left(\begin{array}{c}
\hat{\boldsymbol{E}}\\
\hat{\boldsymbol{H}}
\end{array}\right)
\\
=&
\left(\begin{array}{cccccc}
1 & 0 & -u_{\varepsilon0} & 0 & 0 & -u_{\mu0}\\
0 & 1 & -u_{\varepsilon1} & 0 & 0 & -u_{\mu1}\\
0 & 0 & 1 & 0 & 0 & 0\\
0 & 0 & -u_{\varepsilon3} & 1 & 0 & -u_{\mu3}\\
0 & 0 & -u_{\varepsilon4} & 0 & 1 & -u_{\mu4}\\
0 & 0 & 0 & 0 & 0 & 1
\end{array}\right)
\left(\begin{array}{cccccc}
0 & 0 & 0 &
0 & -i\partial_{u2} & i\partial_{u1}\\
0 & 0 & 0 &
i\partial_{u2} & 0 & -i\partial_{u0}\\
0 & 0 & 0 &
-i\partial_{u1} & i\partial_{u0} & 0\\
0 & i\partial_{u2} & -i\partial_{u1} &
0 & 0 & 0\\
-i\partial_{u2} & 0 & i\partial_{u0} &
0 & 0 & 0\\
i\partial_{u1} & -i\partial_{u0} & 0 &
0 & 0 & 0
\end{array}\right)
\left(1-\boldsymbol{l}_{\varepsilon}-\boldsymbol{l}_{\mu}\right)
\left(\begin{array}{c}
\hat{\boldsymbol{E}}\\
\hat{\boldsymbol{H}}
\end{array}\right)
\\
=&
\left(\begin{array}{cccccc}
-iu_{\mu0}\partial_{u1} & iu_{\mu0}\partial_{u0} & 0 &
iu_{\varepsilon0}\partial_{u1} & -i\partial_{u2}-iu_{\varepsilon0}\partial_{u0} & i\partial_{u1}\\
-iu_{\mu1}\partial_{u1} & iu_{\mu1}\partial_{u0} & 0 &
i\partial_{u2} + iu_{\varepsilon1}\partial_{u1} & -iu_{\varepsilon1}\partial_{u0} & -i\partial_{u0}\\
0 & 0 & 0 &
-i\partial_{u1} & i\partial_{u0} & 0\\
-iu_{\mu3}\partial_{u1} & i\partial_{u2}+iu_{\mu3}\partial_{u0} & -i\partial_{u1} &
iu_{\varepsilon3}\partial_{u1} & -iu_{\varepsilon3}\partial_{u0} & 0\\
-i\partial_{u2}-iu_{\mu4}\partial_{u1} & iu_{\mu4}\partial_{u0} & i\partial_{u0} &
iu_{\varepsilon4}\partial_{u1} & -iu_{\varepsilon4}\partial_{u0} & 0\\
i\partial_{u1} & -i\partial_{u0} & 0 &
0 & 0 & 0
\end{array}\right)
\\
&\times
\left(\begin{array}{cccccc}
1 & 0 & 0 & 0 & 0 & 0\\
0 & 1 & 0 & 0 & 0 & 0\\
-l_{\varepsilon0} & -l_{\varepsilon1} & 1 & -l_{\varepsilon3} & -l_{\varepsilon4} & 0\\
0 & 0 & 0 & 1 & 0 & 0\\
0 & 0 & 0 & 0 & 1 & 0\\
-l_{\mu0} & -l_{\mu1} & 0 & -l_{\mu3} & -l_{\mu4} & 1
\end{array}\right)
\left(\begin{array}{c}
\hat{\boldsymbol{E}}\\
\hat{\boldsymbol{H}}
\end{array}\right)
\\
=&
\left(\begin{array}{ccc}
-iu_{\mu0}\partial_{u1} - i\partial_{u1}l_{\mu0} & iu_{\mu0}\partial_{u0} - i\partial_{u1}l_{\mu1} & 0 
\\
-iu_{\mu1}\partial_{u1} + i\partial_{u0}l_{\mu0} & iu_{\mu1}\partial_{u0} + i\partial_{u0}l_{\mu1} & 0 
\\
0 & 0 & 0 
\\
-iu_{\mu3}\partial_{u1} + i\partial_{u1}l_{\varepsilon0} & i\partial_{u2}+iu_{\mu3}\partial_{u0} + i\partial_{u1}l_{\varepsilon1} & -i\partial_{u1} 
\\
-i\partial_{u2}-iu_{\mu4}\partial_{u1} - i\partial_{u0}l_{\varepsilon0} & iu_{\mu4}\partial_{u0} - i\partial_{u0}l_{\varepsilon1} & i\partial_{u0} 
\\
i\partial_{u1} & -i\partial_{u0} & 0 
\end{array}\right.
\\
&\quad\quad\quad\quad\quad\quad\left.\begin{array}{ccc}
iu_{\varepsilon0}\partial_{u1} - i\partial_{u1}l_{\mu3} & -i\partial_{u2}-iu_{\varepsilon0}\partial_{u0} - i\partial_{u1}l_{\mu4} & i\partial_{u1}\\
i\partial_{u2} + iu_{\varepsilon1}\partial_{u1} + i\partial_{u0}l_{\mu3} & -iu_{\varepsilon1}\partial_{u0} + i\partial_{u0}l_{\mu4} & -i\partial_{u0}\\
-i\partial_{u1} & i\partial_{u0} & 0\\
%
iu_{\varepsilon3}\partial_{u1} + i\partial_{u1}l_{\varepsilon3} & -iu_{\varepsilon3}\partial_{u0} + i\partial_{u1}l_{\varepsilon4} & 0\\
iu_{\varepsilon4}\partial_{u1} - i\partial_{u0}l_{\varepsilon3} & -iu_{\varepsilon4}\partial_{u0} - i\partial_{u0}l_{\varepsilon4} & 0\\
0 & 0 & 0
\end{array}\right)
\left(\begin{array}{c}
h_{0}{E}_{0}\\
h_{1}{E}_{1}\\
\hat{E}_{2}\\
h_{0}{H}_{0}\\
h_{1}{H}_{1}\\
\hat{H}_{2}
\end{array}\right)
\\
= &
\,\omega\left(\begin{array}{cccccc}
\hat\varepsilon_{00} & \hat\varepsilon_{01} & 
0 & 
\hat\alpha_{00} & \hat\alpha_{01} & 
0 \\
\hat\varepsilon_{10} & \hat\varepsilon_{11} & 
0 & 
\hat\alpha_{10} & \hat\alpha_{11} & 
0 \\
0 & 0 & \tilde\varepsilon_{22} & 0 & 0 & \tilde\alpha_{22}\\
\hat\gamma_{00} & \hat\gamma_{01} & 
0 & 
\hat\mu_{00} & \hat\mu_{01} & 
0 \\
\hat\gamma_{10} & \hat\gamma_{11} & 
0 & 
\hat\mu_{10} & \hat\mu_{11} & 
0 \\
0 & 0 & \tilde\gamma_{22} & 0 & 0 & \tilde\mu_{22}
\end{array}\right)
\left(\begin{array}{c}
h_{0}{E}_{0}\\
h_{1}{E}_{1}\\
\hat{E}_{2}\\
h_{0}{H}_{0}\\
h_{1}{H}_{1}\\
\hat{H}_{2}
\end{array}\right)\,.
\end{split}
\label{eq: Details of Maxwell_tilde2}
\end{equation}
From eq. (\ref{eq: Details of Maxwell_tilde2}),
\[
\omega\left(\begin{array}{cc}
\tilde\varepsilon_{22} & \tilde\alpha_{22}\\
\tilde\gamma_{22} & \tilde\mu_{22}
\end{array}\right)
\left(\begin{array}{c}
\hat{E}_{2}\\
\hat{H}_{2}
\end{array}\right)
= 
\left(\begin{array}{cc}
0 & 0\\
i\partial_{u1} & -i\partial_{u0}
\end{array}\right)
\left(\begin{array}{c}
h_{0}{E}_{0}\\
h_{1}{E}_{1}
\end{array}\right)
+
\left(\begin{array}{cc}
-i\partial_{u1} & i\partial_{u0}\\
0 & 0
\end{array}\right)
\left(\begin{array}{c}
h_{0}{H}_{0}\\
h_{1}{H}_{1}
\end{array}\right)
\,.
\]
Note that
\[
\left(\begin{array}{cc}
\tilde\mu_{22} & -\tilde\alpha_{22}\\
-\tilde\gamma_{22} & \tilde\varepsilon_{22}
\end{array}\right)
\left(\begin{array}{cc}
\tilde\varepsilon_{22} & \tilde\alpha_{22}\\
\tilde\gamma_{22} & \tilde\mu_{22}
\end{array}\right)
= \tilde\varepsilon_{22}\tilde\mu_{22} - \tilde\alpha_{22}\tilde\gamma_{22} = c_{22}^{-2}\,.
\]
Then,
\begin{equation}
\left(\begin{array}{c}
\hat{E}_{2}\\
\hat{H}_{2}
\end{array}\right)
= 
\frac{c_{22}^{2}}{\omega}
\left(\begin{array}{cc}
-i\tilde\alpha_{22}\partial_{u1} & i\tilde\alpha_{22}\partial_{u0}\\
i\tilde\varepsilon_{22}\partial_{u1} & -i\tilde\varepsilon_{22}\partial_{u0}
\end{array}\right)
\left(\begin{array}{c}
h_{0}{E}_{0}\\
h_{1}{E}_{1}
\end{array}\right)
+
\frac{c_{22}^{2}}{\omega}
\left(\begin{array}{cc}
-i\tilde\mu_{22}\partial_{u1} & i\tilde\mu_{22}\partial_{u0}\\
i\tilde\gamma_{22}\partial_{u1} & -i\tilde\gamma_{22}\partial_{u0}
\end{array}\right)
\left(\begin{array}{c}
h_{0}{H}_{0}\\
h_{1}{H}_{1}
\end{array}\right)
\,.
\label{eq: hat E2 hat H2}
\end{equation}
From eq. (\ref{eq: Details of Maxwell_tilde2}),
\[\begin{split}
&\left(\begin{array}{cc}
-iu_{\mu0}\partial_{u1} - i\partial_{u1}l_{\mu0} & iu_{\mu0}\partial_{u0} - i\partial_{u1}l_{\mu1}
\\
-iu_{\mu1}\partial_{u1} + i\partial_{u0}l_{\mu0} & iu_{\mu1}\partial_{u0} + i\partial_{u0}l_{\mu1}
\end{array}\right)
\left(\begin{array}{c}
h_{0}{E}_{0}\\
h_{1}{E}_{1}
\end{array}\right)
\\
&+
\left(\begin{array}{cc}
iu_{\varepsilon0}\partial_{u1} - i\partial_{u1}l_{\mu3} & -i\partial_{u2}-iu_{\varepsilon0}\partial_{u0} - i\partial_{u1}l_{\mu4}\\
i\partial_{u2} + iu_{\varepsilon1}\partial_{u1} + i\partial_{u0}l_{\mu3} & -iu_{\varepsilon1}\partial_{u0} + i\partial_{u0}l_{\mu4} 
\end{array}\right)
\left(\begin{array}{c}
h_{0}{H}_{0}\\
h_{1}{H}_{1}
\end{array}\right)
+
\left(\begin{array}{c}
 i\partial_{u1}\\
 -i\partial_{u0}
\end{array}\right)\hat{H}_{2}
\\
=&\,
\omega\left(\begin{array}{cc}
\hat\varepsilon_{00} & \hat\varepsilon_{01}
\\
\hat\varepsilon_{10} & \hat\varepsilon_{11} 
\end{array}\right)
\left(\begin{array}{c}
h_{0}{E}_{0}\\
h_{1}{E}_{1}
\end{array}\right)
+
\omega\left(\begin{array}{cc}
\hat\alpha_{00} & \hat\alpha_{01}  
\\
\hat\alpha_{10} & \hat\alpha_{11} 
\end{array}\right)
\left(\begin{array}{c}
h_{0}{H}_{0}\\
h_{1}{H}_{1}
\end{array}\right).
\end{split}\]
From the above and eq. (\ref{eq: hat E2 hat H2}),
\begin{equation}
\begin{split}
&-i\partial_{u2}
\left(\begin{array}{c}
h_{0}{H}_{0}\\
h_{1}{H}_{1}
\end{array}\right)
\\
=&
\left(\begin{array}{cc}
\partial_{u0}\frac{\tilde\varepsilon_{22}c_{22}^{2}}{\omega}\partial_{u1}-\omega\hat\varepsilon_{10} - iu_{\mu1}\partial_{u1} + i\partial_{u0}l_{\mu0} & -\partial_{u0}\frac{\tilde\varepsilon_{22}c_{22}^{2}}{\omega}\partial_{u0}-\omega\hat\varepsilon_{11} + iu_{\mu1}\partial_{u0} + i\partial_{u0}l_{\mu1}
\\
\partial_{u1}\frac{\tilde\varepsilon_{22}c_{22}^{2}}{\omega}\partial_{u1}+\omega\hat\varepsilon_{00} + iu_{\mu0}\partial_{u1} + i\partial_{u1}l_{\mu0} & -\partial_{u1}\frac{\tilde\varepsilon_{22}c_{22}^{2}}{\omega}\partial_{u0}+\omega\hat\varepsilon_{01} - iu_{\mu0}\partial_{u0} + i\partial_{u1}l_{\mu1}
\end{array}\right)
\left(\begin{array}{c}
h_{0}{E}_{0}\\
h_{1}{E}_{1}
\end{array}\right)
\\
&+
\left(\begin{array}{cc}
\partial_{u0}\frac{\tilde\gamma_{22}c_{22}^{2}}{\omega}\partial_{u1}-\omega\hat\alpha_{10}+iu_{\varepsilon1}\partial_{u1} + i\partial_{u0}l_{\mu3} & -\partial_{u0}\frac{\tilde\gamma_{22}c_{22}^{2}}{\omega}\partial_{u0}-\omega\hat\alpha_{11}-iu_{\varepsilon1}\partial_{u0} + i\partial_{u0}l_{\mu4}  
\\
\partial_{u1}\frac{\tilde\gamma_{22}c_{22}^{2}}{\omega}\partial_{u1}+\omega\hat\alpha_{00}-iu_{\varepsilon0}\partial_{u1} + i\partial_{u1}l_{\mu3} & -\partial_{u1}\frac{\tilde\gamma_{22}c_{22}^{2}}{\omega}\partial_{u0}+\omega\hat\alpha_{01} +iu_{\varepsilon0}\partial_{u0} + i\partial_{u1}l_{\mu4} 
\end{array}\right)
\left(\begin{array}{c}
h_{0}{H}_{0}\\
h_{1}{H}_{1}
\end{array}\right).
\end{split}
\label{eq: check m_bb m_ba}
\end{equation}
We can check the $\boldsymbol{m}_{bb}$ and $\boldsymbol{m}_{ba}$ in eq. (\ref{eq: inv_h m inv_h}) with using the right side of eq. (\ref{eq: check m_bb m_ba}) and the $\boldsymbol{\Psi}$ definition of eq. (\ref{eq: Psi for Maxwell}).

From eq. (\ref{eq: Details of Maxwell_tilde2}),
\[\begin{split}
&\left(\begin{array}{cc}
-iu_{\mu3}\partial_{u1} + i\partial_{u1}l_{\varepsilon0} & i\partial_{u2}+iu_{\mu3}\partial_{u0} + i\partial_{u1}l_{\varepsilon1}
\\
-i\partial_{u2}-iu_{\mu4}\partial_{u1} - i\partial_{u0}l_{\varepsilon0} & iu_{\mu4}\partial_{u0} - i\partial_{u0}l_{\varepsilon1}
\end{array}\right)
\left(\begin{array}{c}
h_{0}{E}_{0}\\
h_{1}{E}_{1}
\end{array}\right)
+
\left(\begin{array}{c}
 -i\partial_{u1}
\\
 i\partial_{u0}
\end{array}\right)\hat{E}_{2}
\\
&+
\left(\begin{array}{cc}
iu_{\varepsilon3}\partial_{u1} + i\partial_{u1}l_{\varepsilon3} & -iu_{\varepsilon3}\partial_{u0} + i\partial_{u1}l_{\varepsilon4}\\
iu_{\varepsilon4}\partial_{u1} - i\partial_{u0}l_{\varepsilon3} & -iu_{\varepsilon4}\partial_{u0} - i\partial_{u0}l_{\varepsilon4}
\end{array}\right)
\left(\begin{array}{c}
h_{0}{H}_{0}\\
h_{1}{H}_{1}
\end{array}\right)
\\
=&
\omega\left(\begin{array}{cc}
\hat\gamma_{00} & \hat\gamma_{01} 
\\
\hat\gamma_{10} & \hat\gamma_{11}
\end{array}\right)
\left(\begin{array}{c}
h_{0}{E}_{0}\\
h_{1}{E}_{1}
\end{array}\right)
+
\omega\left(\begin{array}{cc}
\hat\mu_{00} & \hat\mu_{01}
\\
\hat\mu_{10} & \hat\mu_{11}
\end{array}\right)
\left(\begin{array}{c}
h_{0}{H}_{0}\\
h_{1}{H}_{1}
\end{array}\right).
\end{split}\]
From the above and eq. (\ref{eq: hat E2 hat H2}),
\begin{equation}\begin{split}
&-i\partial_{u2}\left(\begin{array}{cc}
0 & -1
\\
1 & 0
\end{array}\right)
\left(\begin{array}{c}
h_{0}{E}_{0}\\
h_{1}{E}_{1}
\end{array}\right)
\\
=&
\left(\begin{array}{cc}
\partial_{u1}\frac{\tilde\alpha_{22}c_{22}^{2}}{\omega}\partial_{u1}+\omega\hat\gamma_{00} + iu_{\mu3}\partial_{u1} - i\partial_{u1}l_{\varepsilon0} & -\partial_{u1}\frac{\tilde\alpha_{22}c_{22}^{2}}{\omega}\partial_{u0}+\omega\hat\gamma_{01} -iu_{\mu3}\partial_{u0} - i\partial_{u1}l_{\varepsilon1}
\\
-\partial_{u0}\frac{\tilde\alpha_{22}c_{22}^{2}}{\omega}\partial_{u1}+\omega\hat\gamma_{10} + iu_{\mu4}\partial_{u1} + i\partial_{u0}l_{\varepsilon0} & \partial_{u0}\frac{\tilde\alpha_{22}c_{22}^{2}}{\omega}\partial_{u0}+\omega\hat\gamma_{11} -iu_{\mu4}\partial_{u0} + i\partial_{u0}l_{\varepsilon1}
\end{array}\right)
\left(\begin{array}{c}
h_{0}{E}_{0}\\
h_{1}{E}_{1}
\end{array}\right)
\\
&+
\left(\begin{array}{cc}
\partial_{u1}\frac{\tilde\mu_{22}c_{22}^{2}}{\omega}\partial_{u1}+\omega\hat\mu_{00} -iu_{\varepsilon3}\partial_{u1} - i\partial_{u1}l_{\varepsilon3} & -\partial_{u1}\frac{\tilde\mu_{22}c_{22}^{2}}{\omega}\partial_{u0}+\omega\hat\mu_{01} + iu_{\varepsilon3}\partial_{u0} - i\partial_{u1}l_{\varepsilon4}
\\
-\partial_{u0}\frac{\tilde\mu_{22}c_{22}^{2}}{\omega}\partial_{u1}+\omega\hat\mu_{10} -iu_{\varepsilon4}\partial_{u1} + i\partial_{u0}l_{\varepsilon3} & \partial_{u0}\frac{\tilde\mu_{22}c_{22}^{2}}{\omega}\partial_{u0}+\omega\hat\mu_{11} + iu_{\varepsilon4}\partial_{u0} + i\partial_{u0}l_{\varepsilon4}
\end{array}\right)
\left(\begin{array}{c}
h_{0}{H}_{0}\\
h_{1}{H}_{1}
\end{array}\right)
\end{split}
\label{eq: check m_ab m_aa}
\end{equation}
We can check the $\boldsymbol{m}_{ab}$ and $\boldsymbol{m}_{aa}$ in eq. (\ref{eq: inv_h m inv_h}) with using the right side of eq. (\ref{eq: check m_ab m_aa}) and the $\boldsymbol{\Psi}$ definition of eq. (\ref{eq: Psi for Maxwell}).

%% file: App/E_NewtonEq_v3.tex
\chapter{Newton's equation of motion\label{ch: Newton}}

This chapter derives propagation equation \eqref{eq: propagation-equation} for elastic waves based on the Newtonian equation of motion. 
We use two variables which are displacement $\boldsymbol{u}\left(\boldsymbol{x}\right)$ and stress $\boldsymbol{\tau}\left(\boldsymbol{x}\right)$ by Voigt notation.
\[
\boldsymbol{u}^{\mathrm T}=\left(\begin{array}{ccc}
	u_{0} & u_{1} & u_{2}
\end{array}\right)^{\mathrm T}\quad\mathrm{and}\quad
\boldsymbol{\tau}^{\mathrm T}=\left(\begin{array}{cccccc}
	\tau_{00} & \tau_{11} & \tau_{22} & \tau_{12} & \tau_{02} & \tau_{01}
\end{array}\right)^{\mathrm T}.
\]
Hooke's law connects the $\boldsymbol{u}$ and $\boldsymbol{\tau}$ as 
\begin{equation}
\boldsymbol{\tau}  = 
\left(\begin{array}{c}
 \left(\boldsymbol{\Lambda}+2\mu\right)\boldsymbol{D}_{a}\\
 \mu\boldsymbol{D}_{b}
\end{array}\right)\boldsymbol{u}\,,
\label{eq: Hooke law}
\end{equation}
where
\[
\boldsymbol{\Lambda} = 
\left(\begin{array}{ccc}
\lambda & \lambda & \lambda\\
\lambda & \lambda & \lambda\\
\lambda & \lambda & \lambda
\end{array}\right),\quad
\boldsymbol{D}_{a} = 
\left(\begin{array}{ccc}
\frac{\partial}{\partial x_{0}} & 0 & 0\\
0 & \frac{\partial}{\partial x_{1}} & 0\\
0 & 0 & \frac{\partial}{\partial x_{2}}
\end{array}\right)\quad\mathrm{and}\quad
\boldsymbol{D}_{b} = 
\left(\begin{array}{ccc}
0 & \frac{\partial}{\partial x_{2}} & \frac{\partial}{\partial x_{1}}\\
\frac{\partial}{\partial x_{2}} & 0 & \frac{\partial}{\partial x_{0}}\\
\frac{\partial}{\partial x_{1}} & \frac{\partial}{\partial x_{0}} & 0
\end{array}\right).
\]
Here note that the parameters $\lambda\left(\boldsymbol{x}\right)$ and $\mu\left(\boldsymbol{x}\right)$ are Lam\'{e}'s constants.
From Eq. \eqref{eq: Hooke law}, Newtonian equation of motion is shown as
\begin{equation}
\frac{\partial\boldsymbol{\tau}}{\partial t} = 
\left(\begin{array}{c}
 \left(\boldsymbol{\Lambda}+2\mu\right)\boldsymbol{D}_{a}\\
 \mu\boldsymbol{D}_{b}
\end{array}\right)\boldsymbol{v}
\quad\mathrm{and}\quad\rho\frac{\partial \boldsymbol{v}}{\partial t} = -\left(\begin{array}{cc}
\boldsymbol{D}_{a}^{\dagger} & \boldsymbol{D}_{b}^{\dagger}
\end{array}\right) \boldsymbol{\tau},
\label{eq: Newton t}
\end{equation}
where $\boldsymbol{v}=\partial\boldsymbol{u}/\partial t$, and $\rho\left(\boldsymbol{x}\right)$ is density of media.
In the frequency domain, 
Eq. \eqref{eq: Newton t} can be deformed to Eq. \eqref{eq: propagation-equation}, \textit{i.e.} $\boldsymbol{M}\boldsymbol{\Psi}  = -i\left(\partial/\partial x_{2}\right){}
\left(\begin{array}{cc}
\boldsymbol{0} & \boldsymbol{\boldsymbol{1}}\\
\boldsymbol{\boldsymbol{1}} & \boldsymbol{0}
\end{array}\right)\boldsymbol{\Psi}
$ by using
\[
\boldsymbol{M} = \left(\begin{array}{cc}
\boldsymbol{m}_{vv} & \boldsymbol{m}_{v\tau}\\
\boldsymbol{m}_{{\tau}v} & \boldsymbol{m}_{\tau\tau}
\end{array}\right)\quad\mathrm{and}\quad
\boldsymbol{\Psi} = \left(\begin{array}{c}
\boldsymbol{v}\\
\boldsymbol{\tau}_{2}
\end{array}\right) = \left(\begin{array}{c}
-i\omega\boldsymbol{u}\\
\boldsymbol{\tau}_{2}
\end{array}\right),
\]
where
\[
\boldsymbol{m}_{vv} = \frac{1}{\omega}\left(\begin{array}{ccc}
\left(\frac{\partial}{\partial x_{0}}\right)^{\dagger}\frac{4\left(\lambda+\mu\right)\mu}{\lambda+2\mu}\frac{\partial}{\partial x_{0}}+\left(\frac{\partial}{\partial x_{1}}\right)^{\dagger}\mu\frac{\partial}{\partial x_{1}} & \left(\frac{\partial}{\partial x_{0}}\right)^{\dagger}\frac{2\lambda\mu}{\lambda+2\mu}\frac{\partial}{\partial x_{1}}+\left(\frac{\partial}{\partial x_{1}}\right)^{\dagger}\mu\frac{\partial}{\partial x_{0}} & 0\\
\left(\frac{\partial}{\partial x_{1}}\right)^{\dagger}\frac{2\lambda\mu}{\lambda+2\mu}\frac{\partial}{\partial x_{0}}+\left(\frac{\partial}{\partial x_{0}}\right)^{\dagger}\mu\frac{\partial}{\partial x_{1}} & \left(\frac{\partial}{\partial x_{1}}\right)^{\dagger}\frac{4\left(\lambda+\mu\right)\mu}{\lambda+2\mu}\frac{\partial}{\partial x_{1}}+\left(\frac{\partial}{\partial x_{0}}\right)^{\dagger}\mu\frac{\partial}{\partial x_{0}} & 0\\
0 & 0 & 0
\end{array}\right)-\omega\rho\,,
\]
\[
\boldsymbol{m}_{v\tau} = -i\left(\begin{array}{ccc}
0 & 0 & \left(\frac{\partial}{\partial x_{0}}\right)^{\dagger}\frac{\lambda}{\lambda+2\mu}\\
0 & 0 & \left(\frac{\partial}{\partial x_{1}}\right)^{\dagger}\frac{\lambda}{\lambda+2\mu}\\
\left(\frac{\partial}{\partial x_{0}}\right)^{\dagger} & \left(\frac{\partial}{\partial x_{1}}\right)^{\dagger} & 0
\end{array}\right),\quad
\boldsymbol{m}_{{\tau}v} = i\left(\begin{array}{ccc}
0 & 0 & \frac{\partial}{\partial x_{0}}\\
0 & 0 & \frac{\partial}{\partial x_{1}}\\
\frac{\lambda}{\lambda+2\mu}\frac{\partial}{\partial x_{0}} & \frac{\lambda}{\lambda+2\mu}\frac{\partial}{\partial x_{1}} & 0
\end{array}\right),
\]
\[
\boldsymbol{m}_{\tau\tau}  =  -\left(\begin{array}{ccc}
\frac{\omega}{\mu} & 0 & 0\\
0 & \frac{\omega}{\mu} & 0\\
0 & 0 & \frac{\omega}{\lambda+2\mu}
\end{array}\right)\quad\mathrm{and}\quad{}
\boldsymbol{\tau}_{2} = \left(\begin{array}{c}
	\tau_{02}\\
	\tau_{12}\\
	\tau_{22}
\end{array}\right).
\]
The matrix $\boldsymbol{M}$ satisfies $\boldsymbol{M}=\boldsymbol{M}^{\dagger}$ when three parameters $\lambda$, $\mu$ and $\omega\rho$ are real functions.

%% file: App/F_OpticalScattering_v1.tex
\chapter{Optical scattering\label{ch: optical scattering}}

Wave scattering is formally discussed in Chapter \ref{ch: S-matrix BornApprox} and \ref{ch: Roughness}.
This chapter shows more detailed formulas for optical scattering.
We consider Maxwell equation with scalar permittivity $\varepsilon$ and scalar magnetic-permeability $\mu$, and we use the coordinate transform discussed in Chapter \ref{ch: GeneralWaveguides}.
From eq. (\ref{eq: normal Maxwell equation}), the propagation equation (\ref{eq: propagation-equation}) can be reduced to
\begin{equation}
\left\{\begin{aligned}
\boldsymbol{M}\boldsymbol{\Psi} & = -i
	\frac{\partial}{\partial u_{2}}
	\left(
		\begin{array}{cc}
			\boldsymbol{0} & \boldsymbol{1}\\
			\boldsymbol{1} & \boldsymbol{0}
		\end{array}
	\right)
	\boldsymbol{\Psi}\quad \mathrm{as} \quad
\boldsymbol{M}\left(\boldsymbol{V}\right) =
	\left(\begin{array}{cc}
		\boldsymbol{m}_{aa} & \boldsymbol{0}\\
		\boldsymbol{0} & \boldsymbol{m}_{bb}
	\end{array}\right)
\,,
\\
\boldsymbol{m}_{aa} & =\left(\begin{array}{cc}
{V}_{0} + \partial_{u1}{V}_{2} \partial_{u1} &
 -\partial_{u1}{V}_{2}\partial_{u0}\\
 -\partial_{u0}{V}_{2}\partial_{u1} &
 {V}_{1} + \partial_{u0}{V}_{2} \partial_{u0}
\end{array}\right), \quad
\boldsymbol{m}_{bb}  =\left(\begin{array}{cc}
{V}_{3} + \partial_{u0}{V}_{5}\partial_{u0} &
 \partial_{u0}{V}_{5}\partial_{u1}\\
\partial_{u1}{V}_{5}\partial_{u0} &
 {V}_{4} + \partial_{u1}{V}_{5}\partial_{u1}
\end{array}\right)\,,
\\
{V}_{0}&=\frac{h_{2}\omega \mu}{h_{0}},\quad {V}_{1}=h_{0}h_{2}\omega \mu,\quad {V}_{2}= \frac{h_{2}}{h_{0} \omega \varepsilon},\quad {V}_{3}= h_{0}h_{2}\omega \varepsilon,\quad {V}_{4}= \frac{h_{2}\omega \varepsilon}{h_{0}},\quad {V}_{5}= \frac{h_{2}}{h_{0} \omega \mu}.
\end{aligned}\right.\label{eq: the simplest Maxwell equation}
\end{equation}
where $V_{j}$ are elements of the $\boldsymbol{V}$ of eq. \eqref{eq: V0V1}.
Note that $h_{1}$ is always equal to $1$.
The wave function $\boldsymbol{\Psi}$ in eq. (\ref{eq: the simplest Maxwell equation}) is represented by electromagnetic fields and scale factors as  eq. (\ref{eq: Psi for Maxwell}).

\section{Optical scattering except for edge roughness}

We applies the coordinate transformation in chapter \ref{ch: GeneralWaveguides} and the Born approximation in section \ref{sec: BornApprox} to scattering of optical guided waves.
The $\varepsilon$ and $\mu$ in eq. (\ref{eq: the simplest Maxwell equation}) are split to unperturbed terms and perturbed terms:
\[
\varepsilon = {\varepsilon}^{\left(0\right)}\left(u_{0},u_{1}\right)
	+ {\varepsilon}^{\left(1\right)}\left(u_{0},u_{1},u_{2}\right)\quad \mathrm{and} \quad
\mu = {\mu}^{\left(0\right)}\left(u_{0},u_{1}\right)
	+ {\mu}^{\left(1\right)}\left(u_{0},u_{1},u_{2}\right).
\]
From eqs. (\ref{eq: approx h2-h0}) and (\ref{eq: the simplest Maxwell equation}), we can set the diagonal matrices $\boldsymbol{V}^{\left(0\right)}$ and $\boldsymbol{V}^{\left(1\right)}$ of eq. (\ref{eq: V0V1}) as
\begin{equation}
\begin{gathered}
{V}^{\left(0\right)}_{0}=\frac{\omega{\mu}^{\left(0\right)}}{\zeta},\quad {V}^{\left(0\right)}_{1}=\zeta\omega{\mu}^{\left(0\right)},\quad {V}^{\left(0\right)}_{2}= \frac{1}{\zeta\omega{\varepsilon}^{\left(0\right)}},\quad {V}^{\left(0\right)}_{3}= \zeta\omega{\varepsilon}^{\left(0\right)},\quad {V}^{\left(0\right)}_{4}= \frac{\omega{\varepsilon}^{\left(0\right)}}{\zeta},\quad {V}^{\left(0\right)}_{5}= \frac{1}{\zeta\omega{\mu}^{\left(0\right)}},
\\
{V}^{\left(1\right)}_{0}=\frac{\omega\left(\mathit{\Delta}h_{2}{\mu}^{\left(0\right)}+{\mu}^{\left(1\right)}\right)}{\zeta},\quad 
{V}^{\left(1\right)}_{1}=\zeta\omega\left(\mathit{\Delta}h_{2}{\mu}^{\left(0\right)}+{\mu}^{\left(1\right)}\right),\quad 
{V}^{\left(1\right)}_{2}= \frac{\mathit{\Delta}h_{2}-{\varepsilon}^{\left(1\right)}/{\varepsilon}^{\left(0\right)}}{\zeta\omega{\varepsilon}^{\left(0\right)}},
\\
{V}^{\left(1\right)}_{3}= \zeta\omega\left(\mathit{\Delta}h_{2}{\varepsilon}^{\left(0\right)}+{\varepsilon}^{\left(1\right)}\right),\quad 
{V}^{\left(1\right)}_{4}= \frac{\omega\left(\mathit{\Delta}h_{2}{\varepsilon}^{\left(0\right)}+{\varepsilon}^{\left(1\right)}\right)}{\zeta},\quad 
{V}^{\left(1\right)}_{5}= \frac{\mathit{\Delta}h_{2}-{\mu}^{\left(1\right)}/{\mu}^{\left(0\right)}}{\zeta\omega{\mu}^{\left(0\right)}}.
\end{gathered}
\label{eq: optical born approx V0V1}
\end{equation}
Function of $u_{2}$ in eq. (\ref{eq: optical born approx V0V1}) is only $\zeta$, and then
\begin{equation}
\left\{\begin{aligned}
\partial_{u2}\boldsymbol{m}_{aa}^{\left(0\right)} & =
-\kappa_{w}\boldsymbol{m}_{aa}^{\left(0\right)}
+2\kappa_{w}
\left(
\begin{array}{cc}
0 & 0\\
0 & \zeta\omega{\mu}^{\left(0\right)}
\end{array}
\right)\,,
\\
\partial_{u2}\boldsymbol{m}_{bb}^{\left(0\right)} & =
-\kappa_{w}\boldsymbol{m}_{bb}^{\left(0\right)}
+2\kappa_{w}
\left(
\begin{array}{cc}
\zeta\omega{\varepsilon}^{\left(0\right)} & 0\\
0 & 0
\end{array}
\right)
\,.
\end{aligned}\right.
\label{eq: opitcal born approx partial2M0}
\end{equation}

From eqs. (\ref{eq: optical born approx V0V1}) and (\ref{eq: opitcal born approx partial2M0}), we obtain detail of $\hat{M}_{mn}$ of eq. \eqref{eq: Mtild_lm} for $\left|m\right|\leq n_{\max}$:
\begin{equation}
\hat{M}_{mn}\left(u_{2}\right)  =  
 \frac{m}{\left|m\right|}\boldsymbol{\Psi}_{m}^{\left(0\right)\dagger}
\left[
\frac{2\kappa_{w}\left(1-\delta_{mn}\right)}{\beta_{m}-\beta_{n}+0}
\left(\begin{array}{cc}
\begin{array}{cc}
0 & 0\\
0 & \zeta\omega{\mu}^{\left(0\right)}
\end{array} & \boldsymbol{0}\\
\boldsymbol{0} & \begin{array}{cc}
\zeta\omega{\varepsilon}^{\left(0\right)} & 0\\
0 & 0
\end{array}
\end{array}\right)
+
i\left(\begin{array}{cc}
\boldsymbol{m}_{aa}^{\left(1\right)}& \boldsymbol{0}\\
\boldsymbol{0} & \boldsymbol{m}_{bb}^{\left(1\right)}
\end{array}\right)
\right]
\boldsymbol{\Psi}_{n}^{\left(0\right)}
\,.
\label{eq: opitcal born approx gmn}
\end{equation}
From eqs. \eqref{eq: S-matrix app} and \eqref{eq: opitcal born approx gmn}, we can analyze optical scattering caused by the adiabatic tapered waveguide as shown in  Fig. \ref{fig: waveguide}, but we have to add
edge-roughness scattering discussed in Section \ref{sec: roughness} to the scattering properties of optical waveguide.

\section{Edge-roughness scattering of Section \ref{sec: roughness} for optical waveguide\label{sec: edge-roughness for optical WG}}
This section shows details of two approaches in Section \ref{sec: roughness} by using eq. (\ref{eq: the simplest Maxwell equation}).
We consider the straight waveguide, \textit{i.e.} the case that $\zeta = 1$.
The $\boldsymbol{V}^{\left(0\right)}$ of eq. (\ref{eq: optical born approx V0V1}) becomes constant for $u_{2}$, and its components can be reduced to
\begin{gather}
{V}_{0}^{\left(0\right)} = {V}_{1}^{\left(0\right)} = \omega \mu\left(u_{0},u_{1}\right),\quad {V}_{2}^{\left(0\right)}= \frac{1}{ \omega \varepsilon\left(u_{0},u_{1}\right)},\quad {V}_{3}^{\left(0\right)} = {V}_{4}^{\left(0\right)}=  \omega \varepsilon\left(u_{0},u_{1}\right),\quad {V}_{5}^{\left(0\right)} = \frac{1}{\omega \mu\left(u_{0},u_{1}\right)}.
\label{eq: V^(0) for straight WG}
\end{gather}

\subsection{Details of Approach I for \ref{subsec: approach I}\label{subsec: optical approach I}}

Equations (\ref{eq: approach1 V1}) and (\ref{eq: V^(0) for straight WG}) give us the $\boldsymbol{V}^{\left(1\right)}$ of Approach I.
Especially for the abrupt structure in Fig. \ref{fig: waveguide}, the $\boldsymbol{V}_{\mathrm{clad}}$ and $\boldsymbol{V}_{\mathrm{pp}}$ in eq. (\ref{eq: approach1 abrupt V1}) are given by
\begin{equation}
\left\{ \begin{gathered}
V_{0\,\mathrm{clad}} = V_{1\,\mathrm{clad}} = \omega\mu_{\mathrm{clad}}, \quad
V_{2\,\mathrm{clad}} = \frac{1}{\omega \varepsilon_{\mathrm{clad}}}, \quad
V_{3\,\mathrm{clad}} = V_{4\,\mathrm{clad}} = \omega\varepsilon_{\mathrm{clad}}, \quad
V_{5\,\mathrm{clad}} = \frac{1}{\omega \mu_{\mathrm{clad}}}, 
\\
V_{0\,\mathrm{pp}} = V_{1\,\mathrm{pp}} = \omega\left(\mu_{\mathrm{wg}}-\mu_{\mathrm{clad}}\right)H\left(V_{\mathrm{wg}}/{2}-\left|u_{1}\right|\right), \quad
V_{2\,\mathrm{pp}} = \frac{\varepsilon_{\mathrm{wg}}^{-1}-\varepsilon_{\mathrm{clad}}^{-1}}{\omega}H\left(V_{\mathrm{wg}}/{2}-\left|u_{1}\right|\right),
\\
V_{3\,\mathrm{pp}} = V_{4\,\mathrm{pp}} = \omega\left(\varepsilon_{\mathrm{wg}}-\varepsilon_{\mathrm{clad}}\right)H\left(V_{\mathrm{wg}}/{2}-\left|u_{1}\right|\right), \quad
V_{5\,\mathrm{pp}} = \frac{\mu_{\mathrm{wg}}^{-1}-\mu_{\mathrm{clad}}^{-1}}{\omega}H\left(V_{\mathrm{wg}}/{2}-\left|u_{1}\right|\right), 
\end{gathered} \right.
\label{eq: Optical case Vpp}
\end{equation}
where $\varepsilon_{\mathrm{wg}}$ and $\mu_{\mathrm{wg}}$ ($\varepsilon_{\mathrm{clad}}$ and $\mu_{\mathrm{clad}}$) are waveguide (clad) parameters, and $V_{\mathrm{wg}}$ is height of the waveguide as shown in Fig. \ref{fig: waveguide}(b).
By using the normalized roughness parameters of eq. (\ref{eq: roughness A 2 a}), we can represent the  $\widehat{\boldsymbol{V}^{\left(1\right)}}$ in eq. (\ref{eq: approach1 abrupt V1}):
\begin{equation}
\widehat{\boldsymbol{V}^{\left(1\right)}}=\boldsymbol{V}^{\left(1\right)}\left(\widehat{A_{\mathrm{w}}}\left(k\right),\widehat{A_{\mathrm{c}}}\left(k\right)\right) 
 = \sqrt{L_{\mathrm{s}}}
	\left[\frac{1}{2}\widehat{a_{\mathrm{w}}}\left(k\right) + \frac{u_{0}}{W_{\mathrm{wg}}/2}\widehat{a_{\mathrm{c}}}\left(k\right)\right]
	\boldsymbol{V}_{\mathrm{pp}}\left(u_{1}\right)
	\delta\left(W_{\mathrm{wg}}/2-\left|u_{0}\right|\right) .
\label{eq: opitcal roughness 1}
\end{equation}

Next subsection shows another approach for boundary-roughness scattering.

\subsection{Details of Approach II for \ref{subsec: approach II}\label{subsec: optical approach II}}

By using $\mathit{\Delta}h_{0}$ and $\mathit{\Delta}h_{1}$ of eq. (\ref{eq: approx2 h2-h0}), $h_{2}/h_{0}$ and $h_{0}h_{2}$ in the $\boldsymbol{V}$ of eq. (\ref{eq: the simplest Maxwell equation}) are approximated to
\[
\left\{ \begin{aligned}
\frac{h_{2}}{h_{0}} = \left(1+\mathit{\Delta}h_{2}\right)/\left(1+\mathit{\Delta}h_{0}\right) &\simeq 1 + \mathit{\Delta}h_{2} - \mathit{\Delta}h_{0}\,, 
\\
h_{0} h_{2} = \left(1+\mathit{\Delta}h_{0}\right)\left(1+\mathit{\Delta}h_{2}\right) &\simeq 1 + \mathit{\Delta}h_{2} + \mathit{\Delta}h_{0}\,. 
\end{aligned} \right.
\]
From the above equations, let us introduce parameter vector $\boldsymbol{s}$ to Approach II.
Elements of $\boldsymbol{s}$ are defined by
\begin{equation}
{s}_{j+3J}=\mathit{\Delta}h_{2}\left(u_{0},u_{2}\right) - (-1)^{\left[\left(j+J\right)\%3\right]\%2} \mathit{\Delta}h_{0}\left(u_{2}\right) 
\quad \mathrm{for} \quad
\left\{ \begin{aligned}
j&=0,1,2,\\J&=0,1.
\end{aligned} \right.
\label{eq: V^1 for opitcal roughness 2}
\end{equation}
The $\boldsymbol{V}^{\left(1\right)}$ in Subsection \ref{subsec: approach II} can be given as product between the $\boldsymbol{V}^{\left(0\right)}$ of eq. (\ref{eq: V^(0) for straight WG}) and the $\boldsymbol{s}$ of eq. (\ref{eq: V^1 for opitcal roughness 2}): 
\[\boldsymbol{V}^{\left(1\right)}=\boldsymbol{V}^{\left(0\right)}\boldsymbol{s} = \mathrm{diag}\left(V^{\left(0\right)}_{0}{s}_{0},\, \ldots, \, V^{\left(0\right)}_{5}{s}_{5}\right).\]

The $\widehat{\boldsymbol{V}^{\left(1\right)}}\left(k\right)$ of eq. (\ref{eq: approach2 FM1V1}) is also given as $\widehat{\boldsymbol{V}^{\left(1\right)}}\left(k\right)=\boldsymbol{V}^{\left(0\right)} \widehat{\boldsymbol{s}}\left(k\right)$.
By using eqs. (\ref{eq: approach2 FM1V1}), (\ref{eq: roughness A 2 a}) and (\ref{eq: V^1 for opitcal roughness 2}), elements of $\widehat{\boldsymbol{s}}$ is that
\begin{equation}
 \frac{\widehat{s_{j+3J}} \left(k\right)}{\sqrt{L_{\mathrm{s}}}}
 = 
 u_{0} k^{2} \widehat{a_{\mathrm{c}}}\left(k\right)
 + \frac{1}{W_{\mathrm{wg}}} \left( \frac{1}{2} u_{0}^{2} k^{2} - (-1)^{\left[\left(j+J\right)\%3\right]\%2}  \right) \widehat{a_{\mathrm{w}}}\left(k\right)
\quad \mathrm{for} \quad
\left\{ \begin{aligned}
j&=0,1,2,\\J&=0,1.
\end{aligned} \right.
\label{eq: opitcal roughness 2}
\end{equation}
When $a_{\mathrm{w}}$ and $a_{\mathrm{c}}$ are not correlated as mentioned in the end of Section \ref{sec: Auto-correlation function}, we have to use independently each term for $\widehat{a_{\mathrm{w}}}$ and $\widehat{a_{\mathrm{c}}}$ in eqs. (\ref{eq: opitcal roughness 1}) and (\ref{eq: opitcal roughness 2}).

%% file: App/G_NonUniformMesh_v1.tex
\chapter{Non-uniform mesh\label{ch: Non-uniform mesh}}
\begin{figure}[h!] 
\centering{}\includegraphics[bb=0bp 0bp 402bp 570bp,clip,width=0.35\columnwidth]{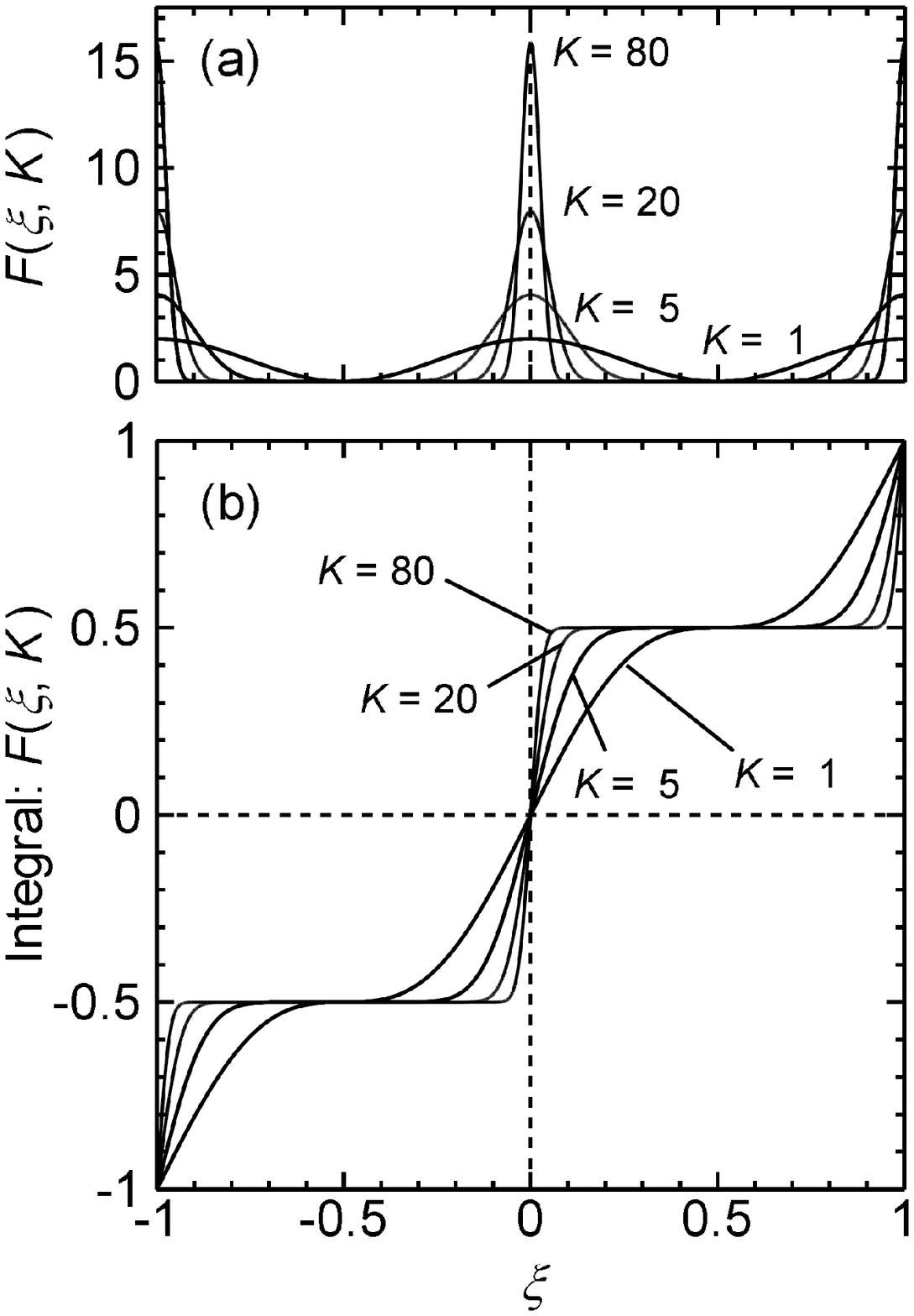}\caption{(a) We plot $F\left(\xi,\,K\right)$ when $K=1$, $5$, $20$, and $80$. 
(b) The staircase of Eq. (\ref{eq: integral of F}).\label{fig: function F}}
\end{figure}

This chapter shows an example of non-uniform mesh techniques discussed in Section \ref{sec: u to xi}.
Then we define a periodic function $F\left({\xi},\,K\right)$, and $F\left({\xi}_{j}/L_{j},\,K_{j}\right)$ is introduced into $du_{j}/d{\xi}_{j}$. 
\begin{equation}
F\left({\xi},\,K\right) =2^{2K-1}\cos^{2K}\left(\pi{\xi}\right)\prod_{I=1}^{K-1}\frac{I}{K+I}\,.
\label{eq: function F}
\end{equation}
Figure \ref{fig: function F} (a) plots the function $F\left(\xi,\,K\right)$ for $K=1$, $5$, $20$ and $80$, and it shows that $F\left(\xi,\,80\right)\simeq0$ when $0.1<\xi<0.9$. 

By using a formula
\[
\cos^{2n}\theta  = \frac{1}{2^{2n-1}}\left[\sum_{r=0}^{n-1}\left(\begin{array}{c}
2n\\
r
\end{array}\right)\cos\left(2n-2r\right)\theta+\frac{1}{2}\left(\begin{array}{c}
2n\\
n
\end{array}\right)\right]\,
\]
from Iwanami Formulae II p. 190,
eq. (\ref{eq: function F}) is deformed into
\begin{equation}
\begin{split}F\left(\xi,\,K\right) & =
\left[\frac{1}{2}\left(\begin{array}{c}
2K\\
K
\end{array}\right)+\sum_{r=0}^{K-1}\left(\begin{array}{c}
2K\\
r
\end{array}\right)\cos\left(2\left(K-r\right)\pi\xi\right)\right]\prod_{I=1}^{K-1}\frac{I}{K+I}
\\ & 
=1+\sum_{r=0}^{K-1}\frac{\left(2K\right)!}{r!\left(2K-r\right)!}\frac{\left(K-1\right)!K!}{\left(2K-1\right)!}\cos\left(2\left(K-r\right)\pi\xi\right)
\\ & 
=1+\sum_{r=0}^{K-1}2K\frac{K!}{\left(2K-r\right)!}\frac{\left(K-1\right)!}{r!}\cos\left(2\left(K-r\right)\pi\xi\right)
\\ & 
=1+\sum_{J=1}^{K}2\frac{K!}{\left(K+J\right)!}\frac{K!}{\left(K-J\right)!}\cos\left(2\pi J\xi\right)
=1+\sum_{J=1}^{K}2\left(\prod_{I=1}^{J}\frac{K+1-I}{K+I}\right)\cos\left(2\pi J\xi\right)
\\ & 
=1+\sum_{J=1}^{K}2\left(\prod_{I=1}^{J}\frac{K-J+I}{K+I}\right)\cos\left(2\pi J\xi\right)\,,
\end{split}
\label{eq: deformed F}
\end{equation}
where
\[
\frac{1}{2}\left(\begin{array}{c}
2K\\
K
\end{array}\right)\prod_{I=1}^{K-1}\frac{I}{K+I}  =  \frac{1}{2}\frac{\left(2K\right)!}{K!K!}\frac{\left(K-1\right)!K!}{\left(2K-1\right)!}
  =  1\,.
\]
Equations (\ref{eq: function F}) and (\ref{eq: deformed F}) for $K = 1,2$ could be checked as
\[\left\{
\begin{aligned}
F\left(\xi,\,1\right) & =2^{2-1}\cos^{2}\left(\pi\xi\right) = 1 + \cos\left(2\pi\xi\right)\,,
\\
F\left(\xi,\,2\right) & =2^{4-1}\cos^{4}\left(\pi\xi\right)\prod_{I=1}^{2-1}\frac{I}{2+I}
= \frac{8}{3}\frac{1+2\cos\left(2\pi\xi\right)+\frac{1+\cos\left(4\pi\xi\right)}{2}}{4}
\\
 & =1+2\frac{2-1+1}{2+1}\cos\left(2\pi\xi\right)+2\left(\prod_{I=1}^{2}\frac{I}{2+I}\right)\cos\left(4\pi\xi\right)
=1+\frac{4}{3}\cos\left(2\pi\xi\right)+\frac{1}{3}\cos\left(4\pi\xi\right)\,.
\end{aligned}
\right.
\]

From eq. (\ref{eq: deformed F}), we obtain a formula for the integral of $F\left(\xi,\,K\right)$. 
\begin{equation}
\int_{0}^{\xi}F\left(x,\,K\right)dx=\xi+\sum_{J=1}^{K}2\left(\prod_{I=1}^{J}\frac{K-J+I}{K+I}\right)\frac{\sin\left(2\pi J\xi\right)}{2\pi J}\,.\label{eq: integral of F}
\end{equation}
Figure \ref{fig: function F}(b) shows eq. (\ref{eq: integral of F}) for $K=1$, $5$, $20$ and $80$. 
The integral of $F$ has staircase shape. 

 The functions (\ref{eq: function F}) and (\ref{eq: integral of F}) have the following properties:
\[
\left\{
\begin{aligned}
F\left(\xi+1,\,K\right) & =  F\left(-\xi,\,K\right)=F\left(\xi,\,K\right)\,,
\\
\max F\left(\xi,\,K\right) & =  F\left(0,\,K\right)=2^{2K-1}\prod_{I=1}^{K-1}\frac{I}{K+I}\,,
\\
\min F\left(\xi,\,K\right) & =  F\left(\frac{1}{2},\,K\right)
=0\,,
\\
\int_{0}^{1}F\left(\xi,\,K\right)d\xi & =  1\,,
\\
\lim_{K\rightarrow\infty}F\left(\xi,\,K\right) & =  \sum_{I=-\infty}^{\infty}\delta\left(\xi-I\right),
\end{aligned}
\right.
\]
where $\delta\left(\xi\right)$ is the Dirac delta function. 

When ten parameters $L_{j}$, $u_{j}\left(0\right)$, $u_{j}\left(L_{j}\right)$, $\min\left(du_{j}/d{\xi}_{j}\right)$ and $K_{j}$ for $j=0,\,1$ are given, we can derive the $u_{j}$ by using eq. (\ref{eq: integral of F}).
\begin{equation}
\left\{
\begin{aligned}
u_{j}\left({\xi}_{j}\right) & = u_{j}\left(0\right) + {\xi}_{j}\min\left(\frac{du_{j}}{d{\xi}_{j}}\right)
+ C_{j} \int^{{\xi}_{j}/L_{j}}_{0} F\left(x,\,K_{j}\right)dx\,,
\\
\frac{du_{j}}{d{\xi}_{j}}\left({\xi}_{j}\right) & = \min\left(\frac{du_{j}}{d{\xi}_{j}}\right)
+ \frac{C_{j}}{L_{j}} F\left({\xi}_{j}/L_{j},\,K_{j}\right)\,,
\\
C_{j} & = u_{j}\left(L_{j}\right) - u_{j}\left(0\right) - L_{j}\min\left(\frac{du_{j}}{d{\xi}_{j}}\right)\,.
\end{aligned}
\right.
\label{eq: function u_j}
\end{equation}
The maximum of ${du_{j}}/{d{\xi}_{j}}$ is given by
\[
\max\left(\frac{du_{j}}{d{\xi}_{j}}\right)  =  \min\left(\frac{du_{j}}{d{\xi}_{j}}\right)
+ \left(2^{2K_{j}-1} \prod_{I=1}^{K_{j}-1}\frac{I}{K_{j}+I}\right) \frac{C_{j}}{L_{j}}\,.
\]

%% file: App/H_DiscDepend_v2.tex
\chapter{Relaxation of discretization dependency\label{ch: RelaxDisc.}}

This chapter shows correction parameter in order to reduce discretization dependency.

\section{Correction factor}

The discretization of differential operator reduces the wavenumber of plane-wave $\exp\left( k x\right)$, because
\[
k > \frac{2}{\mathit{\Delta} x}\sin\left(k\frac{\mathit{\Delta} x}{2}\right) \quad \mathrm{where} \quad k > 0 \quad \mathrm{and} \quad 0 < k\mathit{\Delta} x < \pi\,.
\]
Let us introduce a correction factor $\eta_{x}$ to emphasis the finite difference as
\[
\frac{2}{\mathit{\Delta} x}\sin\left(k\frac{\mathit{\Delta} x}{2}\right) \Rightarrow \frac{2\left(1+\eta_{x}\right)}{\mathit{\Delta} x}\sin\left(k\frac{\mathit{\Delta} x}{2}\right) \,.
\]
We consider $U_{j}$ that surface integral on sphere for squared norm of difference between the $k_{j}$ and the discretized one:
\[
U_{j}=\frac{1}{4\pi}\int_{0}^{\pi}\sin\theta d\theta\int_{0}^{2\pi}d\phi\sum_{m=1}^{2}\left(\frac{2\left(1+\eta_{j}\right)}{\mathit{\Delta} x_{j+m}}\sin\left(k_{j+m}\frac{\mathit{\Delta} x_{j+m}}{2}\right)-k_{j+m}\right)^{2},
\]
where
\begin{gather*}
	k_{0} = k_{r}\sin\theta\cos\phi, \quad k_{1} = k_{r}\sin\theta\sin\phi, \quad k_{2} = k_{r}\cos\theta.
\end{gather*}
Note that $\int_{0}^{\pi}\sin\theta d\theta\int_{0}^{2\pi}d\phi=4\pi$.{}
Variation of the $U_{j}$ is given by 
\[
\frac{\partial U_{j}}{\partial\eta_{j}}	=	\sum_{m=1}^{2}\frac{1}{\pi\left(\mathit{\Delta} x_{j+m}\right)^{2}}\int_{0}^{\pi}\sin\theta d\theta\int_{0}^{2\pi}d\phi\left(2\left(1+\eta_{j}\right)\sin\left(k_{j+m}\frac{\mathit{\Delta} x_{j+m}}{2}\right)-k_{j+m}\mathit{\Delta} x_{j+m}\right)\sin\left(k_{j+m}\frac{\mathit{\Delta} x_{j+m}}{2}\right). 
\]
Then we can obtain $\eta_{j}$ at $\partial U_{j}/\partial\eta_{j}=0$ as follows. 
\begin{align*}
\eta_{j}	&=	\frac{\sum_{m=1}^{2}\left(\mathit{\Delta} x_{j+m}\right)^{-2}\int_{0}^{\pi}\sin\theta d\theta\int_{0}^{2\pi}d\phi\,\left(\sin\left(k_{j+m}\frac{\mathit{\Delta} x_{j+m}}{2}\right)-k_{j+m}\frac{\mathit{\Delta} x_{j+m}}{2}\right)\sin\left(k_{j+m}\frac{\mathit{\Delta} x_{j+m}}{2}\right)}{-\sum_{m=1}^{2}\left(\mathit{\Delta} x_{j+m}\right)^{-2}\int_{0}^{\pi}\sin\theta d\theta\int_{0}^{2\pi}d\phi\,\sin^{2}\left(k_{j+m}\frac{\mathit{\Delta} x_{j+m}}{2}\right)}\\
	&\simeq	\frac{\sum_{m=1}^{2}\left(\mathit{\Delta} x_{j+m}\right)^{2}\int_{0}^{\pi}\sin\theta d\theta\int_{0}^{2\pi}d\phi\,\left(\frac{k_{j+m}}{2}\right)^{4}}
	{3!\sum_{m=1}^{2}\int_{0}^{\pi}\sin\theta d\theta\int_{0}^{2\pi}d\phi\,\left(\frac{k_{j+m}}{2}\right)^{2}},
\end{align*}
where
\[
\int_{0}^{\pi}\sin\theta d\theta\int_{0}^{2\pi}d\phi\,\left(\frac{k_{j}}{2}\right)^{2n}=\left(\frac{k_{r}}{2}\right)^{2n}\frac{4\pi}{2n+1}\,.{}
\]
When $k_{j}\mathit{\Delta} x_{j}/2 \ll 1$, we set
\begin{equation}
\eta_{j}	=\frac{\sum_{m=1}^{2}\left(\mathit{\Delta} x_{j+m}\right)^{2}\left(\frac{k_{r}}{2}\right)^{4}\frac{4\pi}{5}}{3!\sum_{m=1}^{2}\left(\frac{k_{r}}{2}\right)^{2}\frac{4\pi}{3}} 
=	\frac{k_{r}^{2}\left(\mathit{\Delta} x_{j+1}^{2}+\mathit{\Delta} x_{j+2}^{2}\right)}{80}\,.\label{eq: factor eta}
\end{equation}

\section{Definitions of $\mathit{\Delta} x_{j}$ and $k_{r}$}
From section \ref{sec: u to xi}, $\mathit{\Delta} x_{j}$ for $j=0,1,2$ and $k_{r}$ in eq. \eqref{eq: factor eta} are defined as
\begin{equation}
\begin{split}
\mathit{\Delta} x_{j} \left(\boldsymbol{\xi}\right) &= \frac{d{u}_{j}}{d{\xi}_{j}}\,h_{j}
= u_{j}^{'}\left({\xi}_{j}\right)\, h_{j}\left(u_{0}\left({\xi}_{0}\right),u_{1}\left({\xi}_{1}\right),u_{2}\left({\xi}_{2}\right)\right)\, ,\\
k_{r} \left(\boldsymbol{\xi}\right) &= \omega \sqrt{\varepsilon\left(\boldsymbol{\xi}\right)\,\mu\left(\boldsymbol{\xi}\right)}\,.
\end{split}
\label{eq: Dx and kr}
\end{equation}
Note that eq. \eqref{eq: Dx and kr} is referenced in the CFL condition of eq. \eqref{eq: CFL condition}.
From eqs. \eqref{eq: factor eta} and \eqref{eq: Dx and kr}, correction factor $\eta_{j}$ is given by
\begin{equation}
\eta_{j}\left(\boldsymbol{\xi}\right)	
=\frac{\omega^{2}}{80} \varepsilon\left(\boldsymbol{\xi}\right)\,\mu\left(\boldsymbol{\xi}\right){}
\sum_{m=1}^{2}\left(u_{j+m}^{'}\left({\xi}_{j+m}\right)\, h_{j+m}\left(\boldsymbol{u}\left(\boldsymbol{\xi}\right)\right)\right)^{2}\,.\label{eq: disc eta}
\end{equation}
 Note that each point in a Yee cell only has $\varepsilon$ or $\mu$ as shown in Fig. \ref{fig: Yee cell}, and then $\mu$ and $\varepsilon$ in \eqref{eq: disc eta} have to be approximated as
\begin{equation}
\begin{split}
	&\quad \mu \left({l}_{0} + \frac{1-\delta_{0j}}{2},\, {l}_{1} + \frac{1-\delta_{1j}}{2},\, {l}_{2} + \frac{1-\delta_{2j}}{2}\right)\\
	&\simeq \frac{1}{4}
		\sum_{m=0}^{1}\sum_{n=0}^{1}
		\mu\left({l}_{0} + \frac{\delta_{2\,j+m}}{2}+n\delta_{2\,j+1-m},\, {l}_{1} + \frac{\delta_{0\,j+m}}{2}+n\delta_{0\,j+1-m},\, {l}_{2} + \frac{\delta_{1\,j+m}}{2}+n\delta_{1\,j+1-m}\right)\, ,\\
	& \quad \varepsilon \left({l}_{0} + \frac{\delta_{0j}}{2},\, {l}_{1} + \frac{\delta_{1j}}{2},\, {l}_{2} + \frac{\delta_{2j}}{2}\right)\\
	&\simeq \frac{1}{4}
		\sum_{m=0}^{1}\sum_{n=0}^{1}
		\varepsilon\left({l}_{0} + \frac{1-\delta_{2\,j+m}}{2}-n\delta_{2\,j+1-m},\, {l}_{1} + \frac{1-\delta_{0\,j+m}}{2}-n\delta_{0\,j+1-m},\, {l}_{2} + \frac{1-\delta_{1\,j+m}}{2}-n\delta_{1\,j+1-m}\right)\, ,
\end{split}
		\label{eq: average epsilon and mu}
\end{equation}
where $j = 0, 1, 2$.
Equations \eqref{eq: disc eta} and \eqref{eq: average epsilon and mu} give us the $\eta$ in eq. \eqref{eq: epsilon_l mu_l}.

Here, we can check eq. \eqref{eq: average epsilon and mu} as follows.
\begin{align*}
	\mu \left({l}_{0}, {l}_{1} + \frac{1}{2}, {l}_{2} + \frac{1}{2}\right) &\simeq \frac{\mu\left({l}_{0}, {l}_{1} + \frac{1}{2}, {l}_{2}\right)+\mu\left({l}_{0}, {l}_{1}, {l}_{2} + \frac{1}{2}\right)+\mu\left({l}_{0}, {l}_{1} + \frac{1}{2}, {l}_{2} + 1\right)+\mu\left({l}_{0}, {l}_{1} + 1, {l}_{2} + \frac{1}{2}\right)}{4}\\
	\mu \left({l}_{0} + \frac{1}{2}, {l}_{1}, {l}_{2} + \frac{1}{2}\right) &\simeq \frac{\mu\left({l}_{0}, {l}_{1}, {l}_{2} + \frac{1}{2}\right)+\mu\left({l}_{0} + \frac{1}{2}, {l}_{1}, {l}_{2}\right)+\mu\left({l}_{0} + 1, {l}_{1}, {l}_{2} + \frac{1}{2}\right)+\mu\left({l}_{0} + \frac{1}{2}, {l}_{1}, {l}_{2} + 1\right)}{4}\\
	\mu \left({l}_{0} + \frac{1}{2}, {l}_{1} + \frac{1}{2}, {l}_{2}\right) &\simeq \frac{\mu\left({l}_{0} + \frac{1}{2}, {l}_{1}, {l}_{2}\right)+\mu\left({l}_{0}, {l}_{1} + \frac{1}{2}, {l}_{2}\right)+\mu\left({l}_{0} + \frac{1}{2}, {l}_{1} + 1, {l}_{2}\right)+\mu\left({l}_{0} + 1, {l}_{1} + \frac{1}{2}, {l}_{2}\right)}{4}
\end{align*}
and
\begin{align*}
	\varepsilon \left({l}_{0} + \frac{1}{2}, {l}_{1}, {l}_{2}\right) &\simeq \frac{\varepsilon \left({l}_{0} + \frac{1}{2}, {l}_{1}, {l}_{2} + \frac{1}{2}\right)+\varepsilon \left({l}_{0} + \frac{1}{2}, {l}_{1} + \frac{1}{2}, {l}_{2}\right)+\varepsilon \left({l}_{0} + \frac{1}{2}, {l}_{1}, {l}_{2} - \frac{1}{2}\right)+\varepsilon \left({l}_{0} + \frac{1}{2}, {l}_{1} - \frac{1}{2}, {l}_{2}\right)}{4}\\
	\varepsilon \left({l}_{0}, {l}_{1} + \frac{1}{2}, {l}_{2}\right) &\simeq \frac{\varepsilon \left({l}_{0} + \frac{1}{2}, {l}_{1} + \frac{1}{2}, {l}_{2}\right)+\varepsilon \left({l}_{0}, {l}_{1} + \frac{1}{2}, {l}_{2} + \frac{1}{2}\right)+\varepsilon \left({l}_{0} - \frac{1}{2}, {l}_{1} + \frac{1}{2}, {l}_{2}\right)+\varepsilon \left({l}_{0}, {l}_{1} + \frac{1}{2}, {l}_{2} - \frac{1}{2}\right)}{4}\\
	\varepsilon \left({l}_{0}, {l}_{1}, {l}_{2} + \frac{1}{2}\right) &\simeq \frac{\varepsilon \left({l}_{0}, {l}_{1} + \frac{1}{2}, {l}_{2} + \frac{1}{2}\right)+\varepsilon \left({l}_{0} + \frac{1}{2}, {l}_{1}, {l}_{2} + \frac{1}{2}\right)+\varepsilon \left({l}_{0}, {l}_{1} - \frac{1}{2}, {l}_{2} + \frac{1}{2}\right)+\varepsilon \left({l}_{0} - \frac{1}{2}, {l}_{1}, {l}_{2} + \frac{1}{2}\right)}{4}
\end{align*}

%% file: App/I_DiscEquations_v2.tex
\chapter{Details of discrete equations\label{ch: details of disc. eq.}}

This chapter shows detailed derivation of discrete equations.

\section{Derivation of propagation equation (\ref{eq: discrete propagation-eq})\label{sec: detailed disc propagation eq}}

Details of the discrete Maxwell equation (\ref{eq: discrete Maxwell eq}) are given as follows.
\begin{equation}
\left\{
\begin{aligned}
\left(\begin{array}{ccc}
0 & -\bigtriangleup_{2} & \bigtriangleup_{1}\\
\bigtriangleup_{2} & 0 & -\bigtriangleup_{0}\\
-\bigtriangleup_{1} & \bigtriangleup_{0} & 0
\end{array}\right)
\left(\begin{array}{c}
H_{l0}\\H_{l1}\\H_{l2}
\end{array}\right)
& = -i\omega
\left(\begin{array}{ccc}
\varepsilon_{l00} & 0 & 0\\
0 & \varepsilon_{l11} & 0\\
0 & 0 & \varepsilon_{l22}
\end{array}\right)
\left(\begin{array}{c}
E_{l0}\\E_{l1}\\E_{l2}
\end{array}\right),
\\
\left(\begin{array}{ccc}
0 & -\bigtriangledown_{2} & \bigtriangledown_{1}\\
\bigtriangledown_{2} & 0 & -\bigtriangledown_{0}\\
-\bigtriangledown_{1} & \bigtriangledown_{0} & 0
\end{array}\right)
\left(\begin{array}{c}
E_{l0}\\E_{l1}\\E_{l2}
\end{array}\right)
& = i\omega
\left(\begin{array}{ccc}
\mu_{l00} & 0 & 0\\
0 & \mu_{l11} & 0\\
0 & 0 & \mu_{l22}
\end{array}\right)
\left(\begin{array}{c}
H_{l0}\\H_{l1}\\H_{l2}
\end{array}\right).
\end{aligned}\right.
\label{eq: details disc Maxwell eq}
\end{equation}
The four equations in eq. (\ref{eq: details disc Maxwell eq}) are
\begin{equation}
\left\{
\begin{aligned}
-\bigtriangleup_{2}H_{l1} + \bigtriangleup_{1}H_{l2} & = -i\omega \varepsilon_{l00} E_{l0},
\\
\bigtriangleup_{2}H_{l0} - \bigtriangleup_{0}H_{l2} & = -i\omega \varepsilon_{l11} E_{l1},
\\
-\bigtriangledown_{2}E_{l1} + \bigtriangledown_{1}E_{l2} & = i\omega \mu_{l00} H_{l0},
\\
\bigtriangledown_{2}E_{l0} - \bigtriangledown_{0}E_{l2} & = i\omega \mu_{l11} H_{l1}.
\end{aligned}\right.
\label{eq: details 4 equations}
\end{equation}
From eq. (\ref{eq: details disc Maxwell eq}), the $E_{l2}$ and $H_{l2}$ can be represented by other components:
\begin{equation}
E_{l2}  = \frac{i}{\omega\varepsilon_{l22}}
\left( -\bigtriangleup_{1}H_{l0}+\bigtriangleup_{0}H_{l1}\right),
\quad\mathrm{and}\quad
H_{l2}  = \frac{-i}{\omega\mu_{l22}}
\left( -\bigtriangledown_{1}E_{l0}+\bigtriangledown_{0}E_{l1}\right).{}
\label{eq: E2 and H2}
\end{equation}
The above $E_{l2}$ and $H_{l2}$ of \eqref{eq: E2 and H2} can be substituted into eqs. (\ref{eq: details 4 equations}), and then
\[\left\{
\begin{aligned}
-\bigtriangleup_{2}H_{l1} + \bigtriangleup_{1}
\left[
\frac{-i}{\omega\mu_{l22}}
\left( -\bigtriangledown_{1}E_{l0}+\bigtriangledown_{0}E_{l1}\right)
\right]
 & = -i\omega \varepsilon_{l00} E_{l0},
\\
\bigtriangleup_{2}H_{l0} - \bigtriangleup_{0}
\left[
\frac{-i}{\omega\mu_{l22}}
\left( -\bigtriangledown_{1}E_{l0}+\bigtriangledown_{0}E_{l1}\right)
\right]
 & = -i\omega \varepsilon_{l11} E_{l1},
\\
-\bigtriangledown_{2}E_{l1} + \bigtriangledown_{1}
\left[
\frac{i}{\omega\varepsilon_{l22}}
\left( -\bigtriangleup_{1}H_{l0}+\bigtriangleup_{0}H_{l1}\right)
\right]
 & = i\omega \mu_{l00} H_{l0},
\\
\bigtriangledown_{2}E_{l0} - \bigtriangledown_{0}
\left[
\frac{i}{\omega\varepsilon_{l22}}
\left( -\bigtriangleup_{1}H_{l0}+\bigtriangleup_{0}H_{l1}\right)
\right]
 & = i\omega \mu_{l11} H_{l1}.
\end{aligned}\right.
\]
Therefore, the above four equations are deformed into
\begin{equation}\left\{
\begin{aligned}
-i\bigtriangleup_{2}H_{l0} &=  \bigtriangleup_{0}
\left[
\frac{1}{\omega\mu_{l22}}
\left( \bigtriangledown_{1}E_{l0} - \bigtriangledown_{0}E_{l1}\right)
\right]
 - \omega \varepsilon_{l11} E_{l1},
\\
-i\bigtriangleup_{2}H_{l1} &=  \bigtriangleup_{1}
\left[
\frac{1}{\omega\mu_{l22}}
\left( \bigtriangledown_{1}E_{l0}-\bigtriangledown_{0}E_{l1}\right)
\right]
 + \omega \varepsilon_{l00} E_{l0},
\\
i\bigtriangledown_{2}E_{l1} & = \bigtriangledown_{1}
\left[
\frac{1}{\omega\varepsilon_{l22}}
\left( \bigtriangleup_{1}H_{l0}-\bigtriangleup_{0}H_{l1}\right)
\right]
 + \omega \mu_{l00} H_{l0},
\\
-i\bigtriangledown_{2}E_{l0} &=  \bigtriangledown_{0}
\left[
\frac{1}{\omega\varepsilon_{l22}}
\left( -\bigtriangleup_{1}H_{l0}+\bigtriangleup_{0}H_{l1}\right)
\right]
 + \omega \mu_{l11} H_{l1}.
\end{aligned}\right.
\label{eq: details of m_aa and m_bb}
\end{equation}
The four left sides in the above equations are obviously equal to the right side of eq. (\ref{eq: discrete propagation-eq}), and the four right sides give us matrix elements of $\boldsymbol{m}_{bb}$ and $\boldsymbol{m}_{aa}$ in eq. (\ref{eq: discrete m_aa m_bb Psi}).

\section{Details of power flow relations in eq. (\ref{eq: P_+- relations})\label{sec: detailed disc power flow eq}}

From eq. (\ref{eq: discrete propagation-eq}),
\[
\boldsymbol{H}_{2D}^{\dagger}\boldsymbol{m}_{aa}\boldsymbol{H}_{2D}=-i\boldsymbol{H}_{2D}^{\dagger}\bigtriangledown_{2}\boldsymbol{E}_{2D}
= -i\boldsymbol{H}_{2D}^{\dagger}\left[ l_{2} \right]\boldsymbol{E}_{2D}\left[ l_{2} \right]
+i\boldsymbol{H}_{2D}^{\dagger}\left[ l_{2} \right]\boldsymbol{E}_{2D}\left[ l_{2}-1 \right]
\]
and
\[
\boldsymbol{E}_{2D}^{\dagger}\boldsymbol{m}_{bb}\boldsymbol{E}_{2D}=-i\boldsymbol{E}_{2D}^{\dagger}\bigtriangleup_{2}\boldsymbol{H}_{2D}
= -i\boldsymbol{E}_{2D}^{\dagger}\left[ l_{2} \right]\boldsymbol{H}_{2D}\left[ l_{2}+1 \right]
+i\boldsymbol{E}_{2D}^{\dagger}\left[ l_{2} \right]\boldsymbol{H}_{2D}\left[ l_{2} \right]
\]
By using the definition $\boldsymbol{\Psi}_{\pm}$ in eq. (\ref{eq: discrete m_aa m_bb Psi}), the above equations are deformed to
\begin{equation}\left\{
\begin{split}
& i\boldsymbol{H}_{2D}^{\dagger}\left[ l_{2} \right]\left(\boldsymbol{m}_{aa}\left[ l_{2} \right]-\boldsymbol{m}_{aa}^{\dagger}\left[ l_{2} \right]\right)\boldsymbol{H}_{2D}\left[ l_{2} \right]\\
=\, & \boldsymbol{H}_{2D}^{\dagger}\left[ l_{2} \right]\boldsymbol{E}_{2D}\left[ l_{2} \right]
-\boldsymbol{H}_{2D}^{\dagger}\left[ l_{2} \right]\boldsymbol{E}_{2D}\left[ l_{2}-1 \right]
+\boldsymbol{E}_{2D}^{\dagger}\left[ l_{2} \right]\boldsymbol{H}_{2D}\left[ l_{2} \right]
-\boldsymbol{E}_{2D}^{\dagger}\left[ l_{2}-1 \right]\boldsymbol{H}_{2D}\left[ l_{2} \right]
\\
=\, & \boldsymbol{\Psi}_{+}^{\dagger} \left[ l_{2} \right]
\left(\begin{array}{cc} 0&1 \\ 1&0 \end{array}\right)\boldsymbol{\Psi}_{+} \left[ l_{2} \right]
-\boldsymbol{\Psi}_{-}^{\dagger} \left[ l_{2}-1 \right]
\left(\begin{array}{cc} 0&1 \\ 1&0 \end{array}\right)\boldsymbol{\Psi}_{-} \left[ l_{2}-1 \right],
\\
& i\boldsymbol{E}_{2D}^{\dagger}\left[ l_{2} \right]\left(\boldsymbol{m}_{bb}\left[ l_{2} \right]-\boldsymbol{m}_{bb}^{\dagger}\left[ l_{2} \right]\right)\boldsymbol{E}_{2D}\left[ l_{2} \right]\\
=\,& \boldsymbol{E}_{2D}^{\dagger}\left[ l_{2} \right]\boldsymbol{H}_{2D}\left[ l_{2}+1 \right]
-\boldsymbol{E}_{2D}^{\dagger}\left[ l_{2} \right]\boldsymbol{H}_{2D}\left[ l_{2} \right]
+\boldsymbol{H}_{2D}^{\dagger}\left[ l_{2}+1 \right]\boldsymbol{E}_{2D}\left[ l_{2} \right]
-\boldsymbol{H}_{2D}^{\dagger}\left[ l_{2} \right]\boldsymbol{E}_{2D}\left[ l_{2} \right]
\\
=\, & \boldsymbol{\Psi}_{-}^{\dagger} \left[ l_{2} \right]
\left(\begin{array}{cc} 0&1 \\ 1&0 \end{array}\right)\boldsymbol{\Psi}_{-} \left[ l_{2} \right]
-\boldsymbol{\Psi}_{+}^{\dagger} \left[ l_{2} \right]
\left(\begin{array}{cc} 0&1 \\ 1&0 \end{array}\right)\boldsymbol{\Psi}_{+} \left[ l_{2} \right].
\end{split}
\right.\label{eq: details power flow}
\end{equation}

\section{Derivation of transfer matrix of eq. (\ref{eq: discrete transfer matrix})}

From eqs. (\ref{eq: discrete propagation-eq}) and (\ref{eq: discrete m_aa m_bb Psi}),
\[
\left\{\begin{aligned}
i\boldsymbol{m}_{aa}\left[l_{2}\right] \boldsymbol{H}_{2D}\left[l_{2}\right] & = \boldsymbol{E}_{2D}\left[l_{2}\right] - \boldsymbol{E}_{2D}\left[l_{2}-1\right]\,,
\\
i\boldsymbol{m}_{bb}\left[l_{2}\right] \boldsymbol{E}_{2D}\left[l_{2}\right] & = \boldsymbol{H}_{2D}\left[l_{2}+1\right] - \boldsymbol{H}_{2D}\left[l_{2}\right]\,.
\end{aligned}\right.
\]
The above two equations are deformed to
\begin{equation}
\left\{\begin{aligned}
\boldsymbol{H}_{2D}\left[l_{2}+1\right] & =  \boldsymbol{H}_{2D}\left[l_{2}\right] + i\boldsymbol{m}_{bb}\left[l_{2}\right] \boldsymbol{E}_{2D}\left[l_{2}\right]\,,
\\
\boldsymbol{E}_{2D}\left[l_{2}+1\right] & = i\boldsymbol{m}_{aa}\left[l_{2}+1\right] \boldsymbol{H}_{2D}\left[l_{2}+1\right]+\boldsymbol{E}_{2D}\left[l_{2}\right]
\\
& = i\boldsymbol{m}_{aa}\left[l_{2}+1\right]\left(
\boldsymbol{H}_{2D}\left[l_{2}\right] + i\boldsymbol{m}_{bb}\left[l_{2}\right] \boldsymbol{E}_{2D}\left[l_{2}\right]
\right) + \boldsymbol{E}_{2D}\left[l_{2}\right]
\\
& = i\boldsymbol{m}_{aa}\left[l_{2}+1\right]\boldsymbol{H}_{2D}\left[l_{2}\right]
+\left( 1- \boldsymbol{m}_{aa}\left[l_{2}+1\right]\boldsymbol{m}_{bb}\left[l_{2}\right] \right) \boldsymbol{E}_{2D}\left[l_{2}\right]
\,.
\end{aligned}\right.
\label{eq: derivation for transfer matrix}
\end{equation}
Then, we obtain the transfer matrix of eq. (\ref{eq: discrete transfer matrix}).

\section{Check of $\Xi$ definition in eq. (\ref{eq: Xi definition})}

From eq. (\ref{eq: discrete eigenvalue eq}), the expansion in eq. (\ref{eq: Xi definition}) satisfies the discrete propagation equation (\ref{eq: discrete propagation-eq}) as
\begin{equation}
\begin{split}
\boldsymbol{M}_{+}\boldsymbol{\Psi}_{+}\left[l_{2}\right] & =
\left(\begin{array}{cc}
\boldsymbol{m}_{aa} & 0\\
0 & \boldsymbol{m}_{bb}
\end{array}\right)
\sum_{m\neq 0} c_{m} \exp\left(i\theta_{m}l_{2}\right)\boldsymbol{\Xi}_{m}
\\
& = \sum_{m\neq 0} c_{m} \exp\left(i\theta_{m}l_{2}\right)
\left(\begin{array}{cc}
\exp\left(-i\theta_{m}/2\right) & 0\\
0 & 1
\end{array}\right)
 2 \sin \left(
 {\theta_{m}}/{2}
 \right)\left(\begin{array}{cc}
0 & 1\\
1 & 0
\end{array}\right)\boldsymbol{\Phi}_{m}
\\
& = \sum_{m\neq 0} -ic_{m} \exp\left(i\theta_{m}l_{2}\right)
\left(\begin{array}{cc}
0 & \exp\left(-i\theta_{m}/2\right)\\
1 & 0
\end{array}\right)
\left(
 \exp \left( i{\theta_{m}}/{2} \right) - \exp \left( -i{\theta_{m}}/{2} \right)
\right)
\boldsymbol{\Phi}_{m}
\\
& = \sum_{m\neq 0} -ic_{m} \exp\left(i\theta_{m}l_{2}\right)
\left(\begin{array}{cc}
0 & 1 - \exp\left(-i\theta_{m}\right)\\
\exp \left( i{\theta_{m}}/{2} \right) - \exp \left( -i{\theta_{m}}/{2} \right) & 0
\end{array}\right)
\boldsymbol{\Phi}_{m}
\\
& = \sum_{m\neq 0} -i
\left(\begin{array}{cc}
0 & 1 - \exp\left(-i\theta_{m}\right)\\
\exp \left( i{\theta_{m}} \right) - 1 & 0
\end{array}\right)
c_{m} \exp\left(i\theta_{m}l_{2}\right)
\boldsymbol{\Xi}_{m}
= -i
\left(\begin{array}{cc}
0 & \bigtriangledown_{2}\\
\bigtriangleup_{2} & 0
\end{array}\right)
\boldsymbol{\Psi}_{+}\left[l_{2}\right].
\end{split}
\label{eq: discrete propagation eq by Xi}
\end{equation}

\section{ Born approximation for discrete Green's function in eq. (\ref{eq: discrete S-matrix}) \label{sec: discrete Green function}}

Phase shift $\Theta_{n}\left[l_{2}\right]$ from a center of the system is defined by
\begin{equation}
\Theta_{n}\left[l_{2}\right]  = 
\begin{cases}
\sum_{\frac{L_{2}-1}{2} \leq j < l_{2}}\frac{\theta_{n}\left[j\right]}{2} + \sum_{\frac{L_{2}-1}{2} < j \leq l_{2}}\frac{\theta_{n}\left[j\right]}{2}
& \mathrm{for} \quad l_{2} > \frac{L_{2}-1}{2}\,,
\\
0 & \mathrm{if} \quad l_{2} = \frac{L_{2}-1}{2}\,,
\\
-\sum_{l_{2} \leq j < \frac{L_{2}-1}{2}}\frac{\theta_{n}\left[j\right]}{2} - \sum_{l_{2} < j \leq \frac{L_{2}-1}{2}}\frac{\theta_{n}\left[j\right]}{2}
& \mathrm{for} \quad l_{2} < \frac{L_{2}-1}{2}\,.
\end{cases}
\label{eq: Theta(l_2)}
\end{equation}
The above equation is equal to eq. (\ref{eq: Theta(b_m)}) when $l_{2} = b_{m}$.
Equation (\ref{eq: discrete Psi b_m}) can be represented by Green's function which is defined as
\begin{equation}
\begin{split}
\boldsymbol{\Psi}_{n}\left[l_{2}\right]  &= \exp\left( i\Theta_{n}\left[l_{2}\right] \right)\left( \boldsymbol{\Xi}_{n}\left[l_{2}\right] + \sum_{1 \leq j \neq l_{2} < L_{2}-1} \boldsymbol{G}\left[l_{2},j \right]\right),
\\
\boldsymbol{G}\left[l_{2} \gtrless j,j \right] &\triangleq \sum_{m \gtrless 0} g_{m n}\left[j\right]
\exp\left\{ i\left(\Theta_{m}\left[l_{2}\right]-\Theta_{n}\left[l_{2}\right]\right) -i\left(\Theta_{m}\left[j\right]-\Theta_{n}\left[j\right]\right)\right\}
\boldsymbol{\Xi}_{m}\left[l_{2}\right]\,.
\end{split}
\label{eq: discrete Green func}
\end{equation}
This section considers only propagative modes (\textit{i.e.} $ 0 < \left| m \right| \leq n_{\max}$), and it shows derivation of the $\boldsymbol{G}\left[l_{2},j \right]$ from eq. (\ref{eq: discrete propagation-eq}).
We try to separate the $\boldsymbol{M}_{+}$ of eq. (\ref{eq: discrete propagation-eq}) into unperturbed term $\boldsymbol{M}_{+}^{\left(0\right)}$ and perturbed term $\boldsymbol{M}_{+}^{\left(1\right)}$.
\begin{equation}
 i\left(\boldsymbol{M}_{+}^{\left(0\right)}\left[l_{2}\right]+\boldsymbol{M}_{+}^{\left(1\right)}\left[l_{2}\right]\right)\boldsymbol{\Psi}_{n}\left[l_{2}\right] 
= 
 \left(\begin{array}{cc}
\boldsymbol{0} & \bigtriangledown_{2}\\
\bigtriangleup_{2} & \boldsymbol{0}
\end{array}\right)\boldsymbol{\Psi}_{n}\left[l_{2}\right]
\,,
\label{eq: discrete M^0 M^1 Psi_n}
\end{equation}
where the $\boldsymbol{M}_{+}^{\left(j\right)}$ for $j=0,1$ and function of $l_{2}$ are commutative, \textit{e.g.} $\exp\left( i\Theta_{n}\left[l_{2}\right] \right) {\boldsymbol{M}}_{+}^{\left(j\right)}\left[l_{2}\right] = {\boldsymbol{M}}_{+}^{\left(j\right)}\left[l_{2}\right]\exp\left( i\Theta_{n}\left[l_{2}\right] \right)$.
Note that the coefficient $g_{m n}$ in eq. (\ref{eq: discrete Green func}) is set as small value: $g_{m n}\left[j\right]=O\left(\left\| {\boldsymbol{M}}_{+}^{\left(1\right)} \right\|\right)$.
By using the formula of eq. (\ref{eq: Formulas discrete diff}), the right side of eq. (\ref{eq: discrete M^0 M^1 Psi_n}) is deformed to 
\[
\begin{split}
& \quad
 \left(\begin{array}{cc}
\boldsymbol{0} & \bigtriangledown_{2}\\
\bigtriangleup_{2} & \boldsymbol{0}
\end{array}\right)
\exp\left( i\Theta_{n}\left[l_{2}\right] \right)\boldsymbol{\Xi}_{n}\left[l_{2}\right]
\\
&= 
 \left(\begin{array}{cc}
0 & 1 - \exp\left[-i\left(\frac{\theta_{n}\left[l_{2}\right]}{2}+\frac{\theta_{n}\left[l_{2}-1\right]}{2}\right)\right]\\
\exp \left[ i\left(\frac{\theta_{n}\left[l_{2}+1\right]}{2}+\frac{\theta_{n}\left[l_{2}\right]}{2}\right) \right] - 1 & 0
\end{array}\right)
\exp\left( i\Theta_{n}\left[l_{2}\right] \right)
\left(\begin{array}{cc}
1+\frac{\bigtriangleup_{2}}{2} & \boldsymbol{0}\\
\boldsymbol{0} & 1-\frac{\bigtriangledown_{2}}{2}
\end{array}\right)
\boldsymbol{\Xi}_{n}\left[l_{2}\right]
\\
& \quad +\left(\begin{array}{cc}
0 & \frac{1 + \exp\left[-i\left(\frac{\theta_{n}\left[l_{2}\right]}{2}+\frac{\theta_{n}\left[l_{2}-1\right]}{2}\right)\right]}{2}\\
\frac{\exp \left[ i\left(\frac{\theta_{n}\left[l_{2}+1\right]}{2}+\frac{\theta_{n}\left[l_{2}\right]}{2}\right) \right] + 1}{2} & 0
\end{array}\right)
\exp\left( i\Theta_{n}\left[l_{2}\right] \right)
\left(\begin{array}{cc}
\bigtriangleup_{2} & \boldsymbol{0}\\
\boldsymbol{0} & \bigtriangledown_{2}
\end{array}\right)
\boldsymbol{\Xi}_{n}\left[l_{2}\right]
\\
&= 
 \left(\begin{array}{cc}
0 & 1 - \exp\left[-i\left(\frac{\theta_{n}\left[l_{2}\right]}{2}+\frac{\theta_{n}\left[l_{2}-1\right]}{2}\right)\right]\\
\exp \left[ i\left(\frac{\theta_{n}\left[l_{2}+1\right]}{2}+\frac{\theta_{n}\left[l_{2}\right]}{2}\right) \right] - 1 & 0
\end{array}\right)
\exp\left( i\Theta_{n}\left[l_{2}\right] \right)
\boldsymbol{\Xi}_{n}\left[l_{2}\right]
\\
& \quad +\left(\begin{array}{cc}
0 & \exp\left( i\Theta_{n}\left[l_{2}-1\right] \right)\bigtriangledown_{2}\boldsymbol{\Xi}_{n}\left[l_{2}\right]\\
\exp\left( i\Theta_{n}\left[l_{2}+1\right] \right)\bigtriangleup_{2}\boldsymbol{\Xi}_{n}\left[l_{2}\right]  & 0
\end{array}\right).
\end{split}
\]
Furthermore, equation (\ref{eq: discrete propagation eq by Xi}) gives us
\[
i\boldsymbol{M}_{+}^{\left(0\right)}\left[l_{2}\right]
\boldsymbol{\Xi}_{n}\left[l_{2}\right]
= 
 \left(\begin{array}{cc}
0 & 1 - \exp\left(-i\theta_{n}\left[l_{2}\right]\right)\\
\exp \left( i{\theta_{n}}\left[l_{2}\right] \right) - 1 & 0
\end{array}\right)
\boldsymbol{\Xi}_{n}\left[l_{2}\right]
\,.
\]
From the above two relations, a part of the right side in eq. (\ref{eq: discrete M^0 M^1 Psi_n}) is deformed to 
\[
\begin{split}
  \left(\begin{array}{cc}
\boldsymbol{0} & \bigtriangledown_{2}\\
\bigtriangleup_{2} & \boldsymbol{0}
\end{array}\right)
\exp\left( i\Theta_{n}\left[l_{2}\right] \right)\boldsymbol{\Xi}_{n}\left[l_{2}\right]
&= 
i\boldsymbol{M}_{+}^{\left(0\right)}\left[l_{2}\right]
\exp\left( i\Theta_{n}\left[l_{2}\right] \right)\boldsymbol{\Xi}_{n}\left[l_{2}\right]
\\
& \quad +\left(\begin{array}{cc}
0 & \exp\left(-i\frac{\theta_{n}\left[l_{2}\right]}{2}\right) - \exp\left(-i\frac{\theta_{n}\left[l_{2}-1\right]}{2}\right)\\
\exp \left( i\frac{\theta_{n}\left[l_{2}+1\right]}{2} \right) - \exp \left( i\frac{\theta_{n}\left[l_{2}\right]}{2}\right) & 0
\end{array}\right)
\\
& \quad \times
\left(\begin{array}{cc}
\exp\left(i\frac{\theta_{n}\left[l_{2}\right]}{2}\right) & 0\\
0 & \exp\left(-i\frac{\theta_{n}\left[l_{2}\right]}{2}\right)
\end{array}\right)
\exp\left( i\Theta_{n}\left[l_{2}\right] \right)
\boldsymbol{\Xi}_{n}\left[l_{2}\right]
\\
& \quad +\left(\begin{array}{cc}
0 & \exp\left( i\Theta_{n}\left[l_{2}-1\right] \right)\bigtriangledown_{2}\boldsymbol{\Xi}_{n}\left[l_{2}\right]\\
\exp\left( i\Theta_{n}\left[l_{2}+1\right] \right)\bigtriangleup_{2}\boldsymbol{\Xi}_{n}\left[l_{2}\right]  & 0
\end{array}\right).
\end{split}
\]
Then,
\begin{equation}
\begin{split}
& \quad
i\tilde{\boldsymbol{M}}_{+}^{\left(1\right)}\left[l_{2}\right]
\boldsymbol{\Xi}_{n}\left[l_{2}\right]
+ i{\boldsymbol{M}}_{+}^{\left(0\right)}\left[l_{2}\right]
\sum_{1 \leq j \neq l_{2} < L_{2}-1} \boldsymbol{G}\left[l_{2},j \right]
\\
&=
\sum_{1 \leq j \neq l_{2} < L_{2}-1} \exp\left( -i\Theta_{n}\left[l_{2}\right] \right)
\left(\begin{array}{cc}
0 & \bigtriangledown_{2}\\
\bigtriangleup_{2} & 0
\end{array}\right)
\exp\left( i\Theta_{n}\left[l_{2}\right] \right)
\boldsymbol{G}\left[l_{2},j \right]
+ o\left(\left\| \tilde{\boldsymbol{M}}_{+}^{\left(1\right)} \right\|\right)
,
\end{split}
\label{eq: discrete 1st order eq}
\end{equation}
where
\begin{equation}
\begin{split}
i\tilde{\boldsymbol{M}}_{+}^{\left(1\right)}\left[l_{2}\right]
\boldsymbol{\Xi}_{n}\left[l_{2}\right]
&\triangleq
i\boldsymbol{M}_{+}^{\left(1\right)}\left[l_{2}\right]
\boldsymbol{\Xi}_{n}\left[l_{2}\right]
\\
& \quad -
\left(\begin{array}{cc}
0 & \left[\bigtriangledown_{2}\exp\left(-i\frac{\theta_{n}\left[l_{2}\right]}{2}\right)\right]
\exp\left(-i\frac{\theta_{n}\left[l_{2}\right]}{2}\right)
\\
\left[\bigtriangleup_{2}\exp \left( i\frac{\theta_{n}\left[l_{2}\right]}{2} \right)\right] 
\exp\left(i\frac{\theta_{n}\left[l_{2}\right]}{2}\right) & 0
\end{array}\right)
\boldsymbol{\Xi}_{n}\left[l_{2}\right]
\\
& \quad -
\left(\begin{array}{cc}
0 & \exp\left[ -i\left(\bigtriangledown_{2}\Theta_{n}\left[l_{2}\right] \right)\right]\bigtriangledown_{2}\\
\exp\left[ i\left(\bigtriangleup_{2}\Theta_{n}\left[l_{2}\right] \right)\right]\bigtriangleup_{2}  & 0
\end{array}\right)\boldsymbol{\Xi}_{n}\left[l_{2}\right].
\end{split}
\label{eq: discrete tilde M^1}
\end{equation}
It is assumed that $O\left(\left\| \tilde{\boldsymbol{M}}_{+}^{\left(1\right)} \right\|\right) = O\left(\left\| {\boldsymbol{M}}_{+}^{\left(1\right)} \right\|\right)$.
The third term in the right-hand side of eq. (\ref{eq: discrete tilde M^1}) causes that $\exp\left( i\Theta_{n}\left[l_{2}\right] \right)\tilde{\boldsymbol{M}}_{+}^{\left(1\right)}\left[l_{2}\right] \neq \tilde{\boldsymbol{M}}_{+}^{\left(1\right)}\left[l_{2}\right]\exp\left( i\Theta_{n}\left[l_{2}\right] \right)$ even though $\exp\left( i\Theta_{n}\left[l_{2}\right] \right) {\boldsymbol{M}}_{+}^{\left(1\right)}\left[l_{2}\right] = {\boldsymbol{M}}_{+}^{\left(1\right)}\left[l_{2}\right]\exp\left( i\Theta_{n}\left[l_{2}\right] \right)$.
We consider the first term in the right-hand side of eq. (\ref{eq: discrete 1st order eq}), and we can deform it to
\begin{equation}
\begin{split}
&\quad
\exp\left( -i\Theta_{n}\left[l_{2}\right] \right)
\left(\begin{array}{cc}
0 & \bigtriangledown_{2}\\
\bigtriangleup_{2} & 0
\end{array}\right)
\exp\left( i\Theta_{n}\left[l_{2}\right] \right)
\boldsymbol{G}\left[l_{2},j \right]
\\
&=
\left(\begin{array}{cc}
0 & 0\\
\exp\left( i\frac{\theta_{n}\left[l_{2}\right]+\theta_{n}\left[l_{2}+1\right]}{2} \right)
& 0
\end{array}\right)\boldsymbol{G}\left[l_{2}+1,j \right]
 +
\left(\begin{array}{cc}
0 & 1
\\
-1 & 0
\end{array}\right)\boldsymbol{G}\left[l_{2},j \right]
 -
\left(\begin{array}{cc}
0 & \exp\left( -i\frac{\theta_{n}\left[l_{2}-1\right]+\theta_{n}\left[l_{2}\right]}{2} \right)
\\
0 & 0
\end{array}\right)\boldsymbol{G}\left[l_{2}-1,j \right].
\end{split}
\label{eq: differential Green's fun.}
\end{equation}
The above equation (\ref{eq: differential Green's fun.}) can be deformed to the following equation by using the definition of $\boldsymbol{G}$ in eq. (\ref{eq: discrete Green func}):
\begin{equation}
\exp\left( -i\Theta_{n}\left[l_{2}\right] \right)
\left(\begin{array}{cc}
0 & \bigtriangledown_{2}\\
\bigtriangleup_{2} & 0
\end{array}\right)
\exp\left( i\Theta_{n}\left[l_{2}\right] \right)
\boldsymbol{G}\left[l_{2},j \right]
=
i{\boldsymbol{M}}_{+}^{\left(0\right)}\left[l_{2}\right]
\boldsymbol{G}\left[l_{2},j \right]
+ o\left(\left\| \tilde{\boldsymbol{M}}_{+}^{\left(1\right)} \right\|\right)
\quad \mathrm{for} \quad j \lessgtr l_{2}\mp1\,.
\label{eq: differential Green's fun. without pole}
\end{equation}
Here, we can approximate $\theta_{m}\left[l_{2}\pm 1 \right] \simeq \theta_{m}\left[l_{2}\right]$ and $\boldsymbol{\Xi}_{m}\left(l_{2}\pm 1 \right) \simeq \boldsymbol{\Xi}_{m}\left[l_{2}\right]$ for the $\boldsymbol{G}$ in the left-hand side of the above equation.
Equation (\ref{eq: discrete 1st order eq}) can be simplified by eq. (\ref{eq: differential Green's fun. without pole}), and its right-hand side can be expanded by using (\ref{eq: differential Green's fun.}): 
\[
\begin{split}
& \qquad
i\tilde{\boldsymbol{M}}_{+}^{\left(1\right)}\left[l_{2}\right]
\boldsymbol{\Xi}_{n}\left[l_{2}\right]
+ i{\boldsymbol{M}}_{+}^{\left(0\right)}\left[l_{2}\right]
\left(\boldsymbol{G}\left[l_{2},l_{2}-1 \right] + \boldsymbol{G}\left[l_{2},l_{2}+1 \right]\right)
+ o\left(\left\| \tilde{\boldsymbol{M}}_{+}^{\left(1\right)} \right\|\right)
\\
&=
\sum_{j=l_{2}-1,l_{2}+1}
\exp\left( -i\Theta_{n}\left[l_{2}\right] \right)
\left(\begin{array}{cc}
0 & \bigtriangledown_{2}\\
\bigtriangleup_{2} & 0
\end{array}\right)
\exp\left( i\Theta_{n}\left[l_{2}\right] \right)
\boldsymbol{G}\left[l_{2},j \right] 
\\
&=
\left(\begin{array}{cc}
0 & 0\\
\exp\left( i\frac{\theta_{n}\left[l_{2}\right]+\theta_{n}\left[l_{2}+1\right]}{2} \right)
& 0
\end{array}\right)
\left(\boldsymbol{G}\left[l_{2}+1,l_{2}-1 \right] + \boldsymbol{G}\left[l_{2}+1,l_{2}+1 \right]\right)
\\ & \quad 
+
\left(\begin{array}{cc}
0 & 1
\\
-1 & 0
\end{array}\right)
\left(\boldsymbol{G}\left[l_{2},l_{2}-1 \right] + \boldsymbol{G}\left[l_{2},l_{2}+1 \right]\right)
\\ & \quad -
\left(\begin{array}{cc}
0 & \exp\left( -i\frac{\theta_{n}\left[l_{2}-1\right]+\theta_{n}\left[l_{2}\right]}{2} \right)
\\
0 & 0
\end{array}\right)
\left(\boldsymbol{G}\left[l_{2}-1,l_{2}-1 \right] + \boldsymbol{G}\left[l_{2}-1,l_{2}+1 \right]\right).
\end{split}
\]
We can apply the same deformation as eqs. (\ref{eq: differential Green's fun.}) and (\ref{eq: differential Green's fun. without pole}) only to $\bigtriangleup_{2}$ ($\bigtriangledown_{2}$) operation when $j= l_{2} -1$ ($j= l_{2} +1$).
Then, terms for $\boldsymbol{G}\left[l_{2},l_{2}-1 \right]$ and $\boldsymbol{G}\left[l_{2},l_{2}+1 \right]$ can be partially canceled in the above equation:
\begin{equation}
\begin{split}
&\quad 
i\tilde{\boldsymbol{M}}_{+}^{\left(1\right)}\left[l_{2}\right]
\boldsymbol{\Xi}_{n}\left[l_{2}\right]
 +
\left[
\left(\begin{array}{cc}
i & 0
\\
0 & 0
\end{array}\right)
 {\boldsymbol{M}}_{+}^{\left(0\right)}\left[l_{2}\right] 
-\left(\begin{array}{cc}
0 & 1
\\
0 & 0
\end{array}\right)
\right] \boldsymbol{G}\left[l_{2},l_{2}-1 \right]
\\
& \quad  +
\left[
\left(\begin{array}{cc}
0 & 0
\\
0 & i
\end{array}\right)
{\boldsymbol{M}}_{+}^{\left(0\right)}\left[l_{2}\right]
+\left(\begin{array}{cc}
0 & 0
\\
1 & 0
\end{array}\right)
\right] \boldsymbol{G}\left[l_{2},l_{2}+1 \right]
+ o\left(\left\| \tilde{\boldsymbol{M}}_{+}^{\left(1\right)} \right\|\right)
\\
&=
\left(\begin{array}{cc}
0 & 0\\
\exp\left( i\frac{\theta_{n}\left[l_{2}\right]+\theta_{n}\left[l_{2}+1\right]}{2} \right)
& 0
\end{array}\right)
\boldsymbol{G}\left[l_{2}+1,l_{2}+1 \right]
-
\left(\begin{array}{cc}
0 & \exp\left( -i\frac{\theta_{n}\left[l_{2}-1\right]+\theta_{n}\left[l_{2}\right]}{2} \right)
\\
0 & 0
\end{array}\right)
\boldsymbol{G}\left[l_{2}-1,l_{2}-1 \right] .
\end{split}
\label{eq: discrete 1st order eq 2}
\end{equation}
The second and third terms in the left-hand side of eq. (\ref{eq: discrete 1st order eq 2}) are expanded by $\boldsymbol{\Xi}_{m}$:
\[
\begin{split}
&\quad 
\left[
\left(\begin{array}{cc}
i & 0
\\
0 & 0
\end{array}\right)
 {\boldsymbol{M}}_{+}^{\left(0\right)}\left[l_{2}\right] 
-\left(\begin{array}{cc}
0 & 1
\\
0 & 0
\end{array}\right)
\right]\boldsymbol{G}\left[l_{2},l_{2}-1 \right]
+
\left[
\left(\begin{array}{cc}
0 & 0
\\
0 & i
\end{array}\right)
{\boldsymbol{M}}_{+}^{\left(0\right)}\left[l_{2}\right]
+\left(\begin{array}{cc}
0 & 0
\\
1 & 0
\end{array}\right)
\right]\boldsymbol{G}\left[l_{2},l_{2}+1 \right]
\\
&=
\sum_{m>0}
\left[
\left(\begin{array}{cc}
i & 0
\\
0 & 0
\end{array}\right)
 {\boldsymbol{M}}_{+}^{\left(0\right)}\left[l_{2}\right] 
-\left(\begin{array}{cc}
0 & 1
\\
0 & 0
\end{array}\right)
\right]
g_{mn}\left[l_{2}-1\right]\exp\left(
	i\frac{\theta_{m}\left[l_{2}-1\right]+\theta_{m}\left[l_{2}\right]}{2}
	-i\frac{\theta_{n}\left[l_{2}-1\right]+\theta_{n}\left[l_{2}\right]}{2}
\right)\boldsymbol{\Xi}_{m}\left[l_{2}\right]
\\
& \quad +\sum_{m<0}
\left[
\left(\begin{array}{cc}
0 & 0
\\
0 & i
\end{array}\right)
{\boldsymbol{M}}_{+}^{\left(0\right)}\left[l_{2}\right]
+\left(\begin{array}{cc}
0 & 0
\\
1 & 0
\end{array}\right)
\right]
g_{mn}\left[l_{2}+1\right]\exp\left(
	-i\frac{\theta_{m}\left[l_{2}\right]+\theta_{m}\left[l_{2}+1\right]}{2}
	+i\frac{\theta_{n}\left[l_{2}\right]+\theta_{n}\left[l_{2}+1\right]}{2}
\right)\boldsymbol{\Xi}_{m}\left[l_{2}\right]
\\
&=
\sum_{m>0}
g_{mn}\left[l_{2}-1\right]\exp\left(
	i\frac{\theta_{m}\left[l_{2}-1\right]+\theta_{m}\left[l_{2}\right]}{2}
	-i\frac{\theta_{n}\left[l_{2}-1\right]+\theta_{n}\left[l_{2}\right]}{2}
\right)
\left(\begin{array}{cc}
0 & 1-\exp\left(-i\theta_{m}\left[l_{2}\right]\right)-1
\\
0 & 0
\end{array}\right)
\boldsymbol{\Xi}_{m}\left[l_{2}\right]
\\
& \quad +\sum_{m<0}
g_{mn}\left[l_{2}+1\right]\exp\left(
	-i\frac{\theta_{m}\left[l_{2}\right]+\theta_{m}\left[l_{2}+1\right]}{2}
	+i\frac{\theta_{n}\left[l_{2}\right]+\theta_{n}\left[l_{2}+1\right]}{2}
\right)
\left(\begin{array}{cc}
0 & 0
\\
\exp\left(i\theta_{m}\left[l_{2}\right]\right)-1+1 & 0
\end{array}\right)
\boldsymbol{\Xi}_{m}\left[l_{2}\right]
\\
&=
\sum_{m>0}
g_{mn}\left[l_{2}-1\right]\exp\left(
	-i\theta_{n}\left[l_{2}\right]
\right)
\left(\begin{array}{cc}
0 & -1
\\
0 & 0
\end{array}\right)
\boldsymbol{\Xi}_{m}\left[l_{2}\right]
\\
& \quad +\sum_{m<0}
g_{mn}\left[l_{2}+1\right]\exp\left(
	+i\theta_{n}\left[l_{2}\right]
\right)
\left(\begin{array}{cc}
0 & 0
\\
1 & 0
\end{array}\right)
\boldsymbol{\Xi}_{m}\left[l_{2}\right]
+ o\left(\left\| \tilde{\boldsymbol{M}}_{+}^{\left(1\right)} \right\|\right)
.
\end{split}
\]
By using the above relation, equation (\ref{eq: discrete 1st order eq 2}) is deformed to
\begin{equation}
\begin{split}
&\quad 
i\tilde{\boldsymbol{M}}_{+}^{\left(1\right)}\left[l_{2}\right]
\boldsymbol{\Xi}_{n}\left[l_{2}\right]
 +
\sum_{m>0}
g_{mn}\left[l_{2}-1\right]\exp\left(
	-i\theta_{n}\left[l_{2}\right]
\right)
\left(\begin{array}{cc}
0 & -1
\\
0 & 0
\end{array}\right)
\boldsymbol{\Xi}_{m}\left[l_{2}\right]
\\
& \quad +\sum_{m<0}
g_{mn}\left[l_{2}+1\right]\exp\left(
	+i\theta_{n}\left[l_{2}\right]
\right)
\left(\begin{array}{cc}
0 & 0
\\
1 & 0
\end{array}\right)
\boldsymbol{\Xi}_{m}\left[l_{2}\right]
+ o\left(\left\| \tilde{\boldsymbol{M}}_{+}^{\left(1\right)} \right\|\right)
\\
&=
\left(\begin{array}{cc}
0 & 0\\
\exp\left( i\frac{\theta_{n}\left[l_{2}\right]+\theta_{n}\left[l_{2}+1\right]}{2} \right)
& 0
\end{array}\right)
\boldsymbol{G}\left[l_{2}+1,l_{2}+1 \right]
-
\left(\begin{array}{cc}
0 & \exp\left( -i\frac{\theta_{n}\left[l_{2}-1\right]+\theta_{n}\left[l_{2}\right]}{2} \right)
\\
0 & 0
\end{array}\right)
\boldsymbol{G}\left[l_{2}-1,l_{2}-1 \right] .
\end{split}
\label{eq: discrete 1st order eq 3}
\end{equation}
We try to introduce $\boldsymbol{G}\left[l_{2},l_{2} \right]$ into the above equation as
\begin{equation}
\boldsymbol{G}\left[l_{2},l_{2} \right] \triangleq 
\sum_{m>0}
\left(\begin{array}{cc}
g_{mn}\left[l_{2}\right] & 0
\\
0 & 0
\end{array}\right)
\boldsymbol{\Xi}_{m}\left[l_{2}\right]
+\sum_{m<0}
\left(\begin{array}{cc}
0 & 0
\\
0 & g_{mn}\left[l_{2}\right]
\end{array}\right)
\boldsymbol{\Xi}_{m}\left[l_{2}\right]
.
\label{eq: Green func l2 l2}
\end{equation}
If we define other formulation, \textit{e.g.} inverting positive and negative of $m$ instead of eq. (\ref{eq: Green func l2 l2}), there is no consistency with electric and magnetic perturbation-terms and forward and backward scattering-waves.
From eqs. (\ref{eq: discrete 1st order eq 3}) and (\ref{eq: Green func l2 l2}),
\[
\begin{split}
&\quad 
i\tilde{\boldsymbol{M}}_{+}^{\left(1\right)}
\boldsymbol{\Xi}_{n}
 +
\sum_{m>0}
g_{mn}\left[l_{2}-1\right] e^{-i\theta_{n}}
\left(\begin{array}{cc}
0 & -1
\\
0 & 0
\end{array}\right)
\boldsymbol{\Xi}_{m}
+\sum_{m<0}
g_{mn}\left[l_{2}+1\right] e^{i\theta_{n}}
\left(\begin{array}{cc}
0 & 0
\\
1 & 0
\end{array}\right)
\boldsymbol{\Xi}_{m}
+ o\left(\left\| \tilde{\boldsymbol{M}}_{+}^{\left(1\right)} \right\|\right)
\\
&=
\sum_{m>0} g_{mn}\left[l_{2}+1\right]
\left(\begin{array}{cc}
0 & 0\\
e^{i\theta_{n}}
& 0
\end{array}\right)
\boldsymbol{\Xi}_{m}
-
\sum_{m<0} g_{mn}\left[l_{2}-1\right]
\left(\begin{array}{cc}
0 & e^{-i\theta_{n}}
\\
0 & 0
\end{array}\right)
\boldsymbol{\Xi}_{m}
,
\end{split}
\]
where we omit ``$\left[l_{2}\right]$'' from the above notation.
Furthermore, $o\left(\left\| \tilde{\boldsymbol{M}}_{+}^{\left(1\right)} \right\|\right)$ is omitted from the above equation:
\[
\begin{split}
i\tilde{\boldsymbol{M}}_{+}^{\left(1\right)}\left[l_{2}\right]
\boldsymbol{\Xi}_{n}
&=
\sum_{m>0} 
\left(\begin{array}{cc}
0 & e^{-i\theta_{n}} g_{mn}\left[l_{2}-1\right]\\
e^{i\theta_{n}} g_{mn}\left[l_{2}+1\right]
& 0
\end{array}\right)
\boldsymbol{\Xi}_{m}
\\ &\qquad 
-
\sum_{m<0} 
\left(\begin{array}{cc}
0 & e^{-i\theta_{n}} g_{mn}\left[l_{2}-1\right]
\\
e^{i\theta_{n}} g_{mn}\left[l_{2}+1\right] & 0
\end{array}\right)
\boldsymbol{\Xi}_{m}
.
\end{split}
\]
From the special setting of eq. (\ref{eq: h_m e_m special relations}) and the definition of eq. (\ref{eq: Xi definition}),
\[
\begin{split}
i\tilde{\boldsymbol{M}}_{+}^{\left(1\right)}
\boldsymbol{\Xi}_{n}
&=
\sum_{m>0} 
\left(\begin{array}{cc}
0 & e^{-i\theta_{n}} g_{mn}\left[l_{2}-1\right]\\
e^{i\theta_{n}} g_{mn}\left[l_{2}+1\right]
& 0
\end{array}\right)
\left(\begin{array}{c}
e^{-i\theta_{m}/2}\boldsymbol{h}_{m}\\
\boldsymbol{e}_{m}
\end{array}\right)
\\
& \qquad -
\left(\begin{array}{cc}
0 & e^{-i\theta_{n}} g_{-mn}\left[l_{2}-1\right]
\\
e^{i\theta_{n}} g_{-mn}\left[l_{2}+1\right] & 0
\end{array}\right)
\left(\begin{array}{c}
e^{i\theta_{m}/2}\boldsymbol{h}_{m}\\
-\boldsymbol{e}_{m}
\end{array}\right)
\\
& =
\sum_{m>0} 
\left(\begin{array}{c}
e^{-i\theta_{n}} g_{mn}\left[l_{2}-1\right] \boldsymbol{e}_{m}\\
e^{i\theta_{n}} g_{mn}\left[l_{2}+1\right] e^{-i\theta_{m}/2}\boldsymbol{h}_{m}
\end{array}\right)
 -
\left(\begin{array}{cc}
-e^{-i\theta_{n}} g_{-mn}\left[l_{2}-1\right] \boldsymbol{e}_{m}
\\
e^{i\theta_{n}} g_{-mn}\left[l_{2}+1\right] e^{i\theta_{m}/2}\boldsymbol{h}_{m}
\end{array}\right)
\\
& =
\sum_{m>0} 
\left(\begin{array}{c}
e^{-i\theta_{n}}\left[ g_{mn}\left[l_{2}-1\right]
+ g_{-mn}\left[l_{2}-1\right] \right]\boldsymbol{e}_{m}\\
e^{i\theta_{n}}\left[ e^{-i\theta_{m}/2}g_{mn}\left[l_{2}+1\right]
- e^{i\theta_{m}/2} g_{-mn}\left[l_{2}+1\right]\right]\boldsymbol{h}_{m}
\end{array}\right)
,
\end{split}
\]
where $\theta_{-m}=-\theta_{m}$.
From the orthogonality of $\boldsymbol{h}_{m}$ and $\boldsymbol{e}_{m}$ in eq. (\ref{eq: h_n e_m orthogonality}), the above equation gives us the following two equations.
\[
\left\{\begin{aligned}
\left(\begin{array}{cc}
\boldsymbol{h}_{m}^{\mathrm{T}} & 0
\end{array}\right) 
i\tilde{\boldsymbol{M}}_{+}^{\left(1\right)}\left[l_{2}+1\right]
\boldsymbol{\Xi}_{n} &=
\frac{e^{-i\theta_{n}}}{2\cos\left(\theta_{m}/2\right)}\left[ g_{mn} + g_{-mn} \right]\,,
\\
\left(\begin{array}{cc}
0 & \boldsymbol{e}_{m}^{\mathrm{T}}
\end{array}\right) 
i\tilde{\boldsymbol{M}}_{+}^{\left(1\right)}\left[l_{2}-1\right]
\boldsymbol{\Xi}_{n} &=
\frac{e^{i\theta_{n}}}{2\cos\left(\theta_{m}/2\right)}\left[ e^{-i\theta_{m}/2}g_{mn}
- e^{i\theta_{m}/2} g_{-mn}\right]
\end{aligned}\right.
\quad \mathrm{for}\quad m>0\,.
\]
Then, we can solve the above equations for $g_{mn}$ as
\begin{equation}
g_{mn}\left[l_{2}\right] = 
\left(\begin{array}{cc}
e^{i\theta_{m}/2} \boldsymbol{h}_{m}^{\mathrm{T}} & 0
\end{array}\right)i e^{i\theta_{n}}\tilde{\boldsymbol{M}}_{+}^{\left(1\right)}\left[l_{2}+1\right]
\boldsymbol{\Xi}_{n} 
+
\left(\begin{array}{cc}
0 & \boldsymbol{e}_{m}^{\mathrm{T}}
\end{array}\right) 
ie^{-i\theta_{n}}\tilde{\boldsymbol{M}}_{+}^{\left(1\right)}\left[l_{2}-1\right]
\boldsymbol{\Xi}_{n}
+ o\left(\left\| \tilde{\boldsymbol{M}}_{+}^{\left(1\right)} \right\|\right)
\,.
\label{eq: discrete g_mn}
\end{equation}
Note that $g_{mn}\left[l_{2} \leq 0\right] = g_{mn}\left[l_{2} \geq L_{2}-1\right] = 0$ from eq. (\ref{eq: discrete Green func}).
Two terms in the right-hand side of eq. (\ref{eq: discrete g_mn}) then satisfy that
\[
\left\{
\begin{aligned}
\left(\begin{array}{cc}
e^{i\theta_{m}/2} \boldsymbol{h}_{m}^{\mathrm{T}} & 0
\end{array}\right)i e^{i\theta_{n}}\tilde{\boldsymbol{M}}_{+}^{\left(1\right)}\left[l_{2} \leq 1\right] \boldsymbol{\Xi}_{n} = 0 & \quad\mathrm{for}\quad g_{mn}\left[l_{2} \leq 0\right] = 0,
\\
\left(\begin{array}{cc}
0 & \boldsymbol{e}_{m}^{\mathrm{T}}
\end{array}\right) 
ie^{-i\theta_{n}}\tilde{\boldsymbol{M}}_{+}^{\left(1\right)}\left[l_{2} \geq L_{2}-2\right]
\boldsymbol{\Xi}_{n} = 0 & \quad\mathrm{for}\quad g_{mn}\left[l_{2} \geq L_{2}-1\right] = 0.
\end{aligned}
\right.
\]
From the above conditions and the definitions of eqs. (\ref{eq: discrete tilde M^1}), (\ref{eq: discrete propagation-eq}) and (\ref{eq: discrete m_aa m_bb Psi}), the $\varepsilon_{ljj}$ and $\mu_{ljj}$ satisfy that
\[
\left\{
\begin{aligned}
\bigtriangledown_{2}\varepsilon_{l00}\left[l_{2} \leq 0\right] &= \bigtriangledown_{2}\varepsilon_{l11}\left[l_{2} \leq 0\right] = \bigtriangledown_{2}\mu_{l22}\left[l_{2} \leq 0\right] = 0,
\\
\bigtriangledown_{2}\mu_{l00}\left[l_{2} \leq 1\right] &= \bigtriangledown_{2}\mu_{l11}\left[l_{2} \leq 1\right] = \bigtriangledown_{2}\varepsilon_{l22}\left[l_{2} \leq 1\right] = 0,
\\
\bigtriangleup_{2}\varepsilon_{l00}\left[l_{2} \geq L_{2}-2\right] &= \bigtriangleup_{2}\varepsilon_{l11}\left[l_{2} \geq L_{2}-2\right] = \bigtriangleup_{2}\mu_{l22}\left[l_{2} \geq L_{2}-2\right] = 0,
\\
\bigtriangleup_{2}\mu_{l00}\left[l_{2} \geq L_{2}-1\right] &= \bigtriangleup_{2}\mu_{l11}\left[l_{2} \geq L_{2}-1\right] = \bigtriangleup_{2}\varepsilon_{l22}\left[l_{2} \geq L_{2}-1\right] = 0.
\\
\end{aligned}
\right.
\]
If we consider the symmetry for forward and backward directions, we should change the first equation in the above into
\[
\bigtriangledown_{2}\varepsilon_{l00}\left[l_{2} \leq 1\right] = \bigtriangledown_{2}\varepsilon_{l11}\left[l_{2} \leq 1\right] = \bigtriangledown_{2}\mu_{l22}\left[l_{2} \leq 1\right] = 0.
\]

When ${\boldsymbol{M}}_{+}^{\left(0\right)}$ is independent of $l_{2}$, the $\theta_{m}$ and $\boldsymbol{\Xi}_{m}$ are also independent of  $l_{2}$.
Equations (\ref{eq: discrete tilde M^1}) and (\ref{eq: discrete g_mn}) can be then simplified to 
\begin{equation}
\left\{\begin{aligned}
\tilde{\boldsymbol{M}}_{+}^{\left(1\right)}\left[l_{2}\right] & = {\boldsymbol{M}}_{+}^{\left(1\right)}\left[l_{2}\right]
= 
\left(\begin{array}{cc}
{\boldsymbol{m}}_{aa}^{\left(1\right)}\left[l_{2}\right] & 0 \\
0 & {\boldsymbol{m}}_{bb}^{\left(1\right)}\left[l_{2}\right]
\end{array}\right),
\\
g_{mn}\left[l_{2}\right] & = i\boldsymbol{\Xi}_{m}^{\dagger}
\left(\begin{array}{cc}
 e^{i\theta_{n}}{\boldsymbol{m}}_{aa}^{\left(1\right)}\left[l_{2}+1\right] & 0 \\
0 &  e^{-i\theta_{n}}{\boldsymbol{m}}_{bb}^{\left(1\right)}\left[l_{2}-1\right]
\end{array}\right)\boldsymbol{\Xi}_{n}
+ o\left(\left\| {\boldsymbol{M}}_{+}^{\left(1\right)} \right\|\right)
\end{aligned}\right.
\; \mathrm{if}\; 
\bigtriangleup_{2}{\boldsymbol{M}}_{+}^{\left(0\right)}=\bigtriangledown_{2}{\boldsymbol{M}}_{+}^{\left(0\right)}=0.
\label{eq: discrete g_mn for M^(0) independent of l_2}
\end{equation}
Let us apply discrete Fourier transform (DFT) in subsection \ref{sec: DFT definition} to $g_{mn}$.
By using the formula of eq. (\ref{eq: DFT definition}), equation (\ref{eq: discrete g_mn for M^(0) independent of l_2}) is transformed to
\begin{equation}
\widehat{g_{mn}}_{\mathrm{DFT}}\left(\phi\right) =  i\boldsymbol{\Xi}_{m}^{\dagger}
\left(\begin{array}{cc}
 e^{i\left(\theta_{n}+\phi\right)}\widehat{\boldsymbol{m}_{aa}^{\left(1\right)}}_{\mathrm{DFT}}\left(\phi\right) & \boldsymbol{0} \\
\boldsymbol{0} &  e^{-i\left(\theta_{n}+\phi\right)}\widehat{\boldsymbol{m}_{bb}^{\left(1\right)}}_{\mathrm{DFT}}\left(\phi\right)
\end{array}\right)\boldsymbol{\Xi}_{n}
+ o\left(\left\| {\boldsymbol{M}}_{+}^{\left(1\right)} \right\|\right)
.\label{eq: DFT g_mn when M^(0) is independent of l_2}
\end{equation}
The above $\widehat{g_{mn}}_{\mathrm{DFT}}$ can be applied to the discrete $S_{mn}$ of eq. (\ref{eq: discrete S_mn and DFT g_mn}).

\section{Formulas of discrete difference operators}

Note that
\begin{equation}
\left\{
\begin{aligned}
\bigtriangleup A\left[j\right]B\left[j\right]&= A\left[j+1\right]B\left[j+1\right] - A\left[j\right]B\left[j\right]
\\
&= \left( A\left[j+1\right] - A\left[j\right] \right) \frac{B\left[j+1\right] + B\left[j\right]}{2} + \frac{A\left[j+1\right] + A\left[j\right]}{2}\left( B\left[j+1\right] - B\left[j\right] \right)
\\
&= \left( \bigtriangleup A\left[j\right] \right)\left(1+\frac{\bigtriangleup}{2}\right)B\left[j\right] + \left( \left(1+\frac{\bigtriangleup}{2}\right) A\left[j\right] \right)\bigtriangleup B\left[j\right]\,,
\\
\bigtriangledown A\left[j\right]B\left[j\right]&= A\left[j\right]B\left[j\right] - A\left[j-1\right]B\left[j-1\right]
\\
&= \left( A\left[j\right] - A\left[j-1\right] \right) \frac{B\left[j\right] + B\left[j-1\right]}{2} + \frac{A\left[j\right] + A\left[j-1\right]}{2}\left( B\left[j\right] - B\left[j-1\right] \right)
\\
&= \left( \bigtriangledown A\left[j\right] \right)\left(1-\frac{\bigtriangledown}{2}\right)B\left[j\right] + \left( \left(1-\frac{\bigtriangledown}{2}\right) A\left[j\right] \right)\bigtriangledown B\left[j\right]\,.
\end{aligned}\right.
\label{eq: Formulas discrete diff}
\end{equation}

\section{Notation and formulas of discrete Fourier transform (DFT)\label{sec: DFT definition}}

Notation of DFT keeps consistency to one of Fourier transform (FT) in Section \ref{sec: notation}.
We consider a discrete function $f_{d}\left[l_{2}\right]$ for $0 \leq l_{2} < L_{2}$.
The $f_{d}\left[l_{2}\right]$ is discreted from the continuous function $f_{c}\left(u_{2}\right)$ as follows.
\[f_{d}\left[l_{2}\right] = f_{c}\left(u_{2}\left[l_{2}\right]\right)= f_{c}\left(\left(l_{2}-\frac{L_{2}-1}{2}\right){u_{2}^{'}}\right),\]
where we use the notation in Section \ref{sec: u to xi}, and we assume that $u_{2}^{'}$ is constant.

DFT of $f_{d}\left[l_{2}\right]$ is given by
\begin{equation}
{\widehat{f_{d\,}}}_{\mathrm{DFT}}\left(\phi\right) \triangleq \sum_{l_{2}=0}^{L_{2} -1} f_{d}\left[l_{2}\right] \exp\left[-i\phi\left(l_{2}-\frac{L_{2}-1}{2}\right)\right]\quad \mathrm{for}\quad -\pi<\phi<\pi\,.
\label{eq: DFT definition}
\end{equation}
We can approximate the above summation to integration:
\[
{\widehat{f_{d\,}}}_{\mathrm{DFT}}\left(\phi\right) - 
\frac{1}{u_{2}^{'}}\int_{-\frac{L_{2} -1}{2}u_{2}^{'}}^{\frac{L_{2} -1}{2}u_{2}^{'}}f_{c}({u_{2}})\exp\left(-i\frac{\phi u_{2}}{u_{2}^{'}}\right)d u_{2}
<
O\left(\max\left|\frac{df_{c}}{du_{2}}u_{2}^{'}-i\phi f_{c}\right|L_{2}\right).
\]
Therefore, DFT for $l_{2}$ is related to FT for $u_{2}$:
\begin{equation}
\lim_{u_{2}^{'} \rightarrow 0} u_{2}^{'}{\widehat{f_{d\,}}}_{\mathrm{DFT}}\left(\phi\right) = \widehat{f_{c\,}}\left(k\right)
\quad \mathrm{as} \quad 
k=\lim_{u_{2}^{'} \rightarrow 0}\frac{\phi}{u_{2}^{'}}
\,,
\label{eq: DFT FT relations}
\end{equation}
when we assume that $\lim_{u_{2}^{'} \rightarrow 0} {L_{2}u_{2}^{'}} = L_{\mathrm{s}}$, and $f_{c}\left(u_{2} < -L_{\mathrm{s}}/2\right) = f_{c}\left(u_{2} > L_{\mathrm{s}}/2\right) = 0$.

If $\phi$ becomes discrete as $\phi_{n} = 2\pi \left[ n-\left(L_{2}-1\right)/2 \right]/L_{2}$,
we can define inverse discrete Fourier transform (IDFT) as
\[
\begin{split}
f_{d}\left[l\right] &= \frac{1}{L_{2}}\sum_{n=0}^{L_{2}-1}\widehat{f_{d\,}}_{\mathrm{DFT}}\left( \phi_{n}\right) \exp\left(i\phi_{n}\left(l- \frac{L_{2}-1}{2} \right)\right)
\\
&= \frac{1}{L_{2}}\sum_{n=0}^{L_{2}-1}\sum_{l^{'}=0}^{L_{2} -1} f_{d}\left[l^{'}\right] \exp\left(-i\phi_{n}\left(l^{'}-\frac{L_{2}-1}{2}\right)\right) \exp\left(i\phi_{n}\left(l- \frac{L_{2}-1}{2} \right)\right)
\\
&= \frac{1}{L_{2}}\sum_{l^{'}=0}^{L_{2} -1} f_{d}\left[l^{'}\right] \sum_{n=0}^{L_{2}-1} \exp\left(-i\phi_{n}\left(l^{'}-l\right)\right)
= \sum_{l^{'}=0}^{L_{2} -1} f_{d}\left[l^{'}\right] \frac{e^{-\pi i\left(l^{'}-l\right)/L_{2}}\left(e^{\pi i\left(l^{'}-l\right)}-e^{-\pi i\left(l^{'}-l\right)}\right)}{L_{2}\left(1-e^{-2 \pi i\left(l^{'}-l\right)/L_{2}}\right)}\\
&= \sum_{l^{'}=0}^{L_{2} -1} f_{d}\left[l^{'}\right] \delta_{l^{'}l}
=f_{d}\left[l\right]\,.
\end{split}
\]